\documentclass[12pt,lot,lof]{puthesis}

\title{Differentiable Programming for Computational Plasma Physics}

\submitted{May 2024}
\copyrightyear{2024} 
\author{Nicholas Bradley McGreivy}
\advisor{Ammar H. Hakim} 
\department{\\ Astrophysical Sciences \\ Program in Plasma Physics}

\usepackage[T1]{fontenc}

\usepackage{graphicx}
\usepackage{verbatim}
\usepackage{multirow}
\usepackage{longtable}
\usepackage{booktabs}
\usepackage[english]{babel}
\usepackage{siunitx}
\usepackage{amsmath}
\usepackage{layout}

\usepackage[dvipsnames, svgnames]{xcolor}
\colorlet{myIntColor}{DarkRed} 
\colorlet{myCiteColor}{DarkGreen} 
\colorlet{myExtColor}{MidnightBlue} 
\ifdefined\printmode

\usepackage[hyperfootnotes=false, hidelinks]{hyperref}

\else

\ifdefined\proquestmode
\usepackage[hyperfootnotes=false, hidelinks]{hyperref}

\else
\usepackage[hyperfootnotes=false]{hyperref}
\hypersetup{colorlinks,
linkcolor=myIntColor,
citecolor=myCiteColor,
urlcolor =myExtColor,
menucolor=myIntColor,
}

\fi 

\makeatletter
\hypersetup{pdftitle=\@title,pdfauthor=\@author}
\hypersetup{bookmarksnumbered, linktoc=all}
\makeatother

\fi 

\usepackage{cleveref}

\makeatletter
\renewcommand\footnotesize{%
   \@setfontsize\footnotesize\@xipt\@xiipt
   \abovedisplayskip 10\p@ \@plus2\p@ \@minus5\p@
   \abovedisplayshortskip \z@ \@plus3\p@
   \belowdisplayshortskip 6\p@ \@plus3\p@ \@minus3\p@
   \def\@listi{\leftmargin\leftmargini
               \topsep 6\p@ \@plus2\p@ \@minus2\p@
               \parsep 3\p@ \@plus2\p@ \@minus\p@
               \itemsep \parsep}%
   \belowdisplayskip \abovedisplayskip
}
\makeatother

\usepackage{pifont}%
\usepackage{pbox}
\usepackage{stackengine}
\usepackage{scalerel}
\usepackage{amssymb}
\usepackage{booktabs}
\usepackage{longtable}
\usepackage{afterpage}
\usepackage{bm}
\usepackage{nicefrac}
\usepackage{subcaption}
\usepackage{mathtools}
\usepackage{cancel}
\DeclareMathOperator*{\argminC}{\arg\min}   
\DeclareMathOperator*{\argmaxC}{\arg\max}   
\usepackage[version=3]{mhchem}
\usepackage{siunitx}

\usepackage{csquotes}

\newcommand{\curl}{\bm\nabla \times}
\newcommand{\grad}{\bm \nabla}

\newcommand{\code}[1]{{\texttt{#1}}}
\newcommand{\rvect}[1]{\begin{bmatrix} #1 \end{bmatrix}}

\newcommand{\unstretch}[1]{\mbox{#1}}

\usepackage{tikz}
\usetikzlibrary{backgrounds,calc,chains,fit,matrix,positioning,shadows,shapes.misc,circuits.ee}
\tikzset{input/.style={circle, draw, thick, fill=white, minimum size=2.4em, inner sep=0pt}}
\tikzset{int/.style={circle, draw, thick, fill=white, minimum size=2.4em, inner sep=0pt}}
\tikzset{output/.style={circle, draw, thick, fill=white, minimum size=2.4em, inner sep=0pt}}
\tikzset{/tikz/thin/.style={line width=.9pt}}
\tikzset{/tikz/thick/.style={line width=1.4pt}}
\tikzset{every path/.style={thin}}
\tikzset{>=direction ee}

\usetikzlibrary{arrows,petri,topaths,math}
\usepackage{tkz-berge}

\newcommand\nodexonex{0}
\newcommand\nodexoney{0}
\newcommand\nodextwox{4}
\newcommand\nodextwoy{0}
\newcommand\nodeax{0}
\newcommand\nodeay{4}
\newcommand\nodebx{2}
\newcommand\nodeby{2}
\newcommand\nodecx{4}
\newcommand\nodecy{4}
\newcommand\nodedx{4}
\newcommand\nodedy{6}
\newcommand\nodefx{2}
\newcommand\nodefy{8}

\newcommand\extralabel[4][0mm]{\node[label={[label distance=#1]#2:#3}] at (#4){};}

\definecolor{cborange}{RGB}{230,159,0}
\definecolor{cbblue}{RGB}{86,180,233}
\definecolor{cbgreen}{RGB}{0,158,115}
\definecolor{cbred}{RGB}{213,94,0}

\newcommand{\pluseq}{\mathrel{+}=}

\usepackage{pgfmath}
\newcommand\randmin{}
\newcommand\randmax{}
\newcommand\randmultof{}
\newcommand\setrand[4]%
  {\def\randmin{#1}%
   \def\randmax{#2}%
   \def\randmultof{#3}%
   \pgfmathsetseed{#4}%
  }

\newcommand\nextrand
  {\pgfmathparse{int(int((rnd*(\randmax-\randmin+1)+\randmin)/\randmultof)*\randmultof)}%
   \xdef\thisrand{\pgfmathresult}%
  }

\usepackage{algorithm} 
\RequirePackage[noend]{algpseudocode}

\usepackage{mathrsfs}

\DeclareFixedFont{\ttb}{T1}{txtt}{bx}{n}{11} 
\DeclareFixedFont{\ttm}{T1}{txtt}{m}{n}{11}  

\usepackage{color}
\definecolor{deepblue}{rgb}{0,0,0.5}
\definecolor{deepred}{rgb}{0.6,0,0}
\definecolor{deepgreen}{rgb}{0,0.5,0}
\definecolor{verylightgray}{rgb}{0.95,0.95,0.95}

\usepackage{listings}
\newcommand\pythonstyle{\lstset{
language=Python,
backgroundcolor = \color{verylightgray},
breaklines=true,
basicstyle=\ttm,
keywordstyle=\ttm\color{deepgreen},
stringstyle=\color{deepblue},
frame=tb,                         
showstringspaces=false
}}

\lstnewenvironment{python}[1][]
{
\pythonstyle
\lstset{#1}
}
{}

\usepackage{cite}

\ifodd 0
\renewcommand{\maketitlepage}{}
\renewcommand*{\makecopyrightpage}{}
\renewcommand*{\makeabstract}{}

\else

\abstract{
Differentiable programming allows for derivatives of functions implemented via computer code to be calculated automatically.
These derivatives are calculated using automatic differentiation (AD).
This thesis explores two applications of differentiable programming to computational plasma physics.
First, we consider how differentiable programming can be used to simplify and improve stellarator optimization.
We introduce a stellarator coil design code (FOCUSADD) that uses gradient-based optimization to produce stellarator coils with finite build.
Because we use reverse mode AD, which can compute gradients of scalar functions with the same computational complexity as the function, FOCUSADD is simple, flexible, and efficient.
We then discuss two additional applications of AD in stellarator optimization: finding non-axisymmetric magnetic fields that satisfy magnetohydrodynamic (MHD) equilibrium, and optimizing those magnetic fields subject to MHD equilibrium.
Second, we explore how machine learning (ML) can be used to improve or replace the numerical methods used to solve partial differential equations (PDEs), focusing on time-dependent PDEs in fluid mechanics relevant to plasma physics.
Differentiable programming allows neural networks and other techniques from ML to be embedded within numerical methods.
This is a promising, but relatively new, research area.
We focus on two basic questions.
First, can we design ML-based PDE solvers that have the same guarantees of conservation, stability, and positivity that standard numerical methods do?
The answer is yes; we introduce error-correcting algorithms that preserve invariants of time-dependent PDEs.
Second, which types of ML-based solvers work best at solving PDEs?
We perform a systematic review of the scientific literature on solving PDEs with ML.
Unfortunately, we discover two issues, weak baselines and reporting biases, that affect the interpretation reproducibility of a significant majority of published research.
We conclude that using ML to solve PDEs is not as promising as we initially believed.
}

\acknowledgements{
Doing a PhD was a great decision, but not for the reasons I originally thought. When I applied to a PhD at Princeton, I did so because I thought I wanted to become a fusion scientist. Yet the best parts of the PhD have had nothing to do with fusion. Instead, Princeton has given me intellectual freedom, adventure, friendship, community, and a chance to dream big. In my time here, I've been able to take classes in and out of my field, to behave dumb at times as well as smart, to act immature and ultimately to grow up.
I have many people to thank for making the past six and a half years some of the most formative and meaningful of my life.

I first want to thank my advisor, Dr. Ammar Hakim. This dissertation would not have been possible without Ammar, for many reasons. First, Ammar recognized before anyone else did the potential of automatic differentiation in stellarator optimization, and encouraged me to collaborate with Stuart and Coaxiang on AD for stellarator optimization. Second, our work on reproducibility issues in ML-for-PDE research was only possible because of Ammar's extensive knowledge and background in numerical methods, which I did my best to absorb through years of conversations and Slack messages. Third, Ammar taught me a framework for developing numerical methods, which I used in my work on invariant preservation. I am grateful for Ammar's guidance in understanding and implementing numerical methods, for allowing me to stay at his home, meet his family, and collaborate during the COVID pandemic, for proposing a research project for my brother to work on during the COVID lockdowns, and for our many lively and interesting conversations on matters unrelated to research.

I want to thank Dr. Stuart Hudson and Dr. Caoxiang Zhu for collaborating with me on the FOCUSADD coil design code and on efficiently calculating the Biot-Savart law. I have fond memories of our chats in your offices before COVID as well as our zoom calls during COVID lockdowns, which were filled with laughter as well as intense brainstorming and idea generation. Despite the lockdowns, this collaboration made work fun. I also want to thank Lee Gunderson, who joined the collaboration on the Biot-Savart law calculation late in the game but very quickly had a brilliant insight that had eluded Stuart, Caoxiang, and I for months. While I am the first author on that paper, Lee's contribution was clearly the most valuable.

I want to thank Professor Ryan Adams, Deniz Oktay, Joshua Aduol and Alex Beatson for letting me collaborate with you on Randomized Automatic Differentiation. I learned a lot about machine learning and the process of doing machine learning research from this experience. Ryan, in particular, thank you for taking a second chance on me. 

I want to thank Michael Churchill for helpful conversations. While we didn't end up collaborating together on any papers, you encouraged me to focus on issues of numerical stability with machine learned PDE solvers which led to one of the papers I'm most proud of.

I also want to thank the faculty and researchers in my program and department who I've taken courses from or spent significant time with, including Greg Hammett, Amitava Bhattacharjee, Rob Goldston, Matthew Kunz, Nat Fisch, Hong Qin, Sam Cohen, Ilya Dodin, Allan Reiman, William Tang, and Hantao Ji.

I want to thank Beth Leman and Dara Lewis, who helped me facilitate rules and regulations inside the graduate school, the department, the program, the lab, and even HR. Thanks for standing up for me when I needed it, and for helping make exceptions when it allowed me to be the best version of myself. Makia McFarlane, I didn't get to know you during the past year or two but thanks for helping run the program.

I want to thank my family for their love and support, especially to my mom who encouraged me to stick with the PhD even when I was unsure it was the right path, and who took care of me when I tore my ACL and LCL after my first year. Thanks to my brother James for conversations about research, physics, graduate school, and much more.

In the first year of my PhD, Princeton quickly became a happy community for me. I will always cherish the friendships and memories made during this magical first year, memories of laughter and tears and long brunches and loud parties and discovering what F. Scott Fitzgerald called ``the best damn place of all.'' I have many people to thank for making this first year so incredible: Saadia, Tom, Franck, Haamza, Evan, Dan, Karina, Melanie, Haajar, Obi, Maddie, Samantha, Chuk, Jared, Albie, Jackson, Luya, Kim, Francesca, Henry, my friends from Princeton Club Basketball, and my friends in my plasma cohort including Alec, Ben, Eduardo, Sierra, and Yichen. 

Although I tore my ACL and LCL in the late summer after my first year and came back to campus on crutches and a motor scooter, the second year was once again magical thanks to the many wonderful friendships I maintained and the new friendships I made during this time. Thanks to Ruairidh, who I met this second year and who soon became a close friend and someone who occasionally understood me better even than I understood myself. Thanks to Suying, who I can always count on for a good laugh or a funny story or a caring ear to talk to. Thanks also to Connor, Susan, Sori, Danielle, Rachel, John, James, Andres, Kenneth, Aimee, Xita, and Crystal. Finally, thanks to Regina for being my biggest supporter and cheerleader; I'll be cheering so so hard for you in summer 2024.

Towards the end of my third year, when the COVID pandemic started and I began working remotely for two years, I gradually developed a group of friends who became my community. Thanks to Stanley, who quickly welcomed me as his friend and intellectual debate partner, and also helped welcome many more friends into our community. Thanks to JV, with whom I shared many fun adventures and intellectual interests. Thanks also to Jake, Richard, Kallen, Kiana, Ray, Iris, Victoria, Annie, Jackie, Benny, Peter, Danielle, Katie, Matt, Amar, Wookie, Brother, Mia, and Michelle. Your range of backgrounds, interests, and our fascinating conversations helped me develop much broader interests and perspectives; I matured immensely during these two years in part thanks to you and the people you introduced me to.

Finally, thanks to Kayla, my girlfriend for the past three years. Your love, patience, kindness, and wisdom has helped me during the stressful final stretch of my PhD. I've immensely enjoyed and appreciated your intelligence, positive attitude, and easygoing demeanor as we've explored Hawaii, spend a semester in England, and ate our way through New York City.

With love and appreciation to all, Nick
}

\dedication{To Dad, who would've loved to read this.}

\fi

\begin{document}

{\hypersetup{linkcolor=black}
\makefrontmatter
}

\bodyspacing

\chapter{Introduction\label{ch:intro}}

\section{Differentiable programming}

Differentiable programming is a remarkable technology.
Suppose I have a mathematical function $f : \mathbb{R}^m \rightarrow \mathbb{R}^n$, defined via a computer program, which takes an input $\bm x \in \mathbb{R}^n$ and an output $\bm y \in \mathbb{R}^m$.
With only one additional line of code, I can compute the numerical value of the Jacobian (i.e., the derivative) $\bm J = \frac{\partial \bm y}{\partial \bm x}$ for any input $\bm x$.

To illustrate, consider an example. 
Suppose $f: \mathbb{R} \rightarrow \mathbb{R}$, $f(x) = \sin{(\exp{(\sqrt{x})})} + \cos(x^2)$, and I want to compute the derivative $\frac{df}{dx}$.
One option for computing the derivative would be manually, implementing the symbolic derivatives by hand.
\begin{python}
def f(x):
    return sin(exp(sqrt(x))) + cos(x**2)
def dfdx(x): # manual differentiation
    return cos(exp(sqrt(x))) * (1/2) * exp(sqrt(x)) / sqrt(x) - 2 * x * sin(x)
print(dfdx(2.0)) # 2.2071108
\end{python}
Another option is to use differentiable programming. With differentiable programming, computing the derivative of $f$ requires no additional manual work, only a single extra line of code (line 4).
\begin{python}
import differentiable_programming_library as dp
def f(x):
    return dp.sin(dp.exp(dp.sqrt(x))) + dp.cos(x**2)
dfdx = dp.derivative(f)
print(dfdx(2.0)) # 2.2071108
\end{python}
Packages that implement differentiable programming typically compute derivatives using automatic differentiation (AD).\footnote{In the above pseudocode, I used the package \texttt{differentiable\_programming\_library}. Note that this particular package doesn't exist.}
AD is a set of techniques for computing numerical derivatives of composite functions; I discuss how AD works in detail in Chapter \ref{ch:autodiff}. 

Differentiable programming can be an extremely useful technology for certain applications, especially those that compute gradients of scalar objective functions for gradient-based optimization. In the early 2010s, the machine learning (ML) community gradually switched from computing derivatives manually to doing so almost exclusively with AD \cite{Domke_2009,baydin2018automatic}. Because gradients could be computed easily and efficiently, this allowed for rapid prototyping of machine learning models. As a result, AD is considered to be one of the most important factors contributing to the success of ML in the past decade.

The unifying theme of this thesis is exploring how differentiable programming can advance plasma physics research. 
It does so directly, through the use of AD for stellarator optimization, but also indirectly, as the key technology enabling ML to be used in computational physics research.

\section{Motivation: stellarator optimization}


\subsection{Fusion energy}

Matter is made of large numbers of atoms. 1 \unit{\kilogram} of \ce{H2O} contains \num{2.2e25} hydrogen atoms and \num{1.1e25} oxygen atoms. Each atom consists of a small positively charged nucleus made up of protons and neutrons, and may contain negatively charged electrons bonded to and surrounding the nucleus.

Electrons are not necessarily fixed to a particular atom. They can bond with other atoms to form molecules (\ce{H2O} is one such molecule), they can move through matter (especially conductors such as metal), or they can become separated from the nucleus and exist as free electrons.
The energy required to separate an electron from a nucleus (so-called `ionization') is on the order of an electronvolt (1 \unit{\electronvolt} $ = $ \num{1.602e-19} \unit{J}).
For example, the energy required to ionize a single electron from a neutral hydrogen atom is 13.6 \unit{eV}. Thus ionizing 1 \unit{\kilogram} of hydrogen requires 1300 \unit{MJ}, equivalent to about 250 Chipotle burritos.\footnote{This assumes about 1,250 calories per Chipotle burrito.}

Protons and neutrons are not necessarily fixed to a particular nucleus either. Nuclei can spontaneously eject a neutron or an alpha particle, absorb neutrons, split into two or more smaller nuclei (fission), or combine two smaller nuclei into one larger nuclei (fusion). The atom, a word derived from the ancient greek word for indivisible, is therefore not indivisible. The energy associated with changes in atomic nuclei is on the order of mega-electronvolts (\unit{MeV}). The energy released from the fusion of 1 \unit{\kilogram} of \ce{^1H} into \ce{^4_2He} is 679 \unit{TJ}, equivalent to about 130 \textit{million} Chipotle burritos. The energy released from the fission of a \ce{^{235}_{92}U} atom is 202.5 \unit{MeV}; fissioning 1 \unit{\kilogram} of \ce{^{235}_92U} releases about 82 \unit{TJ}. Fusion, it turns out, has higher specific energy than fission.

The energy stored in atomic nuclei can be used for military and industrial applications. Nuclear and thermonuclear explosives rely on energy and neutrons from fission and fusion, while nuclear submarines, aircraft carriers, and power plants all use energy from fission.
Humans, however, are currently unable to build nuclear power plants that rely on fusion energy. Doing so is the primary goal of fusion energy research. 

\subsubsection{Reaction rates}

\begin{figure}
    \centering
    \begin{subfigure}[]{0.49\textwidth}
        \centering
        \includegraphics[width=\textwidth]{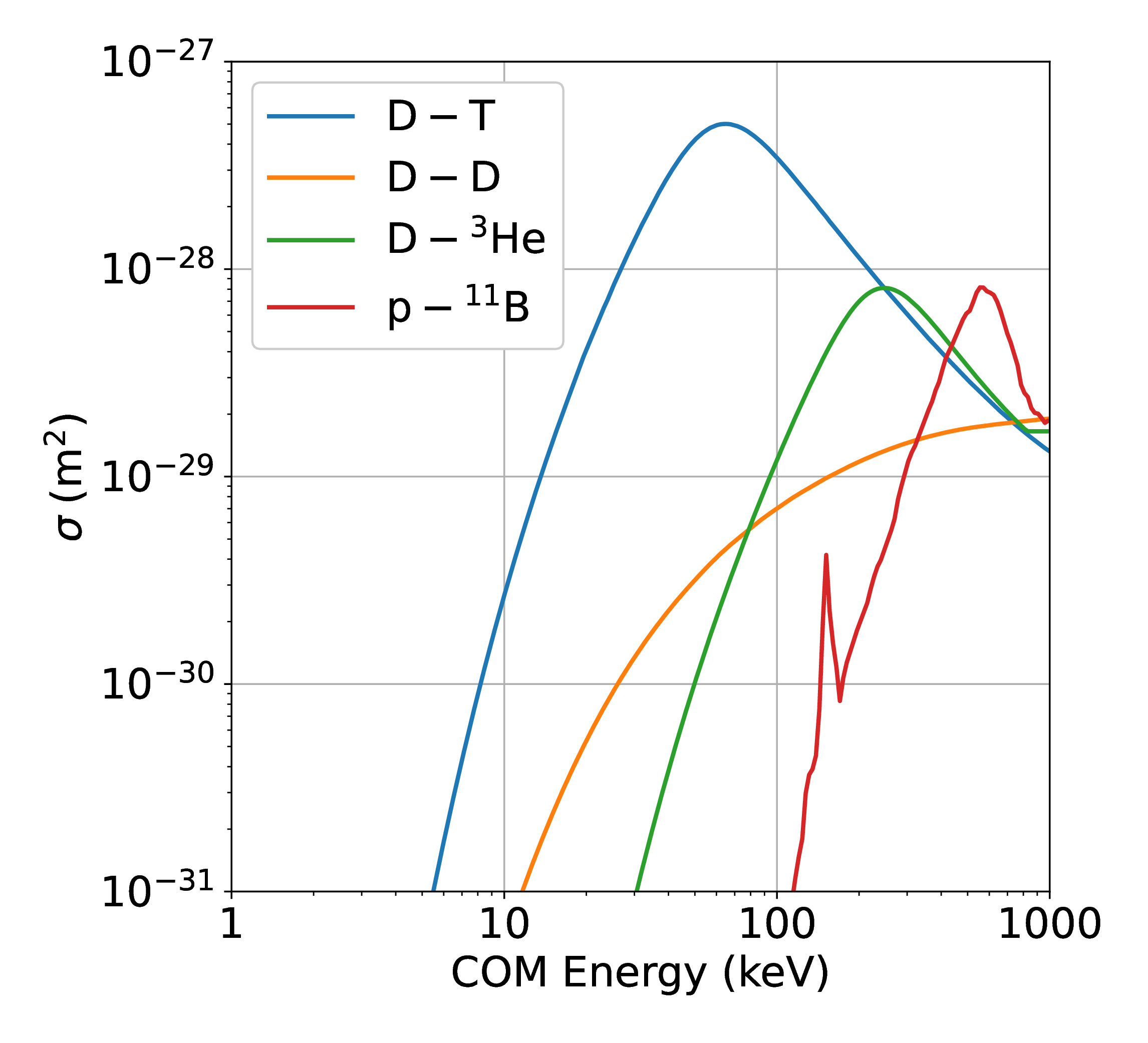}
        \caption{}
        \label{fig:dt_a}
    \end{subfigure}
    \begin{subfigure}[]{0.49\textwidth}
         \centering
         \includegraphics[width=\textwidth]{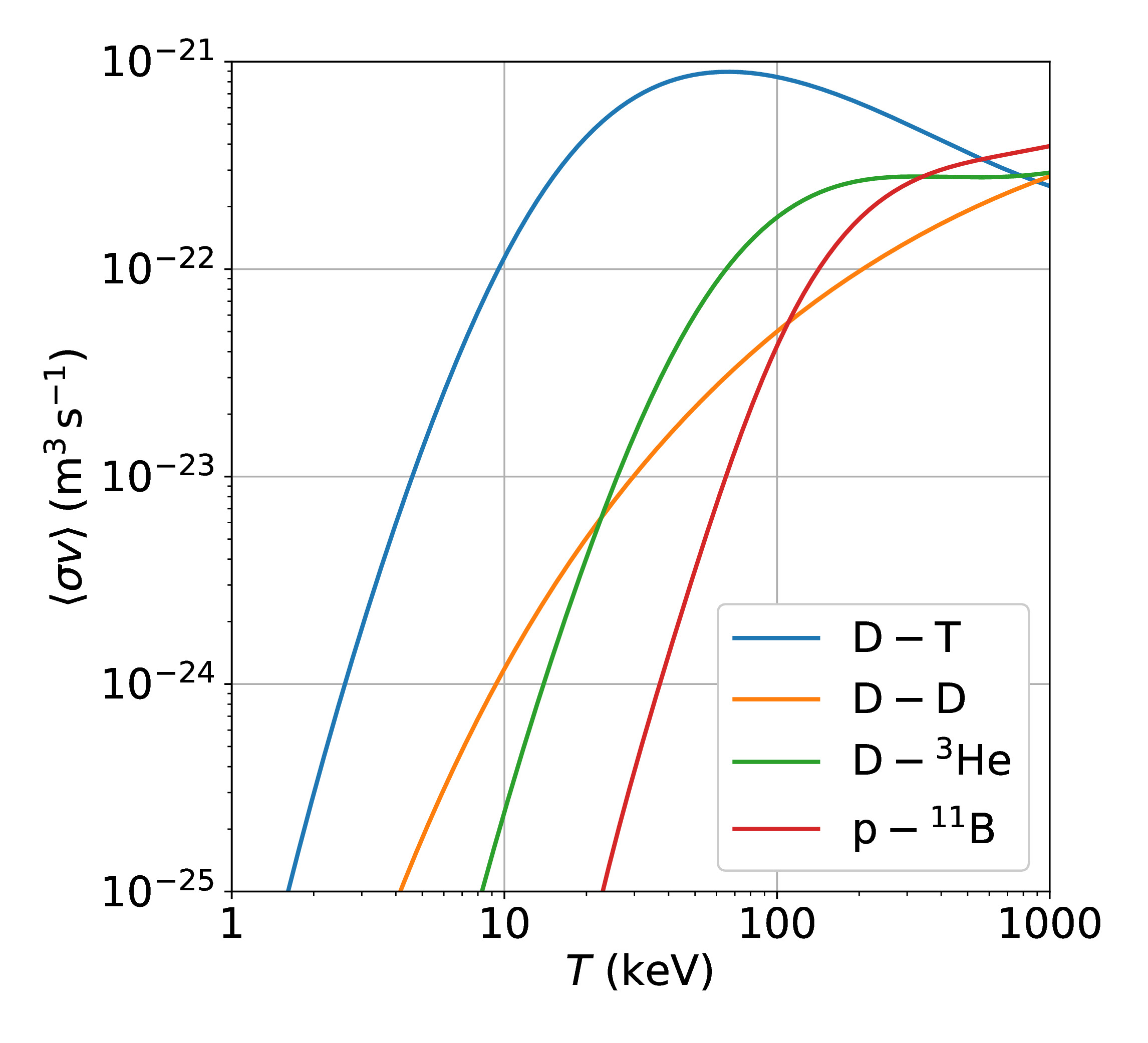}
         \caption{}
         \label{fig:dt_b}
     \end{subfigure}
    \begin{subfigure}[]{0.49\textwidth}
         \centering
         \includegraphics[width=\textwidth]{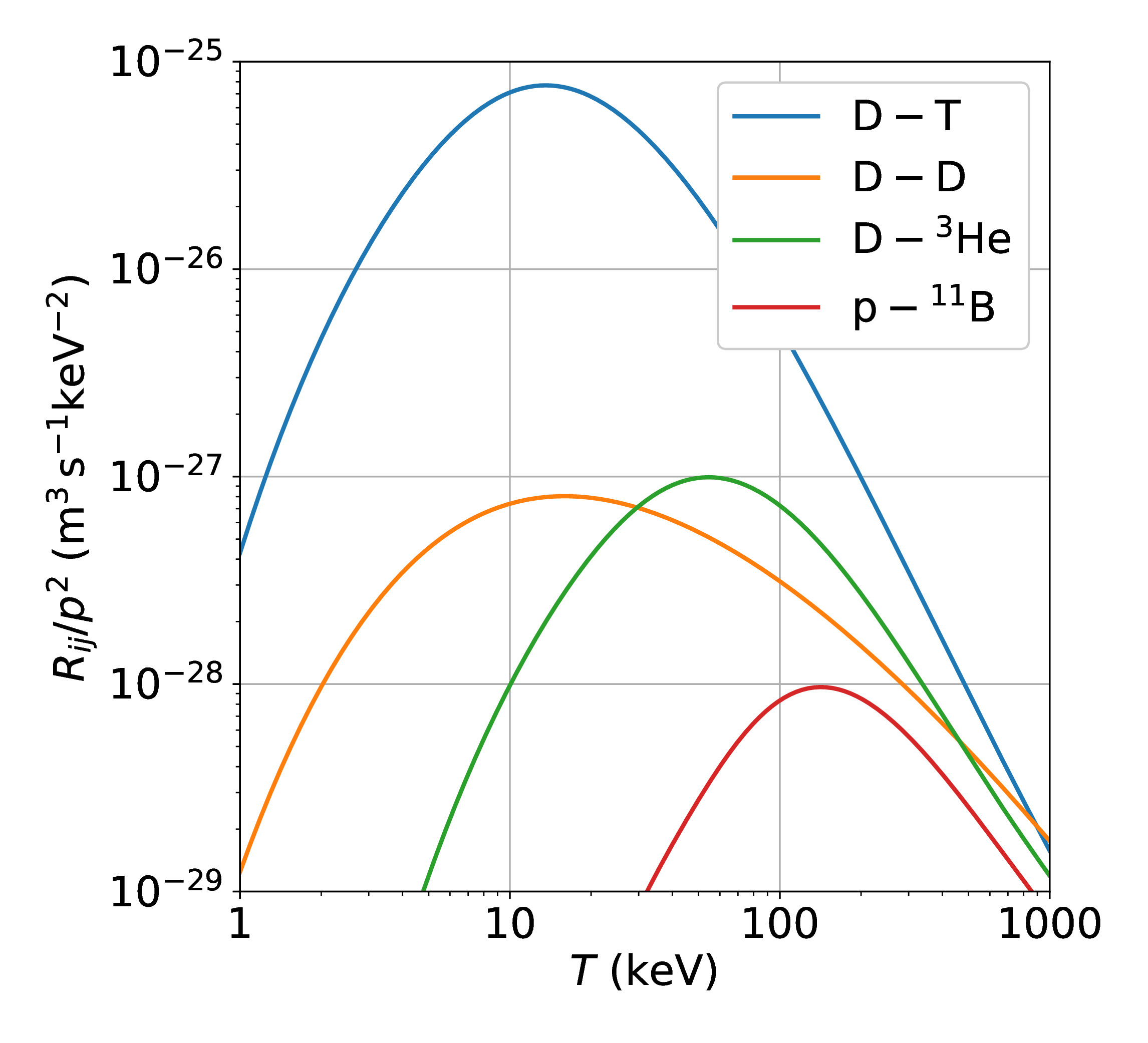}
         \caption{}
         \label{fig:dt_c}
     \end{subfigure}
    \begin{subfigure}[]{0.49\textwidth}
         \centering
         \includegraphics[width=\textwidth]{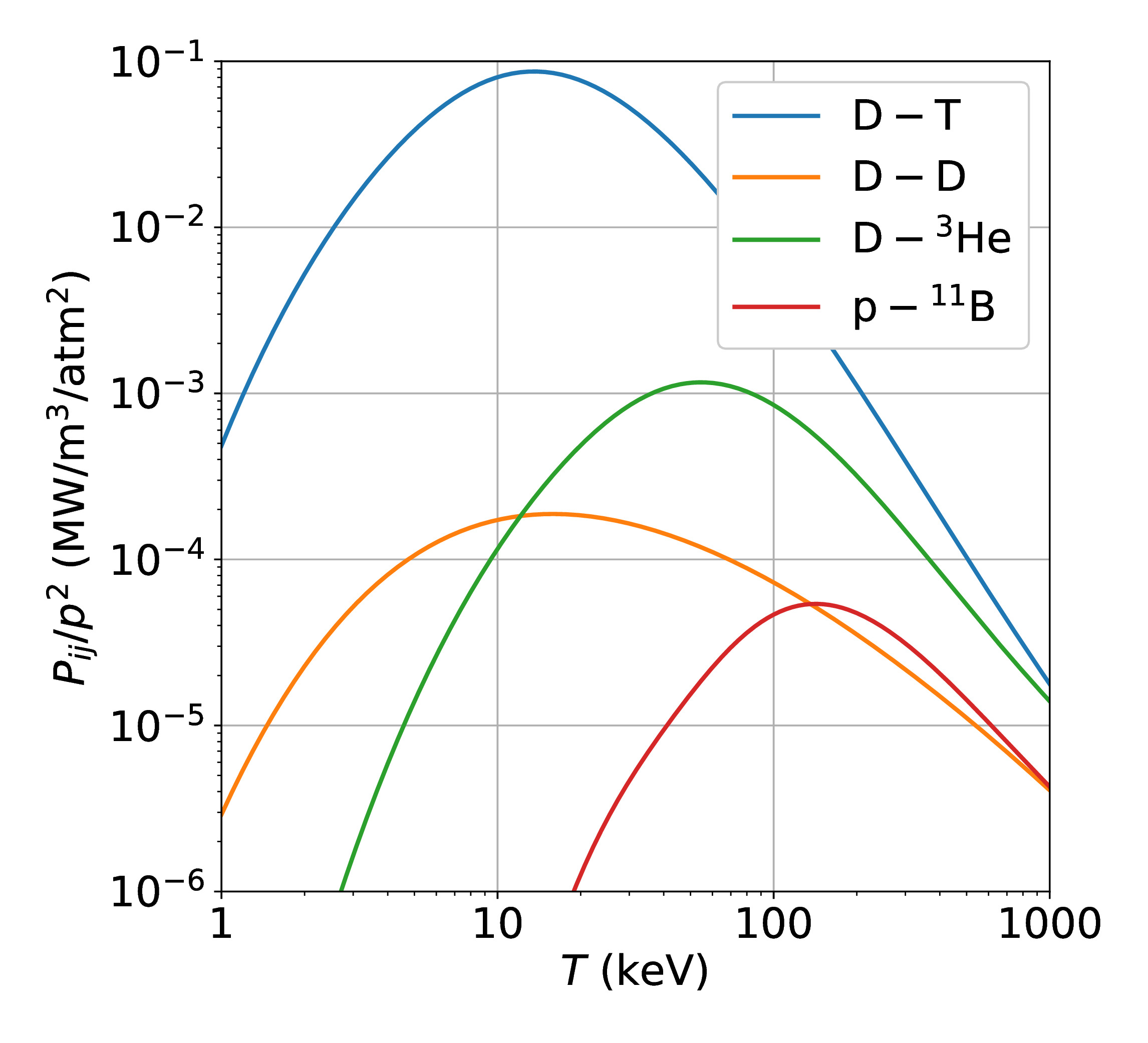}
         \caption{}
         \label{fig:dt_d}
     \end{subfigure}
    \caption{(a) Cross sections of fusion reactions as a function of the center of mass (COM) energy. (b) Reactivities $\langle \sigma v \rangle$ of fusion reactions, assuming that the plasma is in thermodynamic equilibrium. (c) The fusion reaction rate $R_{ij}$ divided by the squared plasma pressure $p^2$, assuming an optimal fuel ratio. (d) The fusion power density (in megawatts per cubic meter) per unit of squared plasma pressure for various fusion reactions. Based on ENDF data from the IAEA.}
        \label{fig:dt_reactions}
\end{figure}

The probability of atomic and nuclear reactions is measured in terms of a so-called `cross section' $\sigma$. Nuclear cross sections depend on quantum mechanics and are measured experimentally in units of \unit{\meter^2} reactions per particle squared.
A beam of particles of species $a$ with number density $n_a$ moving at velocity $v$ through a stationary background of particles of species $b$ with number density $n_b$ will have $n_a n_b v \sigma_{ab}(v)$ reactions per unit time per unit volume. The total reaction rate $R$ in matter with distribution function $f_i(\bm v_i)$ for particle species $i$ will be
\begin{equation}\label{eq:ch1-reactionratesumoverspecies}
R = \sum_{i \le j} R_{ij}
\end{equation}
where (see equations 3.14 and 3.15 of \cite{freidberg2008plasma})
\begin{equation}\label{eq:ch1-reactionratesinglespecies}
R_{ij} = C_{ij}\int f_i(\bm v_i) f_j(\bm v_j) |\bm v_1 - \bm v_2| \sigma(|\bm v_1 - \bm v_2|) \mathop{d\bm v_1}\mathop{d\bm v_2}
\end{equation}
with
\begin{equation}
    \begin{cases}
    C_{ij} = \frac{1}{2}& \text{if } i=j\\
    C_{ij} = 1              & \text{otherwise.}
\end{cases}
\end{equation}
\Cref{fig:dt_a} shows the cross section of different fusion reactions at different center of mass energies.
Because the particle distribution function $f(\bm v)$ is a Maxwellian at thermal equilibrium, the reaction rate $R_{ij}$ at can be calculated using \cref{eq:ch1-reactionratesinglespecies}.
Thus at thermal equilibrium \cref{eq:ch1-reactionratesinglespecies} simplifies to
\begin{equation}\label{eq:ch1-fusionreactivity}
    R_{ij} = n_i n_j \langle\sigma v \rangle
\end{equation}
where $n_i$ is the particle density of species $i$ and $\langle\sigma v \rangle$ is the reactivity.
\Cref{fig:dt_b} plots the reactivity for different fusion reactions in thermal equilibrium at temperature $T$.


The reaction rates in figure \ref{fig:dt_b} are all negligible at temperatures below 0.1 \unit{KeV} (over a million degrees $\unit{C}$). Thus, fusion happens at very high temperatures. Because the typical particle energy at these temperatures is much higher than the energy required to ionize matter, fusion happens in ionized gases called plasmas.
The very high temperatures required is the principal reason that producing energy via fusion is extremely difficult.
Because thermal energy flows from hot to cold, the heat from a fusion reaction will need to be confined in some way to prevent thermal equilibration with the environment outside the reactor.
High temperatures can be confined for very short periods without any confinement (called inertial confinement), or for longer periods by using magnetic fields to reduce the thermal diffusivity of a plasma (called magnetic confinement). We only consider magnetic confinement, and not inertial confinement, in this thesis.

Since the fusion reaction rate $R_{ij}$ is proportional to $n^2$, \cref{eq:ch1-fusionreactivity} suggests that the best way to produce high quantities of energy using fusion power is to increase the plasma density. However, it turns out that increasing the density of a magnetically confined plasma tends to decrease its temperature, such that the pressure $p = \sum_i n_i T_i$ remains approximately constant. Thus it makes sense to rewrite \cref{eq:ch1-fusionreactivity} in terms of $p$ and $T$ instead of $n$ and $T$.

In a fully ionized plasma made up of two ion species $n_i$ and $n_j$ with charge $Z_i e$ and $Z_j e$ and electrons with number density $n_e = n$, assuming equal temperatures $T_i = T_j = T_e = T$, the fusion rate $R_{ij} \propto n_i n_j $ will be maximized subject to the constraint of quasi-neutrality if $n_i = \frac{n}{2 Z_i}$ and $n_j = \frac{n}{2 Z_j}$. These equations can be derived using the method of Lagrange multipliers. Thus $p = nT(1 + \frac{1}{2 Z_i} + \frac{1}{2 Z_j})$, and 
\begin{equation}
    R_{ij} = \frac{n^2}{4 Z_i Z_j} \langle\sigma v \rangle = \frac{1}{4 Z_i Z_j (1 + \frac{1}{2Z_i} + \frac{1}{2Z_j})^2} p^2 \frac{\langle\sigma v \rangle}{T^2}.
\end{equation}
The quantity $R_{ij}/p^2$ is plotted for various fusion reactions in \cref{fig:dt_c}.

Multiplying the number of reactions per volume per second $R_{ij}$ times the average energy per reaction $E_{ij}$ gives the fusion power per volume $P_{ij}$.
\Cref{fig:dt_d} plots the quantity $P_{ij} / p^2$ for various fusion reactions.
At constant pressure, \cref{fig:dt_d} shows that the fusion reaction with the highest power density (by almost two orders of magnitude) is the deuterium-tritium (D-T) reaction 
\begin{equation}\label{eq:ch1-DTreaction}
    \text{\ce{_1^2D + _1^3T -> _2^4He} (3.5 \unit{MeV})\ce{ + ^1_0n} (14.1 \unit{MeV})}
\end{equation}
with the maximum $R_{ij}$ at approximately 15 \unit{KeV}. The \ce{^4_2He} alpha particle is born with 3.5 \unit{MeV}. This particle, through collisions with other plasma particles, can deposit its energy inside a plasma. Alpha heating can, if the rate of heat loss from the plasma is sufficiently low, be used to maintain the plasma temperature in steady state. 

\Cref{fig:dt_reactions} is compiled using Evaluated Nuclear Data File (ENDF) from the IAEA, found at \url{https://www-nds.iaea.org/exfor/endf.htm}. Note that the IAEA data for the \ce{p}-\ce{^11B} peak cross section is lower than other open-source estimates \cite{tentori2023revisiting}, and the reactivities for \ce{D}-\ce{D} and \ce{p}-\ce{^11B} in \cref{fig:dt_b} don't decline at high temperatures as is usually shown. Thus the IAEA data may have some error at large COM energy.

\subsubsection{Fusion power density is much lower than fission}

One important takeaway from \cref{fig:dt_d} is that fusion has much lower power density than fission. Light water fission reactors have power densities on the order of 100 \unit{MW/m^3}, while a fusion reactor with 2.6 \unit{atm} of plasma pressure (the expected plasma pressure of ITER in full operation) would have a power density of only 0.6 \unit{MW/m^3}, about 150 times less than in a typical fission plant.\footnote{ITER has a plasma volume of 840 \unit{m^3} and is designed to produce 500MW of power, about 0.6 \unit{MW/m^3}. This is consistent with \cref{fig:dt_d}. }

\subsubsection{Challenges}

While the D-T reaction has by far the highest power density of any fusion reaction, and thus is by far the easiest to produce, it leads to two major challenges. First, tritium has a radioactive half-life of 12.3 years and so only exists in trace amounts in nature. Tritium would need to be created inside a fusion reactor. Doing so would require that every D-T reaction (\cref{eq:ch1-DTreaction}) produce at least one new tritium atom inside the reactor. This so-called `tritium breeding ratio' must be greater than one because one tritium is lost in in every D-T reaction but also because tritium decays over time and because some amount of tritium will be lost inside a reactor. 
Creating tritium is possible using a three step process. First, using a neutron multiplier (such as Berylium or Lead) to increase the neutron flux. Second, slowing those high-energy neutrons down using a neutron moderator. Third, at low energies allowing the neutrons to react with \ce{^6_3Li} (which is 7.6\% of natural lithium) to breed tritium via the reaction
\begin{equation}\label{eq:ch1-LiLE}
    \text{\ce{^6_3Li +  ^1_0n -> ^4_2He + ^3_1T}} + 4.78{\text{ }}\unit{MeV}.
\end{equation}
Because neutrons have very large mean free paths in low-density plasmas, they will not be slowed down by the hydrogen inside a fusion reactor but will need to be slowed down outside the plasma in what is called the `blanket'.
Some simple calculations show that the width of the blanket must be about 1 \unit{\meter} (more sophisticated calculations show that the width of a blanket would be between 1 \unit{\meter} and 1.5 \unit{\meter} thick) \cite{freidberg2008plasma}. 
All D-T fusion reactors thus require a blanket of roughly 1 \unit{\meter} thickness, to multiply and slow down the neutrons and convert them into tritium. This adds significantly to the cost, size, and complexity of a fusion reactor.

The second major challenge is that neutrons, especially high-energy neutrons, cause metals and other solid materials to degrade over time. Nuclear transmutations due to collisions with neutrons cause materials surrounding fusion reactors to become radioactive, and can lead to the release of gases such as helium and hydrogen that result in voids, swelling, and embrittlement of metals and other crystalline structures \cite{rubel2019fusion}. This neutron damage means that components surrounding the fusion reactor, including the blanket, would have to be replaced multiple times over the lifetime of a fusion reactor \cite{malang1999limitations,beidler2001stellarator}. Due to the radioactivity from neutron activation, the maintenance required for replacing the components inside a fusion reactor would have to be performed remotely using robotic devices \cite{haist2008setting}, also increasing cost and complexity.

We have seen that nuclear physics imposes major challenges to producing fusion energy in a controlled (i.e., non-explosive) manner for energy production. Fusion reactions can only occur at very high energies, implying that fusion reactors will have to contain matter at extremely high temperatures. The only known way of doing so on Earth in steady state is to use magnetic fields to confine a plasma inside a vacuum chamber, greatly increasing the complexity of a reactor. The fusion reaction with the highest power density (by about two orders of magnitude) uses a fuel, tritium, that doesn't exist in nature. Fortunately, it is theoretically possible to create tritium inside a reactor. However, most of the energy in this reaction is released in high-energy neutrons, which rapidly damage (and if not replaced, will eventually destroy) reactors from the inside. The need to both breed tritium and extract energy from neutrons mean that fusion reactors must be surrounded by at least 1 \unit{m} of highly engineered replaceable radioactive components, putting additional constraints on the scale and cost of fusion reactors.

With these challenges in mind, we now consider two of the leading concepts for magnetic confinement fusion, tokamaks and stellarators.

\subsection{Tokamaks}

As we've seen, producing energy from fusion requires very high particle energies -- and thus, in thermodynamic equilibrium, very high temperatures. Because heat tends to flow from hot to cold, the thermal heat loss inside a fusion plasma will need to be kept as low as possible to maintain the high temperature.
Because charged particles in a high-temperature plasma have very large mean free paths, then unless some force is able to confine them, plasma particles inside a reactor will travel in straight lines and rapidly deposit their energy on the reactor walls.
To minimize the heat losses from a steady-state fusion reaction, it is essential to confine plasma particles inside a volume.

One approach for confining plasma particles inside a volume is to use toroidal magnetic fields.
Because a plasma is made up of charged particles, and charged particles travel freely in the direction parallel to magnetic fields but gyrate in the direction perpendicular to magnetic fields, then by wrapping a magnetic field line in a circle, one can imagine that plasma particles moving in the direction parallel to the magnetic field will travel around the circle and arrive back to where they started.
As we'll see, this isn't exactly correct -- toroidal magnetic fields (i.e., purely circular field lines) don't confine charged particles.
Nevertheless, studying why toroidal magnetic fields don't confine plasma particles will allow us to understand the tokamak, which thus far has been the most successful fusion reactor concept.

Suppose a charged particle with mass $m$ and charge $q$ is placed inside an axisymmetric toroidal magnetic field, such that (in a cylindrical coordinate system)
\begin{equation}\label{eq:ch1-toroidallysymmetricB}
    \bm B = B_\phi(r, z) \bm{\hat{\phi}}.
\end{equation}
This magnetic field could be created, for example, by a toroidally-shaped solenoid.
The dynamics of a particle with coordinates $\bm q$ and velocities $\bm{\dot{q}}$ with Lagrangian $\mathcal{L}$ will be determined by the Euler-Lagrange equations,
\begin{equation}
    \frac{\partial \mathcal{L}}{\partial \bm{q}} = \frac{d}{dt} \bigg(\frac{\partial \mathcal{L}}{\partial \bm{\dot{q}}}\bigg).
\end{equation}
If $\frac{\partial \mathcal{L}}{\partial q_i} = 0$, then $\frac{d}{dt}(\frac{\partial \mathcal{L}}{\partial \dot{q}_i}) = 0$ and so $p_i = \frac{\partial \mathcal{L}}{\partial \dot{q}_i}$ will be a conserved quantity.
The Lagrangian for a charged particle in an electromagnetic field is
\begin{equation}
\mathcal{L} = \frac{1}{2}m |\bm{\dot{r}}|^2 + q \bm A \cdot \bm{\dot{r}}  - q \Phi(\bm r, t).
\end{equation}
Here $\bm B = \curl \bm A$ and $\bm E = - \grad \Phi - \frac{\partial \bm A}{\partial t}$. Because $\bm E = 0$ and $\bm B$ is purely toroidal, we can set $\Phi = \frac{\partial \bm A}{\partial t} = 0$, and use the gauge freedom $\bm A \rightarrow \bm A - \grad f$ such that $\bm A = A_z(r, z) \bm{\hat{z}}$ and $B_\phi = - \frac{\partial A_z}{\partial r}$. 
Thus, the Lagrangian for the charged particle is
\begin{equation}
    \mathcal{L} = \frac{1}{2}m (\dot{r}^2 + r^2 \dot{\phi}^2 + \dot{z}^2) + q A_z(r,z) \dot{z}.
\end{equation}
The Lagrangian is independent of $\phi$ ($\frac{\partial \mathcal{L}}{\partial \phi} = 0$) and so $p_\phi = m r^2 \dot{\phi}$ is a conserved quantity.
While $r^2 \dot{\phi}$ is conserved, conservation of $p_\phi$ and conservation of energy put no constraints on the value of $z$.
Thus, particles in toroidal magnetic fields are free (in principal) to drift up or down in the vertical plane.

Charged particles in toroidal magnetic fields, it turns out, do drift vertically. Basic plasma physics theory states that charged particles in electromagnetic fields experience various drifts. In a magnetic field given by \cref{eq:ch1-toroidallysymmetricB}, the curvature and $\grad B$ drifts both point in the ${q} \hat{z}$ direction, and thus charged particles will drift up or down depending on their charge. Toroidal magnetic fields don't confine charged particles in the vertical direction, allowing heat to escape along with the particles.

Tokamaks do confine charged particles. They do so by modifying the magnetic field in \cref{eq:ch1-toroidallysymmetricB} to also have axisymmetric magnetic fields in the $\bm{\hat{r}}$ and $\bm{\hat{z}}$ directions:
\begin{equation}
    \bm B = B_\phi(r,z) \bm{\hat{\phi}} + B_r(r, z) \bm{\hat{r}} + B_z(r,z) \bm{\hat{z}}.
\end{equation}
We'll call $\bm B_t = B_\phi \bm{\hat{\phi}}$ the toroidal magnetic field, and $\bm B_p = B_r(r, z) \bm{\hat{r}} + B_z(r,z) \bm{\hat{z}} = \curl \bm A_p$ the poloidal magnetic field. We can again use gauge freedom to set $\bm A_p = A_\phi(r,z) \bm{\hat{\phi}}$, so that
\begin{equation}
    \bm B_p = -\frac{\partial A_\phi}{\partial z} \bm{\hat{r}} + \frac{1}{r}\frac{\partial}{\partial r}(r A_\phi) \bm{\hat{z}}.
\end{equation}
If $S$ is a 2D surface embedded in 3D space with a normal vector $\bm{\hat{n}}$ at each point, if $\bm B \cdot \bm{\hat{n}}=0$ everywhere on $S$ then $\bm B$ does not cross $S$.
Surfaces with constant $r A_\phi$ have the normal vector $\grad(r A_\phi)$, and we can see that
\begin{equation}
    \bm B \cdot \grad(r A_\phi) = B_r \frac{\partial}{\partial r}(r A_\phi) + B_z \frac{\partial}{\partial z}(r A_\phi) = - \frac{\partial A_\phi}{\partial z}\frac{\partial}{\partial r}(r A_\phi) + \frac{\partial}{\partial r}(r A_\phi) \frac{\partial A_\phi}{\partial z} = 0.
\end{equation}
Thus, $\bm B$ does not cross surfaces of constant $r A_\phi$.
We call these surfaces poloidal surfaces.
You can prove (see section 10.1 of \cite{imbert2019introduction}) that poloidal surfaces are guaranteed to be closed and nested, so long as the magnetic field is axisymmetric.

Unlike toroidal magnetic fields, which can be created by currents outside of the torus, axisymmetric poloidal magnetic fields require toroidal currents inside the torus. To see this, consider the integral $\oint \bm B \cdot \mathop{d\bm \ell}$ around the closed loop formed by the projection of a poloidal surface in a plane of constant $\phi$. By Ampere's law, this integral will be equal to $\mu_0$ times the toroidal current enclosed by the loop.
The vector $d\bm{\ell}$ must be perpendicular to both $\bm{\hat{\phi}}$ and the normal to the surface $\grad(rA_\phi)$, so $d\bm{\ell} \parallel \grad(rA_\phi) \times \bm{\hat{\phi}}$. Since $\grad(rA_\phi) \times \bm{\hat{\phi}} = -r\frac{\partial A_\phi}{\partial z} + \frac{\partial}{\partial r}(r A_\phi) \bm{\hat{z}}$ then $d\bm{\ell} \propto r\bm B_p$.
Thus $\bm B \cdot d\bm{\ell} \propto \bm B_p\cdot \bm B_p$ is non-zero everywhere along the closed loop, and the enclosed toroidal current inside a flux surface must be non-zero.

To see why tokamaks confine charged particles inside a volume, consider the Lagrangian for a particle inside a tokamak
\begin{equation}
    \mathcal{L} = \frac{1}{2}m (\dot{r}^2 + r^2 \dot{\phi}^2 + \dot{z}^2) + q A_z(r,z) \dot{z} + q A_\phi(r, z) r \dot{\phi}.
\end{equation}
The Lagrangian is independent of $\phi$, so
\begin{equation}
    p_\phi = mr^2 \dot{\phi} + q r A_\phi(r, z)
\end{equation}
is a conserved quantity.
As a result, particles will be confined near surfaces of constant $r A_\phi$, which we know are closed and nested. To see this, consider a particle which begins on the flux surface $r A_\phi$. The maximum distance $|\Delta \bm r|$ that the particle can drift normal to its initial flux surface is (to lowest order) given by the expression
\begin{equation}
    q |\grad(r A_\phi)| |\Delta \bm r| = |m r \Delta (r \dot{\phi})|
\end{equation}
which follows from conservation of $p_\phi$. Because $|\grad(r A_\phi)| = r B_p$, and for a typical particle $r \dot \phi \sim v_T$ (the thermal velocity),
\begin{equation}
    |\Delta \bm r| \sim \frac{m r^2 \dot \phi}{q |\grad(r A_\phi)|} = \frac{m v_T}{q B_p} = \frac{v_T}{B_p} = \rho_p.
\end{equation}
$\rho_p$ is the poloidal gyroradius, and will be small relative to the size of the tokamak so long as $B_p$ is sufficiently large.\footnote{Particles with small $v_\parallel \sim r \dot\phi$ can have larger deviations from flux surfaces. These so-called `trapped' particles execute so-called `banana orbits', which lead them to deviate from a flux surface by larger amounts, and thus to transport heat more easily.} Thus, tokamaks confine single particles to stay near flux surfaces.

Tokamaks have been the magnetic confinement concept most successful at producing high temperatures and pressures. As a result, scientists and engineers are considering the tokamak as the basis of possible fusion power plants. One prominent example currently under construction is the ITER tokamak, a more-than \$20B project designed for a physics gain factor $Q \approx 10$. Even this massive device is far from a power plant: ITER is designed for an engineering gain factor $Q_E \approx -0.33$, meaning it is expected to consume more electrical power than it would produce (assuming a conversion efficiency of 40\%).

Although tokamaks are the most successful magnetic confinement devices, they have at least two major problems. First, tokamaks can have what are called `disruptions.' Disruptions involve a rapid loss of thermal energy inside the plasma, and lead to large forces on the structures inside a tokamak and possible runaway electron beams that can deposit their energy locally on the inside of a tokamak. Second, the plasma physics performance of a tokamak is not good enough to scale to a reactor. As Jeffrey Freidberg writes, unfortunately ``a tokamak reactor designed on the basis of standard engineering and nuclear physics constraints does not scale to a reactor. Too much current is required to achieve the necessary confinement time for ignition.'' \cite{freidberg2015designing}
These problems, and many others not mentioned here, have led scientists to consider alternative schemes for confining plasmas at high temperatures to produce fusion energy. One such scheme is the stellarator.

\subsection{Stellarators}

Stellarators have much in common with tokamaks. Both use magnetic fields with a toroidal topology. Both seek to use magnetic fields to confine charged particle orbits to stay within a volume of closed nested flux surfaces. Both require a combination of toroidal and poloidal magnetic fields to produce rotational transform. Both must satisfy magnetohydrodynamic (MHD) force balance and stability.

The key difference between a stellarator and a tokamak is that the magnetic field in a stellarator is not axisymmetric. While tokamaks create toroidal magnetic fields using axisymmetric planar coils and create poloidal magnetic fields using plasma current, stellarators use non-axisymmetric and often non-planar coils to create rotational transform of the magnetic field. Stellarators can use plasma current to create rotational transform, but typically they are designed so that the external magnetic fields produce the dominant portion of the rotational transform.

In non-axisymmetric stellarators, as in tokamaks, one of the most basic and important properties needed to prevent rapid loss of thermal energy from a high-temperature fusion plasma is charged particle confinement. The theory of charged particle confinement in stellarators involves non-orthogonal coordinate systems and is not important to the rest of this thesis. \cite{helander2014theory,imbert2019introduction} are good introductions to this theory. The main idea is to assume that the stellarator contains closed nested flux surfaces, to use a toroidal coordinate system $(\psi, \vartheta, \varphi)$ where $\psi$ is a radial coordinate determined by the flux surface and $\vartheta$ and $\varphi$ are poloidal and toroidal angles relative to the flux surface, and to ensure that the time-averaged radial drift $\bm v_d$ of particles off flux surfaces averages to zero:
\begin{equation}
    \big \langle \bm v_d \cdot \grad \psi  \big \rangle = 0.
\end{equation}
This condition is called \textit{omnigeneity}. There are various ways to satisfy omnigeneity, including $\bm v_d \cdot \grad \psi = 0$ (isodynamicity) and $B(\psi, \vartheta, \varphi) = B(\psi, M \vartheta_B - N \varphi_B)$ (quasisymmetry) \cite{rodriguez2022quasisymmetry}.

Computational optimization is much more important in stellarators than in tokamaks.
Tokamaks only have a few parameters to vary: the toroidal field strength, the strength of the plasma current (and thus the poloidal magnetic field), and the shape of the plasma in the 2D poloidal plane.
Furthermore, the magnitude and effects of the plasma current in a tokamak can be difficult to predict a priori.
In practice, tokamaks rely on experimental validation and extrapolation to accurately predict the performance and properties of new experiments.
Stellarators, by contrast, have 3D magnetic fields, allowing for many more degrees of freedom in the design of stellarator devices.
Because stellarators confine particles using fields produced mostly by external rather than internal currents, the performance of stellarators can be much more reliably predicted from basic theory \cite{boozer2020carbon}.
In practice, all modern stellarators are designed using computational optimization.

Stellarator optimization produces two essential outputs. First, the optimized magnetic field. This non-axisymmetric magnetic field might be optimized for single particle confinement, MHD stability, reduced neoclassical and turbulent transport, and any other desirable properties one might seek. Second, the external current-carrying coils that produce the desired (vacuum) magnetic field. The coil shapes are optimized to produce the desired field as accurately as possible and also as at low cost as possible.
Stellarator design is, therefore, a high-dimensional multi-objective optimization problem.

While it is computationally tractable to exhaustively search low-dimensional spaces to find the global minimum of an objective function, exhaustively searching high-dimensional spaces becomes computationally intractable.
Suppose we want to find the global minimum of $f(\bm x)$, $\bm x \in \mathcal{R}^D$, $x_i \in [0, 1]$, for $i = 1, \dots, D$. We can exhaustively search the design space by choosing $N$ equally spaced points for each dimension, and doing a grid search over all $N^D$ points. However, for even modest $N$ this becomes intractable at large $D$.
High-dimensional optimization instead relies most often on derivative-based optimization methods.
Derivative-based optimization methods typically require the gradient $\grad f$, the Hessian $\partial^2 f$, and/or products of the gradient or Hessian with vectors.

In stellarator optimization, previous works have computed $\grad f$ either using analytic hand-derived expressions or with finite-difference \cite{Zhu_2017,l_singh_preprint}. Other articles have eliminated the need for $\grad f$ entirely by restricting the optimization to only consider convex optimization problems \cite{Merkel_1987_NESCOIL,Landreman_2017}.
Instead, differentiable programming could be used to compute gradients for high-dimensional optimization problems for stellarators.
One motivation of this thesis is to introduce and demonstrate how differentiable programming, by computing gradients using automatic differentiation, can serve as a useful tool in the computational optimization of stellarators.
In \cref{ch:stellarator}, I discuss my own research introducing AD to the stellarator community, through a problem in coil design. I also outline how other researchers are using AD in MHD equilibrium codes for magnetic geometry optimization.

\section{Machine learning (ML)}

\begin{figure}
    \centering
    \includegraphics[width=\textwidth]{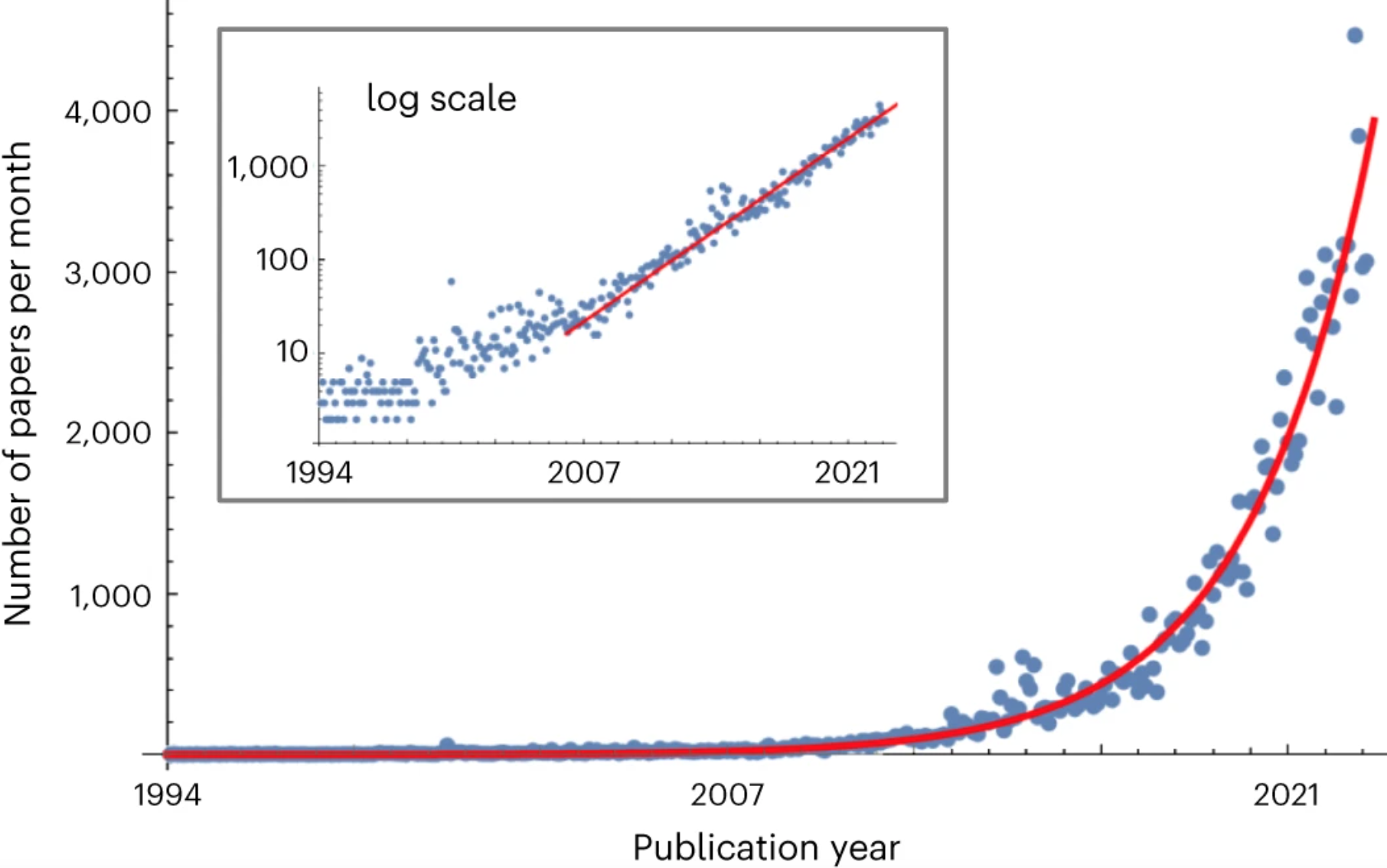}
    \caption{The number of papers published per month in the arXiv categories of AI and ML is growing exponentially. From \cite{krenn2023forecasting}. Obtained via CC BY license.}
    \label{fig:aimlpapers}
\end{figure}

Machine learning (ML) has been a remarkably successful research field. 
Due to its empirical success, ML has become the dominant approach for tackling challenges in artificial intelligence (AI) such as computer vision, natural language processing, and speech recognition.
This has led to enormous interest in ML in academia, industry, and government.
Figure \ref{fig:aimlpapers} shows the number of papers published per month in the ArXiv categories of AI and ML.

\subsection{Supervised learning}

The most common ML paradigm, and the one most relevant to this thesis, is called supervised learning.
Supervised learning assumes that there exists some (deterministic or probabilistic) relationship between inputs $\bm x \in \mathbb{R}^n$ and outputs $\bm y \in \mathbb{R}^m$.
The goal of supervised learning is to output a prediction $\bm{\hat{y}}$ given $\bm x$ which accurately approximates the true relationship between $\bm x$ and $\bm y$.
Doing so requires (a) a parametric model $f_{\bm\theta}$ with parameters $\bm \theta$ such that $\bm{ \hat{y}} = f_{\bm\theta}(\bm x)$, (b) a loss function $\mathcal{L}$, (c) labeled data points $\{\bm x_i, \bm y_i\}_{i=1}^D$, and (d) a learning algorithm which attempts to adjust the parameters $\bm \theta$ towards $\bm \theta^*$ where
\begin{equation}\label{eq:ch1-argminml}
    \bm \theta^* = \argminC_{\bm \theta} \sum_{i=1}^D \mathcal{L}(f_{\bm \theta}(\bm x_i), \bm{y}_i; \bm \theta).
\end{equation}
The parametric models $f_{\bm \theta}$ which have been most successful in AI applications are called neural networks.
Neural networks are flexible function approximators:
universal approximation theorems for neural networks state that, under certain conditions, neural networks with infinitely many parameters can, with appropriately chosen $\bm \theta$, approximate any continuous function \cite{pinkus1999approximation,kidger2020universal}.
Neural networks typically contain repeated `layers', each consisting of a linear transformation (i.e., matrix multiplication) followed by a non-linear transformation.
The parameters of the neural network are the coefficients of the linear transformation; neural networks can have thousands, millions, or (for the largest models) even billions of parameters. 
Additional details of neural networks are not important for this thesis; an excellent introductory text is \cite{goodfellow2016deep}.
Loss functions $\mathcal{L}(f_{\bm \theta}(\bm x_i), \bm{y}_i; \bm \theta)$ are usually chosen such that the function is minimized when $f_{\bm \theta}(\bm x_i) = \bm y_i$; a common choice is the mean squared error (MSE)
\begin{equation}\label{eq:ch1-mse}
    \mathcal{L}(\bm a, \bm b) = \frac{1}{2}\langle (\bm a - \bm b)^2\rangle = \frac{1}{2}\sum_{j=1}^M (a_j - b_j)^2.
\end{equation} 
In some cases, additional so-called `regularization' terms are added to the loss function.
Learning algorithms in machine learning typically perform gradient-based optimization; gradient descent (with learning rate $\eta$)
\begin{equation}
    \bm \theta \longrightarrow \bm \theta - \eta \bm \nabla_{\bm \theta} \mathcal{L}
\end{equation}
and Adam \cite{kingma2014adam} are common gradient-based optimization algorithms in ML.

\subsubsection{Supervised learning as statistical inference}

The supervised learning paradigm in ML can be understood through the lens of statistical inference. To illustrate, consider a simple statistical model. The model assumes that the output data $y \in \mathbb{R}$ is given by a linear sum of the input data $\bm x \in \mathbb{R}^n$ plus some random (Gaussian) noise $\epsilon$:
\begin{equation}\label{eq:ch1-lineardatageneration}
    y = \bm \theta^T \bm x + \epsilon\text{, \hspace{1.0cm} } \epsilon \sim \mathcal{N}(0, \sigma^2).
\end{equation}
(The notation $\sim$ means `a draw from the probability distribution', while $\mathcal{N}(\mu, \sigma^2)$ represents a Gaussian distribution with mean $\mu$ and standard deviation $\sigma$.)
The data is given by a set of $D$ input-output pairs, $\{\bm x_i, y_i\}_{i=1}^D$.
Of course, the data might not be generated according to \ref{eq:ch1-lineardatageneration}. The true relationship between $\bm x$ and $y$ could very well be more complex than a linear model.
Regardless, we'll assume a statistical model of this form and choose the parameters of the model $\bm \theta$ to best fit the data.

There are two ways of choosing $\bm \theta$: maximum likelihood estimation (MLE) and maximum a posteriori (MAP) estimation. MLE chooses the $\bm \theta$ that maximizes the probability of observed data (the likelihood), while MAP uses Bayes' theorem to choose the $\bm \theta$ with the maximum probability given the data (the posterior). We'll consider the MLE estimator here.
The probability of $y$ given $\bm x$ and $\bm \theta$ is (from equation \ref{eq:ch1-lineardatageneration})
\begin{equation}
    p(y | \bm x, \bm \theta) = \mathcal{N}(\bm \theta^T \bm x, \sigma^2) = \frac{1}{({2\pi} \sigma)^{N/2}}\exp\bigg(\frac{-(y-\bm \theta^T \bm x)^2}{2\sigma^2}\bigg).
\end{equation}
Assuming that each of the D input-output pairs is i.i.d., the probability of the data is
\begin{equation}
    p(\{y_i\}_{i=1}^D | \{\bm x_i\}_{i=1}^D, \bm \theta) = \prod_{i=1}^D p(y_i | \bm x_i, \bm \theta).
\end{equation}
According to the MLE estimate, we'll choose the parameters $\bm \theta$ to maximize the probability of the data:
\begin{equation}
    \bm \theta^* = \argmaxC_{\bm \theta} \prod_{i=1}^D p(y_i | \bm x_i, \bm \theta).
\end{equation}
It turns out to be easier to optimize the $\log$ of the probability than the probability, and since the $\log$ function is monotonically increasing, $p$ and $\log{p}$ will have the same $\argmaxC$. 
\begin{equation}\label{eq:ch1-linearloss}
    \bm \theta^* = \argmaxC_{\bm \theta} \prod_{i=1}^D \log{p(y_i | \bm x_i, \bm \theta)} = \argminC_{\bm \theta} \frac{1}{2\sigma^2} \sum_{i=1}^D (y_i - \bm \theta^T \bm x_i)^2 + \text{ constant}
\end{equation}
\Cref{eq:ch1-linearloss} is a convex optimization problem; $\bm \theta^*$ can be computed analytically by setting the gradient with respect to $\bm \theta$ of $\sum_{i=1}^D (y_i - \bm \theta^T \bm x_i)^2$ equal to zero. 

Now compare \cref{eq:ch1-linearloss} to \cref{eq:ch1-argminml,eq:ch1-mse}. They're identical, except with $f_\theta(\bm x_i)$ replaced by $\bm \theta^T \bm x_i$.
Thus, supervised learning with neural networks can be interpreted as maximum likelihood estimation (MLE) of the data, except with a flexible function approximator $\hat{y} = f_{\bm \theta}(\bm x)$ replacing simpler statistical models with tractable likelihood functions.
These flexible function approximators result in high-dimensional non-convex optimization problems that require (or at least empirically perform best with) gradient-based optimization.

\subsection{ML for computational physics}

As interest in machine learning (ML) has grown, more and more scientific fields are exploring the potential of ML to advance science.
One such area is computational physics.
There is immense interest in using ML within computational physics.
At the time of writing (December 17th, 2023), in the Journal of Computational Physics 18 of the 25 (72\%) most cited articles since 2020 and 24 of the 25 (96\%) most downloaded articles in the past 90 days involve neural networks, deep learning, and/or ML.
The most cited of these articles \cite{raissi2019physics}, published by the CRUNCH group at Brown University, has been cited more than 3,300 times in 2023 alone (and over 8,000 times overall) and is on track to become the most cited article in the history of computational physics in fewer than 5 years.
As the time of writing (December 17th, 2023) the CRUNCH research group website expresses a not-unusual level of excitement about the potential of ML, writing:
\begin{displayquote}
    \textit{We are at the cross-roads in Computational Mathematics! The machine learning revolution is real this time around and is changing our field in a fundamental way! We may experience the sudden death of [the finite element method] and other classical numerical methods, and the rise of new and simpler methods using Deep Learning} \cite{crunchgroup}.
\end{displayquote}

\subsubsection{ML-based surrogate models are Department of Energy priority}

There are many ways that ML might be used within computational physics research.
One widely studied approach is to develop so-called `surrogate models'.
Surrogate models are simplified models of more complex physical or computational systems, designed to approximate the behavior of the system at reduced cost.

Two Department of Energy (DOE) Office of Science reports -- the 2020 report \textit{AI for Science} \cite{stevens2020ai} and the 2023 report \textit{AI for Science, Energy, and Security} \cite{carter2023ai} -- reveal the extent to which ML-based surrogate models for applications in high-performance computing (HPC) are expected to be a key part of an ``AI-centric future science landscape'' \cite{stevens2020ai}.

The 2020 report, based on a series of town hall meetings with ``overflowing attendance,'' discusses the potential for faster AI-based surrogate models across many areas of science. The executive summary writes that
\begin{displayquote}
    \textit{AI in HPC has already taken the form of neural networks trained as surrogates to computational functions (or even entire simulations), demonstrating the potential for AI to provide non-linear improvements of multiple orders of magnitude in time-to-solution for HPC applications} \cite{stevens2020ai}.
\end{displayquote}
The 2023 report highlights six key ``expected science, energy, and security outcomes'' related to AI/ML.
The first of the six outcomes is ``AI and Surrogate Models for Scientific Computing,'' which are ``trained by the results of computational models'' and ``demonstrate orders-of-magnitude speedups over the originals'' \cite{carter2023ai}.

\subsubsection{ML for plasma physics}

The field of plasma physics and fusion energy is smaller than the broader field of computational physics, and the interest in using ML is much lower.
Nevertheless, many researchers have explored or are beginning to explore the use of ML for plasma physics.
In experimental magnetic confinement fusion, ML research has focused mainly on tokamak disruption prediction \cite{rea2018disruption,kates2019predicting,churchill2020deep,montes2021interpretable} and plasma control \cite{degrave2022magnetic,char2023offline}.
In computational plasma physics, ML research has focused mainly on the development of surrogate models for computationally intensive plasma models, including for gyrokinetics \cite{dong2021deep,citrin2023fast} and collision operators \cite{miller2021encoder,holloway2021acceleration}.
The 2023 DOE report writes that these surrogate models have ``the potential to revolutionize real-time predictions in magnetic fusion energy, accelerating progress toward favorably modifying the plasma state to a more benign thermodynamic state'' \cite{carter2023ai}.

As I will now discuss, the motivation for the second half of this thesis is to develop surrogate models that more efficiently approximate the solution to partial differential equations (PDEs), and (if possible) to identify a problem that is currently intractable without the use of ML, then to develop surrogate models to help solve that problem.

\section{Motivation: ML for PDEs \label{sec:ch1-motivation-ML-PDEs}}

PDEs are used throughout computational plasma physics. Mathematical models of plasma dynamics are typically given by time-dependent PDEs. Kinetic models (such as the the Vlasov equation), fluid models (such as the Braginskii fluid equations or MHD), and gyrokinetic models are all PDEs.
Thus, one of the most important tasks within computational plasma physics is to develop numerical algorithms for solving PDEs.

Computational scientists have spent decades developing algorithms to approximate the solution of PDEs.
This is a well-developed research area, and much of the computational research at the Princeton Plasma Physics Laboratory (PPPL, where my department is located) involves research and development of these numerical algorithms.
For example, the Gkeyll research group led by my advisor, Ammar Hakim, has developed numerical methods for solving kinetic, gyrokinetic, and fluid PDEs used in plasma physics and fusion energy.

There is a lot of interest in using ML to develop surrogate models for PDEs. The short-term goal of these surrogate models is to find approximate solutions to PDEs at reduced computational cost relative to the `standard' numerical methods that have been developed over the past few decades.
In other words, the short-term goal is to develop faster ML-based solvers.
The long-term goal is to use these faster ML-based surrogate models for downstream applications, such as optimization, uncertainty quantification, 
or inverse problems, and to improve or even replace the standard numerical methods used in research and commercial applications.

ML-based surrogate models for PDEs have, according to the published literature, achieved great success at their short-term goal of accelerating the solution of PDEs \cite{vinuesa2022enhancing}.
\Cref{tab:baselines} in \cref{ch:reproducibility} lists 76 of these articles, each of which claim to be able to output the solution to a PDE faster than a standard numerical method.

Using ML to solve PDEs is, as the famous AI researcher Yann LeCun observed, a ``pretty hot topic, indeed'' \cite{yanntweet}.
Even celebrities are interested in this research area.
Hip-hop artist MC hammer, author of the 1990 smash hit ``U Can't Touch This,'' tweeted about so-called `fourier neural operators' which use specially designed neural networks that learn to solve turbulent PDEs up to 1000$\times$ faster than standard numerical methods \cite{mchammertweet}.

If ML-based surrogate models for PDEs have apparently had success at their short-term goal of being faster, they have not yet achieved their long-term goal of being useful.
To the best of my knowledge, no research paper exists where an ML-based surrogate model for solving a PDE has been used to solve a downstream application or research question that would otherwise be intractable.\footnote{Surrogate models \textit{for macroscopic quantities} have been useful in downstream applications. For example, \cite{rodriguez2022nonlinear} uses Gaussian process regression to predict transport fluxes combined with Bayesian optimization to minimize the required number of gyrokinetic solves to predict transport in the SPARC tokamak, thereby reducing the overall computational cost. However, no surrogate models that output the solution to a PDE have been useful for downstream applications.}
Nor are any ML-based solvers being seriously considered as possible replacements for the numerical methods used in commercial codes for PDEs.

\subsubsection{Why use ML to solve PDEs?}

The previous discussion ignores a more basic question: why use ML to solve PDEs at all? Since the equations are already known, data is not needed to approximate the solution to a PDE. Why would we expect ML-based PDE solvers to be successful compared to standard numerical algorithms?

To illustrate \textit{why} we might expect ML to be useful in developing more efficient numerical methods, below I introduce an example of an ML-based PDE solver. I refer to this example as `machine learned flux prediction'; it could be classified as a so-called `hybrid' solver because it shares many aspects of standard numerical methods.
This type of solver was introduced in \cite{bar2019learning}.
Readers, however, should recognize that ML can be used in many different ways to solve PDEs, and that there is a zoo of different techniques for solving PDEs with ML.

To illustrate how machine learned flux prediction works, and why we might expect it to be successful compared to standard numerical algorithms, suppose we would like to solve for $u(x,t)$ which evolves according to the 1D advection equation
\begin{equation}\label{eq:ch1-1dadvection}
    \frac{\partial u}{\partial t} + c \frac{\partial u}{\partial x} = 0
\end{equation}
with $c \in \mathbb{R}$. Let's assume that $x \in [0, L)$ has periodic boundary conditions and that $t \in \mathbb{R}^+$.
To begin, we'll divide the domain into $N$ equally spaced cells of with $\Delta x = \frac{L}{N}$, so that cell $j$ for $j = 1, \dots, N$ has left boundary $x_{j-\frac{1}{2}} = (j-1)\Delta x$ and right boundary $x_{j+\frac{1}{2}} = j \Delta x$.
Now integrate \cref{eq:ch1-1dadvection} over cell $j$ and divide by $\Delta x$:
\begin{equation}
    \frac{1}{\Delta x} \int_{x_{j-\frac{1}{2}}}^{x_{j+\frac{1}{2}}} \frac{\partial u}{\partial t} \mathop{dx} + \frac{c}{\Delta x} \int_{x_{j-\frac{1}{2}}}^{x_{j+\frac{1}{2}}} \frac{\partial u}{\partial x} \mathop{dx} = 0.
\end{equation}
Setting $u_j(t) = \frac{1}{\Delta x} \int_{x_{j-\frac{1}{2}}}^{x_{j+\frac{1}{2}}} u(x, t) \mathop{dx}$ as the average of $u(x,t)$ in cell $j$, we have
\begin{equation}
    \frac{du_j}{dt} + \frac{c}{\Delta x} \big[u(x,t)\big]_{x_{j-\frac{1}{2}}}^{x_{j+\frac{1}{2}}} = 0
\end{equation}
\begin{equation}\label{eq:ch1-1dadvection-integral}
    \frac{du_j}{dt} + \frac{c}{\Delta x} \big( u_{j+\frac{1}{2}} - u_{j-\frac{1}{2}} \big) = 0
\end{equation}
where $u_{j+\frac{1}{2}}(t) = u(x_{j+\frac{1}{2}}, t)$.
So far, our derivation follows the standard finite volume (FV) method.
\Cref{eq:ch1-1dadvection-integral} represents $N$ coupled ODEs that are an exact integral (weak) form of \cref{eq:ch1-1dadvection}; no approximations have been made.
Standard ODE integration methods exist that can integrate ODEs with low error; given the cell averages $u_j$ for $j = 1, \dots, N$ at time $t$, solving \cref{eq:ch1-1dadvection-integral} requires predicting $u_{j+\frac{1}{2}}(t)$ for all $j$.
Thus, the challenge of computing an accurate approximation to \cref{eq:ch1-1dadvection} boils down to accurately predicting $u_{j+\frac{1}{2}}(t)$ given $\{u_j, u_{j+1}, u_{j-1}, u_{j+2}, \dots\}$.

At this point, standard numerical methods rely on \textit{hand-crafted} rules for predicting the flux $c u_{j+\frac{1}{2}}$ at cell boundaries.
For example, the `centered' scheme 
\begin{equation}
    u_{j+\frac{1}{2}} = \frac{u_j + u_{j+1}}{2}
\end{equation}
is second-order accurate and has phase error but no numerical diffusion, while the `upwind' scheme
\begin{equation}
    u_{j+\frac{1}{2}} = \begin{cases}
u_j \text{ if } c \ge 0\\
u_{j+1} \text{ if } c < 0
\end{cases}
\end{equation}
is first-order accurate and introduces numerical diffusion which decays the magnitude of the solution.
In any case, there are many possible choices for the flux $c u_{j+\frac{1}{2}}$; it is chosen through a combination of theoretical understanding and experimental testing.

However, one could instead imagine replacing hand-crafted rules for the flux $c u_{j+\frac{1}{2}}$ with rules that are learned from data.
If $u_{j+\frac{1}{2}}$ is predicted correctly, then the error in the solution would not be first order or second order but \textit{zero}, or at least very close to it. 
We might therefore expect a data-driven rule for the flux, which attempts to predict $u_{j+\frac{1}{2}}$ as accurately as possible, to perform better than a hand-crafted rule developed by humans.
This is exactly what machine learned flux prediction does: it uses supervised learning to learn $N$ functions $f_{\theta, j+\frac{1}{2}}(u_j, u_{j+1}, u_{j-1}, u_{j+2}, \dots)$ such that
\begin{equation}
    u_{j+\frac{1}{2}} = f_{\theta, j+\frac{1}{2}}(u_j, u_{j+1}, u_{j-1}, u_{j+2}, \dots)
\end{equation}
for $j = 1, \dots, N$. After training on a set of data from simulations that use random draws from a distribution of initial conditions, \cite{bar2019learning} found that machine learned flux prediction is able to solve PDEs with much smaller average error for a given number of gridpoints, or equivalently, to solve PDEs with many fewer gridpoints at equal error.
These findings are consistent with our experiments (see \cref{sec:why_standard_dont_work,sec:verification_burgers}).

In summary, one way of understanding why we might use ML to solve PDEs is that standard numerical methods rely on human-created, hand-crafted rules that work well but are not necessarily optimal. Using supervised learning, we can instead learn data-driven rules that might be more accurate (or have reduced computational cost) relative to standard numerical methods.

\subsubsection{A research direction with the potentially highest payoff}

The 2020 DOE report on AI for Science, within the `{Fusion}' section of the report, writes:
\begin{displayquote}
    \textit{We believe a research direction with the potentially highest payoff may be the integration of our knowledge of physics into ML models. Most existing AI/ML models are either purely data-driven or incorporate very simple physical laws and constraints. Without building the structure of the physical laws into ML methods, it is difficult to interpret the predictions from data-driven models.} \cite{stevens2020ai}
\end{displayquote}
Standard numerical methods for PDEs have a variety of desirable properties -- convergence, stability, conservation, positivity, etc -- that help ensure their results are accurate and reliable.
These properties can usually be proven algebraically or analytically.
However, no existing ML-based solvers do so, or at most they satisfy only very simple constraints.
In my opinion, the outputs of ML-based surrogate models for PDEs are unlikely to ever be trusted unless they maintain these properties as well.
Designing ML-based surrogate models that provably satisfy physical properties is a promising and potentially important research direction.

\subsubsection{Goals}

The latter half of this thesis investigated the use of ML to develop surrogate models for solving PDEs. This line of research had two goals:
\begin{enumerate}
    \item Design surrogate models and/or numerical algorithms for PDEs in plasma physics that are guaranteed to maintain the correct physical properties, thereby pursuing the research direction with the ``potentially highest payoff''.
    \item To identify and solve a problem in computational plasma physics that is currently intractable with standard numerical methods, thereby achieving a long-term goal of this research area.  If we are unable to do so, we want to understand whether ML-based PDE solvers might have any fundamental limitations that prevent them from being useful in research-level problems.
\end{enumerate}

The first goal was successful, as we'll see in \cref{ch:invariant}. However, I was not able to achieve the second goal.
Ultimately, I didn't even attempt to do so.
Instead, we discovered that a majority of the scientific literature, research that had claimed to successfully develop much faster solvers with ML, had made the same errors over and over again.
As a result, the ML-for-PDE solving literature had been systemically overoptimistic.
In \cref{ch:reproducibility}, I perform a systematic review of literature in that area and identify two reproducibility issues (weak baselines and reporting biases) that lead to overoptimism. 
Because the short-term goal of developing faster numerical methods with ML was not as successful as claimed, working on the long-term goals no longer made sense.
I conclude in \cref{ch:conclusion} by discussing the lessons learned from the latter half of my PhD, limitations of ML for numerical methods, and speculate on the future of ML in computational science.

\section{Contributions}

This thesis makes three main contributions and two minor contributions. 

\subsubsection{Main contributions}

The first contribution (\cref{ch:stellarator}) is the development of a new finite-build stellarator coil design code using AD.
This work, in collaboration with Stuart Hudson and Caoxiang Zhu and inspired by conversations with Ammar Hakim, was the first to represent stellarator coils during the optimization process not as infinitely thin filaments or current sheets but as 3D coils with finite extent.
In reality, the main objective of this research was not a new coil design code but rather to introduce AD to the stellarator community.
By demonstrating how AD could simplify the computation of derivatives for first-order optimization, the stellarator community might then adopt AD to easily and efficiently compute gradients of even complex objectives.
Sure enough, follow-up work with the DESC magnetic equilibrium optimization code uses AD for derivative calculations \cite{dudt2020desc,conlin2023desc,panici2023desc}, indicating that this objective was successful.

The second contribution (\cref{ch:invariant}) is to develop algorithms to guarantee that black-box PDE solvers (i.e., data-driven, ML-based solvers) preserve physical properties.
One of the basic goals of computational plasma research is to design algorithms that mimic or inherit the properties of the system they model.
Examples of such properties include conservation of mass, conservation of energy, L2 stability, etc.
While many hundreds of articles have proposed new algorithms that solve PDEs using ML, most of these ML-based PDE solvers have no guarantees of any physical properties. A few articles do conserve linear invariants (such as conservation of mass) using simple constraints such as finite volumes, but none preserve non-linear invariants or guarantee stability.
The key challenge, it turns out, is that standard approaches to preserving non-linear invariants (which put constraints on the flux based on the local solution) are too constraining and significantly degrade the accuracy of black-box solvers.
Preserving invariants without degrading accuracy required developing a new approach. This new approach is based on error correction.
The invariant-preserving error-correcting algorithms we developed aren't silver bullets -- they won't be able to guarantee that a black box algorithm gives qualitatively reasonable solutions -- but they increase the reliability and trustworthiness of ML-based PDE solvers without degrading their accuracy.
It is worth mentioning that there is nothing about these algorithms that requires the use of ML. In particular, these algorithms can be used to preserve invariants in any subgrid model, including Large Eddy Simulation (LES) models of turbulence, or as a correction to standard numerical algorithms.

The third contribution (\cref{ch:reproducibility}) is to diagnose and provide evidence of two reproducibility issues, weak baseline and reporting biases, that affect a significant percentage of ML-for-PDE solving research. Although ML-based PDE solvers are not more reliable than standard solvers, many articles claim to solve PDEs much more efficiently using ML. The efficiency of an ML-based PDE solver is measured relative to a baseline. Performing a systematic review of the ML-for-PDE solving literature, we determine that a significant majority of articles solving fluids-relevant PDEs are comparing to weak baselines. We also determine that articles with larger relative gains with ML are more likely to be comparing to weak baselines.
We find evidence that the second issue, reporting biases, is widespread. Researchers working at the intersection of ML and PDEs appear to be overreporting the successes of ML and underreporting the weaknesses, limitations, and failures of ML when solving PDEs. This issue is certainly not unique to ML, but it does appear to be particularly common in this research area. The main conclusion of this portion of the thesis is that ML not been nearly as successful at solving PDEs as the scientific literature has claimed. At best, ML offers modest improvements in efficiency (at most 10$\times$ faster, often no improvements at all) over standard solvers for fluid-related PDEs, with significant downsides including unpredictable accuracy, worsened reliability, and significant time and effort required to generate training data.

\subsubsection{Minor contributions}

The first minor contribution (\cref{ch:stellarator}) is a modification to a standard technique for computing the Biot-Savart line integral \cite{Hanson_Hirshman}. This modification increases the order of convergence from second to fourth-order accuracy and allows for the Biot-Savart law from a current-carrying filament to be computed much more efficiently. This work was in collaboration with Coaxiang Zhu, Lee Gunderson, and Stuart Hudson.

The second minor contribution (\cref{ch:autodiff}) is an alternative method of computing stochastic gradients with automatic differentiation, which we called randomized automatic differentiation (RAD). In problems with sufficiently simple computational structure (such as simple neural networks), RAD can reduce memory requirements when computing stochastic gradients at the cost of increased variance.
For most problems, however, I believe that RAD has no practical utility and is simply a theoretical framework for developing a deeper understanding of stochastic gradients.
I was a secondary author on this research, which was led by Deniz Oktay and in collaboration with Joshua Aduol and Alex Beatson under the supervision of Professor Ryan Adams in the Computer Science Department.

\chapter{Automatic differentiation (AD) \label{ch:autodiff}}
\noindent\rule{\textwidth}{1pt}

\textit{Part of the contents of this chapter have been published in the following papers: (i) McGreivy, N., Hudson, S. R., \& Zhu, C. ``Optimized finite-build stellarator coils using automatic differentiation.'' Nuclear Fusion 61.2 (2021): 026020 \cite{mcgreivy2020optimized}, (ii) Oktay, D., McGreivy, N., Aduol, J., Beatson, A., \& Adams, R. P. ``Randomized Automatic Differentiation.'' International Conference on Learning Representations (ICLR). 2021 \cite{oktay2021randomized}.}

\noindent\rule{\textwidth}{1pt}

Automatic differentiation, also known as algorithmic differentiation or computational differentiation, is a family of techniques for computing the exact numerical derivatives of a differentiable function specified by a computer program. 
AD is a broadly applicable numerical technique, but is particularly useful for gradient-based optimization in high-dimensional spaces. This is because with reverse mode AD, the gradient of a scalar function of $n$ inputs can be computed at a small multiple of the cost of computing the original function, independent of $n$. 
AD has been extensively studied, and a number of textbooks and review papers exist on the subject \cite{baydin2018automatic,griewank2008evaluating,adimplementation,uwebook,advancesinad}.
It has been used successfully in a number of areas, including machine learning \cite{baydin2018automatic,NIPS2015_5954}, engineering design optimization \cite{racecar}, beam physics \cite{beam,Berz1998VIo}, optimal control \cite{optimal_control}, atmospheric science \cite{mitgcm}, biomagnetic inverse problems \cite{biomagnetic}, a plasma edge code \cite{AD_in_plasma_edge}, and computational finance \cite{greeks}.

While AD can be used to compute numerical derivatives to any order, in many applications only first-order derivatives are computed. For a function $\bm{y} = f(\bm{x})$ from $\bm{x} \in \mathbb{R}^n$ to $\bm{y} \in \mathbb{R}^m$, first-order AD can be used to compute the Jacobian $\bm{J} = \frac{\partial \bm{y}}{\partial \bm{x}}$ at a particular value of $\bm{x}$. However, in practice AD tools usually compute the product of the Jacobian with a vector using either forward mode AD or reverse mode AD. 
In forward mode, AD tools compute the product $\dot{\bm{y}} = \bm{J} \dot{\bm{x}} \in \mathbb{R}^m$ of the Jacobian $\bm J \in \mathbb{R}^{m\times n}$ with a vector $\dot{\bm{x}} \in \mathbb{R}^n$. 
In reverse mode, AD tools compute the product $\overline{\bm{x}} = \overline{\bm{y}}^T\bm{J} \in \mathbb{R}^n$ of a vector $\overline{\bm{y}}\in \mathbb{R}^m$ with the Jacobian $\bm J \in \mathbb{R}^{m\times n}$. 
These are called the Jacobian-vector product (JVP) and vector-Jacobian product (VJP), respectively.

AD works on the principle that arbitrarily complex mathematical functions can be expressed as a sequence of primitive operations; these primitive operations form the building blocks of functions computed by AD tools.
A primitive operation is any function whose derivative of the output of that operation with respect to the input of that operation is known to the AD tool. The derivative of the output with respect to the input of a primitive operation is called the elementary partial derivative or elementary Jacobian matrix; these derivatives are usually derived analytically and must be pre-programmed for each primitive operation in the AD library. The primitive operations can be standard mathematical functions such as \code{divide}, \code{sine}, \code{fft}, and \code{ode\_int}, or they can be user-defined custom operations. An AD tool composes primitive operations together to build a function $f$, then computes the Jacobian of that function by multiplying the elementary Jacobian matrices together as specified by the chain rule. The elementary partial derivatives can be multiplied in any order, and in general choosing the most computationally efficient way to multiply these matrices is an NP-complete problem known as the Jacobian accumulation problem \cite{naumann2008optimal}.

Two ways of multiplying the elementary partial derivatives are forward mode AD and reverse mode AD. Forward mode AD computes the partial derivatives at the same time as the function is being computed, in effect multiplying elementary Jacobian matrices forwards from the beginning of the function to the end. In practice, the result of the computation is the Jacobian-vector product. Reverse mode AD computes the function forwards and then computes the partial derivatives backwards, in effect multiplying elementary Jacobian matrices from the end of the function to the beginning. In practice, the result of the computation is the vector-Jacobian product. Forward mode is sometimes called ``tangent linear mode'' while reverse mode is sometimes called ``adjoint mode'' or ``cotangent mode''. 

For a function $f : \mathbb{R}^n \rightarrow \mathbb{R}^m$ which takes time $\mathcal{O}(1)$ to compute, computing the full Jacobian with forward mode takes time $\mathcal{O}(n)$ and almost no additional memory cost. This can be done by computing $n$ Jacobian-vector products where the (one-hot) vectors are the columns of a $n \times n$ identity matrix. Computing the full Jacobian with reverse mode takes time $\mathcal{O}(m)$, and memory cost proportional to the number of intermediate variables in the computation. This can be done by computing $m$ vector-Jacobian products where the (one-hot) vectors $\bm{\overline{y}}$ are the rows of an $m\times m$ identity matrix. For large computations, the memory cost of reverse mode AD can be extremely large; checkpointing strategies \cite{checkpointing,checkpointingwang} can be used to reduce the memory cost of reverse mode AD at the cost of increased runtime. An important feature of reverse mode AD is that for a scalar function $f: \mathbb{R}^n \rightarrow \mathbb{R}$, the runtime cost of computing the $n$-dimensional gradient is a small multiple of the cost of computing the function itself, independent of $n$. While computing the full Hessian matrix of a scalar function is $\mathcal{O}(n)$ the cost of the original function, Hessian-vector products can be computed in time $\mathcal{O}(1)$ \cite{jax2018github}. These are useful for second-order Hessian-free optimization methods \cite{NumericalOptimizationBook}.
The full Hessian matrix is useful for sensitivity analysis and understanding coil tolerances, an area of research in stellarator coil design \cite{zhu_hessian, Zhu_2019}.

AD is one method of computing derivatives. Other methods include numerical differentiation (finite-difference) and hand-programmed analytic derivatives. Analytic differentiation often results in the fastest derivative computations and, like AD, gives exact derivative values. However, analytic derivatives are both error-prone and time-consuming to derive and program, unlike AD. Numerical differentiation is simple to program, but results in inexact derivative values due to floating-point precision errors and has a computational cost proportional to the number of inputs $n$. 
When computing the gradient of a scalar function, reverse mode AD is comparable in computational efficiency to analytic differentiation and unlike numerical differentiation has a computational cost independent of the number of inputs. Automatic differentiation has significant advantages over other methods of computing derivatives, especially for optimization. The main downsides of using AD are that programs need to be written using an AD software tool, and that for very large computations the memory cost of reverse mode AD can be unwieldy.
For a review of AD software tools and their implementation, see \cite{adimplementation} and \cite{reviewofADimplementation}, and the AD community website\footnote{{www.autodiff.org}}.

In the rest of this chapter, I discuss the theory and principles of automatic differentiation in more detail.

\section{Differentiation \label{sec:ch2-differentiation}}

Suppose I have a scalar function $ f(x) : \mathbb{R} \rightarrow \mathbb{R}$. The definition of the derivative $\frac{df}{dx}$ is
\begin{equation}\label{eq:ch2-definition-derivative}
    \frac{df}{dx} = \lim_{\epsilon \to 0} \frac{f(x+\epsilon) - f(x)}{\epsilon}.
\end{equation}
Higher-order derivatives are defined recursively:
\begin{equation}
    \frac{d^nf}{dx^n} = \lim_{\epsilon \to 0} \frac{\frac{d^{n-1}f}{dx^{n-1}}\big|_{x + \epsilon} - \frac{d^{n-1}f}{dx^{n-1}}\big|_{x}}{\epsilon}.
\end{equation}
The generalization of the scalar derivative $\frac{df}{dx}$ to multivariate functions $f(\bm x): \mathbb{R}^n \to \mathbb{R}^m$ is called the Jacobian. The Jacobian $\bm J$ is an $m \times n$ matrix with the $ij$th entry equal to
\begin{equation}
    J_{ij} = \frac{\partial f_i}{\partial x_j} = \frac{f_i(x_1, \dots, x_j + \epsilon, \dots, x_n) - f(x_1, \dots, x_j, \dots, x_n)}{\epsilon}.
\end{equation}
The full Jacobian matrix can be written as
\begin{equation}\label{eq:ch2-definition-jacobian}
    \bm J = 
\begin{bmatrix}
\frac{\partial f_1}{\partial x_1} & \dots & \frac{\partial f_1}{\partial x_n} \\
\vdots & \ddots & \vdots \\
\frac{\partial f_m}{\partial x_1} & \dots & \frac{\partial f_m}{\partial x_n}
\end{bmatrix}.
\end{equation}
The matrix $\bm J \in \mathbb{R}^{m \times n}$ has $n$ columns, corresponding to the $n$ entries of the input $\bm x$, and $m$ rows, corresponding to the $m$ entries of the output $f$.
Higher-order generalizations of the Jacobian exist; the second derivative matrix, for example, is $\partial^2 f \in \mathbb{R}^{m \times n \times n}$. The $ijk$th entry of $\partial^2f$ is
\begin{equation}
    \partial^2 f_{ijk} = \frac{\partial^2 f_i}{\partial x_j \partial x_k}.
\end{equation}

Of particular importance for applications of automatic differentiation are scalar-valued functions $f(\bm x) : \mathbb{R}^n \to \mathbb{R}$. The derivative of a scalar-valued function is called the gradient, or $\grad f \in \mathbb{R}^n$. The gradient is a column vector, defined as
\begin{equation}\label{eq:ch2-definition-gradient}
    \grad f = \begin{bmatrix}
        \frac{\partial f}{\partial x_1} \\
        \vdots \\
        \frac{\partial f}{\partial x_n}
    \end{bmatrix}.
\end{equation}
The second derivative of a scalar-valued function is known as the Hessian $\partial^2 f \in \mathbb{R}^{n\times n}$. 

The gradient can be thought of as the transpose of the Jacobian (to see this, compare \cref{eq:ch2-definition-gradient} with \cref{eq:ch2-definition-jacobian}).
However, for the purposes of developing a theory of automatic differentiation, the Jacobian of a function $f : \mathbb{R}^n \to \mathbb{R}$ is not a row vector of shape $\mathbb{R}^n$ but rather a matrix of shape $\mathbb{R}^{1\times n}$. 
Obtaining a (transpose) gradient from a Jacobian thus requires calculating the vector-Jacobian product with the vector $\bar{\bm v} = [1]$,
\begin{equation}
    (\grad f)^T = \bar{\bm v}^T \bm J.
\end{equation}

Suppose $f(\bm x) \in  \mathbb{R}^n \to \mathbb{R}^m$.
If we can compute Jacobian-vector products (JVPs) or vector-Jacobian products (VJPs), we can compute $\bm J \in \mathbb{R}^{m \times n}$.
Doing so requires either $n$ JVPs or $m$ VJPs.
If $\dot{\bm v}_j  \in \mathbb{R}^n$ is a one-hot vector with the $i$th entry $\dot{v}_{ji} = \delta_{ij}$ for $j = 1, \dots, n$, then the JVP $\bm J \dot{\bm v}_j$ returns the $j$th column of $\bm J$:
\begin{equation}
    \bm J \dot{\bm v}_j = 
    \begin{bmatrix}
        \frac{\partial f_1}{\partial x_1} & \dots & \frac{\partial f_1}{\partial x_j} & \dots & \frac{\partial f_1}{\partial x_n} \\
        \vdots & & \ddots & & \vdots \\
        \frac{\partial f_m}{\partial x_1} & \dots & \frac{\partial f_m}{\partial x_j} & \dots & \frac{\partial f_m}{\partial x_n}
\end{bmatrix}
    \begin{bmatrix}
        0 \\ \vdots \\ 1 \\ \vdots\\ 0
    \end{bmatrix} = \begin{bmatrix}
        \frac{\partial f_1}{\partial x_j}\\
        \vdots \\
        \frac{\partial f_m}{\partial x_j}
    \end{bmatrix} \in \mathbb{R}^m.
\end{equation}
Computing all $n$ columns requires computing $n$ JVPs, one for each value of $j$.
Likewise, if $\bar{\bm v}_j \in \mathbb{R}^m$ is a one-hot vector with the $i$th entry $\bar{v}_{ji} = \delta_{ij}$ for $j = 1, \dots, m$, then the VJP $\bar{\bm v}_j^T\bm J$ returns the $j$th row of $\bm J$:
\begin{equation}
    \bar{\bm v}_j^T\bm J =  \begin{bmatrix}
        0 & \dots & 1 & \dots & 0
    \end{bmatrix}
    \begin{bmatrix}
\frac{\partial f_1}{\partial x_1} & \dots & \frac{\partial f_1}{\partial x_n} \\
\vdots & & \vdots \\
\frac{\partial f_j}{\partial x_1} & \ddots & \frac{\partial f_j}{\partial x_n}\\
\vdots & & \vdots \\
\frac{\partial f_m}{\partial x_1} & \dots & \frac{\partial f_m}{\partial x_n}
\end{bmatrix}
 = \begin{bmatrix}
        \frac{\partial f_j}{\partial x_1} &
        \dots &
        \frac{\partial f_j}{\partial x_n}
    \end{bmatrix} \in \mathbb{R}^n.
\end{equation}
Computing all $m$ rows requires computing $m$ VJPs, one for each value of $j$.

\subsection{Primitive operations}

The derivative of $\sin{x}$ with respect to $x$ is $\cos{x}$. This can be derived using \cref{eq:ch2-definition-derivative} and the angle addition formulas, but most of us have this fact memorized. Most of us do not, however, have the derivative of $\sin{(x^2e^x)}$ memorized. Determining that its derivative is $(2xe^x + x^2 e^x) \cos{(x^2 e^2)}$ would require some analytic manipulations using the chain rule (see \cref{sec:ch2-chain-rule}) and knowing that $\frac{d}{dx}x^2 = 2x$, $\frac{d}{dx} e^x = e^x$, and $\frac{d}{dx}(ab) = \frac{da}{dx}b + a\frac{db}{dx}$. 

In the terminology of automatic differentiation, primitive operations are those functions whose derivative (or Jacobian) is already pre-programmed by the AD tool.\footnote{Actually, as we'll see in \cref{sec:ch2-implementation}, a primitive operation doesn't need to be able to compute its Jacobian, only a JVP or VJP.}
In the example above, $f(x) = \sin{x}$, $f(x) = x^2$, $f(x) = e^x$, and $f(x_1, x_2) = x_1 * x_2$ are primitive operations.
The derivatives (or Jacobians) of functions made up of arbitrarily complex combinations of primitive operations can be computed using the chain rule, so long as the derivatives (or Jacobians) of each primitive are known.
The derivatives of each primitive operation are called the \textit{elementary partial derivatives} or \textit{elementary Jacobian matrix}.

\subsection{The chain rule \label{sec:ch2-chain-rule}}

Suppose I have a scalar function $f(x)$ which is made up of a composition of two functions $h(x)$ and $g(u)$ such that
\begin{equation}
    f(x) = g \circ h(x) = g(h(x)).
\end{equation}
The chain rule states that if $y = f(x)$, the derivative of $y$ is
\begin{equation}
    \frac{dy}{dx} = g' \circ h(x) \cdot h'(x) = \frac{dy}{du}\Big|_{u=h(x)} \frac{du}{dx}.
\end{equation}
For example, if $f(x) = \sin{(x^2)}$, $u = h(x) = x^2$, and $y = g(u) = \sin{u}$, then according to the chain rule $\frac{dy}{dx} = \frac{dy}{du} \frac{du}{dx} = \cos{(x^2)} 2x$. 

The multivariate chain rule states that the Jacobian of a composition of functions is equal to the product of their Jacobians. Suppose that $f(\bm x): \mathbb{R}^n \to \mathbb{R}^m$ is equal to a composition of $h(\bm x) : \mathbb{R}^n \to \mathbb{R}^d$ and $g(\bm u) : \mathbb{R}^d \to \mathbb{R}^m$, 
\begin{equation}
    f(\bm x) = g \circ h(\bm x) = g(h(\bm x)).
\end{equation} 
If $\bm y = f(\bm x)$, the Jacobian of $\bm y$ equals
\begin{equation}\label{eq:ch2-multivariate-chain-rule}
    \bm J = \frac{\partial \bm y}{\partial \bm u}\Big|_{\bm u = h(\bm x)} \frac{\partial \bm u}{\partial \bm x} = \bm J_g \bm J_h.
\end{equation}
The multivariate chain rule, \cref{eq:ch2-multivariate-chain-rule}, can be rewritten as
\begin{equation}\label{eq:ch2-multivariate-chain-rule-sum}
    J_{ij} = \sum_{k=1}^d \frac{\partial y_i}{\partial u_k} \frac{\partial u_k}{\partial x_j}.
\end{equation}
For a composite function $f(\bm x) : \mathbb{R}^n \to \mathbb{R}^m$ made up of $p$ compositions of functions
\begin{equation}\label{eq:ch2-composite-function}
    f(\bm x) = g_p \circ \dots \circ g_2 \circ g_1(\bm x)
\end{equation}
the chain rule generalizes to
\begin{equation}\label{eq:ch2-multivariate-chain-rule-multiple}
\bm{J} = \bm{J}_{g_p} {\dots} \bm{J}_{g_2} \bm{J}_{g_1}.
\end{equation}
Letting the intermediate variable $\bm u^i = g_i(\bm u^{i-1})$ for $i = 1, \dots, p$ with $\bm u^0 = \bm x$, $\bm u^p = \bm y$, $g_i(\bm u^{i-1}): \mathbb{R}^{d_{i-1}} \to \mathbb{R}^{d_i}$ for $i = 1, \dots, p$, $d_0 = n$, and $d_p = m$, \cref{eq:ch2-multivariate-chain-rule-multiple} can be rewritten as 
\begin{equation}\label{eq:ch2-multivariate-chain-rule-multiple-sum}
    J_{ij} = \sum_{m=1}^{d_{p-1}}\dots \sum_{l=1}^{ d_2}\sum_{k=1}^{d_1} \frac{\partial y_i}{\partial u^{p-1}_m}\dots \frac{\partial u^2_l}{\partial u^1_k}\frac{\partial u^1_k}{\partial x_j}.
\end{equation}
To better understand the multivariate chain rule, consider a simple example. Suppose we want to compute the Jacobian of $f: \mathbb{R}^n \to \mathbb{R}$ where $f(\bm x) = a(\bm x) + b(\bm x)$ and $a$ and $b$ are arbitrary functions.
\Cref{eq:ch2-multivariate-chain-rule} suggests one way of computing the Jacobian $\bm J \in \mathbb{R}^{1\times n}$: writing $f$ as a composition of two functions $h(\bm x)$ and $g(\bm u)$.
To do so, let $\bm u = h(\bm x) : \mathbb{R}^n \to \mathbb{R}^2$, $\bm u = \begin{bmatrix}
    u_1 & u_2
\end{bmatrix}^T$ and $h(\bm x) = \begin{bmatrix}
    a(\bm x) & b(\bm x)
\end{bmatrix}^T$. 
Then, let $g(\bm u) = u_1 + u_2 : \mathbb{R}^2 \to \mathbb{R}$.
$g$ is simply the \code{add} function whose Jacobian $\bm J_g \in \mathbb{R}^{1 \times 2}$ is
\begin{equation}
    \bm J_g = \begin{bmatrix}
        1 & 1
    \end{bmatrix}.
\end{equation}
The Jacobian of $h$, $\bm J_h \in \mathbb{R}^{2 \times n}$ is
\begin{equation}
    \begin{bmatrix}
        \frac{\partial a}{\partial x_1} & \dots & \frac{\partial a}{\partial x_n} \\
        \frac{\partial b}{\partial x_1} & \dots & \frac{\partial b}{\partial x_n}
    \end{bmatrix}.
\end{equation}
Using \cref{eq:ch2-multivariate-chain-rule},
\begin{equation}
    \bm J = \bm J_g \bm J_h = \begin{bmatrix}
        1 & 1
    \end{bmatrix} \begin{bmatrix}
        \frac{\partial a}{\partial x_1} & \dots & \frac{\partial a}{\partial x_n} \\
        \frac{\partial b}{\partial x_1} & \dots & \frac{\partial b}{\partial x_n}
    \end{bmatrix} = \begin{bmatrix}
        \frac{\partial a}{\partial x_1} + \frac{\partial b}{\partial x_1} & \dots & \frac{\partial a}{\partial x_n} + \frac{\partial b}{\partial x_n}
    \end{bmatrix} = \bm J_a + \bm J_b.
\end{equation}
As expected, the Jacobian of a sum is the sum of the Jacobians.

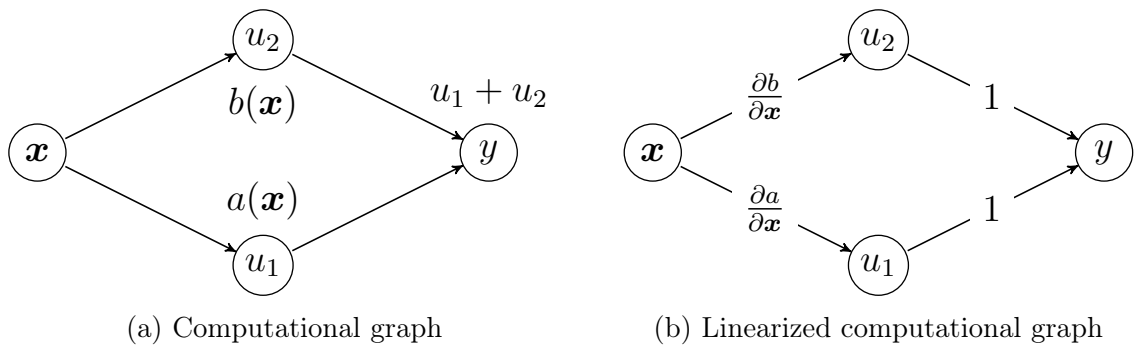
\begin{figure}
    \centering
    \begin{subfigure}[]{0.49\textwidth}
        \centering

\begin{tikzpicture}[scale=1.2,transform shape]
		
  			\tikzstyle{every node}=[node distance = 4cm,%
            		                bend angle    = 45]
			\Vertex[x=0, y=1.25, L=\texttt{$\bm x$}]{x1}

			\Vertex[x=2.5, y=0, L=\texttt{$u_1$}]{u1} 
                \extralabel[2mm]{90}{\texttt{$a(\bm x)$}}{u1}
			\Vertex[x=2.5, y=2.5, L=\texttt{$u_2$}]{u2}
                \extralabel[2mm]{270}{\texttt{$b(\bm x)$}}{u2}

			\Vertex[x=5, y=1.25, L=\texttt{$y$}]{y}
                \extralabel[2mm]{90}{\texttt{$u_1 + u_2$}}{y}
			\tikzstyle{EdgeStyle}=[post]
			\Edge(x1)(u1)
			\Edge(x1)(u2)
			\Edge(u1)(y)
			\Edge(u2)(y)
                
		\end{tikzpicture}

        \caption{Computational graph}
        \label{fig:ch2-compgraph-add-ab}
    \end{subfigure}
    \begin{subfigure}[]{0.45\textwidth}
        \centering

\begin{tikzpicture}[scale=1.2,transform shape]
		
  			\tikzstyle{every node}=[node distance = 4cm,%
            		                bend angle    = 45]
			\Vertex[x=0, y=1.25, L=\texttt{$\bm x$}]{x1}

			\Vertex[x=2.5, y=0, L=\texttt{$u_1$}]{u1} 
			\Vertex[x=2.5, y=2.5, L=\texttt{$u_2$}]{u2}
			\Vertex[x=5, y=1.25, L=\texttt{$y$}]{y}
			\tikzstyle{EdgeStyle}=[post]
			\Edge[label=\texttt{$\frac{\partial a}{\partial \bm x}$}](x1)(u1)
			\Edge[label=\texttt{$\frac{\partial b}{\partial \bm x}$}](x1)(u2)
			\Edge[label=\texttt{$1$}](u1)(y)
			\Edge[label=\texttt{$1$}](u2)(y)
                
		\end{tikzpicture}

        \caption{Linearized computational graph}
        \label{fig:ch2-lin-compgraph-add-ab}
    \end{subfigure}

    \caption{(a) The computational graph for $f(\bm x) = a(\bm x) + b(\bm x)$. Here $u_1 = a(\bm x)$ and $u_2 = b(\bm x)$ are intermediate variables corresponding to the output of the primitive operations $a$ and $b$, while $y = u_1 + u_2$ is an output variable corresponding to the output of the primitive operation \code{add}. (b) The linearized computational graph for $f$. Here the edge weight on an edge from node $v_i$ to $v_j$ represents the value of the partial derivative $\frac{\partial v_j}{\partial v_i}$.}
    \label{fig:ch2-compgraph-example}
\end{figure}

\Cref{eq:ch2-multivariate-chain-rule-sum,eq:ch2-multivariate-chain-rule-multiple-sum} suggest another way of computing the Jacobian using the chain rule: writing $f(\bm x)$ as a directed acyclic graph (DAG) with the input variables $\bm x$, intermediate variables $\bm u$, and output variable $\bm y$ as nodes in the graph, and each primitive operation represented as one or more edges in the graph. An edge $e_{ij}$ from node $v_i$ to node $v_j$ implies that the variable $v_i$ is an input to some primitive operation for which the variable $v_j$ is an output. This DAG is typically known as the `computational graph'. The computational graph for $f(\bm x) = a(\bm x) + b(\bm x)$ is sketched in \cref{fig:ch2-compgraph-add-ab}.
The so-called `linearized computational graph' is a DAG with the same set of nodes $V$ and edges $E$ as the computational graph, except edges from node $v_i$ to $v_j$ have edge weights equal to the value of the partial derivative $\frac{\partial v_j}{\partial v_i}$.
The linearized computational graph is sketched in \cref{fig:ch2-lin-compgraph-add-ab}.
Using \cref{eq:ch2-multivariate-chain-rule-sum}, we see that if $y = f(\bm x)$ the Jacobian $\bm J$ is equal to the sum of the product of the edge weights of the linearized computational graph along each of the two paths from $\bm x$ to $y$:
\begin{equation}
    \bm J = \sum_{i=1}^2 \frac{\partial y}{\partial u_i}\frac{\partial u_i}{\partial \bm x} = 1 \cdot \frac{\partial a}{\partial \bm x} + 1 \cdot \frac{\partial b}{\partial \bm x} = \bm J_a + \bm J_b.
\end{equation}

Using the simple example of $f(\bm x) = a(\bm x) + b(\bm x)$, we've interpreted the multivariate chain rule either as function composition or as primitive operations creating graphs. As we'll see in \cref{sec:ch2-ad-as-jacobian,sec:ch2-ad-as-dp}, these two interpretations are different and more general ways of understanding AD.
The former interpretation is simpler and mathematically more elegant, but the latter is closer to how AD is actually implemented in practice.

\section{AD as Jacobians of composite functions \label{sec:ch2-ad-as-jacobian}}

In this section, we'll discuss an informal theory of AD based purely on composite functions. This theory isn't sufficiently general to include all functions -- composite functions of the form \cref{eq:ch2-composite-function} only allow for sequential operations, and many functions are not sequential as they involve branching (fan-in and fan-out) and/or running multiple sequential operations in parallel -- but it is sufficiently simple to illustrate most of the key ideas of AD.
For a formal theory of AD based on the chain rule for sequential composition and two other rules, see \cite{Elliott-2018-ad-extended}.

Suppose I have a function $f(\bm x)  : \mathbb{R}^n \to \mathbb{R}^m$, made up of a composition of functions as in \cref{eq:ch2-composite-function}. For concreteness, I'll assume that $f$ is made up of four compositions
\begin{equation}\label{eq:ch2-ja-example}
    f(\bm x) = D \circ C \circ B \circ A(\bm x)
\end{equation}
but I could choose any number of compositions to illustrate the essential ideas.
Using the chain rule (\cref{eq:ch2-multivariate-chain-rule-multiple}),
\begin{equation}\label{eq:ch2-mv-cr-ex}
    \bm J = \bm J_{D} \bm J_{C} \bm J_{B} \bm J_{A}.
\end{equation}
Suppose that $\bm J_A = \mathbb{R}^{d_1 \times n}$, $\bm J_B = \mathbb{R}^{d_2 \times d_1}$, $\bm J_C = \mathbb{R}^{d_3 \times d_2}$, and $\bm J_D = \mathbb{R}^{m \times d_3}$.

\subsection{Jacobian accumulation \label{sec:ch2-ad-as-comp-jac-acc}}

According to \cref{eq:ch2-mv-cr-ex}, computing the Jacobian $\bm J$ requires three matrix multiplications. These matrices can be multiplied in any order. The order in which they are multiplied determines the efficiency of the computation of $\bm J$. 
There are six options for the order of matrix multiplication:
\begin{enumerate}
    \item Forward accumulation: $\bm J = \bm J_{D} (\bm J_{C} (\bm J_{B} \bm J_{A}))$. $\mathcal{O}(n(d_2 d_1 + d_3 d_2 + m d_3))$ multiply-add operations.
    \item Reverse accumulation: $\bm J = ((\bm J_{D} \bm J_{C}) \bm J_{B}) \bm J_{A}$. $\mathcal{O}(m(d_3 d_2 + d_2d_1 + d_1 n))$ multiply-add operations.
    \item Middle-left accumulation: $\bm J = (\bm J_{D} (\bm J_{C} \bm J_{B})) \bm J_{A}$. $\mathcal{O}(d_1(d_3 d_2 + m d_3 + m n))$ multiply-add operations.
    \item Middle-right accumulation: $\bm J = \bm J_{D} ((\bm J_{C} \bm J_{B}) \bm J_{A})$. $\mathcal{O}(d_3(d_2d_1 + d_1 n + m n))$ multiply-add operations.
    \item Left-right accumulation: $\bm J = (\bm J_{D} \bm J_{C}) (\bm J_{B} \bm J_{A})$. $\mathcal{O}(d_2 (d_3 m + d_1 n + nm))$ multiply-add operations. 
    \item Right-left accumulation: same as left-right accumulation. 
\end{enumerate}
The process of computing a Jacobian $\bm J$ by multiplying the elementary partial derivatives according the chain rule is called either \textit{derivative accumulation} or \textit{Jacobian accumulation} \cite{griewank2008evaluating}.
The end result of Jacobian accumulation is predetermined, but the computational efficiency is not.
In the example above, the method of computing the Jacobian with the optimal efficiency depends on the values of $n$, $d_1$, $d_2$, $d_3$, and $m$.
\Cref{eq:ch2-ja-example} has only four compositions and so the method of accumulating the Jacobian with the optimal efficiency can be determined by brute force; for functions of $n$ compositions, the best known algorithm to compute the optimal order of $n$ matrix chain multiplications has complexity $\mathcal{O}(n \log n)$ \cite{hu1982computation}.

In principal, differentiable programming libraries could implement AD by developing tools to determine the most efficient method for Jacobian accumulation for a given $f$. However, doing so for non-composite functions is an NP-complete problem (as we'll discuss in \cref{sec:ch2-ad-as-dp-jac-acc}).
In practice, AD methods tend to implement only the first two orders in the example listed above: forward accumulation and reverse accumulation.
Actually, most of the time what they do is even simpler: they accumulate products of vectors with the Jacobian, computing either JVPs (forward mode) or VJPs (reverse mode).

\subsection{Forward mode}

Forward mode AD computes the JVP $\dot{\bm{y}} = \bm J \dot{\bm x}$ of a function $f(\bm x): \mathbb{R}^n \to \mathbb{R}^m$. $\dot{\bm{x}} \in \mathbb{R}^n$ and $\dot{\bm y} \in \mathbb{R}^m$ are called tangent vectors.

For the function $f$ given by \cref{eq:ch2-ja-example}, the JVP with $\dot{\bm{x}}$ is $\bm J \dot{\bm{x}} = \bm J_{D} \bm J_{C} \bm J_{B} \bm J_{A} \dot{\bm{x}}$.
The most efficient way to compute this JVP is using forward accumulation:
\begin{equation}
    \bm J \dot{\bm{x}} = \bm J_{D} (\bm J_{C} (\bm J_{B} (\bm J_{A} \dot{\bm{x}}))).
\end{equation}
Forward accumulation only requires matrix-vector products; it is the only order of accumulating a JVP which doesn't require matrix-matrix products. 
Thus, forward accumulation for computing a JVP will be more efficient than any other order of accumulation.\footnote{This statement assumes the Jacobian matrices are dense. It also assumes that $d_1$, $d_2$, and $d_3$ are all greater than 1.}
In this case, forward accumulation requires $\mathcal{O}(nd_1 + d_1 d_2 + d_2 d_3 + d_3 m)$ multiply-add operations, which is $\mathcal{O}(1)$ times the cost of computing the original function $f$. 

As discussed in \cref{sec:ch2-differentiation}, computing the full Jacobian $\bm J$ using JVPs requires $n$ JVPs with $n$ one-hot vectors $\bm v_j$, each returning one column of $\bm J$. For functions $f(x) : \mathbb{R} \to \mathbb{R}^m$, one JVP is sufficient to compute the entire Jacobian matrix.

\subsection{Reverse mode}

Reverse mode AD computes the VJP $\bar{\bm{x}}^T = \bar{\bm y}^T\bm J$ of a function $f(\bm x): \mathbb{R}^n \to \mathbb{R}^m$. $\bar{\bm{y}} \in \mathbb{R}^m$ and $\bar{\bm x} \in \mathbb{R}^n$ are called adjoint vectors or cotangent vectors.

For the function $f$ given by \cref{eq:ch2-ja-example}, the VJP with $\bar{\bm{y}}$ is $\bar{\bm y}^T\bm J = \bar{\bm y}^T\bm J_{D} \bm J_{C} \bm J_{B} \bm J_{A}$.
The most efficient way to compute this VJP is using reverse accumulation:
\begin{equation}
    \bar{\bm y}^T\bm J = (((\bar{\bm y}^T\bm J_{D}) \bm J_{C}) \bm J_{B}) \bm J_{A}.
\end{equation}
Reverse accumulation only requires matrix-vector products; it is the only order of accumulating a VJP which doesn't require matrix-matrix products. 
Thus, reverse accumulation for computing a VJP will be more efficient than any other order of accumulation.
In this case, reverse accumulation requires $\mathcal{O}(md_3 + d_3 d_2 + d_2 d_1 + d_1 n)$ multiply-add operations, which is $\mathcal{O}(1)$ times the cost of computing the original function $f$. 

As discussed in \cref{sec:ch2-differentiation}, computing the full Jacobian $\bm J$ using VJPs requires $m$ VJPs with $m$ one-hot vectors $\bm v_j$, each returning one row of $\bm J$. For scalar-valued functions $f(x) : \mathbb{R}^n \to \mathbb{R}$, one VJP is sufficient to compute the entire Jacobian matrix.

\section{AD as dynamic programming on graphs \label{sec:ch2-ad-as-dp}}

The most common way of understanding AD is as a family of algorithms for computing derivatives of functions built of primitive operations and represented by graphs. We briefly introduced the graph interpretation of AD in \cref{sec:ch2-chain-rule}, and will expand on that discussion in this section.

\subsection{Computational graphs \label{sec:ch2-comp-graph}}

Suppose I have a mathematical function $f : \mathbb{R}^n \to \mathbb{R}^m$ defined through a computer program as a sequence of primitive operations. The function has $n$ input variables $x_i$, $i = 1, \dots, n$, $p$ intermediate variables $v_i$, $i = 1, \dots, p$ and $m$ output variables $y_i$, $i = 1, \dots, m$.
To simplify notation, let $v_{i-n} = x_i$ for $i = 1, \dots, n$ and $v_{p+i} = y_i$ for $i = 1, \dots, m$.
The intermediate and output variables are defined as
\begin{equation}
    v_j = \varphi_j(v_i)_{i \prec j}
\end{equation}
for $j = 1, \dots, p+m$.
The notation $i \prec j$ means that $v_j$ depends on $v_i$ directly through $\varphi_j$. $\varphi_j$ is a primitive operation which with inputs $v_i$ for all $i \prec j$ and scalar output $v_j$.

A computational graph for $f$ can be constructed by adding a node for each variable $v_j$, then adding edges $e_{ij}$ from node $v_i$ to $v_j$ for all $i \prec j$.
The result is a DAG, with paths between the input and output variables.
A linearized computational graph (LCG) for $f$ has the same structure as the computational graph, except with edge weights corresponding to the elementary partial derivative $\frac{\partial v_j}{\partial v_j}$.

\begin{figure}[t]
		\begin{subfigure}{0.49\textwidth}
	\begin{lstlisting}[language=Python,columns=flexible]
from math import sin, exp
 
def f(x1, x2):
  a = exp(x1)
  b = sin(x2)
  c = b * x2
  d = a * c
  return a * d
	\end{lstlisting}
		\caption{Pseudocode}
		\label{fig:ch2-rad-ex-a}
		\end{subfigure}
		\hfill%
		\begin{subfigure}[b]{0.49\textwidth}
		\centering%
		\begin{tikzpicture}[scale=0.8,transform shape]
		
  			\tikzstyle{every node}=[node distance = 4cm,%
            		                bend angle    = 45]
			\Vertex[x=\nodexonex, y=\nodexoney, L=\texttt{x1}]{x1}
			\Vertex[x=\nodextwox, y=\nodextwoy, L=\texttt{x2}]{x2}

			\Vertex[x=\nodeax, y=\nodeay, L=\texttt{a}]{a} 
			\extralabel[2mm]{0}{\texttt{exp(x1)}}{a}

			\Vertex[x=\nodebx, y=\nodeby, L=\texttt{b}]{b}
			\extralabel[2mm]{0}{\texttt{sin(x2)}}{b}

			\Vertex[x=\nodecx, y=\nodecy, L=\texttt{c}]{c}
			\extralabel[2mm]{0}{\texttt{b * x2}}{c}
			
			\Vertex[x=\nodedx, y=\nodedy, L=\texttt{d}]{d}
			\extralabel[2mm]{0}{\texttt{a * c}}{d}

			\Vertex[x=\nodefx, y=\nodefy, L=\texttt{f}]{f}
			\extralabel[2mm]{0}{\texttt{a * d}}{f}

			\tikzstyle{EdgeStyle}=[post]
			\Edge(x1)(a)
			\Edge(x2)(b)
			\Edge(b)(c)
			\Edge(x2)(c)
			\Edge(a)(d)
			\Edge(c)(d)
			\Edge(a)(f)
			\Edge(d)(f)
		\end{tikzpicture}
		\caption{Computational graph}
		\label{fig:ch2-rad-ex-b}
		\end{subfigure}
		\hfill%
		\begin{subfigure}[b]{0.49\textwidth}	
		\centering%
		\begin{tikzpicture}[scale=0.8,transform shape]
  			\tikzstyle{every node}=[node distance = 4cm,%
            		                bend angle    = 45]
			\Vertex[x=\nodexonex, y=\nodexoney, L=\texttt{x1}]{x1}
			\Vertex[x=\nodextwox, y=\nodextwoy, L=\texttt{x2}]{x2}
			\Vertex[x=\nodeax, y=\nodeay, L=\texttt{a}]{a} 
			\Vertex[x=\nodebx, y=\nodeby, L=\texttt{b}]{b}
			\Vertex[x=\nodecx, y=\nodecy, L=\texttt{c}]{c}
			\Vertex[x=\nodedx, y=\nodedy, L=\texttt{d}]{d}
			\Vertex[x=\nodefx, y=\nodefy, L=\texttt{f}]{f}

			\tikzstyle{EdgeStyle}=[post]
			\Edge[label=\texttt{exp(x1)}](x1)(a)
			\Edge[label=\texttt{cos(x2)}](x2)(b)
			\Edge[label=\texttt{x2}](b)(c)
			\Edge[label=\texttt{b}](x2)(c)
			\Edge[label=\texttt{c}](a)(d)
			\Edge[label=\texttt{a}](c)(d)
			\Edge[label=\texttt{d}](a)(f)
			\Edge[label=\texttt{a}](d)(f)
		\end{tikzpicture}
		\caption{Linearized computational graph}
		\label{fig:ch2-rad-ex-c}
		\end{subfigure}
		\hfill%
		\begin{subfigure}[b]{0.49\textwidth}	
		\centering%
		\begin{tikzpicture}[scale=0.8,transform shape]
  			\tikzstyle{every node}=[node distance = 4cm,%
            		                bend angle    = 45]
			\Vertex[x=\nodexonex, y=\nodexoney, L=\texttt{x1}]{x1}
			\Vertex[x=\nodextwox, y=\nodextwoy, L=\texttt{x2}]{x2}
			\Vertex[x=\nodeax, y=\nodeay, L=\texttt{a}]{a} 
			\Vertex[x=\nodebx, y=\nodeby, L=\texttt{b}]{b}
			\Vertex[x=\nodecx, y=\nodecy, L=\texttt{c}]{c}
			\Vertex[x=\nodedx, y=\nodedy, L=\texttt{d}]{d}
			\Vertex[x=\nodefx, y=\nodefy, L=\texttt{f}]{f}
			
		\draw[line width=2,rounded corners, color=cborange] 
			([xshift=-5mm,yshift=0mm]\nodexonex,\nodexoney) -- 
			([xshift=-5mm,yshift=1mm]\nodeax,\nodeay) -- 
			([xshift=-5mm,yshift=0mm]\nodefx,\nodefy);
		\draw[line width=2,rounded corners, color=cbblue] 
			([xshift=5mm,yshift=0mm]\nodexonex,\nodexoney) -- 
			([xshift=5mm,yshift=-1mm]\nodeax,\nodeay) --
			([xshift=-7mm,yshift=-6mm]\nodedx,\nodedy) --
			([xshift=-7mm,yshift=0mm]\nodedx,\nodedy) --
			([xshift=0mm,yshift=-6.5mm]\nodefx,\nodefy);
		\draw[line width=2,rounded corners, color=cbred] 
			([xshift=-6mm,yshift=0mm]\nodextwox,\nodextwoy) -- 
			([xshift=-6mm,yshift=0mm]\nodebx,\nodeby) --
			([xshift=-5mm,yshift=0mm]\nodecx,\nodecy) --
			([xshift=-5mm,yshift=0.5mm]\nodedx,\nodedy) --
			([xshift=0mm,yshift=-4mm]\nodefx,\nodefy);
		\draw[line width=2,rounded corners, color=cbgreen] 
			([xshift=5mm]\nodextwox,\nodextwoy) -- 
			([xshift=5mm]\nodecx,\nodecy) --
			([xshift=5mm,yshift=2mm]\nodedx,\nodedy) --
			([xshift=5mm,yshift=0mm]\nodefx,\nodefy);
			
			\tikzstyle{EdgeStyle}=[post]
			\Edge[label=\texttt{exp(x1)}](x1)(a)
			\Edge[label=\texttt{cos(x2)}](x2)(b)
			\Edge[label=\texttt{x2}](b)(c)
			\Edge[label=\texttt{b}](x2)(c)
			\Edge[label=\texttt{c}](a)(d)
			\Edge[label=\texttt{a}](c)(d)
			\Edge[label=\texttt{d}](a)(f)
			\Edge[label=\texttt{a}](d)(f)
			
		\end{tikzpicture}
		\caption{Bauer paths}
		\label{fig:ch2-rad-ex-d}
		\end{subfigure}
		\caption{Illustration of the basic concepts of the linearized computational graph (LCG) and Bauer's formula.
		(a) pseudocode for a simple function with intermediate variables;
		(b) the primal computational graph, a DAG with variables as vertices and flow moving upwards to the output;
		(c) the linearized computational graph (LCG) in which the edges are labeled with the values of the local derivatives;
		(d) illustration of the four paths that must be evaluated to compute the Jacobian.
		(Example from Paul D.\ Hovland.)
		}
		\label{fig:ch2-rad-ex}
	
	\end{figure}
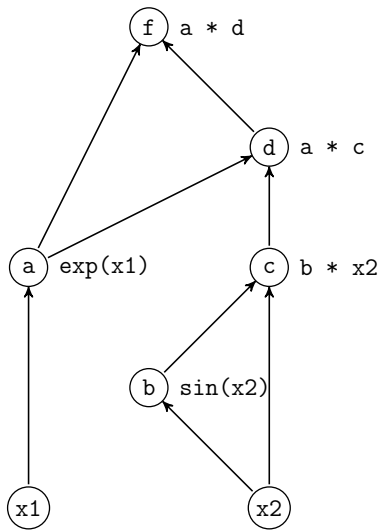
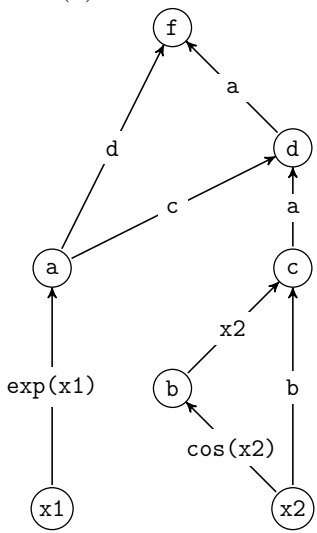
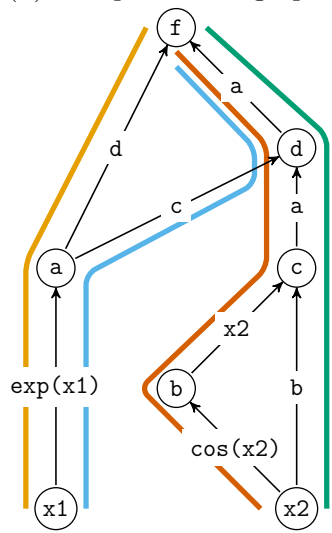

To illustrate, consider an example. Suppose $f(x_1, x_2) = e^{2x_1} x_2 \sin{x_2}$. Psuedo-code for $f$ is shown in \cref{fig:ch2-rad-ex-a}.
This pseudo-code defines a sequence of operations that set intermediate variables equal to the output of primitive operations, eventually returning $f$.
The computational graph representing that sequence of operations is shown in \cref{fig:ch2-rad-ex-b}.
The LCG of $f$ is shown in \cref{fig:ch2-rad-ex-c}.
The edge weights of the LCG are equal to the elementary partial derivatives.

\subsection{Bauer's formula \label{sec:ch2-bauers}}

Consider a function $f : \mathbb{R} \to \mathbb{R}$ made up of four compositions:
\begin{equation}\label{eq:ch2-dp-ex}
    f(x) = D \circ C \circ B \circ A(x).
\end{equation}
$f$ can be computed by a sequence of primitive operations: $\bm a = A(x)$, $\bm b = B(\bm a)$, and $\bm c = C(\bm b)$, and $y = D(\bm c)$.
Here $x$ is an input variable, $\bm a \in \mathbb{R}^{d_1}$, $\bm b \in \mathbb{R}^{d_2}$, and $\bm c \in \mathbb{R}^{d_3}$ are intermediate variables, and $y$ is an output variable.
By the chain rule (\cref{eq:ch2-multivariate-chain-rule-multiple-sum}), the derivative $\frac{\partial y}{\partial x}$ is
\begin{equation}\label{eq:ch2-dp-ex-cr}
    \frac{\partial y}{\partial x} = \sum_{k=1}^{d_3}\sum_{j=1}^{d_2}\sum_{i=1}^{d_1} \frac{\partial y}{\partial c_k}\frac{\partial c_k}{\partial b_j}\frac{\partial b_j}{\partial a_i} \frac{\partial a_i}{\partial x}.
\end{equation}
Notice that there are $d_1 d_2 d_3$ terms in the summation, each of which consists of the product of four elementary partial derivatives.

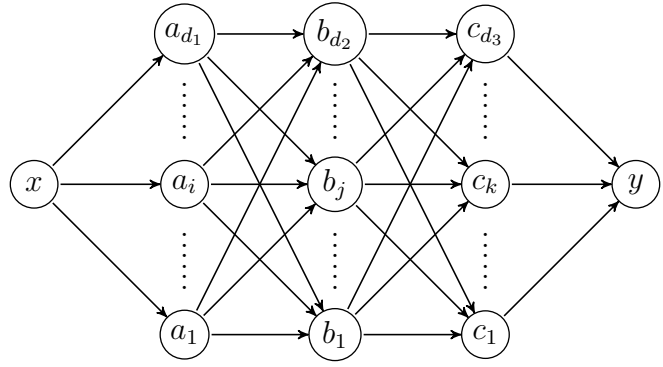
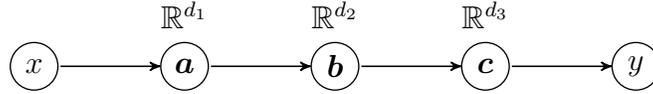
\begin{figure}
\centering
	\begin{subfigure}{\textwidth}
	\centering%
		\begin{tikzpicture}[scale=1.0,transform shape]
		
  			\tikzstyle{every node}=[node distance = 4cm,%
            		                bend angle    = 45]
			\Vertex[x=0, y=2, L=\texttt{$x$}]{x1}
			\Vertex[x=2, y=0, L=\texttt{$a_1$}]{a1} 
			\Vertex[x=2, y=2, L=\texttt{$a_i$}]{a2}
			\Vertex[x=2, y=4, L=\texttt{$a_{d_1}$}]{a3} 
   
                \extralabel[3.0mm]{90}{{$\vdots$}}{a1}
                \extralabel[7.2mm]{90}{{$\vdots$}}{a1}
                \extralabel[3.0mm]{90}{{$\vdots$}}{a2}
                \extralabel[7.2mm]{90}{{$\vdots$}}{a2}

			\Vertex[x=4, y=0, L=\texttt{$b_1$}]{b1} 
			\Vertex[x=4, y=2, L=\texttt{$b_j$}]{b2} 
			\Vertex[x=4, y=4, L=\texttt{$b_{d_2}$}]{b3} 

                \extralabel[3.0mm]{90}{{$\vdots$}}{b1}
                \extralabel[7.2mm]{90}{{$\vdots$}}{b1}
                \extralabel[3.0mm]{90}{{$\vdots$}}{b2}
                \extralabel[7.2mm]{90}{{$\vdots$}}{b2}

			\Vertex[x=6, y=0, L=\texttt{$c_1$}]{c1} 
			\Vertex[x=6, y=2, L=\texttt{$c_k$}]{c2} 
			\Vertex[x=6, y=4, L=\texttt{$c_{d_3}$}]{c3} 

                \extralabel[3.0mm]{90}{{$\vdots$}}{c1}
                \extralabel[7.2mm]{90}{{$\vdots$}}{c1}
                \extralabel[3.0mm]{90}{{$\vdots$}}{c2}
                \extralabel[7.2mm]{90}{{$\vdots$}}{c2}

			\Vertex[x=\nodefy, y=\nodefx, L=\texttt{$y$}]{y}

			\tikzstyle{EdgeStyle}=[post]
			\Edge(x1)(a1)
			\Edge(x1)(a2)
			\Edge(x1)(a3)
			\Edge(a1)(b1)
			\Edge(a1)(b2)
			\Edge(a1)(b3)
			\Edge(a2)(b1)
			\Edge(a2)(b2)
			\Edge(a2)(b3)
			\Edge(a3)(b1)
			\Edge(a3)(b2)
			\Edge(a3)(b3)
			\Edge(b1)(c1)
			\Edge(b1)(c2)
			\Edge(b1)(c3)
			\Edge(b2)(c1)
			\Edge(b2)(c2)
			\Edge(b2)(c3)
			\Edge(b3)(c1)
			\Edge(b3)(c2)
			\Edge(b3)(c3)
			\Edge(c1)(y)
			\Edge(c2)(y)
			\Edge(c3)(y)
		\end{tikzpicture}
		\caption{Computational graph for \cref{eq:ch2-dp-ex}}
		\label{fig:ch2-rad-ex2-a}
	\end{subfigure}
	\begin{subfigure}{\textwidth}
		\centering%
		\vspace{0.2cm}
		\begin{tikzpicture}[scale=1.0,transform shape]
		
  			\tikzstyle{every node}=[node distance = 4cm,%
            		                bend angle    = 45]
			\Vertex[x=0, y=2, L=\texttt{$x$}]{x1}

			\Vertex[x=2, y=2, L=\texttt{$\bm a$}]{a}
			\extralabel[2mm]{90}{\texttt{$\mathbb{R}^{d_1}$}}{a}

			\Vertex[x=4, y=2, L=\texttt{$\bm b$}]{b} 
			\extralabel[2mm]{90}{\texttt{$\mathbb{R}^{d_2}$}}{b}

			\Vertex[x=6, y=2, L=\texttt{$\bm c$}]{c}
			\extralabel[2mm]{90}{\texttt{$\mathbb{R}^{d_3}$}}{c}

			\Vertex[x=\nodefy, y=\nodefx, L=\texttt{$y$}]{y}

			\tikzstyle{EdgeStyle}=[post]
			\Edge(x1)(a)
			\Edge(a)(b)
			\Edge(b)(c)
			\Edge(c)(y)
		\end{tikzpicture}

		\caption{Vector graph for (a)
		}
		\label{fig:ch2-rad-ex2-b}
	\end{subfigure}

		\caption{(a) The underlying computational graph for \cref{eq:ch2-dp-ex}. A comparison with \cref{eq:ch2-dp-ex-cr} reveals that the derivative $\frac{\partial y}{\partial x}$ equals the sum over paths of the products of the edge weights in the linearized computational graph. (b) A vectorized representation of the computational graph, simplifying the presentation.   
		}
		\label{fig:ch2-rad-ex2}
	\end{figure}

A computational graph for $f$ is shown in \cref{fig:ch2-rad-ex2-a}.
(\Cref{fig:ch2-rad-ex2-b} is a vector representation of \cref{fig:ch2-rad-ex2-a}. For functions consisting of vector operations, the vectorized computation graph is a more compact representation. A single path in the vectorized computation graph represents many paths in the underlying scalar computation graph.)
The LCG is not shown for simplicity.
Notice that there are $d_1 d_2 d_3$ paths from $x$ to $y$, that each path consists of four edges, and that the product of the edge weights of each path in the LCG is $\frac{\partial y}{\partial c_k}\frac{\partial c_k}{\partial b_j}\frac{\partial b_j}{\partial a_i} \frac{\partial a_i}{\partial x}$.
Thus, by inspection, we see that \cref{eq:ch2-dp-ex-cr} can be rewritten as a sum over paths of the product of the edge weights of each path in the LCG:
\begin{equation}\label{eq:ch2-bauers-simple}
    \frac{dy}{dx} = \sum_{[x \to y]} \prod_{(k, l) \in [x \to y]} \frac{\partial v_l}{\partial v_k}.
\end{equation}
where~$[x\to y]$ indexes paths from node~$x$ to node~$y$ and~${(k,l)\in[x\to y]}$ denotes the set of edges in that path.
In fact, a generalization of \cref{eq:ch2-bauers-simple} holds for any LCG representing a function $\bm y = f(\bm x)$:
\begin{equation}\label{eq:ch2-bauers}
    J_{ij} = \frac{dy_i}{dx_j} = \sum_{[i \to j]} \prod_{(k, l) \in [i \to j]} \frac{\partial v_l}{\partial v_k}.
\end{equation}
This equation is known as Bauer's formula \cite{bauer1974computational}.
\Cref{fig:ch2-rad-ex-d} shows all of the Bauer paths required to compute the Jacobian of $f(x_1,x_2) = e^{2x_1} x_2 \sin{x_2}$ using \cref{eq:ch2-bauers}.

\subsection{Jacobian accumulation \label{sec:ch2-ad-as-dp-jac-acc}}

As mentioned in \cref{sec:ch2-ad-as-comp-jac-acc}, Jacobian accumulation is the process of computing a Jacobian $\bm J$ by multiplying the elementary partial derivatives according the chain rule.
Performing Jacobian accumulation using a naive implementation of Bauer's formula (\cref{eq:ch2-bauers}) would take time exponential in the depth of the computational graph because the number of paths in a computational graph of width $d$ and depth $p$ goes as $\mathcal{O}(d^p)$. For a visual illustration of this exponential increase, see \cref{fig:ch2-rad-ex2-a}.

AD can be understood as a family of dynamic programming (DP) algorithms for computing \cref{eq:ch2-bauers}. DP algorithms find overlapping substructure within a problem to reduce the complexity of the original problem. Often, this ends up reducing a problem with exponential complexity into one with polynomial complexity.
The overlapping structure in \cref{eq:ch2-bauers} comes from the fact that different paths can contain the same product of edge weights.
For example, in \cref{fig:ch2-rad-ex2-a}, the product of edge weights in the paths $x \to a_1 \to b_1 \to c_1 \to y$ and $x \to a_1 \to b_1 \to c_2 \to y$ both contain the product $\frac{\partial b_1}{\partial a_1} \frac{\partial a_1}{\partial x}$.

The two most commonly used DP algorithms for reducing the runtime complexity of \cref{eq:ch2-bauers} are forward mode (\cref{sec:ch2-ad-as-dp-forward}) and reverse mode (\cref{sec:ch2-ad-as-dp-reverse}). 
However, these are not the only DP algorithms to reduce the runtime of Bauer's formula.
More general DP algorithms, such as vertex, edge, and face elimination, manipulate the LCG by removing vertices, edges, or faces until the graph becomes bipartite with edges only between input and output variables \cite{naumann2002elimination,naumann2004optimal}. These remaining edges are the entries of the Jacobian matrix.
In practice, determining the optimal strategy for Jacobian accumulation is NP-complete \cite{naumann2008optimal}.
Most implementations of AD don't attempt to perform these more general DP algorithms, and instead only implement the special cases of forward and reverse mode.

\subsection{Forward mode \label{sec:ch2-ad-as-dp-forward}}

Forward mode AD computes the JVP $\dot{\bm y} = \bm J \dot{\bm x}$ for a function $f : \mathbb{R}^n \to \mathbb{R}^m$. Forward mode's inputs are the input vector $\bm x \in \mathbb{R}^n$ and the tangent vector $\dot{\bm x} \in \mathbb{R}^n$.
Using the computational graph notation introduced in \cref{sec:ch2-comp-graph}, forward mode AD performs the following steps:
\begin{enumerate}
    \item Assign $v_{i-n} = x_i$ and $\dot v_{i-n} = x_i$ for $i = 1, \dots, n$.
    \item Starting with $j=1$, perform $v_j = \varphi(v_i)_{i \prec j}$. Add node $v_j$ to the computational graph. Add edges $e_{ij}$ for all $i \prec j$. Then perform 
    \begin{equation}
        \dot v_j = \sum_{\substack{i \in \text{parents} \\ \text{of } j}} \frac{\partial v_j}{\partial v_i} \dot v_i.
    \end{equation}
    Repeat for $j = 2, \dots, p + m$.
    \item Assign $y_i = v_{p+i}$ and $\dot y_i = \dot v_{p+i}$ for $i = 1, \dots, m$.
\end{enumerate}
Step 2 can be understood as computing the JVP $\bm J_\varphi \bm \dot{v}_i$ for each primitive operation $\varphi$.
The overall result of computing the JVP for each primitive is to accumulate the JVP of the entire function defined by the computational graph.

\subsection{Reverse mode \label{sec:ch2-ad-as-dp-reverse}}

Reverse mode AD computes the VJP $\bar{\bm x}^T = \bar{\bm y}^T \bm J$ for a function $f : \mathbb{R}^n \to \mathbb{R}^m$. Reverse mode's inputs are the input vector $\bm x \in \mathbb{R}^n$ and the cotangent vector $\bar{\bm y} \in \mathbb{R}^m$. Using the computational graph notation introduced in \cref{sec:ch2-comp-graph}, reverse mode AD performs the following steps:
\begin{enumerate}
    \item Assign $v_{i-n} = x_i$ for $i = 1, \dots, n$.
    \item Starting with $j=1$, perform $v_j = \varphi(v_i)_{i \prec j}$. Add node $v_j$ to the computational graph. Add edges $e_{ij}$ for all $i \prec j$. Repeat for $j = 2, \dots, p+m$.
    \item Assign $\bar v_{p+i} = \bar y_i$ for $i = 1, \dots, m$.
    \item Starting with $j = p$, perform 
    \begin{equation}\label{eq:ch2-reverse-mode-update}
        \bar v_i = \sum_{\substack{j \in \text{children} \\ \text{of } i}} \bar v_j \frac{\partial v_j}{\partial v_i}.
    \end{equation}
    Repeat for $j = p-1, \dots, 1, \dots, -n+1$.
    \item Assign $\bar x_i = \bar v_{i-n}$ for $i = 1, \dots, n$.
\end{enumerate}
Step 2 builds the computational graph. Step 4, by contrast, traverses the computational graph in reverse order. Step 4 can be understood as computing the VJP $\bar{\bm v}_j^T\bm J_\varphi$ for each primitive operation $\varphi$ and adding the result to the cotangent of the inputs to the primitive.
The overall result of computing the VJP for each primitive in reverse order through the computational graph is to accumulate the VJP of the entire function defined by the computational graph.

\subsection{Differentiation with AD: an example}

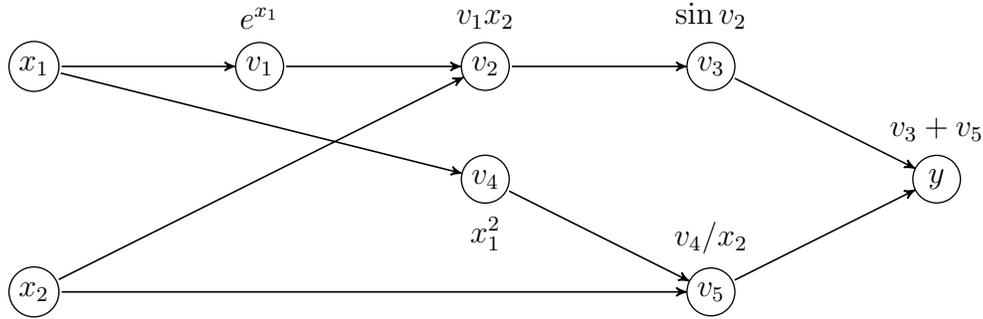
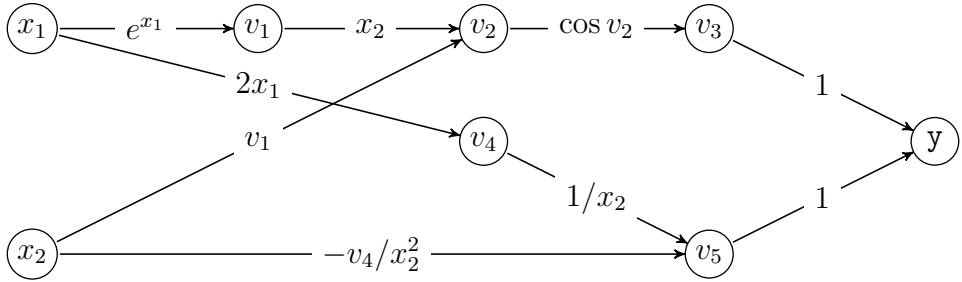
\begin{figure}
\centering

\begin{subfigure}{0.8\textwidth}
    \centering 

\begin{tikzpicture}[scale=1,transform shape]
		
\tikzstyle{every node}=[node distance = 4cm, bend angle = 45]

\Vertex[x=0, y=3, L={\texttt{$x_1$}}]{x1}
\Vertex[x=0, y=0, L=\texttt{$x_2$}]{x2}

\Vertex[x=3, y=3, L=\texttt{$v_1$}]{v1}
\extralabel[2mm]{90}{\texttt{$e^{x_1}$}}{v1}

\Vertex[x=6, y=3, L=\texttt{$v_2$}]{v2}
\extralabel[2mm]{90}{\texttt{$v_1 x_2$}}{v2}

\Vertex[x=9, y=3, L=\texttt{$v_3$}]{v3}
\extralabel[2mm]{90}{\texttt{$\sin{v_2}$}}{v3}

\Vertex[x=6, y=1.5, L=\texttt{$v_4$}]{v4}
\extralabel[2mm]{270}{\texttt{$x_1^2$}}{v4}

\Vertex[x=9, y=0, L=\texttt{$v_5$}]{v5}
\extralabel[2mm]{90}{\texttt{$v_4/x_2$}}{v5}

\Vertex[x=12, y=1.5, L=\texttt{$y$}]{y}
\extralabel[2mm]{90}{\texttt{$v_3 + v_5$}}{y}

\tikzstyle{EdgeStyle}=[post]
\Edge(x1)(v1)
\Edge(x1)(v4)
\Edge(x2)(v2)
\Edge(x2)(v5)
\Edge(v1)(v2)
\Edge(v2)(v3)
\Edge(v3)(y)
\Edge(v4)(v5)
\Edge(v5)(y)
\end{tikzpicture}

    \caption{Computational graph}
\label{fig:compgraphAD-CG}
\end{subfigure}

\vspace{0.2cm}

\begin{subfigure}{0.8\textwidth}
    \centering 
\begin{tikzpicture}[scale=1,transform shape]
		
\tikzstyle{every node}=[node distance = 4cm, bend angle = 45]

\Vertex[x=0, y=3, L=\texttt{$x_1$}]{x1}
\Vertex[x=0, y=0, L=\texttt{$x_2$}]{x2}
\Vertex[x=3, y=3, L=\texttt{$v_1$}]{v1}
\Vertex[x=6, y=3, L=\texttt{$v_2$}]{v2}
\Vertex[x=9, y=3, L=\texttt{$v_3$}]{v3}
\Vertex[x=6, y=1.5, L=\texttt{$v_4$}]{v4}
\Vertex[x=9, y=0, L=\texttt{$v_5$}]{v5}
\Vertex[x=12, y=1.5, L=\texttt{y}]{y}

\tikzstyle{EdgeStyle}=[post]
\Edge[label=\texttt{$e^{x_1}$}](x1)(v1)
\Edge[label=\texttt{$2x_1$}](x1)(v4)
\Edge[label=\texttt{$v_1$}](x2)(v2)
\Edge[label=\texttt{$-v_4/x_2^2$}](x2)(v5)
\Edge[label=\texttt{$x_2$}](v1)(v2)
\Edge[label=\texttt{$\cos{v_2}$}](v2)(v3)
\Edge[label=\texttt{$1$}](v3)(y)
\Edge[label=\texttt{$1/x_2$}](v4)(v5)
\Edge[label=\texttt{$1$}](v5)(y)
\end{tikzpicture}

    \caption{Linearized computational graph}
\label{fig:compgraphAD-LCG}
\end{subfigure}

\caption{(a) A computational graph for the function defined in equation \ref{eq:examplefunc}. (b) The LCG. The variables on the left, $x_1$ and $x_2$, are the inputs to the function $f(x_1,x_2)$. The intermediate variables $v_1$, $v_2$, $v_3$, $v_4$, and $v_5$ are produced during the computation of $f$. The function $f$ is computed from left to right across the graph.
In forward mode, the tangents (derivatives) are also computed from left to right across the graph, at the same time as the intermediate variables are being computed. In reverse mode, the cotangents (derivatives) are computed from right to left after the function is computed from left to right.} 
    \label{fig:compgraphAD}
\end{figure}

Let us illustrate how forward and reverse mode work with an example. Suppose we want to compute the derivatives of the function $f$ given by
\begin{equation} \label{eq:examplefunc}
    y = f(x_1,x_2) = \sin(e^{x_1} x_2) + x_1^2/x_2{.}
\end{equation}
A computational graph and LCG for $f$ is shown in figure \ref{fig:compgraphAD}. 
Since $f$ is a scalar function of two variables, then forward mode AD should be able to compute the full Jacobian in two forward mode computations, while reverse mode AD should be able to compute the full Jacobian in one reverse mode computation.

\subsubsection{Forward mode}

We now perform forward mode AD on $f$ to compute the Jacobian-vector product $\bm{J}\dot{\bm x}$ at some value of $\bm x$. For concreteness, we arbitrarily set $x_1=2$ and $x_2=4$. Since $f: \mathbb{R}^2 \rightarrow \mathbb{R}$, then $\bm{J} \in \mathbb{R}^{1\times2}$. Here we will set $\bm{\dot{x}} = \rvect{1,0}^T$ which will give us the partial derivative $\frac{\partial f}{\partial x_1}$ at $x_1=2$ and $x_2=4$:
\begin{equation}
    \dot{y}
     = 
    \begin{bmatrix}
    \frac{\partial f}{\partial x_1} & \frac{\partial f}{\partial x_2} 
    \end{bmatrix}
    \begin{bmatrix} 1 \\
    0
    \end{bmatrix} = \frac{\partial f}{\partial x_1}{.}
\end{equation}
Because $\bm{\dot{x}} = \rvect{1,0}^T$, the tangent variable $\dot{v}_i$ equals $\frac{\partial v_i}{\partial x_1}$. 
If $\bm{\dot{x}}$ were $\rvect{0,1}^T$, then $\dot{v}_i$ would equal $\frac{\partial v_i}{\partial x_2}$. If $\bm{\dot{x}}$ were $\rvect{\alpha,\beta}^T$, then $\dot{v}_i$ would equal $\alpha \frac{\partial v_i}{\partial x_1} + \beta  \frac{\partial v_i}{\partial x_2}$.

\begin{table}
    \centering
    \begin{tabular}{|c|c|c|c|c|}
        \hline Step &
        Variable & Value & Tangent  &  Value \\ \hline
        1 & $x_1$ & 2.0 & $\dot{x}_1 = \frac{\partial x_1}{\partial x_1}$ & 1.0 \\ \hline
        2 & $x_2$ & 4.0 & $\dot{x}_2 =\frac{\partial x_2}{\partial x_1}$ & 0.0 \\ \hline
        3 & $v_1$ & 7.389 & $\dot{v}_1 =\dot{x}_1 \frac{\partial v_1}{\partial x_1}$ &  7.389  \\ \hline
        4 & $v_2$ & {29.556} & $\dot{v}_2 =\dot{v}_1 \frac{\partial v_2}{\partial v_1} + \dot{x}_2 \frac{\partial v_2}{\partial x_2}$ & {29.556} \\ \hline
        5 & $v_3$ & -0.959 & $\dot{v}_3 =\dot{v}_2 \frac{\partial v_3}{\partial v_2}$ & -8.42 \\ \hline
        6 & $v_4$ & 4.0 & $\dot{v}_4 =\dot{x}_1\frac{\partial v_4}{\partial x_1}$ & 4.0 \\ \hline
        7 & $v_5$ & 1.0 & $\dot{v}_5 =\dot{x}_2 \frac{\partial v_5}{\partial x_2}+ \dot{v}_4 \frac{\partial v_5}{\partial v_4}$ & 1.0 \\ \hline
        8 & $y$ & 0.0414 & $\dot{y} = \dot{v}_3\frac{\partial y}{\partial v_3} + \dot{v}_5 \frac{\partial y}{\partial v_5}$ & -7.421 \\ \hline
    \end{tabular}
    \caption{The steps of the forward mode computation which compute the function from equation \ref{eq:examplefunc} and its derivative $\dot{y} = \frac{\partial f}{\partial x_1}$. In steps 1 and 2, the input variables $x_1$ and $x_2$ and their tangents are set. In steps 3-7, the intermediate variables in figure \ref{fig:compgraphAD} are computed and their tangents are computed using the update rule in equation \ref{eq:forwardupdate}. In step 8, the output variable $y$ is set, along with its tangent. }
    \label{tab:forwardmode}
\end{table}

We can use the computational graph in figure \ref{fig:compgraphAD} along with table \ref{tab:forwardmode} to understand how $\frac{\partial f}{\partial x_1}$ is computed with forward mode AD. In steps 1 and 2 of table \ref{tab:forwardmode}, $x_1$ and $x_2$ are set to $2$ and $4$, while their tangents $\dot{x}_1$ and $\dot{x}_2$ are set to 1 and 0 to match $\bm{\dot{x}} = \rvect{1,0}^T$. In steps 3-8, the computational graph in figure \ref{fig:compgraphAD} is traversed in topological order. For each vertex, $v_j$ is computed along with its tangent $\dot{v}_j$. $\dot{v}_j$ is computed using the update rule
\begin{equation} \label{eq:forwardupdate}
    \dot{v}_j = \sum_{\substack{i \in \text{parents} \\ \text{of } j}} \frac{\partial v_j}{\partial v_i} \dot{v}_i {.}
\end{equation}
{This update rule is equivalent to the chain rule.} The elementary partial derivatives $\frac{\partial v_j}{\partial v_i}$ are the edge weights in the LCG.

In the last row of table \ref{tab:forwardmode} we have the result of the computation $y = 0.0414$, as well as the tangent $\dot{y} = \frac{\partial f}{\partial x_1} = - 7.421$. Notice that this gave us one column of the Jacobian at a computational cost equal to a small multiple of the cost of evaluating the function itself. To compute both columns of the Jacobian would require two forward mode evaluations.  

\subsubsection{Reverse mode}

We now perform reverse mode AD on $f$ to compute the vector-Jacobian product $\bm{\overline{y}^T J}$ at $x_1=2$ and $x_2=4$. Since $f: \mathbb{R}^2 \rightarrow \mathbb{R}$, then $\bm{J} \in \mathbb{R}^{1\times2}$. Here we will set $\bm{\overline{y}} = [1]$ which will give us the transpose gradient {$\rvect{\frac{\partial f}{\partial x_1},\frac{\partial f}{\partial x_2}}$} at $x_1=2$ and $x_2=4$:
\begin{equation}
    \overline{\bm{x}}^T = 
    \begin{bmatrix}
    1
    \end{bmatrix}
    \begin{bmatrix}
    \frac{\partial y}{\partial x_1} & \frac{\partial y}{\partial x_2} 
    \end{bmatrix} = {
    \begin{bmatrix}
    \frac{\partial y}{\partial x_1} & \frac{\partial y}{\partial x_2}
    \end{bmatrix}}{.}
\end{equation}
Reverse mode AD associates a derivative value $\overline{v}_i$ for each intermediate variable $v_i$. $\overline{v}_i$ is called the cotangent variable for $v_i$. $\overline{v}_i$ is computed after the function evaluates $y$, by combining partial derivatives from end to beginning. Because $\bm{\overline{y}} = [1]$, then the cotangent variable $\overline{v}_i \equiv \frac{\partial y}{\partial v_i}$ for each variable.

\begin{table}
    \centering
    \begin{tabular}{|c|c|c|}\hline
        Step & Variable & Value \\ \hline
        1 & $x_1$ & 2.0 \\ \hline
        2 & $x_2$ & 4.0 \\ \hline
        3 & $v_1$ & 7.389 \\ \hline
        4 & $v_2$ & 29.556 \\ \hline
        5 & $v_3$ & -0.959 \\ \hline
        6 & $v_4$ & 4.0 \\ \hline
        7 & $v_5$ & 1.0 \\ \hline
        8 & $y$ & 0.0414 \\ \hline
    \end{tabular}
    \quad
    \begin{tabular}{|c|c|c|}
    \hline
    Step & Cotangent & Value \\ \hline
    9 & $\overline{y} = \frac{\partial y}{\partial y}$ & 1.0 \\ \hline
    10 & $\overline{v}_5 = {\overline{y}}\frac{\partial y}{\partial v_5}$  & 1.0 \\ \hline
    11 & $\overline{v}_4 = \overline{v}_5 \frac{\partial v_5}{\partial v_4}$ & 0.25   \\ \hline
    12 & $\overline{v}_3 = {\overline{y}} \frac{\partial y}{\partial v_3}$ & 1.0 \\ \hline
    13 & $\overline{v}_2 = \overline{v}_3 \frac{\partial v_3}{\partial v_2}$ & -0.285   \\ \hline
    14 & $\overline{v}_1 = \overline{v_2}\frac{\partial v_2}{\partial v_1}$  & -1.140 \\ \hline
    15 & $\overline{x}_2 = \overline{v}_5 \frac{\partial v_5}{\partial x_2} + \overline{v}_2 \frac{\partial v_2}{\partial x_2}$ & -2.355 \\ \hline 
    16 & $\overline{x}_1 = \overline{v}_4 \frac{\partial v_4}{\partial x_1} + \overline{v}_1 \frac{\partial v_1}{\partial x_1}$ & -7.421 \\ \hline
    \end{tabular}
    \caption{The steps of the reverse mode computation which compute the function from equation \ref{eq:examplefunc} and its derivatives $\frac{\partial f}{\partial x_1}$ and $\frac{\partial f}{\partial x_2}$. In steps 1 and 2, the input variables $x_1$ and $x_2$ are set. In steps 3-8, the forward pass of the computation is performed and the value of each intermediate variable is stored for use in the backwards pass. In step 9, the cotangent variable $\overline{y}$ is set to 1. In steps 10-16, the computational graph in figure \ref{fig:compgraphAD} is traversed in reverse topological order and the cotangent variables of each variable in the graph are computed using the update rule in equation \ref{eq:reverseupdate}. }
    \label{tab:reversemode}
\end{table}

We can use the computational graph and LCG in figure \ref{fig:compgraphAD} along with table \ref{tab:reversemode} to understand the detailed computations performed by reverse mode AD. 
On the left side of table \ref{tab:reversemode}, the forward pass is performed.
In the forward pass, the computational graph in figure \ref{fig:compgraphAD} is traversed in topological order.
For each edge, the elementary partial derivatives $\frac{\partial v_j}{\partial v_i}$ are computed and stored in memory.
On the right side of table \ref{tab:reversemode}, the backwards pass is performed.
In the backwards pass, the LCG is traversed in reverse topological order.
The cotangent variable $\overline{y}$ is first set to 1.
Then, for each vertex the cotangent variable $\overline{v}_i$ is computed using the update rule 
\begin{equation}\label{eq:reverseupdate}
    \overline{v}_i \equiv \frac{\partial y}{\partial v_i} = \sum_{\substack{j \in \text{children} \\ \text{of } i}} \overline{v}_j \frac{\partial v_j}{\partial v_i}{.}
\end{equation}
{This update rule is equivalent to the chain rule.} 
The elementary partial derivatives $\frac{\partial v_j}{\partial v_i}$ are the edge weights in the LCG.

At the end of the backwards pass, we have the result of the computation $y = 0.0414$, as well as the cotangent variables $\overline{x}_1 \equiv \frac{\partial f}{\partial x_1} = - 7.421$ and $\overline{x}_2 \equiv \frac{\partial f}{\partial x_2} = - 2.355$. Notice that reverse mode AD gives the full Jacobian of a scalar function at a computational cost equal to a small multiple of the cost of evaluating the function itself.

\section{AD in practice \label{sec:ch2-implementation}}

\subsection{Implementing AD}

The implementation of AD can be more complex than the theory of AD.
While forward mode can be relatively straightforward to implement, reverse mode can be quite complex.
Fortunately, we don't need to understand perfectly how AD tools are implemented in order to use them. 
A good AD tool will hide all of the implementation details from the user, allowing them to program naturally in a high-level language yet still be able to compute derivatives automatically.

A few common approaches to implementing AD are operator overloading, explicit construction of computational graphs, and program transformation.
There are two main approaches to operator overloading. The first is to define new data types (such as, for example, \texttt{adouble} instead of \texttt{double}) that contain hidden information, then overloading existing primitives (such as, for example, \texttt{adouble add(adouble x, adouble y)} instead of \texttt{double add(double x, double y)}) to operate with these new data types.
The second is to use a so-called `tape' or `Wengert list' to record the sequence of primitive operations and intermediate variables computed by the program, then (in reverse mode) reverse the order of the primitive operations in the tape and add the result of each VJP to the cotangents of the inputs to the primitive.
An example is {Tapenade} \cite{hascoet2013tapenade}.
{Theano} \cite{bergstra2011theano} and {TensorFlow 1.0} (2017-2019) explicitly build a computational graph of symbolic expressions, then pass values and cotangents through the computational graph. 
Program transformation involves a custom interpreter which translates source code into an intermediate language, transforms the intermediate code to compute derivatives, then either translates the representation back into source code in the original language or executes the intermediate representation using some low-level language.
Source code transformation is typically done in FORTRAN \cite{bischof1996adifor}, while the latter approach is sometimes done in high-level languages like Python. 
In a functional language such as Haskell, program transformation can be done using a compiler plugin \cite{Elliott-2018-ad-extended}.

If the JVP and VJP rules for each primitive operation are written in terms of primitive operations, it is possible for an AD tool to compute higher-order derivatives.
A discussion of how higher-order derivatives are implemented is outside the scope of this thesis.

\subsubsection{Forward mode implementation via program transformation}

We now demonstrate a simple and minimal implementation of forward mode AD in Python based on the strategy of program transformation.
The goal of this implementation is to implement \texttt{jvp}, a program transformation that takes a function $f : \mathbb{R}^n \to \mathbb{R}^m$ and outputs a function $\textnormal{jvp}(f) : (\mathbb{R}^n, \mathbb{R}^n) \to (\mathbb{R}^m, \mathbb{R}^m)$. In Haskell type signature notation, this transformation can be written as \texttt{jvp : (a -> b) -> (a, Ta) -> (b, Tb)} where \texttt{Ta} is the tangent type of \texttt{a}. 

Our minimal implementation uses two classes, \texttt{Node} and \texttt{Primitive}. Transformed functions use \texttt{Node} datatypes, rather than floats, as inputs and outputs. Nodes contain both the value $x$ and the tangent $\dot{x}$. Each \texttt{Primitive} simply contains a string referring to its name.
\begin{python}
from typing import NamedTuple
class Node(NamedTuple):
	value: float
	tangent: float
class Primitive(NamedTuple):
	name: str
\end{python}
Next, we implement a library of primitive operations. In this minimal implementation, our library only contains two primitives, \texttt{add} and \texttt{multiply}.
For each primitive, we implement evaluation rules and JVP rules. Evaluation rules take variables of type \texttt{float} as inputs and outputs, while JVP rules take variables of type \texttt{Node} as inputs and outputs.
Each primitives \texttt{name} is used as a lookup key in a dictionary of functions to evaluate primitives and their derivative rules.
\begin{python}
add_p = Primitive('add')
mul_p = Primitive('mul')
eval_rules = {}
eval_rules[add_p] = lambda x, y: x + y
eval_rules[mul_p] = lambda x, y: x * y
jvp_rules = {}
jvp_rules[add_p] = lambda x, y: Node(x.value + y.value, x.tangent + y.tangent) 
jvp_rules[mul_p] = lambda x, y: Node(x.value * y.value, x.tangent * y.value + x.value * y.tangent) 
def add(x, y): return call_prim(add_p, x, y)
def multiply(x, y): return call_prim(mul_p, x, y)
\end{python}
All primitive calls are wrapped calls to the \texttt{call\_prim} function. \texttt{call\_prim} uses the evaluation rule for the primitive if all of the inputs are of type \texttt{float}, and the JVP rule for the primitive if any of the inputs are of type \texttt{Node}.
\begin{python}
def call_prim(prim, *args, **kwargs):
	if any(isinstance(arg, Node) for arg in args):
		args = [arg if isinstance(arg, Node) else Node(arg, 0.0) for arg in args]
		return jvp_rules[prim](*args, **kwargs)
	else:
		return eval_rules[prim](*args, **kwargs)
\end{python}
All that remains is to implement the \texttt{jvp} transformation.
\begin{python}
def jvp(func):
	def transformed_func(*args):
		nodes = [Node(*arg) for arg in args]
		output = func(*nodes)
		if isinstance(output, Node):
			return output.tangent
		else:
			return tuple([node.tangent for node in output])
	return transformed_func
\end{python}
Now, users can program arbitrary functions composed of any primitive in the library. For example, we can program the function $f(x,y) = x^2 + y^2, 2y^2 : \mathbb{R}^2 \to \mathbb{R}^2$. \texttt{f} takes inputs of type \texttt{float} and returns outputs of type \texttt{float}.
\begin{python}
def f(x, y):
	a = multiply(x, x)
	b = multiply(y, y)
	c = add(a, b)
	d = multiply(2, b)
	return c, d
x = 2.0
y = 3.0
outputs = f(x, y)
print(outputs) # (13.0, 18.0)
\end{python}
Applying the transformation \texttt{jvp} to \texttt{f} returns a function that takes $n=2$ tuples of inputs $x$ and tangents $\dot{x}$, and returns $m=2$ outputs of type \texttt{float} corresponding to the tangents $\dot{y}$. 
\begin{python}
jvp_f = jvp(f)
tx = (x, 1.0)
ty = (y, 0.0)
tangents = jvp_f(tx, ty)
print(tangents) # (4.0, 0.0)
\end{python}

\subsubsection{Reverse mode implementation using program transformation}

Simple and minimal implementations of reverse mode AD are also possible. One example is {MiniGrad}, hosted at \url{https://github.com/kennysong/minigrad/tree/master}, which uses operator overloading in Python.
A full implementation of reverse-mode AD based on program transformation is too complex to be contained within this thesis. One such implementation is explained at \url{https://jax.readthedocs.io/en/latest/autodidax.html}.

The key idea with reverse mode is to implement the (forward) evaluation and (reverse) VJP rules for each primitive. The pseudocode below sketches how a primitive might implement these rules.
The forward evaluation rule takes $n$ inputs and returns two tuples: the $m$ output variables as well as whichever intermediate variables need to be stored in memory for computing the VJP.
\begin{python}
def example_fwd(inputs): # inputs is tuple of n variables
	... # perform intermediate computation
	outputs = ... # tuple of m outputs
	save_for_reverse = ... # intermediate variables needed for vjp
	return outputs, save_for_reverse
vjp_rules_fwd[example_p] = example_fwd
\end{python}
The reverse VJP rule takes the saved variables as well as $m$ cotangent variables, and returns $n$ cotangents with respect to the input variables.
\begin{python}
def example_rev(saved, cotangents): # cotangents is tuple of m variables
	v1, v2, ... = saved # unpack intermediate variables
	... # perform intermediate computations
	return input_cotangents # tuple of n variables
vjp_rules_rev[example_p] = example_rev
\end{python}

Reverse mode AD tools first compute the so-called `forward pass' of the function, computing \texttt{vjp\_rules\_fwd[prim]} for each primitive \texttt{prim}. Next, once the entire function $f$ is evaluated, the AD tool starts with the cotangent vector $\bar{\bm y}$ of the output $\bm y$ and, following \cref{eq:reverseupdate}, computes the so-called `backwards pass' by adding the result of \texttt{vjp\_rules\_rev[prim]} to the cotangent vector of each input of each primitive \texttt{prim} in the reverse order of evaluation.

\subsection{Using AD}

There are many AD implementations. Each implementation tends to have different syntax for computing derivatives.
Users who program with AD tools will need to learn the unique syntax of each implementation in order to compute derivatives of mathematical functions.
Thus, it is impossible to discuss in general terms how to program with AD tools.
Readers interested in using a particular AD tool will have to find an example or tutorial explaining how to use that tool.

Researchers in ML have developed new, and in many ways improved, tools for computing derivatives with AD. At present, TensorFlow, PyTorch, and JAX are the most widely used AD tools.
Each allows for a wide range of mathematical functions to be computed on both CPU and GPU.
However, PyTorch and TensorFlow are intended more to be neural network libraries with AD implementations, while JAX is more of an AD library with neural network extensions.
JAX is particularly well-suited for scientific computing and is used in this dissertation in \cref{ch:stellarator,ch:invariant,ch:reproducibility}.
JAX has an extremely simple and intuitive syntax. Functions are programmed using the syntax of the Python library \texttt{numpy}, then gradients are computed using the transformation \texttt{grad}.
Computing the gradient of an arbitrarily complex scalar function -- for example, the \texttt{loss} function in logistic regression -- is as simple as \texttt{grad(loss)}: 
\begin{python}
from jax import grad, numpy as jnp
def sigmoid(z):
	return 1 / (1 + jnp.exp(-z))
def predict(params, x):
    return sigmoid(jnp.dot(x, params['w']) + params['b'])
def loss(params): # scalar logistic regression loss function
    preds = predict(params, x)
    return -jnp.sum(jnp.log(preds * y + (1 - preds) * (1 - y)))
x = jnp.array([0.52, 1.12,  0.77]) # random length-3 input
y = jnp.array([True]) # random binary output
params = {'w': jnp.zeros(x.shape[0]), 'b': 0.}
print(loss(params)) # 0.6931472
grad_loss = grad(loss) # computes gradient of loss
print(grad_loss(params)) # {'b': DeviceArray(-0.5), 'w': DeviceArray([-0.26 , -0.56 , -0.385])}
\end{python}
\noindent JAX also allows for higher-order derivatives to be computed.
For instance, using the transformations \texttt{jacfwd} or \texttt{jacrev} to compute the Jacobian of \texttt{grad(f)} returns the Hessian of \texttt{f}:
\begin{python}
from jax import grad, jacfwd, numpy as jnp
def f(x): # f : R^2 -> R
	return x[0]**2 * jnp.sin(x[0] * x[1]**3) + 3 * x[1]
hessian_f = jacfwd(grad(f)) # jacrev(grad(f)) could also be used
x = jnp.array([-2.,1.5])
print(hessian_f(x)) # [[ -4.5061264  -9.687016 ][ -9.687019  263.78568  ]]
\end{python}
\noindent Further information on how to use JAX can be found at \url{https://jax.readthedocs.io/}.

\section{Runtime and memory costs of AD \label{sec:ch2-runtime-memory}}

Computing the JVP and/or VJP of a primitive operation can be performed with a runtime cost equal to, at most, a small multiple of the cost of evaluating the primitive. This is true for scalar primitives (for example, the JVP of $y=\sin{x}$ is $\dot{y} = \cos{(x)}\dot{x}$ and the VJP gives $\bar{x} \pluseq \bar{y}\cos{x}$), vector primitives (for example, the JVP of the matrix-matrix product $\bm C = \bm A \cdot \bm B$ is $\dot{\bm{C}} = \dot{\bm{{A}}} \bm{B} + \bm{A} \dot{\bm{B}}$ and the VJP gives $\bar{\bm{A}} \pluseq \bar{\bm{C}}\bm{B}^T$, $\bar{\bm{B}} \pluseq \bar{\bm{C}} \bm{A}^T$), implicitly-defined primitive functions (as we will see in \cref{sec:ch2-implicit-differentiation}), and any other primitive we might implement.
Chaining JVPs or VJPs of primitive operations to compute the overall JVP or VJP of an arbitrarily complex function $f$ requires one JVP or VJP per primitive operation and thus can be performed at a runtime cost equal to a small multiple of evaluating $f$.
Using a simple measure of runtime complexity, \cite{griewank2008evaluating} calculates that computing a JVP in forward mode requires no more than $2.5\times$ the cost of evaluating the original function, while computing a VJP in reverse mode requires no more than $5\times$ the cost of evaluating the original function.
In practice, the runtime ratio in reverse mode can be higher depending on the time required to store and retrieve variables saved in memory.

As already discussed, AD can compute the full Jacobian of $f : \mathbb{R}^n \to \mathbb{R}^m$ using $n$ forward mode evaluations or $m$ reverse mode evaluations.
Thus forward mode is more efficient for `fan-out' functions where $n \ll m$, while reverse mode is more efficient for `fan-in' functions where $n \gg m$.
Of particular importance is the case $m=1$, for which a \textit{cheap gradient principle} applies. As defined by \cite{griewank2008evaluating}, the cheap gradient principle states that ``operation counts and random access memory requirements for gradients are bounded multiples of those for the functions.''
This fortunate result implies that computing gradients for optimization using reverse-mode AD can be done without a significant increase in runtime.

Unfortunately, reverse-mode AD has memory requirements proportional to the number of intermediate variables in the computation. Thus, optimization may require a significant increase in memory usage.
This increased memory requirement is due to the fact that when computing the VJP of a primitive operation, the cotangent vectors are only computed after the forwards pass of the function is completed, and so the variables required for computing the VJP of each primitive must be stored in memory.
For example, the scalar primitive $y = \sin{x}$ must save either $x$ or $\cos{x}$ in memory to compute the VJP $\bar{x} \pluseq \bar{y} \cos{x}$, and the vector primitive $\bm C = \bm A \cdot \bm B$ must save both $\bm A$ and $\bm B$ in memory to compute the VJP $\bar{\bm{A}} \pluseq \bar{\bm{C}}\bm{B}^T$, $\bar{\bm{B}} \pluseq \bar{\bm{C}} \bm{A}^T$.

For large computations, the memory requirements of reverse-mode AD can be significant.
The memory required to compute gradients of, for example, large neural networks or long scientific computations can easily surpass the typical 8 to 32 GB of memory on a GPU.

Fortunately, so-called `checkpointing' or `rematerialization' strategies can be used to reduce the memory consumption of reverse-mode AD at the cost of increased runtime.
The idea of checkpointing is, on the forwards pass, to only store intermediate variables in memory at certain `checkpoint' nodes in the computational graph.
Then, on the backwards pass, intermediate variables can be recomputed by performing one or more additional forward passes starting from the checkpoints.
Whether or not checkpointing can be used and useful depends on the structure of the computational graph; checkpointing strategies have been developed for common computational graph structures such as neural networks and iterative numerical computations.
For computational graphs with a regular structure of repeated `layers', such as neural networks or time-dependent simulations, checkpointing strategies can be used to reduce memory consumption from $\mathcal{O}(n)$ to $\mathcal{O}(\sqrt{n})$ at the cost of only a single additional forwards pass \cite{chen2016training}.
Alternative strategies can reduce memory consuption to $\mathcal{O}(\log{n})$ at the cost of $\mathcal{O}(\log{n})$ additional forward passes \cite{griewank2000algorithm}.
Checkpointing is generally most effective when the number of `layers' in a computational graph is
much larger than the memory required at each layer.

Computing higher-order derivatives with AD can be significantly more expensive than first-order derivatives.
Computing the Hessian $\partial^2 f$ of a function $f : \mathbb{R}^n \to \mathbb{R}$ requires computing the Jacobian of $\grad f : \mathbb{R}^n \to \mathbb{R}^n$, which results in a function $\partial^2 f : \mathbb{R}^n \to \mathbb{R}^{n\times n}$. Computing the Jacobian of a function from $\mathbb{R}^n$ to $\mathbb{R}^n$ takes time $\mathcal{O}(n)$ the cost of the original function, regardless of whether forward or reverse mode is used to compute the Jacobian.
Since $\grad f$ can be computed in time $\mathcal{O}(1)$ using reverse mode, computing the Hessian using nested calls to reverse mode (to compute $\grad f$) and forward mode (to compute the Jacobian of $\grad f$) takes time $\mathcal{O}(n)$, with the same memory requirements as reverse mode AD.

For optimization problems with large $n$ (such as neural networks), a runtime of $\mathcal{O}(n)$ times the cost of evaluating the scalar objective function can be prohibitively expensive.
Thus, AD is rarely used to compute the Hessian matrix for second-order optimization.
It is possible, however, to perform second-order optimization using only Hessian-vector products (HVPs), which can be computed in time $\mathcal{O}(1)$.
The trick to doing so is to use the fact that $\partial^2{f} \bm v = \grad((\grad f)^T \bm v)$.
Since $(\grad f)^T \bm v$ is a scalar function that takes time $\mathcal{O}(1)$ to compute using reverse-mode AD, then computing its gradient also takes time $\mathcal{O}(1)$ using reverse-mode AD.
In JAX, this can be done as follows:
\begin{python}
from jax import grad, numpy as jnp
def hvp(f):
    def transformed_func(x, v):
        return grad(lambda x: jnp.dot(grad(f)(x), v))(x)
    return transformed_func
\end{python}
While the runtime cost of computing an HVP does not scale with number of input variables $n$, the runtime and memory cost of computing a HVP both are larger than computing a VJP by constant factors.

\section{Implicit differentiation \label{sec:ch2-implicit-differentiation}}

So far, we've discussed how to compute derivatives of functions composed of primitives defined as explicit functions of the inputs, such as $f(x) = x^2$ and $f(x) = \sin{x}$.
Yet in many applications relevant to scientific computing (and in some applications involving machine learning), derivatives of implicitly defined functions are desired.
For example, $\bm u \in \mathbb{R}^d$ might be defined through the solution to the implicit function $\bm g(\bm u, \bm p) = 0$, where $\bm g : \mathbb{R}^d \times \mathbb{R}^p \to \mathbb{R}^d$.
Examples of implicitly defined functions include systems of linear equations $\bm A \bm u = \bm b$ (in which case $\bm g(\bm u(\bm p), \bm p) = \bm A(\bm p) \bm u - \bm b(\bm p)$), systems of non-linear equations $\bm F(\bm u, \bm p) = 0$ (in which case $\bm g = \bm F)$, and the solutions to attracting fixed point iterations $\bm u = \bm \Phi(\bm u, \bm p)$ (in which case $\bm g = \bm \Phi(\bm u, \bm p) - \bm u$).
AD tools should be able to compute derivatives with respect to the results of implicitly defined functions, even if the elementary Jacobian matrix cannot be explicitly derived.

To be concrete, suppose that we want to use reverse mode AD to compute the gradient of the scalar objective function $f(\bm u(\bm p), \bm p)$ with respect to a vector of parameters $\bm p \in \mathbb{R}^p$ where
$\bm u$ is defined through the implicit function $\bm g(\bm u(\bm p), \bm p) = 0$.
The following discussion can easily be generalized to forward mode AD and non-scalar functions.
Using the chain rule, we have
\begin{equation}\label{eq:ch2-implicit-first}
    \frac{df}{d\bm p} = \frac{\partial f}{\partial \bm p} + \frac{\partial f}{\partial \bm u} \frac{\partial \bm u}{\partial \bm p}.
\end{equation}
Computing $\frac{\partial f}{\partial \bm p}$ and $\frac{\partial f}{\partial \bm u}$ with AD is straightforward.
However, $\bm u$ is the result of an implicitly defined function and thus $\frac{\partial \bm u}{\partial \bm p}$ cannot be computed directly.
Instead we'll eliminate $\frac{\partial \bm u}{\partial \bm p}$, which we can do using the analytic form for the Jacobian $\frac{d \bm g}{d \bm p}$. Since $\bm g = 0$ for any value of $\bm p$, then if $\bm p$ changes, $\bm g$ will not change either, implying that $\frac{d \bm g}{d \bm p} = 0$.
Assuming that $\frac{\partial \bm g}{\partial \bm u}$ is continuously differentiable and non-singular, then according to the implicit function theorem there exists an implicitly defined function $\bm u(\bm p)$ with continuous derivatives. Using the chain rule,
\begin{equation}
    \frac{d \bm g}{d \bm p} = \frac{\partial \bm g}{\partial \bm p} + \frac{\partial \bm g}{\partial \bm u} \frac{\partial \bm u}{\partial \bm p} = 0.
\end{equation}
Note that these partial derivatives are all evaluated at the solution $\bm u = \bm u^*$ to the implicit equation $\bm g(\bm u^*(\bm p), \bm p) = 0$. Inverting gives
\begin{equation}
    \frac{\partial \bm u}{\partial \bm p} = - \bigg(\frac{\partial \bm g}{\partial \bm u}\bigg)^{-1}\frac{\partial \bm g}{\partial \bm p}
\end{equation}
which plugged into \cref{eq:ch2-implicit-first} gives
\begin{equation}\label{eq:ch2-implicit-second}
    \frac{df}{d\bm p} = \frac{\partial f}{\partial \bm p} - \frac{\partial f}{\partial \bm u} \bigg(\frac{\partial \bm g}{\partial \bm u}\bigg )^{-1}\frac{\partial \bm g}{\partial \bm p}.
\end{equation}
The so-called `adjoint' $A^\dagger$ of an operator $A$ is defined via the inner product relation $\langle c | A b \rangle = \langle A^\dagger c | b \rangle$. For matrices, $A^\dagger = A^T$. Thus
\begin{equation}
    \frac{\partial f}{\partial \bm u} \bigg(\frac{\partial \bm g}{\partial \bm u}\bigg )^{-1}\frac{\partial \bm g}{\partial \bm p} = \bigg(\Big(\big(\frac{\partial \bm g}{\partial \bm u}\big )^{-1}\Big)^T\frac{\partial f}{\partial \bm u}\bigg)\frac{\partial \bm g}{\partial \bm p} = \bigg(\Big(\big(\frac{\partial \bm g}{\partial \bm u}\big )^{T}\Big)^{-1}\frac{\partial f}{\partial \bm u}\bigg)\frac{\partial \bm g}{\partial \bm p} 
\end{equation}
using $(A^T)^{-1} = (A^{-1})^T$. \Cref{eq:ch2-implicit-second} can now be written as
\begin{equation}\label{eq:ch2-implicit-gradient}
    \frac{df}{d\bm p} = \frac{\partial f}{\partial \bm p} + \bm \lambda^T \frac{\partial \bm g}{\partial \bm p}
\end{equation}
with the so-called `adjoint' variable $\bm \lambda$ defined via the linear equation
\begin{equation}\label{eq:ch2-adjoint-variable}
    \bigg(\frac{\partial \bm g}{\partial \bm u}\bigg)^T \bm \lambda = - \frac{\partial f}{\partial \bm u}.
\end{equation}
\Cref{eq:ch2-adjoint-variable} is an equation for the adjoint variable $\bm \lambda$, which can be plugged into \cref{eq:ch2-implicit-gradient} to give an equation for the gradient $\frac{df}{d\bm p}$.
Note that the derivatives $\frac{\partial \bm g}{\partial \bm u}$, $\frac{\partial f}{\partial \bm u}$, and $\frac{\partial f}{\partial \bm p}$ are typically either already evaluated in the course of solving $\bm g = 0$ or can be derived using reverse-mode AD.
Once $\bm \lambda$ is known, the VJP $\bm \lambda^T \frac{\partial \bm g}{\partial \bm p}$ can be computed using reverse-mode AD.

\Cref{eq:ch2-adjoint-variable,eq:ch2-implicit-gradient} are often referred to as `the adjoint method'.
Readers should be aware that multiple equations and methods can be described as adjoint methods.
`Continuous' adjoint methods refers to the analytic manipulation of (ordinary or partial) differential equations to derive a different set of differential equations that compute gradients with respect to the solution of those equations \cite{kidger2022neural}.
The `discrete' adjoint method usually refers to the derivatives of discretized equations approximating the solution to differential equations; these derivatives can be computed either analytically or with AD \cite{nadarajah2000comparison}.
\Cref{eq:ch2-adjoint-variable,eq:ch2-implicit-gradient}, by contrast, can be used to compute the derivative of any implicitly defined function, though they are used in the discrete adjoint method.

The results of this subsection have a simple interpretation. The implicit equation $\bm g = 0$ can be understood as a primitive operation with input $\bm p \in \mathbb{R}^p$ and output $\bm u \in \mathbb{R}^d$.
\Cref{eq:ch2-implicit-gradient,eq:ch2-adjoint-variable} are how we compute the VJP of the primitive $\bm g = 0$.
$\frac{\partial f}{\partial \bm u}$ is the cotangent vector (the `v' in VJP), $\bm \lambda$ is the cotangent vector which gives the result of the VJP, and the linear system for $\bm \lambda$ (\cref{eq:ch2-adjoint-variable}), sometimes known as the `adjoint equation', is the linear equation whose solution gives the VJP of the primitive $\bm g = 0$.

Several examples of implicit functions are worth examining. Additional details and examples are discussed in \cite{blondel2022efficient}.

\subsubsection{Linear systems}

The linear system of equations $\bm g = \bm A(\bm p) \bm u - \bm b(\bm p)$ has $\frac{\partial \bm g}{\partial \bm u} = \bm A(\bm p)$, and thus the adjoint variable $\bm \lambda$ requires a linear solve with the transpose of $\bm A$:
\begin{equation}
    \bm A^T \bm \lambda = - \frac{\partial f}{\partial \bm u}.
\end{equation}

\subsubsection{Non-linear systems}

The non-linear system of equations $\bm g = \bm F$ can be solved using the Newton iteration $\bm u^{n+1} = \bm u^n + \Delta \bm u$ with $\bm F(\bm u^n + \Delta \bm u, \bm p) \approx \bm F(\bm u^n, \bm p) + \frac{\partial \bm F}{\partial \bm u}\Delta \bm u = 0$, which requires repeatedly finding solutions to the system of linear equations $\frac{\partial \bm F}{\partial \bm u}\Delta \bm u = - \bm F(\bm u^n, \bm p)$.
Solving for the adjoint variable 
only requires an additional linear solve:
\begin{equation}
    \bigg(\frac{\partial \bm F}{\partial \bm u}\bigg)^T \bm \lambda = - \frac{\partial f}{\partial \bm u}.
\end{equation}
The matrix $\frac{\partial \bm F}{\partial \bm u}$ is already constructed for the Newton iterations and thus can usually be reconstructed for the adjoint equation.
Notably, even though the non-linear solve for $\bm F = 0$ requires an iterative computation with repeated linear solves, computing the derivative of $\bm F = 0$ only requires a single linear solve.

\subsubsection{Fixed point iterations}

The fixed point iteration $\bm u = \bm \Phi(\bm u, \bm p)$ is another implicitly defined function. 
Fixed point iterations can be solved using iterative techniques. 
Setting $\bm g = \bm \Phi - \bm u$, then $\frac{\partial \bm g}{\partial \bm u} = \frac{\partial \bm \Phi}{\partial \bm u} - {I}$. 
Thus \cref{eq:ch2-adjoint-variable} then becomes
\begin{equation}
    \bm \lambda = \bigg(\frac{\partial \bm \Phi}{\partial \bm u}\bigg)^T\bm \lambda + \frac{\partial f}{\partial \bm u}
\end{equation}
which is also a fixed point iteration for $\bm \lambda$. 
In other words, the adjoint equation for a fixed point iteration is also a fixed point iteration.

\subsubsection{Stationary points}

A stationary point is where a derivative $\frac{\partial h}{\partial \bm u}$ of a scalar function $h(\bm u) : \mathbb{R}^n \to \mathbb{R}$ is zero.
This could arise, for example, as the result of an optimization problem
\begin{equation}
    \bm u^* = \argminC_{\bm u} h(\bm u, \bm p).
\end{equation}
In this case, $\bm g(\bm u(\bm p), \bm p) = \frac{\partial h}{\partial \bm u}$ and the gradient of a scalar function $\frac{\partial f}{\partial \bm u}$ can be computed using \cref{eq:ch2-implicit-gradient,eq:ch2-adjoint-variable}, replacing $\bm g$ with $\frac{\partial h}{\partial \bm u}$.

\subsubsection{Overdetermined and underdetermined systems of equations}

We want to solve $\bm G(\bm u(\bm p), \bm p) = 0$ where $\bm G : \mathbb{R}^n \times \mathbb{R}^p \to \mathbb{R}^m$. $\bm G = 0$ may be overdetermined or underdetermined and may have no solution or infinitely many solutions. Since $\bm G(\bm u(\bm p), \bm p) = 0 $ may not have a unique solution, $\bm u$ is typically chosen to be the solution to an optimization problem
\begin{equation}
    \bm u^* = \argminC_{\bm u} \frac{1}{2} ||\bm G||^2.
\end{equation}
The solution will be at a stationary point of $h(\bm u, \bm p) = \frac{1}{2}||\bm G ||^2$, where $\frac{\partial h}{\partial \bm u} = \bm G^T \frac{\partial \bm G}{\partial \bm u}$.
The matrices $\frac{\partial \bm g}{\partial \bm p} : \mathbb{R}^{n \times p}$ and $\frac{\partial \bm g}{\partial \bm u} : \mathbb{R}^{n \times n}$ in \cref{eq:ch2-implicit-gradient,eq:ch2-adjoint-variable} have $\bm g = \frac{\partial h}{\partial \bm u} = \bm G^T \frac{\partial \bm G}{\partial \bm u}$ and are given by
\begin{equation}
    \frac{\partial \bm g}{\partial \bm p} = \Big( \frac{\partial \bm G}{\partial \bm u}\Big)^T  \frac{\partial \bm G}{\partial \bm p} + \bm G^T \frac{\partial^2 \bm G}{\partial \bm u \partial \bm p}
\end{equation}
\begin{equation}
    \frac{\partial \bm g}{\partial \bm u} = \Big( \frac{\partial \bm G}{\partial \bm u}\Big)^T \frac{\partial \bm G}{\partial \bm u} + \bm G^T \frac{\partial^2 \bm G}{\partial \bm u^2}.
\end{equation}
The gradient $\frac{df}{d\bm p}$ can be written as
\begin{equation}\label{eq:ch2-implicit-gradient-overdetermined}
    \frac{df}{d\bm p} = \frac{\partial f}{\partial \bm p} + \bm \lambda^T \bigg(\Big( \frac{\partial \bm G}{\partial \bm u} \Big)^T\frac{\partial \bm G}{\partial \bm p} + \bm G^T \frac{\partial^2 \bm G}{\partial \bm u \partial \bm p}\bigg)
\end{equation}
with the adjoint variable given by
\begin{equation}\label{eq:ch2-adjoint-variable-overdetermined}
    \bigg(\Big( \frac{\partial \bm G}{\partial \bm u}\Big)^T\frac{\partial \bm G}{\partial \bm u} + \bm G^T \frac{\partial^2 \bm G}{\partial \bm u^2}\bigg)^T \bm \lambda = - \frac{\partial f}{\partial \bm u}.
\end{equation}
Notice that if $\bm G = 0$ and $\frac{\partial \bm G}{\partial \bm u}$ is square (so $m = n$) and non-singular, then a short calculation reveals that \cref{eq:ch2-adjoint-variable-overdetermined,eq:ch2-implicit-gradient-overdetermined} reduce to the usual formulas for implicit differentiation, \cref{eq:ch2-implicit-gradient,eq:ch2-adjoint-variable}.

\section{Stochastic automatic differentiation \label{sec:ch2-stochastic-gradients}}

Deep neural networks have taken center stage as a powerful way to construct and train massively-parametric ML models for supervised, unsupervised, and reinforcement learning tasks.
AD tools like PyTorch, TensorFlow, and JAX provide a computational substrate for rapidly exploring a wide variety of differentiable architectures without performing tedious and error-prone gradient derivations.
In modern ML workflows, first-order stochastic optimization is used to train these differentiable architectures.
Stochastic gradient descent variants such as AdaGrad \cite{duchi2011adaptive} and Adam \cite{kingma2014adam} form the core of almost all successful optimization techniques for these models, performing so-called `minibatch stochastic gradient descent' by using small subsets of the data to compute noisy but unbiased gradient estimates.

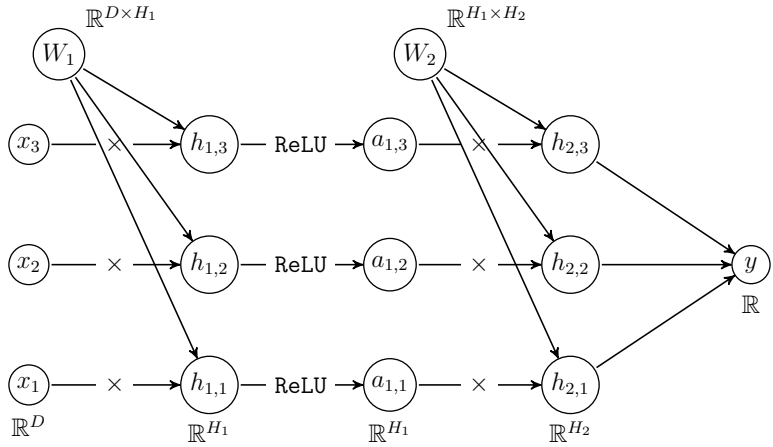
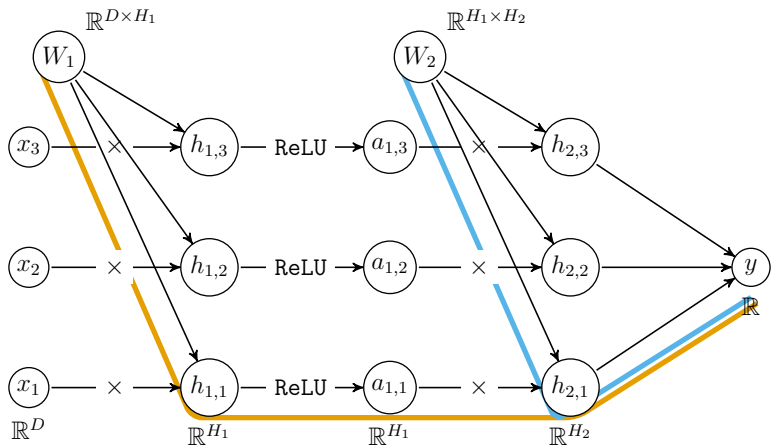
\begin{figure}
\centering
	\begin{subfigure}{0.7\textwidth}
		\centering
		\begin{tikzpicture}[scale=0.8,transform shape]
		
  			\tikzstyle{every node}=[node distance = 4cm,%
            		                bend angle    = 45]
			\Vertex[x=0.5, y=5.5, L=\texttt{$W_1$}]{W1}
			\extralabel[2.0mm]{45}{\texttt{$\mathbb{R}^{D \times H_1}$}}{W1}

			\Vertex[x=0, y=0, L=\texttt{$x_1$}]{x1}
			\Vertex[x=0, y=2, L=\texttt{$x_2$}]{x2}
			\Vertex[x=0, y=4, L=\texttt{$x_3$}]{x3}
			\extralabel[2mm]{270}{\texttt{$\mathbb{R}^D$}}{x1}

			\Vertex[x=3, y=0, L=\texttt{$h_{1,1}$}]{h11}
			\Vertex[x=3, y=2, L=\texttt{$h_{1,2}$}]{h12}
			\Vertex[x=3, y=4, L=\texttt{$h_{1,3}$}]{h13}
			\extralabel[3.0mm]{270}{\texttt{$\mathbb{R}^{H_1}$}}{h11}

			\Vertex[x=6, y=0, L=\texttt{$a_{1,1}$}]{a11} 
			\Vertex[x=6, y=2, L=\texttt{$a_{1,2}$}]{a12} 
			\Vertex[x=6, y=4, L=\texttt{$a_{1,3}$}]{a13} 
			\extralabel[3.0mm]{270}{\texttt{$\mathbb{R}^{H_1}$}}{a11}

			\Vertex[x=6.5, y=5.5, L=\texttt{$W_2$}]{W2}
			\extralabel[2.0mm]{45}{\texttt{$\mathbb{R}^{H_1 \times H_2}$}}{W2}

			\Vertex[x=9, y=0, L=\texttt{$h_{2,1}$}]{h21} 
			\Vertex[x=9, y=2, L=\texttt{$h_{2,2}$}]{h22} 
			\Vertex[x=9, y=4, L=\texttt{$h_{2,3}$}]{h23}
			\extralabel[3.0mm]{270}{\texttt{$\mathbb{R}^{H_2}$}}{h21}

			\Vertex[x=12, y=\nodefx, L=\texttt{$y$}]{y}
			\extralabel[2.0mm]{270}{\texttt{$\mathbb{R}$}}{y}

			\tikzstyle{EdgeStyle}=[post]
			\Edge[label=\texttt{$\times$}](x1)(h11)
			\Edge[label=\texttt{$\times$}](x2)(h12)
			\Edge[label=\texttt{$\times$}](x3)(h13)
			\Edge(W1)(h11)
			\Edge(W1)(h12)
			\Edge(W1)(h13)

			\Edge[label=\texttt{ReLU}](h11)(a11)
			\Edge[label=\texttt{ReLU}](h12)(a12)
			\Edge[label=\texttt{ReLU}](h13)(a13)

			\Edge[label=\texttt{$\times$}](a11)(h21)
			\Edge[label=\texttt{$\times$}](a12)(h22)
			\Edge[label=\texttt{$\times$}](a13)(h23)
			\Edge(W2)(h21)
			\Edge(W2)(h22)
			\Edge(W2)(h23)

			\Edge(h21)(y)
			\Edge(h22)(y)
			\Edge(h23)(y)
		\end{tikzpicture}
		\caption{Neural network computational graph}
		\label{fig:ch2-nn-graph}
		\end{subfigure}
		\begin{subfigure}{0.7\textwidth}	
		\centering%
		\begin{tikzpicture}[scale=0.8, transform shape]
		
  			\tikzstyle{every node}=[node distance = 4cm,%
            		                bend angle    = 45]
    		\draw[line width=2,rounded corners, color=cborange] 
    			([xshift=-4mm,yshift=0mm]0.5,5.5) -- 
    			([xshift=-3mm,yshift=-5mm]3,0) -- 
    			([xshift=-5mm,yshift=-5mm]6,0) -- 
    			([xshift=1mm,yshift=-5mm]9,0) -- 
    			([xshift=1mm,yshift=-6mm]12,2);

    		\draw[line width=2,rounded corners, color=cbblue] 
    			([xshift=-4mm,yshift=-0mm]6.5,5.5) -- 
    			([xshift=-2.5mm,yshift=-5.5mm]9,0) -- 
    			([xshift=0mm,yshift=-5mm]12,2);

			\Vertex[x=0.5, y=5.5, L=\texttt{$W_1$}]{W1}
			\extralabel[2.0mm]{45}{\texttt{$\mathbb{R}^{D \times H_1}$}}{W1}

			\Vertex[x=0, y=0, L=\texttt{$x_1$}]{x1}
			\Vertex[x=0, y=2, L=\texttt{$x_2$}]{x2}
			\Vertex[x=0, y=4, L=\texttt{$x_3$}]{x3}
			\extralabel[2mm]{270}{\texttt{$\mathbb{R}^D$}}{x1}

			\Vertex[x=3, y=0, L=\texttt{$h_{1,1}$}]{h11}
			\Vertex[x=3, y=2, L=\texttt{$h_{1,2}$}]{h12}
			\Vertex[x=3, y=4, L=\texttt{$h_{1,3}$}]{h13}
			\extralabel[3.0mm]{270}{\texttt{$\mathbb{R}^{H_1}$}}{h11}

			\Vertex[x=6, y=0, L=\texttt{$a_{1,1}$}]{a11} 
			\Vertex[x=6, y=2, L=\texttt{$a_{1,2}$}]{a12} 
			\Vertex[x=6, y=4, L=\texttt{$a_{1,3}$}]{a13} 
			\extralabel[3.0mm]{270}{\texttt{$\mathbb{R}^{H_1}$}}{a11}

			\Vertex[x=6.5, y=5.5, L=\texttt{$W_2$}]{W2}
			\extralabel[2.0mm]{45}{\texttt{$\mathbb{R}^{H_1 \times H_2}$}}{W2}

			\Vertex[x=9, y=0, L=\texttt{$h_{2,1}$}]{h21} 
			\Vertex[x=9, y=2, L=\texttt{$h_{2,2}$}]{h22} 
			\Vertex[x=9, y=4, L=\texttt{$h_{2,3}$}]{h23}
			\extralabel[3.0mm]{270}{\texttt{$\mathbb{R}^{H_2}$}}{h21}

			\Vertex[x=12, y=\nodefx, L=\texttt{$y$}]{y}
			\extralabel[2.0mm]{270}{\texttt{$\mathbb{R}$}}{y}

			\tikzstyle{EdgeStyle}=[post]
			\Edge[label=\texttt{$\times$}](x1)(h11)
			\Edge[label=\texttt{$\times$}](x2)(h12)
			\Edge[label=\texttt{$\times$}](x3)(h13)
			\Edge(W1)(h11)
			\Edge(W1)(h12)
			\Edge(W1)(h13)

			\Edge[label=\texttt{ReLU}](h11)(a11)
			\Edge[label=\texttt{ReLU}](h12)(a12)
			\Edge[label=\texttt{ReLU}](h13)(a13)

			\Edge[label=\texttt{$\times$}](a11)(h21)
			\Edge[label=\texttt{$\times$}](a12)(h22)
			\Edge[label=\texttt{$\times$}](a13)(h23)
			\Edge(W2)(h21)
			\Edge(W2)(h22)
			\Edge(W2)(h23)

			\Edge(h21)(y)
			\Edge(h22)(y)
			\Edge(h23)(y)
		\end{tikzpicture}
		\caption{Computational Graph with Mini-batching}
		\label{fig:nnpaths}

		\end{subfigure}
		\caption{NN computation graphs.}
		\label{fig:nnvisual}
	\end{figure}

An illustration of a computational graph for a simple two-layer ReLU neural network is shown in \cref{fig:ch2-nn-graph}.
In this example, the dataset is size 3 (individual datapoints are $x_1$, $x_2$, and $x_3$) and the parameters of the neural network are $W_1$ and $W_2$.

\Cref{fig:nnpaths} illustrates the concept of minibatch stochastic gradient descent, where a minibatch of size 1 is used to compute the stochastic gradient with respect to the loss.
Only the paths in the LCG corresponding to the input variable $x_1$, marked in blue and orange in \cref{fig:nnpaths}, are needed to compute stochastic estimates of $\frac{\partial y}{\partial W_1}$ and $\frac{\partial y}{\partial W_2}$.
Since the scalar objective function
\begin{equation}
    y = \frac{1}{N}\sum_{i=1}^N L(x_i)
\end{equation}
is a sum of losses for each individual datapoint, then the stochastic gradient estimate
\begin{equation}
    \grad \hat{y} = \frac{1}{|B|} \sum_{j \in B} \grad L(x_j)
\end{equation}
with minibatch $B$ of size $|B|$ consisting of datapoints $x_j$ sampled uniformly from the full dataset is an unbiased Monte Carlo estimate of the full gradient $\grad y$.

Note: the rest of \cref{ch:autodiff} is advanced material. Most readers may prefer to skip to \cref{ch:stellarator}.

\subsection{Alternative methods for stochastic gradients}

Minibatch stochastic gradient descent trades off memory and runtime for increased variance.
The question considered in this subsection is: on what theoretical principle does minibatch stochastic gradient descent work? 
Is it possible to use this theoretical principle to develop alternative methods besides minibatching for computing unbiased gradient estimators? Ideally, we would like to find alternative methods to reduce memory and/or runtime at the cost of increased variance.

As we'll see in \cref{sec:ch2-stochastic-case-studies}, except for certain computational graph structures, the framework for randomized automatic differentiation developed in this section cannot be used effectively as it leads to exponentially increasing variance (`exploding variance') as the depth of the computational graph increases.
Nevertheless, we consider two special computational graph structures (ReLU neural networks and a linear PDE) for which these alternative methods don't lead to exploding variance and construct reduced-memory unbiased estimators for them.

\subsubsection{Path sampling}

Baeur's formula (reproduced below) gives the Jacobian of a function computed using AD:
\begin{equation}
        J_{ij} = \frac{dy_i}{dx_j} = \sum_{[i \to j]} \prod_{(k, l) \in [i \to j]} \frac{\partial v_l}{\partial v_k}.
\end{equation}
Observe that in Bauer's formula each Jacobian entry is expressed as a sum over paths in the LCG.

A simple strategy for developing an unbiased estimator $\hat{\bm J}$ such that $\mathbb{E} \big[\hat{\bm J}\big] = \bm J$ is to sample paths from the computational graph, forming a Monte Carlo estimate of \cref{eq:ch2-bauers}.
Minibatch stochastic gradient descent is one such Monte Carlo estimator; it samples a subset of paths in the computational graph (as illustrated in \cref{fig:nnpaths}) and, because each path in the computational graph is equally likely to be sampled, results in an unbiased estimate of the gradient.
Minibatch stochastic gradient descent reduces memory and compute in exchange for increased variance, but requires knowledge of the structure of the computational graph and cannot be used in all computational graph structures. 

A general strategy for developing a Monte Carlo estimate of \cref{eq:ch2-bauers} which applies to all computational graph structures is to sample paths (rather than datapoints) at random from the computational graph.
Multiple paths can be sampled without significant computation overhead by performing a topological sort of the vertices and iterating through vertices, sampling multiple outgoing edges for
each vertex, as explained in \cref{alg:pathsampling}.
The main storage savings from \cref{alg:pathsampling} comes from Line 9, where we only consider a vertex if it has an incoming edge that has been sampled. In computational graphs with a large number of independent paths, this will significantly reduce memory required, whether we record intermediate variables and recompute the LCG or store entries of the LCG directly.
Note that the original computational graph is used for the forward pass, and randomization is used to determine a LCG to use for the backward pass in place of the original computation graph.

\begin{algorithm}
	\caption{Stochastic reverse mode AD with path sampling} 
	\begin{algorithmic}[1]
	    \State \textbf{Inputs}: 
	    \State \quad $\mathcal G = (V, E)$ - Computational Graph. $d_v$ denotes outdegree, $v.succ$ successor set of vertex $v$.
	    \State \quad $y$ - Output vertex
	    \State \quad $\Theta = (\theta_1, \theta_2, \hdots, \theta_m) \subset V$ - Input vertices
	    \State \quad $k > 0$ - Number of samples per vertex 
	    \State \textbf{Initialization}: 
	    \State \quad $Q(e) = 0, \forall e \in E$
		\For {$v$ in topological order; synchronous with forward computation}
		    \If {No incoming edge of $v$ has been sampled}
		        \State Continue
		    \EndIf
		    \For {$k$ times}
		        \State Sample $i$ from $[d_v]$ uniformly.
		        \State $Q(v, v.succ[i]) \gets Q(v, v.succ[i]) + \frac{d_v}{k} \frac{\partial v.succ[i]}{\partial v}$
		    \EndFor
		\EndFor
		\State Run reverse mode AD from $y$ to $\Theta$ using $Q$ as intermediate partials.
		\State \textbf{Output}: $\nabla_\Theta y$
	\end{algorithmic}
	\label{alg:pathsampling}
\end{algorithm}

To see that path sampling gives an unbiased gradient estimate, we use proof by induction. We traverse the vertices in reverse topological order. For every vertex $z$, we denote $\hat z$ as our stochastic estimate of $\bar z = \frac{\partial y}{\partial z}$. For our base case, we let $\hat y = \frac{d y}{d y} = 1$, so $\mathbb E \hat y = \bar y$. For the induction step on vertex $z$, the reverse mode AD cotangent update rule (\cref{eq:ch2-reverse-mode-update}) gives
\begin{equation}
    \hat{z} = \sum_{\substack{v \in \text{children} \\ \text{of } z}} \hat{v} \frac{\partial \hat{v}}{\partial z}
\end{equation}
where $\frac{\partial \hat{v}}{\partial z}$ is the stochastic edge weight $Q(z, v)$.
Thus
\begin{equation}
    \mathbb{E}[\hat{z}] =  \sum_{\substack{v \in \text{children} \\ \text{of } z}} \mathbb{E}[\hat{v}\frac{\partial \hat{v}}{\partial z}] = \sum_{\substack{v \in \text{children} \\ \text{of } z}} \mathbb{E}[\hat{v}] \mathbb{E}[\frac{\partial \hat{v}}{\partial z}] = \sum_{\substack{v \in \text{children} \\ \text{of } z}} \bar{v}\mathbb{E}[\frac{\partial \hat{v}}{\partial z}]
\end{equation}
using (a) the inductive hypothesis that $\mathbb{E}[\hat{v}] = \bar{v}$ and (b) the fact that $\hat{v}$ doesn't depend on $\frac{\partial \hat{v}}{\partial z}$ (due to the reverse topological ordering of vertices) and thus they are independent random variables.
We also have that
\begin{equation}
    \mathbb{E}[\frac{\partial \hat{v}}{\partial z}] = \sum_{i=1}^k p(v_i = v) \frac{d_z}{k} \frac{\partial v}{\partial z}
\end{equation}
where $d_z$ is the out-degree of $z$ and $p(v_i = v)$ is the probability that the sampled vertex $v_i$ on the $i$th draw is vertex $v$. Using line 12 of \cref{alg:pathsampling}, we see that $p(v_i = v) = \frac{1}{d_z}$.
This then simplifies to
\begin{equation}
    \mathbb{E}[\frac{\partial \hat{v}}{\partial z}] = \sum_{i=1}^k \frac{1}{d_z} \frac{d_z}{k} \frac{\partial v}{\partial z} = \frac{\partial v}{\partial z},
\end{equation}
which implies that
\begin{equation}
    \mathbb{E}[\hat{z}] = \sum_{\substack{v \in \text{children} \\ \text{of } z}} \bar{v}\mathbb{E}[\frac{\partial \hat{v}}{\partial z}] = \bar{v} \frac{\partial v}{\partial z} = \bar{z}
\end{equation}
which completes the proof.

\subsubsection{Random matrix injection}

In computation graphs consisting of vector operations, the vectorized computation graph is a more compact representation (see, for example, \cref{fig:ch2-rad-ex-b}). 
A single path in the vectorized computation graph represents many paths in the underlying scalar computation graph.
We introduce an alternative view on sampling paths in this case which we refer to as `random matrix injection'.

As an example of random matrix injection, consider once again the function given by \cref{eq:ch2-dp-ex} whose computational graph is sketched in \cref{fig:ch2-rad-ex}.
For concreteness, we set $d_1 = d_2 = d_3 = 3$.
The gradient is given by
\begin{align}
    \frac{\partial y}{\partial \theta} = \frac{\partial y}{\partial \bm c}\frac{\partial \bm c}{\partial \bm b} \frac{\partial \bm b}{\partial \bm a} \frac{\partial \bm a}{\partial \theta}
\end{align}
where $\bm a \in \mathbb{R}^3$, $\bm b \in \mathbb{R}^3$ and $\bm c \in \mathbb{R}^3$, $\frac{\partial \bm c}{\partial \bm b}, \frac{\partial \bm b}{\partial \bm a}$ are Jacobian matrices of shape $\mathbb{R}^{3\times 3}$, $\frac{\partial y}{\partial \bm c} \in \mathbb{R}^{1 \times 3}$, and $\frac{\partial \bm a}{\partial \theta} \in \mathbb{R}^{3 \times 1}$.

We now note that the contribution of the path $ p = \theta \to a_1 \to b_2 \to c_2 \to y$ to the gradient is,
\begin{align}
    \frac{\partial y}{\partial \bm c}P_2\frac{\partial \bm c}{\partial \bm c} P_2 \frac{\partial \bm b}{\partial \bm a} P_1 \frac{\partial \bm a}{\partial \theta}
\end{align}
where~${P_i = e_i e_i^T}$ (outer product of standard basis vectors). Sampling from~$\{P_1, P_2, P_3\}$ and right multiplying a Jacobian is equivalent to sampling the paths passing through a vertex in the scalar graph.

In general, if we have transition~${\bm b \to \bm c}$ in a vectorized computational graph, where~${\bm b \in \mathbb R^d, \bm c \in \mathbb R^m}$, we can insert a random matrix~${P = \nicefrac{d}{k} \sum_{s=1}^k P_{s}}$ where each~$P_s$ is sampled uniformly from~$\{P_1, P_2, \hdots, P_d\}$. With this construction,~${\mathbb{E}[P] = I_d}$, so
\begin{align}
    \mathbb E \left[\frac{\partial \bm c}{\partial \bm b} P \right] = \frac{\partial \bm c}{\partial \bm b}\,.
\end{align}

If we have a matrix chain product, we can use the fact that the expectation of a product of independent random variables is equal to the product of their expectations, so drawing independent random matrices $P_B$, $P_C$ would give
\begin{equation}
    \mathbb{E} \left[\frac{\partial y}{\partial \bm c}P_C\frac{\partial \bm c}{\partial \bm b} P_B\right] = \frac{\partial y}{\partial \bm c}\mathbb{E} \left[P_C\right]\frac{\partial \bm c}{\partial \bm b} \mathbb{E} \left[P_B\right] = \frac{\partial y}{\partial \bm c} \frac{\partial \bm c}{\partial \bm b}.
\end{equation}

Right multiplication by $P$ may be achieved by sampling the intermediate Jacobian: one does not need to actually assemble and multiply the two matrices. For clarity we adopt the notation ${\mathcal S_P \left[\frac{\partial \bm c}{\partial \bm b}\right] = \frac{\partial \bm c}{\partial \bm b} P}$. This is sampling (with replacement) $k$ out of the $d$ vertices represented by $\bm b$, and only considering paths that pass from those vertices.

The important properties of $P$ that enable memory savings with an unbiased approximation are
\begin{align}
    \mathbb E P &= I_d & &\text{and}&
    P = RR^T, R &\in \mathbb R^{d \times k}, k < d\,.
\end{align}
We could therefore consider other matrices with the same properties. 

In vectorized computational graphs, we can imagine a two-level sampling scheme. We can both sample paths from the computational graph where each vertex on the path corresponds to a vector. We can also sample within each vector path, with sampling performed via matrix injection as above.

How random matrix injection can actually create memory savings in practice depends on the primitive operation.
In many situations the full intermediate Jacobian for a vector operation is unreasonable to store. Consider the operation $B \to C$ where $B, C \in \mathbb R^d$. The Jacobian is $d \times d$. Thankfully many common operations are element-wise, leading to a diagonal Jacobian that can be stored as a $d$-vector.
In our implementations of elementwise operations, we do not directly construct and sparsify the Jacobians. We instead sparsify the input vectors or the compact version of the Jacobian in a way that has the same effect.
Another common operation is matrix-vector products. Consider $Ab = c$, $\nicefrac{\partial c}{\partial b} = A$. Although $A$ has many more entries than $c$ or $b$, in many applications $A$ is either a parameter to be optimized (and thus already stored in memory) or is easily recomputed.
In these cases, random matrix injection is unnecessary.
Unfortunately, some primitive operations do not have a compactly-representable Jacobian.
In these cases, it is possible to assemble the full Jacobian and project it into a lower dimensional space, but doing so will not save memory relative to the standard VJP update rule.
The softmax function $\sigma_i(\bm z) = \frac{e^{z_i}}{\sum_{j=1}^k z_i}$ is one such primitive operation, as it has the Jacobian $J_{ij} = \sigma_i \delta_{ij} - \sigma_i \sigma_j$.
It is possible, however, to implement softmax using elementwise primitives, each of which has their own compactly-representable Jacobian and for which matrix injection can be used.

\subsection{Case studies in stochastic AD \label{sec:ch2-stochastic-case-studies}}

\subsubsection{The exploding variance problem}

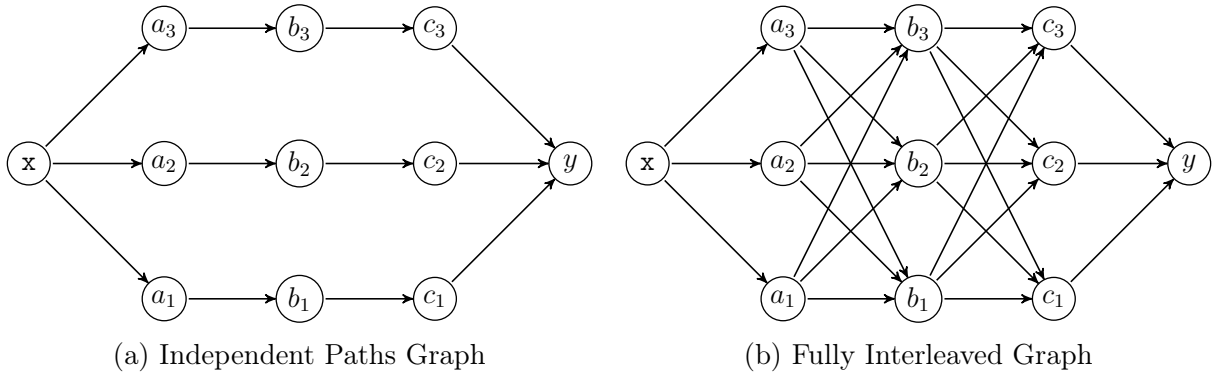
\begin{figure}
\centering
	\begin{subfigure}{0.49\textwidth}
		\centering%
		\begin{tikzpicture}[scale=0.9,transform shape]
  			\tikzstyle{every node}=[node distance = 4cm, bend angle    = 45]
			\Vertex[x=0, y=2, L=\texttt{x}]{x1}

			\Vertex[x=2, y=0, L=\texttt{$a_1$}]{a1} 
			\Vertex[x=2, y=2, L=\texttt{$a_2$}]{a2} 
			\Vertex[x=2, y=4, L=\texttt{$a_3$}]{a3} 

			\Vertex[x=4, y=0, L=\texttt{$b_1$}]{b1} 
			\Vertex[x=4, y=2, L=\texttt{$b_2$}]{b2} 
			\Vertex[x=4, y=4, L=\texttt{$b_3$}]{b3} 

			\Vertex[x=6, y=0, L=\texttt{$c_1$}]{c1} 
			\Vertex[x=6, y=2, L=\texttt{$c_2$}]{c2} 
			\Vertex[x=6, y=4, L=\texttt{$c_3$}]{c3} 

			\Vertex[x=\nodefy, y=\nodefx, L=\texttt{$y$}]{y}
			\tikzstyle{EdgeStyle}=[post]
			\Edge(x1)(a1)
			\Edge(x1)(a2)
			\Edge(x1)(a3)
			\Edge(a1)(b1)
			\Edge(a2)(b2)
			\Edge(a3)(b3)
			\Edge(b1)(c1)
			\Edge(b2)(c2)
			\Edge(b3)(c3)
			\Edge(c1)(y)
			\Edge(c2)(y)
			\Edge(c3)(y)
		\end{tikzpicture}
		\caption{Independent Paths Graph}
		\label{fig:good-graph}
	\end{subfigure}
	\begin{subfigure}{0.49\textwidth}
	\centering%
		\begin{tikzpicture}[scale=0.9,transform shape]
		
  			\tikzstyle{every node}=[node distance = 4cm,%
            		                bend angle    = 45]
			\Vertex[x=0, y=2, L=\texttt{x}]{x1}

			\Vertex[x=2, y=0, L=\texttt{$a_1$}]{a1} 
			\Vertex[x=2, y=2, L=\texttt{$a_2$}]{a2} 
			\Vertex[x=2, y=4, L=\texttt{$a_3$}]{a3} 

			\Vertex[x=4, y=0, L=\texttt{$b_1$}]{b1} 
			\Vertex[x=4, y=2, L=\texttt{$b_2$}]{b2} 
			\Vertex[x=4, y=4, L=\texttt{$b_3$}]{b3} 

			\Vertex[x=6, y=0, L=\texttt{$c_1$}]{c1} 
			\Vertex[x=6, y=2, L=\texttt{$c_2$}]{c2} 
			\Vertex[x=6, y=4, L=\texttt{$c_3$}]{c3} 

			\Vertex[x=\nodefy, y=\nodefx, L=\texttt{$y$}]{y}

			\tikzstyle{EdgeStyle}=[post]
			\Edge(x1)(a1)
			\Edge(x1)(a2)
			\Edge(x1)(a3)
			\Edge(a1)(b1)
			\Edge(a1)(b2)
			\Edge(a1)(b3)
			\Edge(a2)(b1)
			\Edge(a2)(b2)
			\Edge(a2)(b3)
			\Edge(a3)(b1)
			\Edge(a3)(b2)
			\Edge(a3)(b3)
			\Edge(b1)(c1)
			\Edge(b1)(c2)
			\Edge(b1)(c3)
			\Edge(b2)(c1)
			\Edge(b2)(c2)
			\Edge(b2)(c3)
			\Edge(b3)(c1)
			\Edge(b3)(c2)
			\Edge(b3)(c3)
			\Edge(c1)(y)
			\Edge(c2)(y)
			\Edge(c3)(y)
		\end{tikzpicture}
		\caption{Fully Interleaved Graph}
		\label{fig:bad-graph}
	\end{subfigure}
		\caption{\small Common computational graph patterns. The graphs may be arbitrarily deep and wide. (a) A small number of independent paths. Path sampling has constant variance with depth. (b) The number of paths increases exponentially with depth; path sampling and matrix injection give exponentially increasing variance with depth.
		}
		\label{fig:graphcompare}
\end{figure}

The variance incurred by path sampling and random matrix injection will depend on the structure of the LCG. We present two extremes in \cref{fig:graphcompare}. In \cref{fig:good-graph}, each path is independent and there are a small number of paths. If we sample a fixed fraction of all paths, variance will be constant in the depth of the graph. In contrast, in \cref{fig:bad-graph}, the paths overlap, and the number of paths increases exponentially with depth. Sampling a fixed fraction of all paths would require almost all edges in the graph, and sampling a fixed fraction of vertices at each layer (using random matrix injection, as an example) would lead to exponentially increasing variance with depth.
We refer to this tendency for exponentially increasing variance as the \textit{exploding variance problem}.

Thus, unfortunately, stochastic gradient methods will in general lead to exploding variance.
Indeed, our initial efforts to apply random matrix injection schemes to neural network graphs resulted in variance exponential with depth of the network, which prevented stochastic optimization from converging.

For certain computational graph structures, however, using a priori knowledge of the computational graph it is possible to apply stochastic gradient methods to reduce memory consumption without exploding variance.
In the rest of this subsection, we develop tailored sampling strategies for computation graphs corresponding to problems of common interest, exploiting properties of these graphs to avoid the exploding variance problem.
It is worth noting that these tailored sampling strategies rely on unique features of these computational graphs, in particular the use of ReLU activations in case study 1 and a linear explicit timestepping update in case study 2.
In the author's opinion, due to the limited set of computational graphs which do not have exploding variance, these alternative stochastic gradient methods are unlikely to be useful for practical optimization problems.

\subsubsection{Case study 1: ReLU neural networks}

We consider neural networks composed of fully connected layers, convolution layers, ReLU nonlinearities, and pooling layers. We take advantage of the important property that many of the intermediate Jacobians can be compactly stored, and the memory required during reverse-mode is often bottlenecked by a few operations. We draw a vectorized computational graph for a typical simple neural network in figure \ref{fig:nnvisual}. Although the diagram depicts a dataset of size of $3$, mini-batch size of size $1$, and $2$ hidden layers, we assume the dataset size is $N$. Our analysis is valid for any number of hidden layers, and also recurrent networks. We are interested in the gradients $\nicefrac{\partial y}{\partial W_1}$ and $\nicefrac{\partial y}{\partial W_2}$.

We wish to use our principles to derive a randomization scheme that can be used on top of mini-batch SGD. We ensure our estimator is unbiased as we randomize by applying random matrix injection independently to various intermediate Jacobians. Consider a path corresponding to data point $1$. The contribution to the gradient $\nicefrac{\partial y}{\partial W_1}$ is
\begin{align}
    \frac{\partial y}{\partial h_{2, 1}} \frac{\partial h_{2, 1}}{\partial a_{1, 1}} \frac{\partial a_{1, 1}}{\partial h_{1, 1}} \frac{\partial h_{1, 1}}{\partial W_1}.
\end{align}
Using random matrix injection to sample every Jacobian would lead to exploding variance. Instead, we analyze each term to see which are memory bottlenecks. 

$\nicefrac{\partial y}{\partial h_{2, 1}}$ is the Jacobian with respect to (typically) the loss. Memory requirements for this Jacobian are independent of depth of the network. The dimension of the classifier is usually smaller ($10-1000$) than the other layers (which can have dimension $10,000$ or more in convolutional networks). Therefore, the Jacobian at the output layer is not a memory bottleneck.

$\nicefrac{\partial h_{2, 1}}{\partial a_{1, 1}}$ is the Jacobian of the hidden layer with respect to the previous layer activation. This can be constructed from $W_2$, which must be stored in memory, with memory cost independent of mini-batch size. In convnets, due to weight sharing, the effective dimensionality is much smaller than $H_1 \times H_2$. In recurrent networks, it is shared across timesteps. Therefore, these are not a memory bottleneck.

$\nicefrac{\partial a_{1, 1}}{\partial h_{1, 1}}$ contains the Jacobian of the ReLU activation function. This can be compactly stored using $1$-bit per entry, as the gradient can only be $1$ or $0$. Note that this is true for ReLU activations in particular, and not true for general activation functions, although ReLU is widely used in deep learning. For ReLU activations, these partials are not a memory bottleneck.

$\nicefrac{\partial h_{1, 1}}{\partial W_1}$ contains the memory bottleneck for typical ReLU neural networks. This is the Jacobian of the hidden layer output with respect to $W_1$, which, in a multi-layer perceptron, is equal to $x_1$. For $B$ data points, this is a $B \times D$ dimensional matrix.

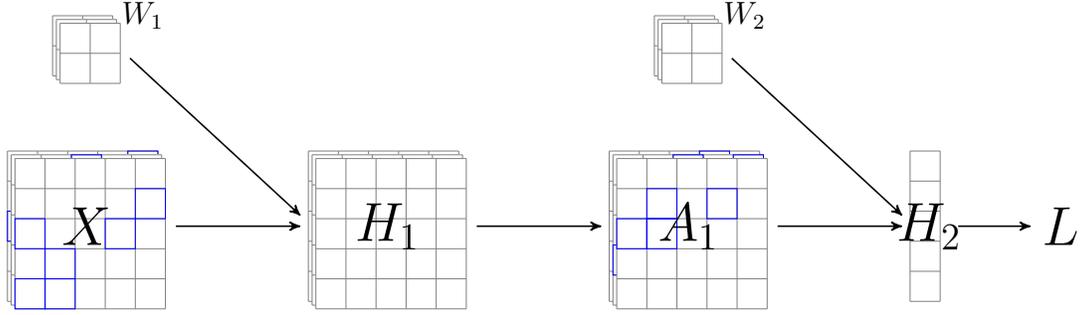
\begin{figure}
\centering
\begin{tikzpicture}[scale=1.0,transform shape]
	{
		\newcommand*{\xMin}{0}%
		\newcommand*{\xMax}{5}%
		\newcommand*{\xMaxm}{4}%
		\newcommand*{\yMin}{0}%
		\newcommand*{\yMax}{5}%
		\newcommand*{\yMaxm}{4}%
		\newcommand*{\tick}{0.4}

		\newcommand*{\xstart}{0.0}
		\newcommand*{\ystart}{0.0}
		\node at (\xstart + \xMin * \tick +0.05, \ystart + \yMax * \tick / 2.0) (l1i) {};

		\foreach \j in {0, ..., 2} {
		    \newcommand*{\xoffset}{0.05 * \j}
		    \newcommand*{\yoffset}{-0.05 * \j}
			\fill [white] (\xstart + \xoffset,\ystart + \yoffset) rectangle (\xstart + \tick * \xMax + \xoffset, \ystart + \tick * \yMax + \yoffset);
		    \foreach \i in {\xMin,...,\xMax} {
		        \draw [very thin,gray] (\xstart + \i * \tick + \xoffset,\ystart + \yMin * \tick + \yoffset) -- (\xstart + \i * \tick + \xoffset,\ystart + \yMax * \tick + \yoffset);
		    }
		    \foreach \i in {\yMin,...,\yMax} {
		        \draw [very thin,gray] (\xstart + \xMin * \tick + \xoffset,\ystart + \i * \tick + \yoffset) -- (\xstart + \xMax * \tick + \xoffset,\ystart + \i * \tick + \yoffset);
		    }
			\setrand{1}{4}{1}{18 + \j}
			\foreach \i in {\xMin, ..., \xMaxm} {
				\foreach \k in {\yMin, ..., \yMaxm} {
					\nextrand
					\ifnum \thisrand>3 {
						\newcommand*{\cornerx}{\xoffset + \i * \tick + \xstart}
						\newcommand*{\cornery}{\yoffset + \k * \tick + \ystart}
						\draw [blue, very thin] (\cornerx,\cornery) rectangle (\cornerx + \tick, \cornery + \tick);

					} \else {}\fi
				}
			}

		}
		\node at (\xstart + \xMax * \tick +0.05, \ystart + \yMax * \tick / 2.0) (l1o) {};
		\node at (\xstart + \xMax * \tick / 2.0 +0.05, \ystart + \yMax * \tick / 2.0) (x) {\LARGE $X$};

	}

	{
		\newcommand*{\xMin}{0}%
		\newcommand*{\xMax}{2}%
		\newcommand*{\xMaxm}{2}%
		\newcommand*{\yMin}{0}%
		\newcommand*{\yMax}{2}%
		\newcommand*{\yMaxm}{2}%
		\newcommand*{\tick}{0.4}

		\newcommand*{\xstart}{0.6}
		\newcommand*{\ystart}{3.0}
		\node at (\xstart + \xMin * \tick +0.05, \ystart + \yMax * \tick / 2.0) (w1i) {};

		\foreach \j in {0, ..., 2} {
		    \newcommand*{\xoffset}{0.05 * \j}
		    \newcommand*{\yoffset}{-0.05 * \j}
			\fill [white] (\xstart + \xoffset,\ystart + \yoffset) rectangle (\xstart + \tick * \xMax + \xoffset, \ystart + \tick * \yMax + \yoffset);
		    \foreach \i in {\xMin,...,\xMax} {
		        \draw [very thin,gray] (\xstart + \i * \tick + \xoffset,\ystart + \yMin * \tick + \yoffset) -- (\xstart + \i * \tick + \xoffset,\ystart + \yMax * \tick + \yoffset);
		    }
		    \foreach \i in {\yMin,...,\yMax} {
		        \draw [very thin,gray] (\xstart + \xMin * \tick + \xoffset,\ystart + \i * \tick + \yoffset) -- (\xstart + \xMax * \tick + \xoffset,\ystart + \i * \tick + \yoffset);
		    }
		}
		\node at (\xstart + \xMax * \tick +0.05, \ystart + \yMax * \tick / 2.0) (w1o) {};
		\node at (\xstart + \xMax * \tick +0.4, \ystart + \yMax * \tick) (w1l) {$W_1$};
	}

	{
		\newcommand*{\xMin}{0}%
		\newcommand*{\xMax}{5}%
		\newcommand*{\xMaxm}{4}%
		\newcommand*{\yMin}{0}%
		\newcommand*{\yMax}{5}%
		\newcommand*{\yMaxm}{4}%
		\newcommand*{\tick}{0.4}

		\newcommand*{\xstart}{4.0}
		\newcommand*{\ystart}{0.0}
		\node at (\xstart + \xMin * \tick +0.05, \ystart + \yMax * \tick / 2.0) (l2i) {};

		\foreach \j in {0, ..., 2} {
		    \newcommand*{\xoffset}{0.05 * \j}
		    \newcommand*{\yoffset}{-0.05 * \j}
			\fill [white] (\xstart + \xoffset,\ystart + \yoffset) rectangle (\xstart + \tick * \xMax + \xoffset, \ystart + \tick * \yMax + \yoffset);
		    \foreach \i in {\xMin,...,\xMax} {
		        \draw [very thin,gray] (\xstart + \i * \tick + \xoffset,\ystart + \yMin * \tick + \yoffset) -- (\xstart + \i * \tick + \xoffset,\ystart + \yMax * \tick + \yoffset);
		    }
		    \foreach \i in {\yMin,...,\yMax} {
		        \draw [very thin,gray] (\xstart + \xMin * \tick + \xoffset,\ystart + \i * \tick + \yoffset) -- (\xstart + \xMax * \tick + \xoffset,\ystart + \i * \tick + \yoffset);
		    }
		}
		\node at (\xstart + \xMax * \tick +0.05, \ystart + \yMax * \tick / 2.0) (l2o) {};
		\node at (\xstart + \xMax * \tick / 2.0 +0.05, \ystart + \yMax * \tick / 2.0) (x) {\LARGE $H_1$};

	}

	{
		\newcommand*{\xMin}{0}%
		\newcommand*{\xMax}{5}%
		\newcommand*{\xMaxm}{4}%
		\newcommand*{\yMin}{0}%
		\newcommand*{\yMax}{5}%
		\newcommand*{\yMaxm}{4}%
		\newcommand*{\tick}{0.4}

		\newcommand*{\xstart}{8.0}
		\newcommand*{\ystart}{0.0}
		\node at (\xstart + \xMin * \tick +0.05, \ystart + \yMax * \tick / 2.0) (l3i) {};

		\foreach \j in {0, ..., 2} {
		    \newcommand*{\xoffset}{0.05 * \j}
		    \newcommand*{\yoffset}{-0.05 * \j}
			\fill [white] (\xstart + \xoffset,\ystart + \yoffset) rectangle (\xstart + \tick * \xMax + \xoffset, \ystart + \tick * \yMax + \yoffset);
		    \foreach \i in {\xMin,...,\xMax} {
		        \draw [very thin,gray] (\xstart + \i * \tick + \xoffset,\ystart + \yMin * \tick + \yoffset) -- (\xstart + \i * \tick + \xoffset,\ystart + \yMax * \tick + \yoffset);
		    }
		    \foreach \i in {\yMin,...,\yMax} {
		        \draw [very thin,gray] (\xstart + \xMin * \tick + \xoffset,\ystart + \i * \tick + \yoffset) -- (\xstart + \xMax * \tick + \xoffset,\ystart + \i * \tick + \yoffset);
		    }
			\setrand{1}{4}{1}{14 + \j}
			\foreach \i in {\xMin, ..., \xMaxm} {
				\foreach \k in {\yMin, ..., \yMaxm} {
					\nextrand
					\ifnum \thisrand>3 {
						\newcommand*{\cornerx}{\xoffset + \i * \tick + \xstart}
						\newcommand*{\cornery}{\yoffset + \k * \tick + \ystart}
						\draw [blue, very thin] (\cornerx,\cornery) rectangle (\cornerx + \tick, \cornery + \tick);

					} \else {}\fi
				}
			}

		}
		\node at (\xstart + \xMax * \tick +0.05, \ystart + \yMax * \tick / 2.0) (l3o) {};
		\node at (\xstart + \xMax * \tick / 2.0 +0.05, \ystart + \yMax * \tick / 2.0) (x) {\LARGE $A_1$};
	}

	{
		\newcommand*{\xMin}{0}%
		\newcommand*{\xMax}{2}%
		\newcommand*{\xMaxm}{2}%
		\newcommand*{\yMin}{0}%
		\newcommand*{\yMax}{2}%
		\newcommand*{\yMaxm}{2}%
		\newcommand*{\tick}{0.4}

		\newcommand*{\xstart}{8.6}
		\newcommand*{\ystart}{3.0}
		\node at (\xstart + \xMin * \tick +0.05, \ystart + \yMax * \tick / 2.0) (w2i) {};

		\foreach \j in {0, ..., 2} {
		    \newcommand*{\xoffset}{0.05 * \j}
		    \newcommand*{\yoffset}{-0.05 * \j}
			\fill [white] (\xstart + \xoffset,\ystart + \yoffset) rectangle (\xstart + \tick * \xMax + \xoffset, \ystart + \tick * \yMax + \yoffset);
		    \foreach \i in {\xMin,...,\xMax} {
		        \draw [very thin,gray] (\xstart + \i * \tick + \xoffset,\ystart + \yMin * \tick + \yoffset) -- (\xstart + \i * \tick + \xoffset,\ystart + \yMax * \tick + \yoffset);
		    }
		    \foreach \i in {\yMin,...,\yMax} {
		        \draw [very thin,gray] (\xstart + \xMin * \tick + \xoffset,\ystart + \i * \tick + \yoffset) -- (\xstart + \xMax * \tick + \xoffset,\ystart + \i * \tick + \yoffset);
		    }
		}
		\node at (\xstart + \xMax * \tick +0.05, \ystart + \yMax * \tick / 2.0) (w2o) {};
		\node at (\xstart + \xMax * \tick +0.4, \ystart + \yMax * \tick) (w2l) {$W_2$};
	}

	{
		\newcommand*{\xMin}{0}%
		\newcommand*{\xMax}{1}%
		\newcommand*{\xMaxm}{0}%
		\newcommand*{\yMin}{0}%
		\newcommand*{\yMax}{5}%
		\newcommand*{\yMaxm}{4}%
		\newcommand*{\tick}{0.4}

		\newcommand*{\xstart}{12.0}
		\newcommand*{\ystart}{0.0}
		\node at (\xstart + \xMin * \tick +0.05, \ystart + \yMax * \tick / 2.0) (l4i) {};

		\foreach \j in {0} {
		    \newcommand*{\xoffset}{0.05 * \j}
		    \newcommand*{\yoffset}{-0.05 * \j}
			\fill [white] (\xstart + \xoffset,\ystart + \yoffset) rectangle (\xstart + \tick * \xMax + \xoffset, \ystart + \tick * \yMax + \yoffset);
		    \foreach \i in {\xMin,...,\xMax} {
		        \draw [very thin,gray] (\xstart + \i * \tick + \xoffset,\ystart + \yMin * \tick + \yoffset) -- (\xstart + \i * \tick + \xoffset,\ystart + \yMax * \tick + \yoffset);
		    }
		    \foreach \i in {\yMin,...,\yMax} {
		        \draw [very thin,gray] (\xstart + \xMin * \tick + \xoffset,\ystart + \i * \tick + \yoffset) -- (\xstart + \xMax * \tick + \xoffset,\ystart + \i * \tick + \yoffset);
		    }
		}
		\node at (\xstart + \xMax * \tick +0.05, \ystart + \yMax * \tick / 2.0) (l4o) {};
		\node at (\xstart + \xMax * \tick / 2.0 +0.05, \ystart + \yMax * \tick / 2.0) (x) {\LARGE $H_2$};
	}

	\node at (14, 0.4 * 5/2) (loss) {\LARGE $L$};

	\tikzstyle{EdgeStyle}=[post]
	\Edge(l1o)(l2i)
	\Edge(w1o)(l2i)
	\Edge(l2o)(l3i)
	\Edge(w2o)(l4i)
	\Edge(l3o)(l4i)
	\Edge(l4o)(loss)

	\end{tikzpicture}
	\caption{Convnet activation sampling for one mini-batch element. $X$ is the image, $H$ is the pre-activation, and $A$ is the activation. $A$ is the output of a ReLU, so we can store the Jacobian $\nicefrac{\partial A_1}{\partial H_1}$ with $1$ bit per entry. For $X$ and $H$ we sample spatial elements and compute the Jacobians $\nicefrac{\partial H_1}{\partial W_1}$ and $\nicefrac{\partial H_2}{\partial W_2}$ with the sparse tensors.}
	\label{fig:convnetsample}
\end{figure}

Accordingly, we choose to sample $\nicefrac{\partial h_{1, 1}}{\partial W_1}$, replacing the matrix chain with~${\frac{\partial y}{\partial h_{2, 1}} \frac{\partial h_{2, 1}}{\partial a_{1, 1}} \frac{\partial a_{1, 1}}{\partial h_{1, 1}} \mathcal S_{P_{W_1}} \left[\frac{\partial h_{1, 1}}{\partial W_1}\right]}$. For an arbitrarily deep NN, this can be generalized:
\begin{align}
    &\frac{\partial y}{\partial h_{d, 1}} \frac{\partial h_{d, 1}}{\partial a_{d-1, 1}} \frac{\partial a_{d-1, 1}}{\partial h_{d-1, 1}} \hdots \frac{\partial a_{1, 1}}{\partial h_{1, 1}} \mathcal S_{P^{W_1}}\!\!\! \left[\frac{\partial h_{1, 1}}{\partial W_1}\right], & 
    &\frac{\partial y}{\partial h_{d, 1}} \frac{\partial h_{d, 1}}{\partial a_{d-1, 1}} \frac{\partial a_{d-1, 1}}{\partial h_{d-1, 1}} \hdots \frac{\partial a_{2, 1}}{\partial h_{2, 1}} \mathcal S_{P^{W_2}}\!\!\! \left[\frac{\partial h_{2, 1}}{\partial W_2}\right]
\end{align}
This can be interpreted as sampling activations on the backward pass. This is our proposed alternative SGD scheme for neural networks: along with sampling data points, we can also sample activations, while maintaining an unbiased approximation to the gradient. This does not lead to exploding variance, as along any path from a given neural network parameter to the loss, the sampling operation is only applied to a single Jacobian. Sampling for convolutional networks is visualized in Figure \ref{fig:convnetsample}.

Experiments using this randomization scheme to (in theory) reduce the memory usage of a small fully connected network trained on MNIST and a small convolutional network trained on CIFAR-10 can be found in \cite{oktay2021randomized}.

\subsubsection{Case study 2: reaction-diffusion PDE}

Our second application is motivated by the observation that many scientific computing problems involve a repeated or iterative computation resulting in a layered computational graph. We may apply RAD to get a stochastic estimate of the gradient by subsampling paths through the computational graph. For certain problems, we can leverage problem structure to develop a low-memory stochastic gradient estimator without exploding variance. To illustrate this possibility we consider the optimization of a linear reaction-diffusion PDE on a square domain with Dirichlet boundary conditions, representing the production and diffusion of neutrons in a fission reactor \cite{McClarren2018}. Simulating this process involves solving for a potential~$\phi(x,y,t)$ varying in two spatial coordinates and in time. The solution obeys the partial differential equation:
\begin{align}
    \frac{\partial \phi(x,y,t)}{\partial t } = D \bm\nabla^2 \phi(x,y,t) + C(x,y,t,\bm\theta) \phi(x,y,t)
\end{align}
We discretize the PDE in time and space and solve on a spatial grid using an explicit update rule~${\bm\phi_{t+1} = \bm{M} \bm{\phi}_{t} + \Delta t\bm C_t \odot \bm \phi_t}$, where $M$ summarizes the discretization of the PDE in space. The exact form is available in the appendix. The initial condition is~${\bm\phi_0=\sin{(\pi x)}\sin{(\pi y)}}$, with~${\bm\phi = 0}$ on the boundary of the domain. The loss function is the time-averaged squared error between~$\bm \phi$ and a time-dependent target,~${L = \nicefrac{1}{T}\sum_{t} ||\bm\phi_t(\bm\theta) - \bm \phi^{\text{target}}_t||^2_2}$.
The target is~${\bm\phi^{\text{target}}_t = \bm\phi_0 + \nicefrac{1}{4}\sin{(\pi t)}\sin{(2\pi x)}\sin{(\pi y)}}$. The source $C$ is given by a seven-term Fourier series in $x$ and $t$, with coefficients given by ${\bm\theta \in \mathbb{R}^7}$, where $\bm\theta$ is the control parameter to be optimized. 
The parameters $\bm\theta$, which are to be optimized, are the coefficients of that Fourier series.

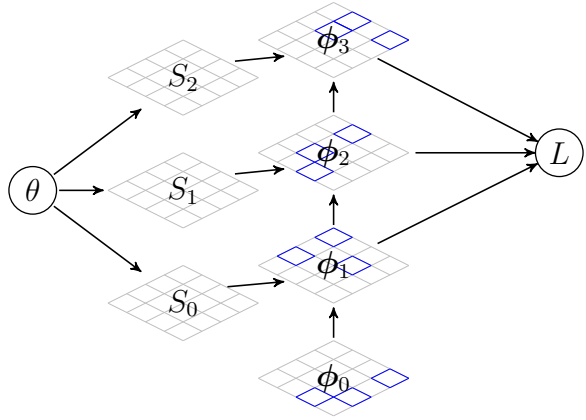
\begin{figure}
\centering
		\begin{tikzpicture}[scale=1.0,transform shape]
\tikzstyle{every node}=[node distance = 4cm, bend angle    = 45]
\Vertex[x=0, y=2.5, L=\texttt{$\theta$}]{t}
\draw[lightgray, very thin] (-1.0+2,0.0+1) -- (0.0+2,0.5+1);
\draw[lightgray, very thin] (0.0+2,0.5+1) -- (1.0+2,0.0+1);
\draw[lightgray, very thin] (1.0+2,0.0+1) -- (0.0+2,-0.5+1);
\draw[lightgray, very thin] (0.0+2,-0.5+1) -- (-1.0+2,0.0+1);
\draw[lightgray, very thin] (-0.5+2,-0.25+1) -- (0.5+2,0.25+1);
\draw[lightgray, very thin] (0.5+2,-0.25+1) -- (-0.5+2,0.25+1);
\draw[lightgray, very thin] (-0.75+2,-0.125+1) -- (0.25+2,0.375+1);
\draw[lightgray, very thin] (-0.25+2,-0.375+1) -- (0.75+2,0.125+1);
\draw[lightgray, very thin] (0.75+2,-0.125+1) -- (-0.25+2,0.375+1);
\draw[lightgray, very thin] (0.25+2,-0.375+1) -- (-0.75+2,0.125+1);
\draw[lightgray, very thin] (-1.0+2,0.0+2.5) -- (0.0+2,0.5+2.5);
\draw[lightgray, very thin] (0.0+2,0.5+2.5) -- (1.0+2,0.0+2.5);
\draw[lightgray, very thin] (1.0+2,0.0+2.5) -- (0.0+2,-0.5+2.5);
\draw[lightgray, very thin] (0.0+2,-0.5+2.5) -- (-1.0+2,0.0+2.5);
\draw[lightgray, very thin] (-0.5+2,-0.25+2.5) -- (0.5+2,0.25+2.5);
\draw[lightgray, very thin] (0.5+2,-0.25+2.5) -- (-0.5+2,0.25+2.5);
\draw[lightgray, very thin] (-0.75+2,-0.125+2.5) -- (0.25+2,0.375+2.5);
\draw[lightgray, very thin] (-0.25+2,-0.375+2.5) -- (0.75+2,0.125+2.5);
\draw[lightgray, very thin] (0.75+2,-0.125+2.5) -- (-0.25+2,0.375+2.5);
\draw[lightgray, very thin] (0.25+2,-0.375+2.5) -- (-0.75+2,0.125+2.5);
\draw[lightgray, very thin] (-1.0+2,0.0+4) -- (0.0+2,0.5+4);
\draw[lightgray, very thin] (0.0+2,0.5+4) -- (1.0+2,0.0+4);
\draw[lightgray, very thin] (1.0+2,0.0+4) -- (0.0+2,-0.5+4);
\draw[lightgray, very thin] (0.0+2,-0.5+4) -- (-1.0+2,0.0+4);
\draw[lightgray, very thin] (-0.5+2,-0.25+4) -- (0.5+2,0.25+4);
\draw[lightgray, very thin] (0.5+2,-0.25+4) -- (-0.5+2,0.25+4);
\draw[lightgray, very thin] (-0.75+2,-0.125+4) -- (0.25+2,0.375+4);
\draw[lightgray, very thin] (-0.25+2,-0.375+4) -- (0.75+2,0.125+4);
\draw[lightgray, very thin] (0.75+2,-0.125+4) -- (-0.25+2,0.375+4);
\draw[lightgray, very thin] (0.25+2,-0.375+4) -- (-0.75+2,0.125+4);
\node at (2,1) (S0) {$S_0$};
\node at (1.6,1.3) (S0_f) {};
\node at (2.4,1.2) (S0_f2) {};
\node at (2,2.5) (S1) {$S_1$};
\node at (1.1,2.5) (S1_f) {};
\node at (2.5,2.7) (S1_f2) {};
\node at (2,4) (S2) {$S_2$};
\node at (1.6,3.7) (S2_f) {};
\node at (2.5,4.2) (S2_f2) {};
\draw[lightgray, very thin] (-1.0+4,0.0+4.5) -- (0.0+4,0.5+4.5);
\draw[lightgray, very thin] (0.0+4,0.5+4.5) -- (1.0+4,0.0+4.5);
\draw[lightgray, very thin] (1.0+4,0.0+4.5) -- (0.0+4,-0.5+4.5);
\draw[lightgray, very thin] (0.0+4,-0.5+4.5) -- (-1.0+4,0.0+4.5);
\draw[lightgray, very thin] (-0.5+4,-0.25+4.5) -- (0.5+4,0.25+4.5);
\draw[lightgray, very thin] (0.5+4,-0.25+4.5) -- (-0.5+4,0.25+4.5);
\draw[lightgray, very thin] (-0.75+4,-0.125+4.5) -- (0.25+4,0.375+4.5);
\draw[lightgray, very thin] (-0.25+4,-0.375+4.5) -- (0.75+4,0.125+4.5);
\draw[lightgray, very thin] (0.75+4,-0.125+4.5) -- (-0.25+4,0.375+4.5);
\draw[lightgray, very thin] (0.25+4,-0.375+4.5) -- (-0.75+4,0.125+4.5);
\draw[blue, very thin] (0.5+4,0.0+4.5) -- (0.75+4,0.125+4.5);
\draw[blue, very thin] (0.75+4,0.125+4.5) -- (1.0+4,0.0+4.5);
\draw[blue, very thin] (1.0+4,0.0+4.5) -- (0.75+4,-0.125+4.5);
\draw[blue, very thin] (0.75+4,-0.125+4.5) --(0.5+4,0.0+4.5);
\draw[blue, very thin] (0.0+4,0.25+4.5) -- (0.25+4,0.375+4.5);
\draw[blue, very thin] (0.25+4,0.375+4.5) -- (0.5+4,0.25+4.5);
\draw[blue, very thin] (0.5+4,0.25+4.5) -- (0.25+4,0.125+4.5);
\draw[blue, very thin] (0.25+4,0.125+4.5) --(0.0+4,0.25+4.5);
\draw[blue, very thin] (0.0 - 0.25 + 4, 0.125 + 0.0 + 4.5) -- (0.0 + 0.0 + 4,0.125 + 0.125 + 4.5);
\draw[blue, very thin] (0.0 + 0.0 +4, 0.125 + 0.125 + 4.5) -- (0.0 + 0.25 +4,0.125 + 0.0 + 4.5);
\draw[blue, very thin] (0.0 + 0.25 +4 , 0.125 + 0.0 + 4.5) -- (0.0 + 0.0 + 4, 0.125 -0.125 + 4.5);
\draw[blue, very thin] (0.0 + 0.0 + 4, 0.125 - 0.125 + 4.5) --(0.0 - 0.25 +4 , 0.125 + 0.0 + 4.5);
\draw[lightgray, very thin] (-1.0+4,0.0+3) -- (0.0+4,0.5+3);
\draw[lightgray, very thin] (0.0+4,0.5+3) -- (1.0+4,0.0+3);
\draw[lightgray, very thin] (1.0+4,0.0+3) -- (0.0+4,-0.5+3);
\draw[lightgray, very thin] (0.0+4,-0.5+3) -- (-1.0+4,0.0+3);
\draw[lightgray, very thin] (-0.5+4,-0.25+3) -- (0.5+4,0.25+3);
\draw[lightgray, very thin] (0.5+4,-0.25+3) -- (-0.5+4,0.25+3);
\draw[lightgray, very thin] (-0.75+4,-0.125+3) -- (0.25+4,0.375+3);
\draw[lightgray, very thin] (-0.25+4,-0.375+3) -- (0.75+4,0.125+3);
\draw[lightgray, very thin] (0.75+4,-0.125+3) -- (-0.25+4,0.375+3);
\draw[lightgray, very thin] (0.25+4,-0.375+3) -- (-0.75+4,0.125+3);
\draw[blue, very thin] (-0.25 - 0.25 + 4, 0.0 + 0.0 + 3) -- (-0.25 + 0.0 + 4, 0.0 + 0.125 + 3);
\draw[blue, very thin] (-0.25 + 0.0 +4, 0.0 + 0.125 + 3) -- (-0.25 + 0.25 +4, 0.0 + 0.0 + 3);
\draw[blue, very thin] (-0.25 + 0.25 +4 , 0.0 + 0.0 + 3) -- (-0.25 + 0.0 + 4, 0.0 -0.125 + 3);
\draw[blue, very thin] (-0.25 + 0.0 + 4, 0.0 - 0.125 + 3) --(-0.25 - 0.25 +4 , 0.0 + 0.0 + 3);
\draw[blue, very thin] (-0.25 - 0.25 + 4, -0.25 + 0.0 + 3) -- (-0.25 + 0.0 + 4, -0.25 + 0.125 + 3);
\draw[blue, very thin] (-0.25 + 0.0 +4, -0.25 + 0.125 + 3) -- (-0.25 + 0.25 +4, -0.25 + 0.0 + 3);
\draw[blue, very thin] (-0.25 + 0.25 +4 , -0.25 + 0.0 + 3) -- (-0.25 + 0.0 + 4, -0.25 -0.125 + 3);
\draw[blue, very thin] (-0.25 + 0.0 + 4, -0.25 - 0.125 + 3) --(-0.25 - 0.25 +4 , -0.25 + 0.0 + 3);
\draw[blue, very thin] (0.25 - 0.25 + 4, 0.25 + 0.0 + 3) -- (0.25 + 0.0 + 4, 0.25 + 0.125 + 3);
\draw[blue, very thin] (0.25 + 0.0 +4, 0.25 + 0.125 + 3) -- (0.25 + 0.25 +4, 0.25 + 0.0 + 3);
\draw[blue, very thin] (0.25 + 0.25 +4 , 0.25 + 0.0 + 3) -- (0.25 + 0.0 + 4, 0.25 -0.125 + 3);
\draw[blue, very thin] (0.25 + 0.0 + 4, 0.25 - 0.125 + 3) --(0.25 - 0.25 +4 , 0.25 + 0.0 + 3);
\draw[lightgray, very thin] (-1.0+4,0.0+1.5) -- (0.0+4,0.5+1.5);
\draw[lightgray, very thin] (0.0+4,0.5+1.5) -- (1.0+4,0.0+1.5);
\draw[lightgray, very thin] (1.0+4,0.0+1.5) -- (0.0+4,-0.5+1.5);
\draw[lightgray, very thin] (0.0+4,-0.5+1.5) -- (-1.0+4,0.0+1.5);
\draw[lightgray, very thin] (-0.5+4,-0.25+1.5) -- (0.5+4,0.25+1.5);
\draw[lightgray, very thin] (0.5+4,-0.25+1.5) -- (-0.5+4,0.25+1.5);
\draw[lightgray, very thin] (-0.75+4,-0.125+1.5) -- (0.25+4,0.375+1.5);
\draw[lightgray, very thin] (-0.25+4,-0.375+1.5) -- (0.75+4,0.125+1.5);
\draw[lightgray, very thin] (0.75+4,-0.125+1.5) -- (-0.25+4,0.375+1.5);
\draw[lightgray, very thin] (0.25+4,-0.375+1.5) -- (-0.75+4,0.125+1.5);
\draw[blue, very thin] (0.25 - 0.25 + 4, 0.0 + 0.0 + 1.5) -- (0.25 + 0.0 + 4, 0.0 + 0.125 + 1.5);
\draw[blue, very thin] (0.25 + 0.0 +4, 0.0 + 0.125 + 1.5) -- (0.25 + 0.25 +4, 0.0 + 0.0 + 1.5);
\draw[blue, very thin] (0.25 + 0.25 +4 , 0.0 + 0.0 + 1.5) -- (0.25 + 0.0 + 4, 0.0 -0.125 + 1.5);
\draw[blue, very thin] (0.25 + 0.0 + 4, 0.0 - 0.125 + 1.5) --(0.25 - 0.25 +4 , 0.0 + 0.0 + 1.5);
\draw[blue, very thin] (0.0 - 0.25 + 4, 0.375 + 0.0 + 1.5) -- (0.0 + 0.0 + 4, 0.375 + 0.125 + 1.5);
\draw[blue, very thin] (0.0 + 0.0 +4, 0.375 + 0.125 + 1.5) -- (0.0 + 0.25 +4, 0.375 + 0.0 + 1.5);
\draw[blue, very thin] (0.0 + 0.25 +4 , 0.375 + 0.0 + 1.5) -- (0.0 + 0.0 + 4, 0.375 -0.125 + 1.5);
\draw[blue, very thin] (0.0 + 0.0 + 4, 0.375 - 0.125 + 1.5) --(0.0 - 0.25 +4 , 0.375 + 0.0 + 1.5);
\draw[blue, very thin] (-0.5 - 0.25 + 4, 0.125 + 0.0 + 1.5) -- (-0.5 + 0.0 + 4, 0.125 + 0.125 + 1.5);
\draw[blue, very thin] (-0.5 + 0.0 +4, 0.125 + 0.125 + 1.5) -- (-0.5 + 0.25 +4, 0.125 + 0.0 + 1.5);
\draw[blue, very thin] (-0.5 + 0.25 +4 , 0.125 + 0.0 + 1.5) -- (-0.5 + 0.0 + 4, 0.125 -0.125 + 1.5);
\draw[blue, very thin] (-0.5 + 0.0 + 4, 0.125 - 0.125 + 1.5) --(-0.5 - 0.25 +4 , 0.125 + 0.0 + 1.5);
\draw[lightgray, very thin] (-1.0+4,0.0+0) -- (0.0+4,0.5+0);
\draw[lightgray, very thin] (0.0+4,0.5+0) -- (1.0+4,0.0+0);
\draw[lightgray, very thin] (1.0+4,0.0+0) -- (0.0+4,-0.5+0);
\draw[lightgray, very thin] (0.0+4,-0.5+0) -- (-1.0+4,0.0+0);
\draw[lightgray, very thin] (-0.5+4,-0.25+0) -- (0.5+4,0.25+0);
\draw[lightgray, very thin] (0.5+4,-0.25+0) -- (-0.5+4,0.25+0);
\draw[lightgray, very thin] (-0.75+4,-0.125+0) -- (0.25+4,0.375+0);
\draw[lightgray, very thin] (-0.25+4,-0.375+0) -- (0.75+4,0.125+0);
\draw[lightgray, very thin] (0.75+4,-0.125+0) -- (-0.25+4,0.375+0);
\draw[lightgray, very thin] (0.25+4,-0.375+0) -- (-0.75+4,0.125+0);
\draw[blue, very thin] (-0.25 - 0.25 + 4, -0.25 + 0.0 + 0) -- (-0.25 + 0.0 + 4, -0.25 + 0.125 + 0);
\draw[blue, very thin] (-0.25 + 0.0 +4, -0.25 + 0.125 + 0) -- (-0.25 + 0.25 +4, -0.25 + 0.0 + 0);
\draw[blue, very thin] (-0.25 + 0.25 +4 , -0.25 + 0.0 + 0) -- (-0.25 + 0.0 + 4, -0.25 -0.125 + 0);
\draw[blue, very thin] (-0.25 + 0.0 + 4, -0.25 - 0.125 + 0) --(-0.25 - 0.25 +4 , -0.25 + 0.0 + 0);
\draw[blue, very thin] (0.25 - 0.25 + 4, -0.25 + 0.0 + 0) -- (0.25 + 0.0 + 4, -0.25 + 0.125 + 0);
\draw[blue, very thin] (0.25 + 0.0 +4, -0.25 + 0.125 + 0) -- (0.25 + 0.25 +4, -0.25 + 0.0 + 0);
\draw[blue, very thin] (0.25 + 0.25 +4 , -0.25 + 0.0 + 0) -- (0.25 + 0.0 + 4, -0.25 -0.125 + 0);
\draw[blue, very thin] (0.25 + 0.0 + 4, -0.25 - 0.125 + 0) --(0.25 - 0.25 +4 , -0.25 + 0.0 + 0);
\draw[blue, very thin] (0.5+4,0.0+0) -- (0.75+4,0.125+0);
\draw[blue, very thin] (0.75+4,0.125+0) -- (1.0+4,0.0+0);
\draw[blue, very thin] (1.0+4,0.0+0) -- (0.75+4,-0.125+0);
\draw[blue, very thin] (0.75+4,-0.125+0) --(0.5+4,0.0+0);
\node at (4,0) (p0) {$\bm\phi_0$};
\node at (4,0.35) (p0_f4) {};
\node at (4,1.5) (p1) {$\bm\phi_1$};
\node at (3.5,1.3) (p1_f) {};
\node at (4.4,1.7) (p1_f2) {};
\node at (4,1.1) (p1_f3) {};
\node at (4,1.85) (p1_f4) {};
\node at (4,3) (p2) {$\bm\phi_2$};
\node at (3.5,2.8) (p2_f) {};
\node at (4.9,3) (p2_f2) {};
\node at (4,2.6) (p2_f3) {};
\node at (4,3.35) (p2_f4) {};
\node at (4,4.5) (p3) {$\bm\phi_3$};
\node at (3.5,4.3) (p3_f) {};
\node at (4.4,4.35) (p3_f2) {};
\node at (4,4.1) (p3_f3) {};
\Vertex[x=7, y=3, L=\texttt{$L$}]{L}
		\tikzstyle{EdgeStyle}=[post]
		\Edge(p0_f4)(p1_f3)
		\Edge(p1_f4)(p2_f3)
		\Edge(p2_f4)(p3_f3)
		\Edge(t)(S0_f)
		\Edge(t)(S1_f)
		\Edge(t)(S2_f)
		\Edge(S0_f2)(p1_f)
		\Edge(S1_f2)(p2_f)
		\Edge(S2_f2)(p3_f)
		\Edge(p1_f2)(L)
		\Edge(p2_f2)(L)
        \Edge(p3_f2)(L)
		\end{tikzpicture}
    \caption{Visualization of sampling scheme for reaction-diffusion PDE.}
	\label{fig:pdefigs}
\end{figure}

The gradient is~$
{\frac{\partial L}{\partial \bm \theta} = \sum_{t=1}^T \frac{\partial L}{\partial \bm \phi_t} \sum_{i=1}^t \big(\prod_{j=i}^{t-1} \frac{\partial \bm \phi_{j+1}}{\partial \bm \phi_j}\big) \frac{\partial \bm \phi_i}{\partial \bm C_{i-1}} \frac{\partial \bm C_{i-1}}{\partial \bm\theta}}$.
As the reaction-diffusion PDE is linear and explicit,~${\nicefrac{\partial \bm\phi_{j+1}}{\partial \bm\phi_j} \in \mathbb{R}^{N_x^2 \times N_x^2}}$ is known and independent of~$\bm\phi$. We avoid storing $\bm C$ at each timestep by recomputing $\bm C$ from $\bm\theta$ and $t$.
This permits a low-memory stochastic gradient estimate without exploding variance by sampling from~${\nicefrac{\partial L}{\partial \bm \phi_t} \in \mathbb{R}^{N_x^2}}$ and the diagonal matrix~$\nicefrac{\partial \bm \phi_{i}}{\partial \bm C_{i-1}}$, replacing~${\frac{\partial L}{\partial \theta}}$ with the unbiased estimator
\begin{align}
{\sum_{t=1}^T  \mathcal S_{P^{\bm\phi_t}} \left[\frac{\partial L}{\partial \bm\phi_{t}}\right] \sum_{i=1}^{t} \big(\prod_{j = i}^{t-1} \frac{\partial \bm\phi_{j+1}}{\partial \bm\phi_j}\big) \mathcal S_{P^{\bm\phi_{i-1}}} \left[\frac{\partial\bm \phi_i}  {\partial \bm C_{i-1}}\right]\frac{\partial \bm C_{i-1}}{\partial \bm\theta}\,.}
\end{align}
An experiment with this reduced-memory gradient estimator can be found in \cite{oktay2021randomized}.

\chapter{AD for stellarator optimization \label{ch:stellarator}}
\noindent\rule{\textwidth}{1pt}

\textit{Part of the contents of this chapter have been published in the following papers: (i) McGreivy, N., Hudson, S. R., \& Zhu, C. ``Optimized finite-build stellarator coils using automatic differentiation.'' Nuclear Fusion 61.2 (2021): 026020 \cite{mcgreivy2020optimized}, (ii) McGreivy, N., Zhu, C., Gunderson, L. M., \& Hudson, S. R. ``Computation of the Biot–Savart line integral with higher-order convergence using straight segments.'' Physics of Plasmas 28.8 (2021) \cite{mcgreivy2021computation}.}

\noindent\rule{\textwidth}{1pt}

The stellarator, as we know, is a toroidal magnetic fusion concept which confines plasma using a rotational transform of the vacuum magnetic field \cite{helander2014theory}. 
The rotational transform of the vacuum magnetic field in a stellarator device is created by non-axisymmetric current-carrying coils. The non-axisymmetry of the current-carrying coils and magnetic field allow for a large number of degrees of freedom in the design of a stellarator device. Stellarator design therefore can be formulated as an optimization problem over these degrees of freedom \cite{boozer2020carbon}. The objective of this optimization problem is to simultaneously maximize the plasma performance and minimize the engineering and construction costs of the device.

\section{Stellarator coil optimization with FOCUSADD \label{sec:ch3-coil}} 

Well-designed coils are prerequisites to achieving the performance and cost goals of a stellarator device. 
Usually, stellarator coils are designed to reproduce a given target magnetic field. 
Because inverting the Biot-Savart law is an ill-posed problem, there is no coil set with a finite number of coils that can exactly reproduce an arbitrary magnetic field {throughout a volume}.
Therefore, the goal of stellarator coil design is to find a set of coils which produces the target magnetic field well enough to accomplish the performance goals of the experiment and which can be built and assembled at the cheapest possible cost. 

In practice, achieving the goals of stellarator coil design means successfully optimizing a well-crafted objective function which includes a number of complex physics and engineering objectives. {One term in this objective function usually encourages minimizing the surface integral of the normal magnetic field squared on the outer plasma surface. Other terms could be added; f}or example, a{n} objective function {could be crafted which gives} resonant error fields greater weight in the objective function relative to less damaging non-resonant error fields. {Explicitly targeting resonant error fields was implemented in the island healing techniques developed for reducing chaos in the NCSX stellarator \cite{PhysRevLett.89.275003}.
More recently, a method for identifying the important error fields was presented by Zhu {\em et al.} \cite{Zhu_2019}.} In addition, the coil-coil spacing should be as large as possible to allow for increased access to the plasma for maintenance, diagnostics, and beams. In a power plant, the coils should be at least one meter from the plasma to allow for the tritium breeding blanket and neutron shielding. Each of these objectives could be converted to a potentially complex scalar objective function or possibly a constraint on the optimization which allows for a set of coils to be found which best satisfy the desired engineering constraints.  

Successful optimization of a scalar objective function in non-convex, high-dimensional spaces often relies on the use of derivative information to inform the optimization process. Many stellarator coil design codes have performed gradient-based optimization by computing finite-difference derivatives \cite{strickler_2002}, while recent work has allowed for efficient derivative computations by computing analytic derivatives \cite{Paul_2018,Zhu_2018,paul_abel_landreman_dorland_2019,antonsen_paul_landreman_2019}. In this section, we instead use automatic differentiation to efficiently compute the required derivatives. 

\begin{figure}
    \centering
    \includegraphics[width=0.6\textwidth]{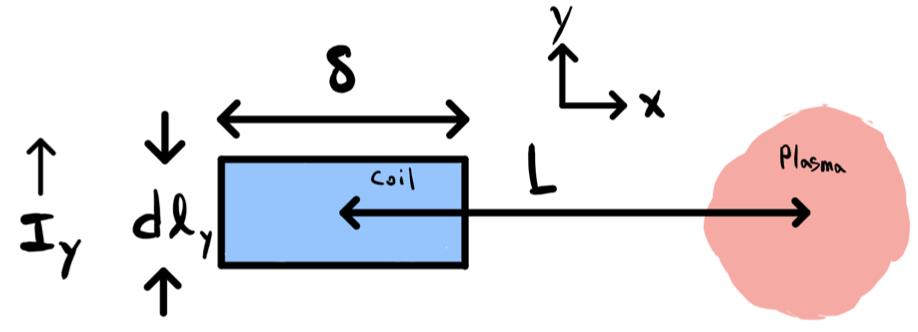}
    \caption{A sketch of an infinitesimal finite-build coil segment carrying current in the $y$-direction. The coil is assumed to have zero thickness in the $z$-direction, and thickness $\delta$ in the $x$-direction. The center of the coil is a distance $L$ from the plasma. The lowest-order correction of the magnetic field $B_z$ due to the finite-build of the coil is second-order in $\delta/L$. }
    \label{fig:biot-savart}
\end{figure}

Existing coil design codes have ultimately optimized the positions of filamentary (zero thickness) coils. However, any real coil will be made up of a winding pack that carries current over a non-zero volume. While such a coil can be approximated as a single filament in space, this approximation is only valid in the limit that the coil thickness is much less than the distance from the coil to the plasma. The correction to the magnetic field due to coil finite build is second-order in the coil thickness $\delta$ divided by the coil-plasma distance $L$, which can be shown using a Taylor expansion of the magnetic field due to an infinitesimal finite-build coil segment as sketched in figure \ref{fig:biot-savart}. Assuming the coil has zero thickness in the $z$-direction, the magnetic field at the plasma is given by
\begin{equation}\label{eq:one}
        \mathrm{d}B_z = -\frac{\mu_0 I_y \mathrm{d} \ell_y}{4 \pi \delta }\int_{-\delta/2}^{\delta/2} \frac{\mathrm{d}x}{(L+x)^2}{.}
\end{equation}
Taylor expanding the denominator to lowest order gives a second-order correction in the ratio $\nicefrac{\delta}{L}$:
\begin{equation}\label{eq:expansion}
    \mathrm{d}B_z \approx -\frac{\mu_0 I_y \mathrm{d}\ell_y}{4\pi \delta L^2}\int_{-\delta/2}^{\delta/2} \Big(1 - \cancelto{0}{\frac{2x}{L}} + \frac{3x^2}{L^2}\Big)\mathrm{d}x \approx \mathrm{d}B_{\text{filament}}\Big(1 + \frac{\delta^2}{4L^2}\Big){.}
\end{equation}

{Previous experiments have constructed the geometry of finite-build coils in different ways. One approach, used by the Wendelstein 7-X (W7-X) experiment, is to define an orthonormal frame around a central filament consisting of the tangent to the filament, the normal to a winding surface, and the binormal perpendicular to both, and then defining the edges of rectangular cross-section coils by extending the edges outwards from the central filament in the normal and binormal directions \cite{physics_engineering_design_w7x}.}

In this section, we introduce a new stellarator coil design code which directly optimizes finite-build (non-zero thickness) coils, thereby accounting for the finite-build correction to the produced magnetic field. This code is called FOCUSADD (\textbf{F}lexible \textbf{O}ptimized \textbf{Cu}rves in \textbf{S}pace using \textbf{A}utomatic \textbf{D}ifferentiation and Finite Buil\textbf{d}). 
One key motivation of this work is to introduce AD to the stellarator community. Our results (\cref{sec:ch3-results}) focus on comparing optimized finite-build coils to optimized filamentary coils.

Code for FOCUSADD can be found at \href{https://github.com/nickmcgreivy/FOCUSADD}{https://github.com/nickmcgreivy/FOCUSADD}. 
Originally, FOCUSADD was adapted from FOCUS and written in FORTRAN with the source transformation tool OpenAD/F \cite{OpenAD/F}. FOCUSADD was then rewritten in JIT-compiled Python using \href{https://github.com/google/jax}{JAX} \cite{jax2018github}, a research project under development by Google.
JAX is a library for composable transformations of Python+NumPy programs, including automatic differentiation, vectorization, JIT-compilation to GPU/TPU, and SPMD parallelization.

Recent work using the OMIC code \cite{l_singh_preprint} has, for the first time, optimized coils which have a finite build. OMIC starts with a set of filamentary coils, defines a winding pack around each coil, then optimizes the {orientation} of the coil winding packs using gradient-based optimization and gradients computed using finite-difference. While FOCUS \cite{Zhu_2017} optimizes the position of filamentary coils and OMIC optimizes the {orientation} of finite-build coils, neither code can optimize both position and {winding pack orientation} at once. Both quantities can be optimized at once by FOCUSADD. While we demonstrate this capability (section \cref{sec:ch3-results}), optimizing the {winding pack orientation} of finite-build coils is not a primary focus of this research. 

OMIC uses finite-difference derivatives to compute the gradient of a scalar objective function. One limitation of this approach is that the computational cost of computing the gradient increases linearly with the number of optimization parameters. Analytic derivatives could, in principle, be used to efficiently compute the required gradient, at a computational cost independent of the number of optimization parameters. However, finding and computing analytic derivatives for finite-build stellarator coils is challenging due to the significantly increased complexity of the objective function. Both of these challenges -- computational efficiency and computing analytic derivatives -- are solved here by using reverse mode AD to develop a new finite-build coil design code (\cref{sec:ch3-focusadd,sec:ch3-objective}).

AD is particularly well-suited to the problem of stellarator coil design. This is because in order to satisfy the many physics and engineering objectives of stellarator coil design, we need to optimize a high-dimensional ({200}-5000 optimization parameters) objective function (\cref{sec:ch3-objective}) whose analytic derivatives may be extremely difficult to write down and program analytically. AD allows us to efficiently perform gradient-based optimization of such an objective function; efficiently computing the required derivatives would be a significant challenge if AD were not used. We would also like to be able to efficiently optimize many different objective functions, without taking the time and effort to derive and program analytic derivatives for each objective function that we might consider. Automatic differentiation allows us to neither derive nor program these analytic derivatives; we only need to compute the value of an objective function and its derivatives are computed automatically and efficiently.

\subsection{Multi-filament coil parametrization \label{sec:ch3-focusadd}}

Existing coil design codes have ultimately optimized the positions of zero-thickness or filamentary coils.
A filamentary approximation to a finite-build coil is valid in the limit that the coil thickness is much less than the minimum distance between the coils and the plasma. If the coil thickness is non-zero, which will be true of any real set of coils, then the errors in this approximation are second-order in the ratio of the coil thickness to the coil-plasma distance as shown in equation \ref{eq:expansion}. To account for these finite-build corrections, we need to directly optimize coils with non-zero thickness. 

\begin{figure}
    \centering
    \includegraphics[width=0.5\textwidth]{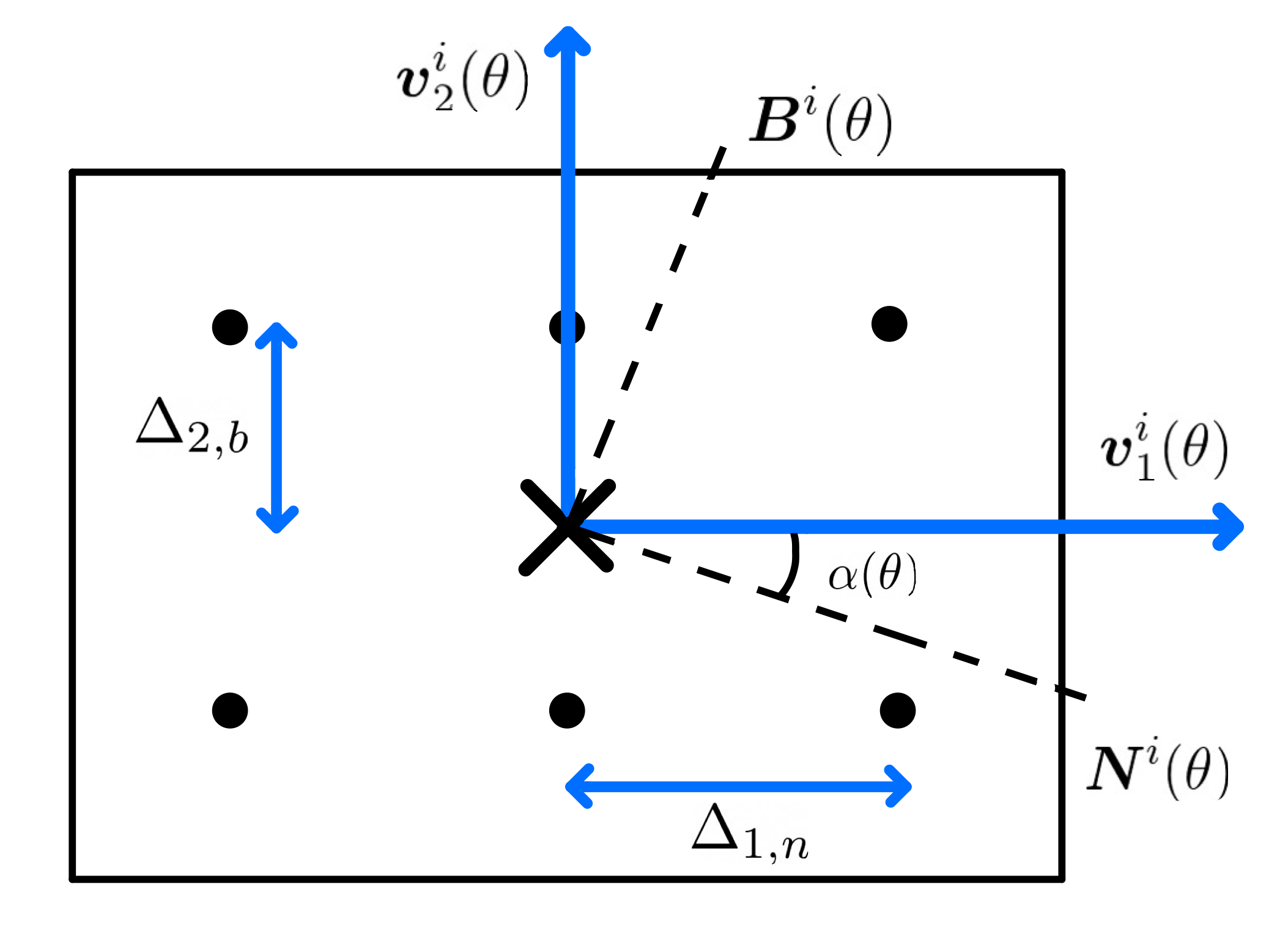}
    \caption{A coil cross-section showing the multi-filament approximation to the coil winding pack used by FOCUSADD{; the coil tangent vector $\bm T^i$ (not shown) is normal to the plane}. The winding pack centroid is shown as a black X. The current-carrying filaments are shown as black circles. The filaments are placed on a rectangular grid whose coordinate axes are defined by the vectors $\bm{v}_1$ and $\bm{v}_2$; the coordinate axes are rotated by an angle $\alpha^i$ with respect to normal and binormal vectors $\bm{N}^i$ and $\bm{B}^i$. The indices of the rectangular grid are $n$ and $b$ {and are counted from the bottom left corner}; filaments at index $n$ are displaced relative to the winding pack centroid in the $\bm{v}_1$ direction by a distance $\Delta_{1,n}$ and filaments at index $b$ are displaced relative to the winding pack centroid in the $\bm{v}_2$ direction by a distance $\Delta_{2,b}$.}
    \label{fig:multifilament}
\end{figure}

In this section, we directly optimize finite-build modular coils using a multi-filament approximation to the winding pack. 
{The filaments used in the multi-filament approximation are, at each poloidal angle $\theta$, placed on a rectangular grid centered on the winding pack centroid $\bm r^i$.}\footnote{{What we call the ``winding pack centroid", other authors have called the ``central filament". We choose this term because our multi-filament approximation may have no central current-carrying filament, see figure \ref{fig:multifilament}.}}
{The position of the winding pack centroid $\bm{r}^i = \{x^i, y^i, z^i\}$ is parametrized as a function of $\theta$ with a Fourier series given by}
\begin{equation}
     \label{eq:coil_r}
     \begin{split}
        x^i(\theta) &= \sum_{m=0}^{N_F-1} \left[ X^i_{c,m} \cos(m \theta) + X^i_{s,m} \sin(m \theta) \right] \\
        y^i(\theta) &= \sum_{m=0}^{N_F-1} \left[Y^i_{c,m} \cos(m \theta) + Y^i_{s,m} \sin(m \theta) \right]\\
        z^i(\theta) &= \sum_{m=0}^{N_F-1} \left[Z^i_{c,m} \cos(m \theta) + Z^i_{s,m} \sin(m \theta) \right]
     \end{split}{.}
\end{equation}
{$N_F$ is an integer describing the number of modes in the Fourier series.} The parameters of this Fourier series can be combined into a single vector $\bm{R}$, where $\bm{R} \equiv \big\{\bm{X}^i_c,\bm{X}^i_s, \bm{Y}_c^i,\bm{Y}_s^i, \bm{Z}_c^i,\bm{Z}_s^i\big\}, \ i=1, \cdots, N_c${, where $N_c$ is the number of coils}.

Like the OMIC code \cite{l_singh_preprint}, we use a multi-filament approximation to the coil winding pack and have the freedom to allow the winding pack to rotate. Figure \ref{fig:multifilament} displays a cross-section of the multi-filament approximation to a coil winding pack at a particular poloidal angle $\theta$. The centroid of the coil winding pack{, $\bm r^i$,} is shown with a black X in figure \ref{fig:multifilament}; the black circles{, $\bm{r}^i_{n,b}$, are current-carrying filaments indexed by $n$ and $b$, starting from the bottom left corner}. In figure \ref{fig:multifilament}, six filaments are placed on a 3 by 2 grid, but any number of filaments are allowed. 

The vectors $\bm{v}^i_1(\theta)$ and $\bm{v}^i_2(\theta)$ in figure \ref{fig:multifilament} {form} the coordinate axes of the rectangular grid upon which the coil filaments are placed. $\bm{v}^i_1$ and $\bm{v}_2^i$ are rotated relative to a normal vector $\bm{N}^i$ and binormal vector $\bm{B}^i$ by an angle $\alpha^i(\theta)$, given by 
\begin{equation}\label{eq:rotate_coil}
    \begin{bmatrix}
        \bm{v}^i_1(\theta) \\ 
        \bm{v}^i_2(\theta) 
    \end{bmatrix} = 
    \begin{bmatrix}
        \cos\alpha^i & -\sin\alpha^i \\
        \sin\alpha^i & \cos\alpha^i
    \end{bmatrix} 
    \begin{bmatrix}
        \bm{N}^i(\theta) \\
        \bm{B}^i(\theta) 
    \end{bmatrix}{.}
\end{equation}
$\alpha^i$ is parametrized by another Fourier series, given by
\begin{equation}\label{eq:rotfourier}
    \alpha^i(\theta) = \frac{N_R \theta}{2} +  \sum_{m=0}^{N_{FR}-1
    } \left[ A_{c,m}^i \cos{(m\theta)}  + A_{s,m}^i \sin{(m\theta)} \right]{.}
\end{equation}
{$N_{FR}$ is an integer describing the number of modes describing this Fourier series; if the winding packs are not free to rotate it is set to zero.} $N_R$ in equation \ref{eq:rotfourier} is an integer describing the number of {half} rotations of the coil winding pack; $N_R$ is normally set to zero. The parameters of this Fourier series can be combined into a single vector, $\bm{A}$ where $\bm{A} \equiv \big\{\bm{A}^i_c,\bm{A}^i_s\big\}, \ i=1, \cdots, N_c$.  {The tangent vector $\bm T^i$ is defined by the Frenet-Serret equations for the winding pack centroid $\bm r^i$, while $\bm{N}^i$ and $\bm{B}^i$ are defined by the so-called ``center of mass frame", introduced in \cite{l_singh_preprint}. The center of mass frame defines the normal vector $\bm N^i$ using
\begin{equation}\label{eq:normal_vec}
\bm N^i \equiv \frac{\bm \delta^i - (\bm \delta^i \cdot \bm T^i)}{||\bm \delta^i - (\bm \delta^i \cdot \bm T^i)||},
\end{equation}
the normalized component of $\bm \delta^i$ perpendicular to $\bm T^i$, where $\bm\delta^i(\theta) \equiv \bm r^i(\theta) - \bm R^i_{c,0}$ and $\bm R^i_{c,0}$ is the center of mass of the $i$th coil. The binormal vector $\bm B^i$ is defined as $\bm T^i \times \bm N^i$. $\bm T^i(\theta)$, $\bm N^i(\theta)$, $\bm B^i(\theta)$ define an orthonormal coordinate system around each coil at each poloidal angle $\theta$; this coordinate system defines the orientation of the winding pack at zero rotation.}

The position of $\bm{r}^i_{n,b}$, the filament in the $i$th coil with indices $n$ and $b$, is given by
\begin{equation}\label{eq:spacing}
    \bm{r}^i_{n,b}(\theta) = \bm{r}^i + \Delta_{1,n} \bm{v}_1^i(\theta) + \Delta_{2,b} \bm{v}_2^i(\theta){.}
\end{equation}
$n$ and $b$ are indices of the filaments on the rectangular grid{, counting up from 1}. $\Delta_{1,n} \equiv (n - \frac{N_1+1}{2})l_1$ and $\Delta_{2,b} \equiv (b - \frac{N_2+1}{2}) l_2$ where $l_1$ and $l_2$ are the spacing between gridpoints in the $\bm{v}_1$ and $\bm{v}_2$ directions and $N_1$ and $N_2$ are the number of gridpoints in the $\bm{v}_1$ and $\bm{v}_2$ directions.

Each finite-build coil is ultimately parametrized by four Fourier series, three for the centroid of the coil winding pack and one for the {orientation} of the winding pack in space. The parameters of these Fourier series can be combined into a single vector $\bm{p}$, where $\bm{p} \equiv \big\{ \bm{R},\bm{A} \big\}$.

\subsection{Objective function \label{sec:ch3-objective}}

The goal of stellarator coil design is to find a set of coils which reproduces the target magnetic field well enough to accomplish the performance goals of the experiment and which can be built and assembled at the cheapest possible cost. In practice, this design problem is formulated as an optimization problem. The goal is to find optimal coil parameters $\bm{p}^*$ which minimize an objective function $f$. 

\begin{equation}
    \bm{p}^* = \argminC_{\bm{p}} f(\bm{p})
\end{equation}
This objective function should be chosen to meet the goals of the experiment, and therefore should incorporate both physics and engineering objectives. The standard way of incorporating multiple objectives in an optimization problem is to sum the multiple objectives into a total objective function $f_{total}$. 

\begin{equation}\label{eq:ftotal}
    f_{total}(\bm{p}) = f_{Phys}(\bm{p}) + f_{Eng}(\bm{p})
\end{equation}

Gradient-based optimization is commonly used to optimize high-dimensional non-convex objective functions. Performing gradient-based optimization requires the computation of derivatives of the objective function. Reverse mode AD allows for the gradient to be computed efficiently -- independent of the number of parameters, and at a runtime cost of {a small multiple of} the cost of computing the objective function -- for arbitrary differentiable functions. In practice, AD makes gradient-based optimization of high-dimensional objective functions easy and efficient. The job of the stellarator coil designer is then to craft an objective function which accounts for the complex physics and engineering objectives required of the stellarator coils and which can be effectively optimized. 

For simplicity we choose to optimize the same objective function as FOCUS \cite{Zhu_2017}. The physics objective function is the so-called ``quadratic flux'', given by the integral over the outer toroidal surface of the square of the dot product between the magnetic field $\bm{B}$ and the surface normal vector $\bm{n}${:}
\begin{equation}\label{eq:quadratic_flux}
    f_{Phys} \equiv \int_S (\bm{B} \bm{\cdot} \bm{n})^2 dA{.}
\end{equation}
For simplicity, FOCUSADD ignores the magnetic field generated due to currents in the plasma. The vacuum magnetic field is given by the Biot-Savart law integrated over the filaments in each multi-filament coil:
\begin{equation}\label{eq:biot_savart}
    \bm{B}(\bm{r}) = \sum_{i=1}^{N_c} \sum_{n=1}^{N_1} \sum_{b=1}^{N_2} \frac{\mu_0  I^i_{n,b}}{4 \pi} \oint \frac{d\bm{l}^i_{n,b} \times (\bm{r} - \bm{r}^{i}_{n,b})}{|\bm{r} - \bm{r}^{i}_{n,b}|^3}{.}
\end{equation}
For simplicity, FOCUSADD assumes each filament carries the same current and does not treat each coil's current as an optimization parameter. 

Here we choose $f_{Eng}$ to be proportional to the average length of the coil centroids, as given by equation
\begin{equation}\label{eq:feng}
    f_{Eng} \equiv \lambda_L \frac{1}{N_c} \sum_{i=1}^{N_c} L_i{.}
\end{equation} 
We weigh the length penalty with a regularization coefficient $\lambda_L$. Although the raw material cost of the coils will be approximately proportional to the length of the coils, this simplified engineering objective function is likely a poor approximation to the true cost of manufacturing and assembling real stellarator coils. Future coil design codes can and should account for the many complex engineering requirements on the coils; automatic differentiation makes optimizing these objective functions easy and efficient.

\begin{figure}
    \centering
    \includegraphics[width=0.8\textwidth]{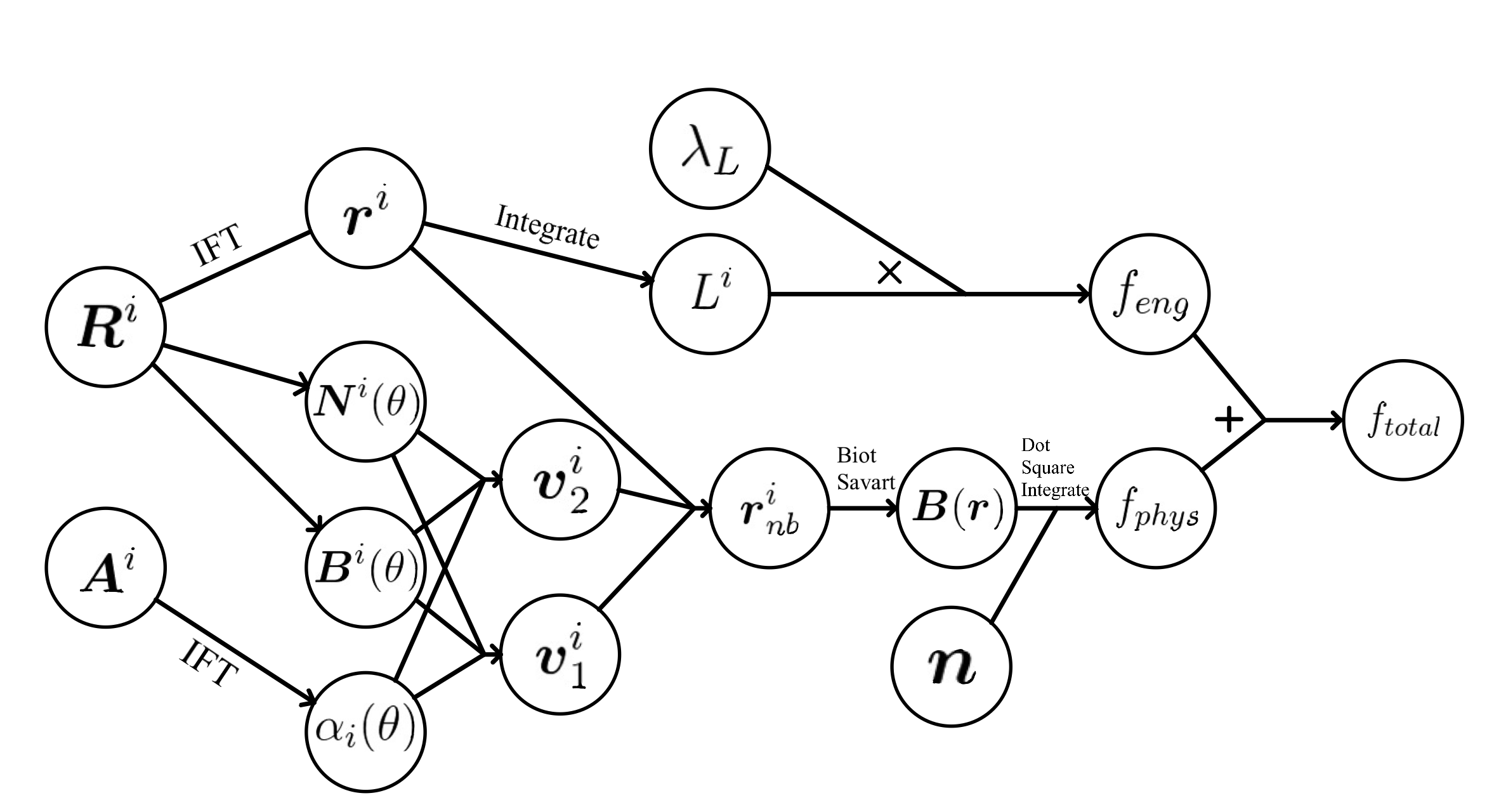}
    \caption{A computational graph for the function defined in equation \ref{eq:ftotal}. The variables on the left, $\{\bm{R}^i\}_{i=1}^{N_c}$ and $\{\bm{A}^i\}_{i=1}^{N_c}$, represent the coil parameters we want the gradient of $f_{total}$ with respect to. ``IFT" stands for ``inverse Fourier transform''. To compute this gradient, we use reverse mode AD by computing the intermediate variables in topological order and then the cotangent variables in reverse topological order. }
    \label{fig:compgraphftotal}
\end{figure}

A computational graph which represents the calculation of the objective function in equation \ref{eq:ftotal} is shown in figure \ref{fig:compgraphftotal}. The vertices are the variables computed as a result of the computation and the edges are the operations performed on those variables. Not all intermediate variables or operations are shown in figure \ref{fig:compgraphftotal}. We would like to compute the gradient of $f_{total}$ with respect to $\bm{p}\equiv \big\{ \bm{R},\bm{A} \big\}$. To compute this gradient, an AD tool computes $f_{total}$ by traversing the graph from beginning to end, then computes the cotangent variables by traversing the graph from end to beginning. At the end of the computation, we have the gradient $\frac{d f}{d \bm{p}}$. Gradient descent with momentum {\cite{momentum}} is used to minimize the objective function.

\subsection{Finite-build coil optimization results \label{sec:ch3-results}}

To understand the effect of optimizing finite-build coils, we investigate two configurations: (i) a rotating elliptical stellarator, and (ii) a boundary for {a} W7-X{-like stellarator}. We first use the rotating elliptical stellarator to both establish the viability of the method and to demonstrate the {optimization} of the coil winding pack {orientation}. We then use the W7-X{-like} boundary to test the algorithm and to investigate the effects of including finite-build on the resulting coils. {In each case, the center of mass frame was used to define the zero-rotation frame, and nine filaments were placed on a square 3 by 3 grid in a multi-filament approximation to the coil finite build. Although a real conductor or superconductor will typically have more than nine filaments, adding more filaments is an increasingly small correction. For our purposes of testing the algorithm and understanding the finite-build effects, nine filaments is sufficient. }

\subsubsection{Elliptical stellarator}

\begin{figure}
	\centering%
    \includegraphics[width=0.65\textwidth]{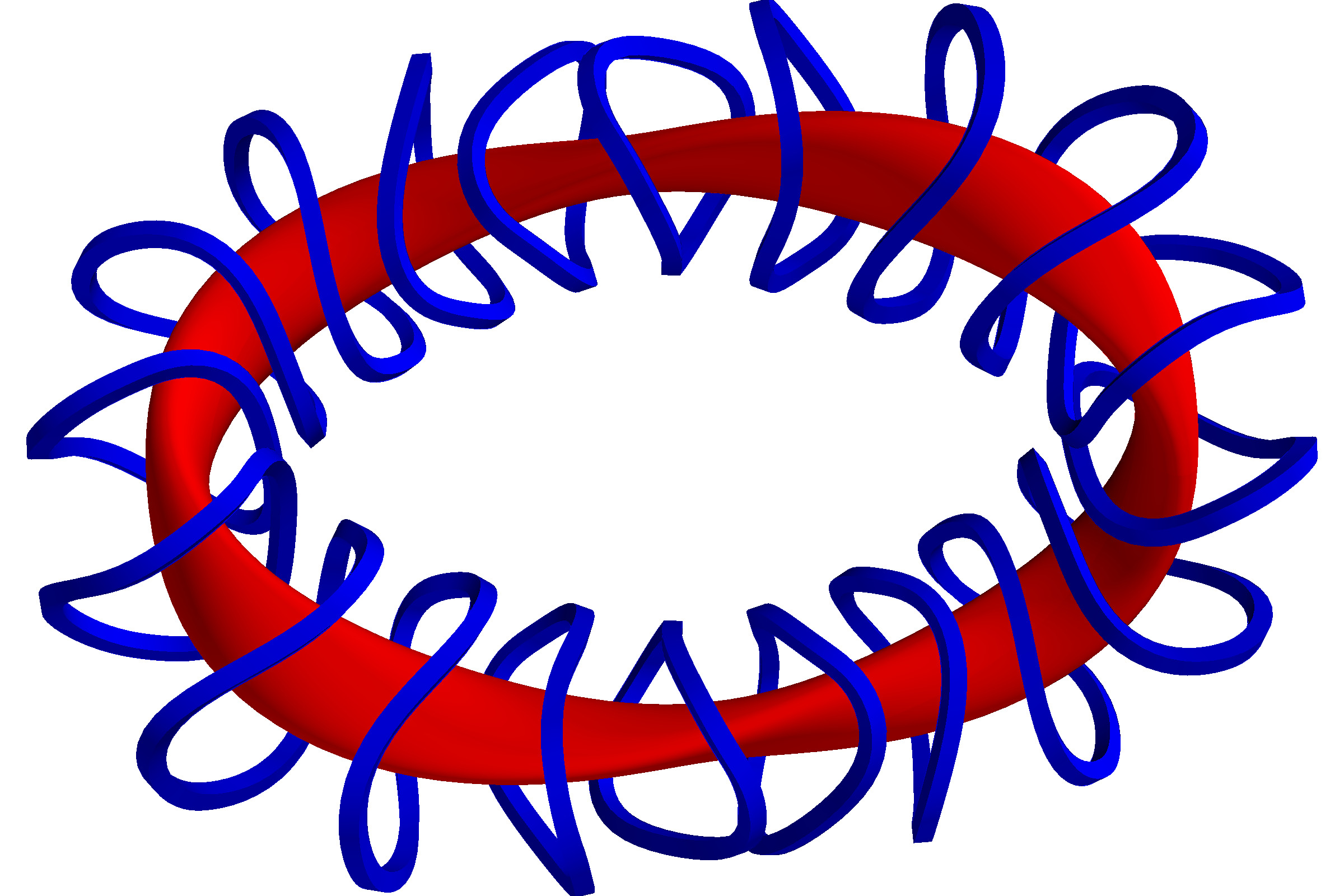}
    \caption{The outer toroidal surface and optimized zero-rotation coils of the four-period elliptical rotating stellarator used to test the FOCUSADD method. A length penalty of $\lambda_L = 0.1$ is chosen to force the coils to remain close to the plasma. The coil winding pack centroids overlay filamentary optimized coils (not shown), demonstrating that the finite-build is a small correction. }
    \label{fig:elliptical_stellarator}
\end{figure}

\begin{figure}
    \centering%
    \includegraphics[width=0.3\textwidth]{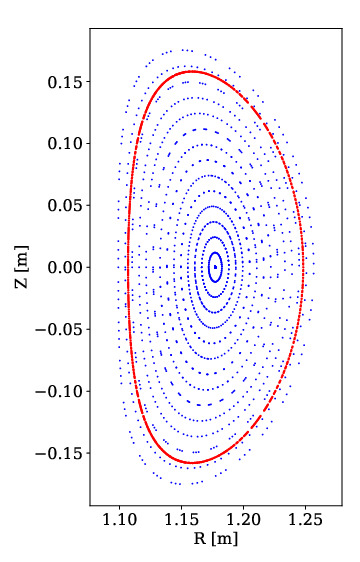}
        \caption{A poincare plot of the magnetic field lines of the optimized finite-build coils in figure \ref{fig:elliptical_stellarator}, showing nested magnetic surfaces. The target surface is shown in red. }
    \label{fig:poincare}
\end{figure}

The FOCUSADD code was used to find optimized coils for a rotating elliptical stellarator. This allowed us to establish the viability of the method and to demonstrate the {optimization} of the coil winding pack {orientation}. The outer surface of the four-period, elliptical cross-section stellarator is shown in figure \ref{fig:elliptical_stellarator}. Also shown in figure \ref{fig:elliptical_stellarator} are optimized finite-build coils with zero rotation of the winding pack relative to the center of mass frame. A Poincare plot showing nested magnetic surfaces for optimized finite-build coils is shown in figure \ref{fig:poincare}. 


\subsubsection{Winding pack {orientation optimization}}

\begin{figure}
    \centering
    \includegraphics[width=0.6\textwidth, angle=90]{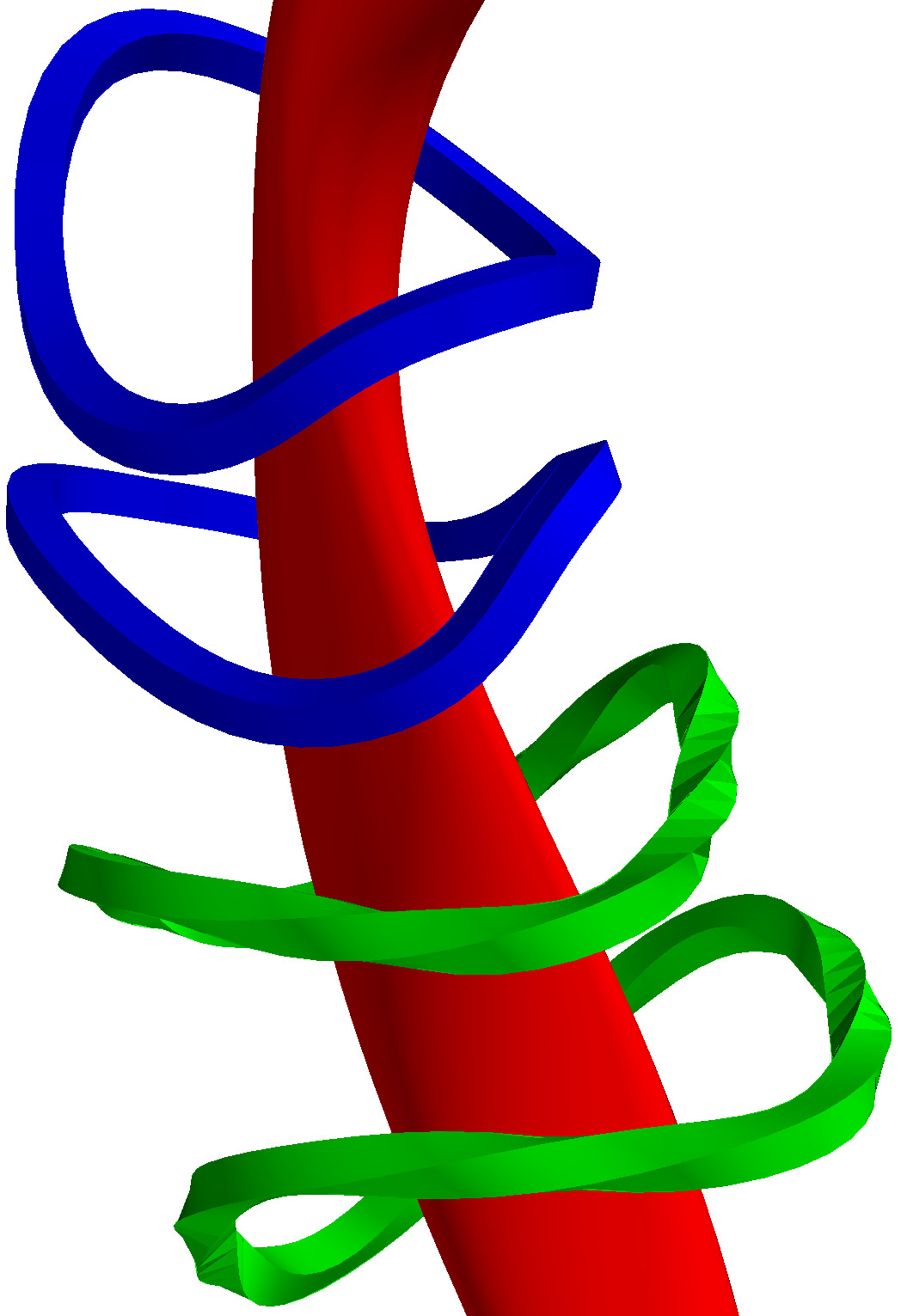}
    \caption{Finite-build coils are compared for the rotating elliptical stellarator. The leftmost two coils (blue) have zero winding pack rotation {relative to the center of mass frame}, while the rightmost two coils (green) are allowed to rotate freely. There are 9 filaments which are placed on a 3 by 3 square grid.}
    \label{fig:coil_comparison}
\end{figure}

We examine the case where the coil winding pack {orientation} and centroid position are simultaneously optimized. Such optimized square cross-section finite-build coils are shown in figure \ref{fig:coil_comparison}; the right two coils (green) are allowed to rotate freely while the left two coils (blue) have zero winding pack rotation. Allowing the coils to freely rotate decreases the quadratic flux by 20\% relative to the zero-rotation coils. However, due to the {rapidly twisting} winding pack profile, these coils would be extremely challenging to engineer; we conclude that the winding pack cannot simply be allowed to rotate freely. This conclusion is consistent with the results from Singh, et al. \cite{l_singh_preprint}, who find that regularizing the coil rotation profiles leads to coils with less rotation relative to the center of mass frame. While a regularization penalty on the coil rotation {profile} could easily be added to FOCUSADD, we have chosen not to focus on rotation profile optimization in this research. 

The remainder of this section uses coils with zero winding pack rotation in the center of mass frame; this particular choice of frame is arbitrary but results in simple winding packs. Further discussion of the coil winding pack {orientation} can be found in \cref{sec:ch3-coilopt-conclusion}.

\subsubsection{W7-X{-like stellarator} }

To investigate the effect of directly optimizing finite-build coils, we find optimized coils for a W7-X{-like stellarator} surface. {The particular surface is geometrically similar to the standard configuration of W7-X, although no known reference for it exists in the literature. We make no attempt to exactly match the standard configuration surface.} {Wendelstein 7-X (W7-X) is a 5.5m major radius, five-period experimental stellarator device in Greifswald, Germany. The W7-X magnetic field is produced by 50 non-planar and 20 planar modular superconducting coils. Each of the five periods consists of ten coils; due to stellarator symmetry there are only five unique coil shapes. We use the phrase ``W7-X-like" to emphasize that} the goal at this stage is not to find coils as if designing an actual experiment{, nor is the goal to compare to any experimental design. R}ather{,} the goal is to test the method and determine approximately the importance of the finite-build effects {on a realistic stellarator configuration}; this goal can most simply be phrased with the question \textit{``Does finite-build matter?"}.

{To investigate these finite-build effects we focus on two quantities. The first quantity, $\Delta r$, is (roughly speaking) the distance the coils shift due to finite-build. $\Delta r$ is defined as the mean minimum distance between the winding pack centroid of optimized zero-rotation finite-build coils and optimized filamentary coils. The second quantity, $\Delta e$, is (roughly speaking) by what percent does the normalized field error change if finite-build is not accounted for in the optimization. The normalized field error $e$ is defined as
\begin{equation}\label{eq:normalized_field_error}
    e \equiv \frac{1}{\int_S dA }\int_S \frac{|\bm B \cdot \bm n|}{|B|} dA.
\end{equation}
$\Delta e \equiv (\nicefrac{e'_{fil}}{e_{fb}} - 1) * 100$ is defined as the ratio of $e'_{fil}$, the normalized field error of coils which are optimized with a filamentary approximation but then are built with zero rotation and finite-build, and $e_{fb}$, the normalized field error of zero-rotation coils which are optimized with a finite-build approximation and then built with zero rotation and finite-build, minus 1, all times 100.}
\begin{figure}
    \centering
    \includegraphics[width=0.6\textwidth]{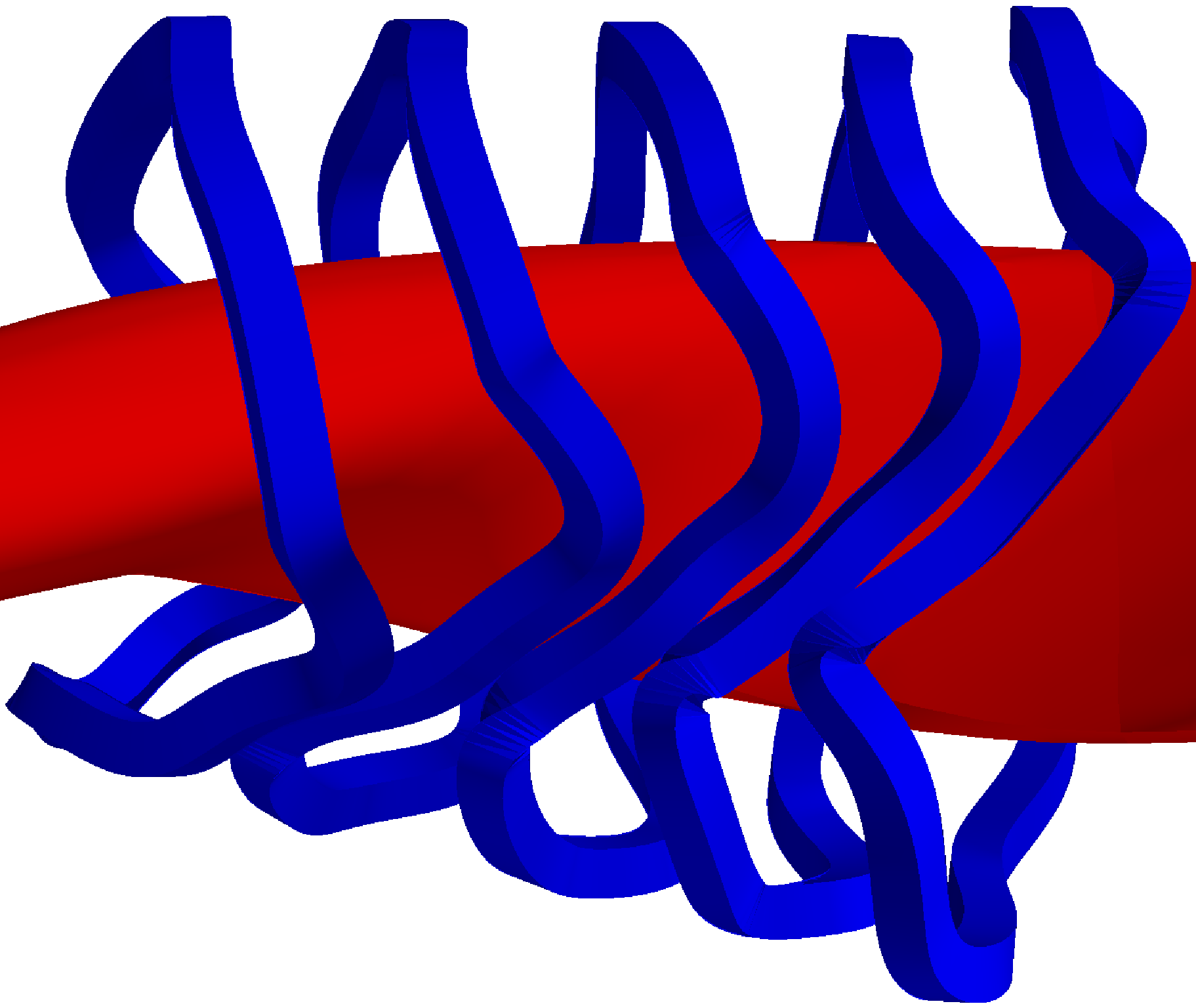}
    \caption{Half-period consisting of five W7-X finite-build coils. Coils shown are $18$cm by 18cm in cross-section.}
    \label{fig:w7x}
\end{figure}

A half-period of a W7-X{-like} outer plasma surface and optimized zero-rotation finite-build coils are shown in figure \ref{fig:w7x}. The finite-build coil dimensions (18cm by 18cm) in figure \ref{fig:w7x} are approximately the same dimensions as the true W7-X conductor ({19.2cm by 16cm}) {\cite{w7x_coilsize}}. The regularization penalty on the coil lengths, $\lambda_L$, is set to 0.25. {Each optimization run is performed for 10,000 iterations, which ensures convergence.}

\begin{figure}
		\centering%
		\begin{subfigure}{0.49\textwidth}
		\centering%
        \includegraphics[width=0.98\textwidth]{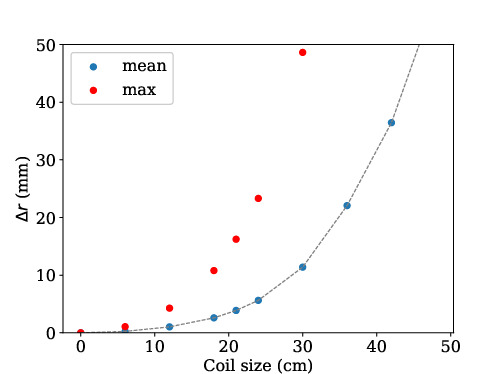}
		\caption{$\Delta r(\delta)$ for W7-X.}
		\label{fig:w7x_r}
		\end{subfigure}
		\begin{subfigure}{0.49\textwidth}	
		\centering%
        \includegraphics[width=0.98\textwidth]{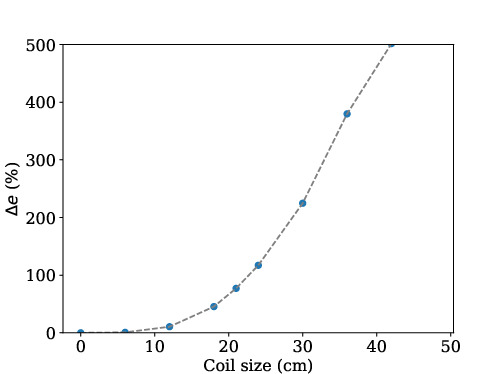}
		\caption{$\Delta e(\delta)$ for W7-X.}
		\label{fig:w7x_e}
		\end{subfigure}
		\caption{W7-X: (a) $\Delta r$, the mean shift in the coil position between finite-build and filamentary coils, is plotted in blue as a function of coil size $\delta$. The maximum shift is plotted in red. (b) $\Delta e$, the percent change in the normalized field error if finite-build is not accounted for, as a function of coil size $\delta$. For 18cm by 18cm coils, of comparable size to the W7-X conductor cross-section ({19.2cm by 16cm}), the mean shift $\Delta r \approx 2.5$mm, the maximum shift is about 11mm, and the normalized field error $\Delta e \approx 50\%$.}
		\label{fig:curves_w7x}
	\end{figure}

To investigate the effect{s} of including finite build, we vary the size of the $\delta$ by $\delta$ {winding pack} cross-section from $\delta = 0${cm} to $\delta = 50$cm while keeping all {other} quantities fixed and calculate the effects on $\Delta r$ and $\Delta e${; the results} are shown in figures \ref{fig:w7x_r} and \ref{fig:w7x_e}. {Although eventually the coils will overlap as $\delta$ increases, this is not an issue for our purpose of better understanding the finite build effects. We extend $\delta$ to such a large value for the purposes of illustration.}

In figure \ref{fig:w7x_r}, we can see that as the coil size $\delta$ is increased, the mean shift in the coil positions increases. For coils of approximately the same dimensions as the W7-X coils (18cm by 18cm), we have the result that the mean shift $\Delta r \approx 2.5$mm, while the maximum shift is about 11mm. The coil tolerances in the W7-X experimental design {were} {$\pm 3$mm during the manufacturing process and $\pm 2$mm during the assembly process}{; once the machine was assembled, the maximum deviation of any reference point was measured to be 5.7mm from its manufacture value} \cite{Andreeva_2015}. Because the mean coil shift is of the same magnitude as the coil tolerances, and the maximum shift much larger, we can reasonably conclude that finite-build effects should be accounted for in the optimization of coils for a stellarator such as W7-X.

\begin{figure}
		\centering
		\begin{subfigure}{0.45\textwidth}
		\centering
        \includegraphics[width=0.6\textwidth]{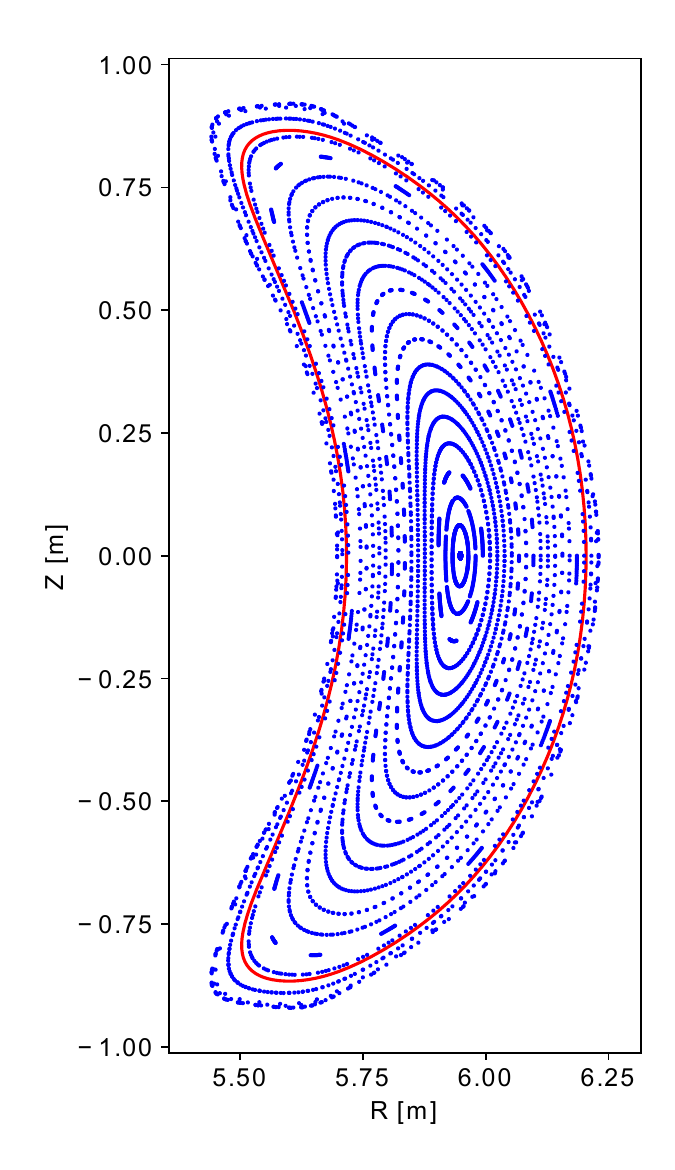}
		\caption{}
		\label{fig:poincare_fb}
		\end{subfigure}
		\begin{subfigure}{0.45\textwidth}
		\centering
        \includegraphics[width=0.6\textwidth]{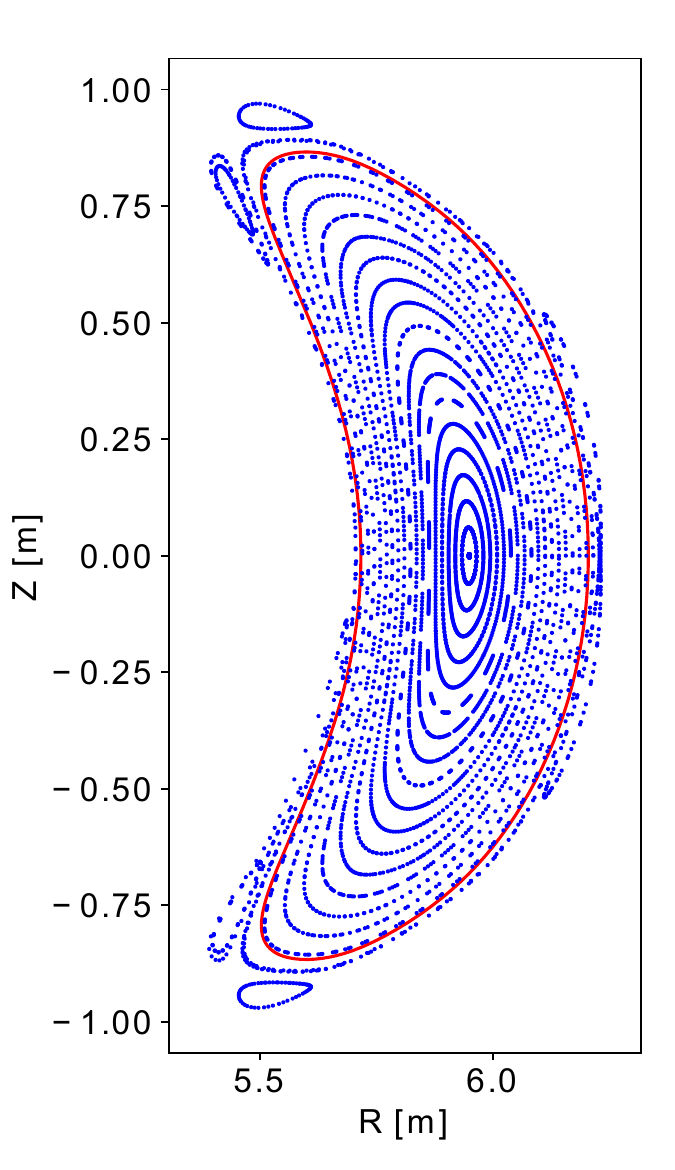}
		\caption{}
		\label{fig:poincare_18}
		\end{subfigure}
		\begin{subfigure}{0.45\textwidth}
		\centering
        \includegraphics[width=0.6\textwidth]{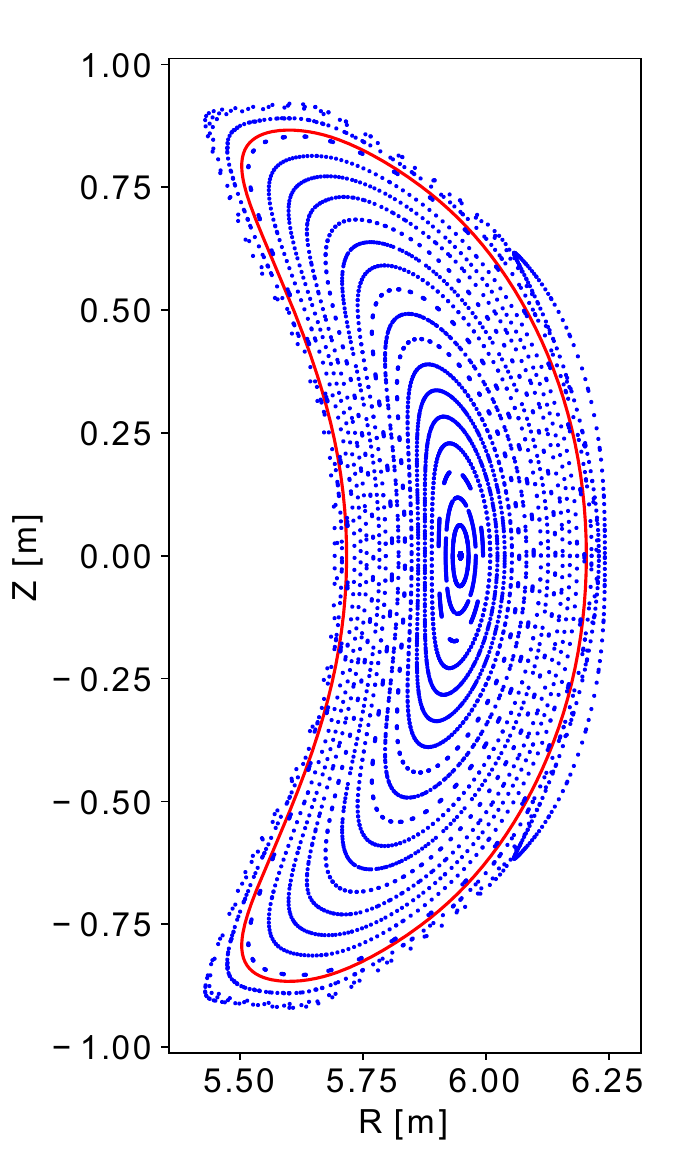}
		\caption{}
		\label{fig:poincare_fb36}
		\end{subfigure}
		\begin{subfigure}{0.45\textwidth}
		\centering
        \includegraphics[width=0.6\textwidth]{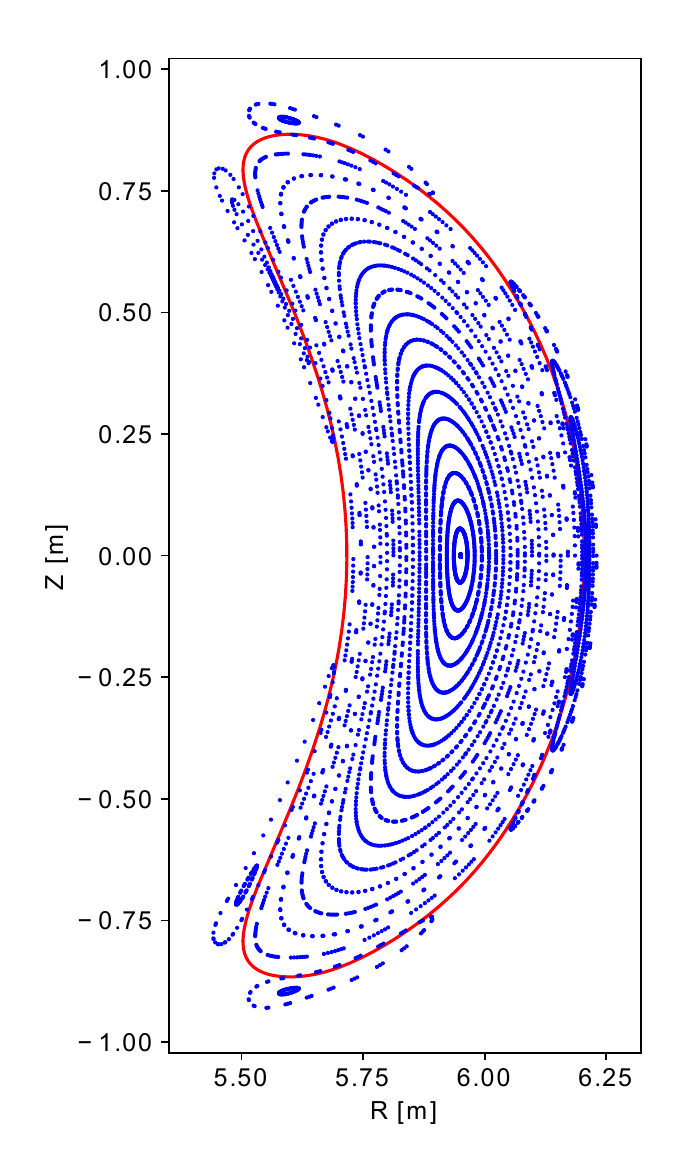}
		\caption{}
		\label{fig:poincare_36}
		\end{subfigure}
		
		\caption{Poincare plots for the W7-X{-like} stellarator surface, with (a) finite build coils (18cm by 18cm), optimized using a multi-filament approximation. (b) finite build coils (18cm by 18cm), optimized using a filamentary approximation. {(c) finite build coils (36cm by 36cm), optimized using a multi-filament approximation.} (d) finite build coils (36cm by 36cm), optimized using a filamentary approximation. Comparing (b) to (a) {as well as (d) to (c)}, we see that optimizing finite-build coils with a filamentary approximation leads to a different island structure relative to directly optimizing finite-build coils; for this particular surface, as the field error increases, the (5,5) magnetic islands move radially inwards towards the target surface. }
		\label{fig:poincare_w7x}
\end{figure}

In figure \ref{fig:w7x_e}, we similarly see that as the coil size $\delta$ is increased, that $\Delta e$ increases. For coils of approximately the same dimensions as the W7-X coils, the normalized field error $e$ increases by approximately 50\% {from its initial value of $4 * 10^{-4}$} if the finite-build is not accounted for. Determining whether this 50\% increase will lead to a significant degradation in the properties we care about (such as radial transport, MHD stability, energetic particle confinement, etc) would require a detailed analysis of the equilibrium that is outside the scope of this research. 

{Lastly, we look at} Poincare plots of the magnetic field lines {to get a qualitative} understand{ing} {of how finite-build effects might affect the magnetic field}. 
Figure{s} \ref{fig:poincare_fb} {and \ref{fig:poincare_fb36}} show Poincare plots of zero-rotation finite-build coils optimized using the multi-filament model; {figure \ref{fig:poincare_fb} has an 18cm by 18cm cross-section, while figure \ref{fig:poincare_fb36} has a 36cm by 36cm cross-section}. Figures \ref{fig:poincare_18} and \ref{fig:poincare_36} show Poincare plots of the magnetic field produced by finite-build coils which are optimized using a filamentary approximation, figure \ref{fig:poincare_18} has an 18cm by 18cm cross-section and $\Delta e \approx 50\%$, while figure \ref{fig:poincare_36} has a 36cm by 36cm cross-section and $\Delta e \approx 400\%$. Comparing figure \ref{fig:poincare_18} to \ref{fig:poincare_fb} {as well as figure \ref{fig:poincare_36} to \ref{fig:poincare_fb36}}, we see a small qualitative difference in the magnetic field structure from finite-build coils which are optimized using a filamentary approximation. 
{The structure of the magnetic field in the core shows little change, which is consistent with the understanding that finite-build effects diminish in magnitude with distance from the coils. The structure of the magnetic field in the edge shows some small changes in the size of the (5,5) magnetic island and the chaotic zones. This is also consistent with our understanding: closer to the coils, the finite build effects are larger. 
Also, because the width of the magnetic islands is proportional to the square root of the magnitude of the resonant perturbation and inversely proportional to the shear \cite{ISI:A1991FD80100018}, and overlapping islands creates chaotic fieldlines \cite{ISI:A1979HD59000001}, we expect that small changes in the magnetic field can result in noticeable changes in the island size and structure of the magnetic field in the edge region.
This} provides further evidence for the conclusion that finite build effects should be accounted for in the optimization {of coils for} a stellarator such as W7-X. {Finite-build effects are an important consideration for the optimization of the divertor, particularly for low-shear stellarators.}

\subsection{Conclusion \label{sec:ch3-coilopt-conclusion}}

We have introduced a new finite-build stellarator coil design code called FOCUSADD which represents coils as filamentary closed curves in space surrounded by a multi-filament approximation to the coil winding pack. The optimization parameters are the Fourier series describing the winding pack centroid positions and optionally the winding pack {orientation}. These parameters are optimized using a simple gradient descent with momentum algorithm. The required derivatives are calculated using the reverse mode automatic differentiation (AD) tool JAX. By using reverse mode AD, the derivatives are calculated easily and efficiently, at a cost independent of the number of parameters. 

We found that by allowing coil winding packs to rotate freely in space, the quadratic flux can be further decreased, but at the cost of finding coils whose rotation profiles would be extremely difficult to engineer. Singh et al. \cite{l_singh_preprint} come to the same conclusion, but find that regularizing the coil rotation {profiles} results in more feasible coils. However, each of these approaches uses the arbitrary center of mass frame; neither is based on a rigorous or quantitative understanding of what makes a coil challenging or expensive to build. Future work in finite-build stellarator coil design should develop such understanding to design either a new winding pack frame or a penalty on the rotation profile which would result in a winding pack rotation profile which is easier to engineer. An open question in finite-build stellarator coil design is whether there will be any freedom in the rotation profile to optimize for physics objectives, or whether engineering feasibility alone will determine the {winding pack orientation}. 

We compared zero-rotation finite-build coils to filamentary coils for a W7-X{-like} surface, and found that the optimized finite-build coils were shifted relative to the optimized filamentary coils. For coils of approximately the same sizes as were built in the experiment, the mean shift in coil positions was approximately 2.5mm. We also found that optimizing filamentary rather than finite-build coils leads to a{n increase in the normalized field error $e$ as well as a} qualitatively different magnetic island structure. These calculations suggest that the finite build of stellarator coils is a non-negligible effect which should be included in the optimization of coils for real stellarator experiments.

Automatic differentiation has allowed us to compute the gradient of our objective function efficiently and easily. Although this objective function consists of the standard quadratic flux plus a simple penalty term, this objective function could easily be modified in future work to include other objective functions, such as coil-coil spacing, inter-coil electromagnetic forces, curvature, or torsion.

\section{Magnetic field optimization \label{sec:ch3-magnetic-field}}

In the previous section, we discussed how AD could be used to optimize stellarator coils to produce a desired magnetic field defined by the Biot-Savart law. In this section, we'll discuss how AD can be used to solve for non-axisymmetric magnetic fields that satisfy magnetohydrodynamic (MHD) equilibrium. We'll then discuss how AD can be used to optimize those magnetic fields, subject to the constraint that MHD equilibrium is satisfied.

\subsection{Optimization with AD \label{sec:ch3-optimization-ad}}

Before discussing how to solve for magnetic fields that satisfy MHD equilibrium, we'll introduce a few basic methods used in derivative-based optimization.

The unconstrained optimization problem is typically formulated as finding $\bm x^* \in \mathbb{R}^n$ which minimizes a scalar objective function or loss function $L(\bm x) : \mathbb{R}^n \to \mathbb{R}$,
\begin{equation}\label{eq:ch3-scalar-optimization}
    \bm x^* = \argminC_{\bm{x}} L(\bm{x}).
\end{equation}
Typically, the goal is to find the global minimum of $L$, though if $L$ is non-convex then optimization methods can usually only be guaranteed to find a local minimum.
At a local minimum,
\begin{equation}
    \grad_{\bm x} L = 0.
\end{equation}
Thus, optimization problems that involve minimizing a scalar objective function can also be formulated as finding the solution to a non-linear vector-valued function $\grad_{\bm x} L = 0$ where $\grad_{\bm x} L : \mathbb{R}^n \to \mathbb{R}^n$. 

Finding the solution to $\grad_{\bm x} L = 0$ is a special case of the more general problem of root-finding for the possibly non-linear equation 
\begin{equation}\label{eq:ch3-f-equals-0}
    \bm F(\bm x) = 0
\end{equation}
where $\bm F : \mathbb{R}^n \to \mathbb{R}^m$.
In general, \cref{eq:ch3-f-equals-0} may have no solutions (overdetermined) or infinitely many solutions (underdetermined).
In these cases, $\bm x$ can be chosen as the solution to an optimization problem
\begin{equation}\label{eq:ch3-nlls-optimization}
    \bm x^* = \argminC_{\bm{x}} \frac{1}{2}||\bm F||^2
\end{equation}
known as non-linear least squares.

\subsubsection{Gradient descent}

The simplest gradient-based optimization method is gradient descent. Gradient descent iteratively updates $\bm x$ until it is sufficiently close to a local minimum. Each update is given by
\begin{equation}
    \bm x \longleftarrow \bm x - \eta \grad_{\bm x} L
\end{equation}
where $\eta \in \mathbb{R}$ is a learning rate. Gradient descent (and variations on gradient descent such as Adam \cite{kingma2014adam}) are the most common optimization methods for neural networks, but these methods are often suboptimal for lower-dimensional problems.
Reverse mode AD can be used to compute $\grad_{\bm x} L$ at a cost $\mathcal{O}(1)$ times that of computing $L(\bm x)$. 

\subsubsection{Newton's method}

Newton's method involves Taylor expanding $L(\bm x + \Delta \bm x)$ to second order around $\bm x$
\begin{equation}
    L(\bm x + \Delta \bm x) = L(\bm x) + \Delta \bm x^T \grad_{\bm x} L + \frac{1}{2}\Delta \bm x^T \bigg(\frac{\partial^2 L}{\partial \bm x^2}\bigg) \Delta \bm x + \mathcal{O}(\Delta \bm x^3).
\end{equation}
and choosing $\Delta \bm x$ to minimize $L(\bm x + \Delta \bm x)$. This happens when $\grad_{\bm x} L = 0$,
\begin{equation}
    \grad_{\bm x} L(\bm x + \Delta \bm x) \approx \grad_{\bm x} L + \bigg(\frac{\partial^2 L}{\partial \bm x^2}\bigg) \Delta \bm x = 0,
\end{equation}
so $\Delta \bm x$ given by
\begin{equation}\label{eq:ch3-nm-scalar}
    \Delta \bm x = -\bigg(\frac{\partial^2 L}{\partial \bm x^2}\bigg)^{-1} \grad_{\bm x} L. 
\end{equation}
Each update is given by
\begin{equation}\label{eq:ch3-iterativeupdate}
    \bm x \longleftarrow \bm x + \Delta \bm x
\end{equation}
AD can be used to compute $\frac{\partial^2 L}{\partial \bm x^2}$ at cost $\mathcal{O}(n)$ times that of computing $L(\bm x)$.
Thus, each iteration of Newton's method is $\mathcal{O}(n)$ times slower than gradient descent, which can be slow for large $n$. However, near a local minimum Newton's method is known to converge in many fewer iterations than gradient descent.

Newton's method can also be used as a root-finding algorithm to solve $\bm F = 0$. In this case, a Taylor expansion to first order of $\bm F(\bm x + \Delta \bm x)$ is used
\begin{equation}
    \bm F(\bm x + \Delta \bm x) = \bm F(\bm x) + \bm J_{\bm F} \Delta \bm x + \mathcal{O}(\Delta \bm x^2)
\end{equation}
and the linear approximation is set to zero:
\begin{equation}
    \bm F(\bm x) + \bm J_{F} \Delta \bm x = 0
\end{equation}
\Cref{eq:ch3-iterativeupdate} is again used as an iterative update equation, except with
\begin{equation}\label{eq:ch3-nm-root}
    \Delta \bm x = - \big(\bm J_{\bm F}\big)^{-1} \bm F(\bm x).    
\end{equation}
AD can be used to compute $\bm J_{\bm F}$ in time $\mathcal{O}(n)$ the cost of computing $\bm F$. Notice that if $\bm F = \grad_{\bm x}L$, then $\bm J_{\bm F} = \frac{\partial^2 L}{\partial \bm x^2}$ and \cref{eq:ch3-nm-root} becomes identical to \cref{eq:ch3-nm-scalar}.

If $m \ne n$, or if $n=m$ but the Jacobian $(J_{\bm F})$ is not invertible, $\bm F = 0$ can instead formulated as a non-linear least squares optimization problem with $L = \frac{1}{2}||\bm F||^2$. Newton's method for non-linear least squares is called the Gauss-Newton method.
In the Gauss-Newton method, a Taylor expansion to first order of $\bm F(\bm x + \Delta \bm x)$ is again used, giving
\begin{equation}
    L \approx \frac{1}{2}|| \bm F(\bm x) + \bm J_{F} \Delta \bm x ||^2.
\end{equation}
Setting $\grad_{\Delta \bm x} L = 0$ gives
\begin{equation}
    \grad_{\Delta \bm x} L = \bm J_{\bm F}^T(\bm F(\bm x) + \bm J_{\bm F} \Delta \bm x) = 0.
\end{equation}
If $m > n$, then $(\bm J_{\bm F}^T\bm J_{\bm F})$ is non-singular and 
$\Delta \bm x$ is given by
\begin{equation}\label{eq:ch3-newton-overdetermined}
    \Delta \bm x = -\bm J_{\bm F}^{+} \bm F(\bm x)
\end{equation}
where $\bm J_{\bm F}^{+} = (\bm J_{\bm F}^T\bm J_{\bm F})^{-1} \bm J_{\bm F}^T \bm F(\bm x)$ is the (left) Moore-Penrose pseudoinverse of $\bm J_{\bm F}$.
If $n < m$, then a lengthier calculation comes to the same result, except that $\bm J_{\bm F}^+ = \bm J_{\bm F}^T(\bm J_{\bm F}\bm J_{\bm F}^T)^{-1}$ is the (right) Moore-Penrose pseudoinverse of $\bm J_{\bm F}$.
$\bm J_{\bm F}$ can be computed with cost $\mathcal{O}(n)$ or $\mathcal{O}(m)$ depending on whether forward- or reverse-mode AD is used.
In practice, the Gauss-Newton method is often replaced with the Levenberg-Marquardt method, which has
\begin{equation}
    \Delta \bm x = -(\bm J_{\bm F}^T\bm J_{\bm F} + \lambda \bm I)^{-1} \bm J_{\bm F}^T \bm F(\bm x)
\end{equation}
for some scalar regularization parameter $\bm \lambda$.

\subsubsection{PDE-constrained optimization}

Suppose that the (approximate) solution $\bm x^*$ to a PDE can be written as the solution to the non-linear equation $\bm F = 0$.
$\bm x^*$ may be the solution to an optimization problem of the form \cref{eq:ch3-scalar-optimization}, in which case $\bm F = \grad_{\bm x} L(\bm x)$.
$\bm x^*$ might also be the solution to a non-linear least squares problem of the form \cref{eq:ch3-nlls-optimization}.
For example, when solving Poisson's equation $\nabla^2 \phi = \chi$ using the finite element method, the discrete solution $\bm x^*$ is given by the solution to a linear equation $\bm A \bm x - \bm b = 0$.
When, for example, solving the MHD equilibrium equation $(\bm \nabla \times \bm B) \times \bm B = \mu_0 \grad p$ (discussed in \cref{sec:ch3-mhd-equilibrium}), the solution $\bm x^*$ is given by the solution to the non-linear equation $\bm F (\bm x) = 0$.

Suppose also that we would like to find the parameters $\bm p$ that minimize a scalar cost function $f(\bm x, \bm p)$ where $\bm x$ is constrained to be the solution to the PDE equation $\bm F(\bm x, \bm p) = 0$.
For example, the parameters $\bm p$ might be the geometry of the outer flux surface of a stellarator magnetic field, the cost function $f(\bm x, \bm p)$ might measure the omnigeneity or quasisymmetry of the magnetic fields (and thus the loss of charged particles in a stellarator), and the parameters $\bm x$ describing the magnetic field might be constrained to satisfy the MHD equilibrium equation. We'll discuss $\bm x$ and $\bm p$ in more detail in \cref{sec:ch3-mhd-equilibrium}.

Fortunately, this problem setup is identical to the discussion of implicit differentiation in \cref{sec:ch2-implicit-differentiation}. $f(\bm x, \bm p)$ can be optimized using gradient descent, with $\grad_{\bm p} f$ given by \cref{eq:ch2-implicit-gradient}:
\begin{equation}\label{eq:ch3-pde-constrained-optimization}
    \grad_{\bm p} f = \frac{\partial f}{\partial \bm p} + \bm \lambda^T \frac{\partial \bm F}{\partial \bm p}.
\end{equation}
The adjoint vector $\bm \lambda$ is given by \cref{eq:ch2-adjoint-variable}
\begin{equation}\label{eq:ch3-pde-constrained-adjoint}
    \bigg(\frac{\partial \bm F}{\partial \bm x}\bigg)^T \bm \lambda = - \frac{\partial f}{\partial \bm x}.
\end{equation}
Using AD, $\frac{\partial f}{\partial \bm p}$ and $\frac{\partial f}{\partial \bm x}$ can be computed using reverse mode AD in time $\mathcal{O}(1)$, while the vector-Jacobian product $\bm \lambda^T \frac{\partial \bm F}{\partial \bm p}$ can be computed in time $\mathcal{O}(1)$ using reverse mode AD once the the adjoint vector $\bm \lambda$ is computed using \cref{eq:ch3-pde-constrained-adjoint}.
Using either forward mode AD or reverse mode AD, $\frac{\partial \bm F}{\partial \bm x}$ can be computed either in $\mathcal{O}(m)$ or $\mathcal{O}(n)$ times the cost of computing $\bm F$;
however, if the solution to the PDE equation $\bm F = 0$ is found using Newton's method, then $\bm J_{\bm F} = \frac{\partial \bm F}{\partial \bm x}$ is already calculated in the process of solving $\bm F = 0$, and the only additional cost is another linear solve.

In practice, $\frac{\partial \bm F}{\partial \bm x}$ is often not square and/or not invertible, and $\bm x$ is typically given by the solution to a non-linear least squares optimization problem. Fortunately, this problem setup is identical to the discussion of overdetermined and underdetermined implicit differentiation in \cref{sec:ch2-implicit-differentiation}. If the PDE solution is given by a non-linear least-squares optimization problem, the gradient of the cost function can be computed using \cref{eq:ch2-adjoint-variable-overdetermined,eq:ch2-implicit-gradient-overdetermined}.

\subsection{Computing MHD equilibrium \label{sec:ch3-mhd-equilibrium}}

We'd like to solve the MHD equilibrium equation
\begin{equation}\label{eq:ch3-mhd-equilibrium}
    \bm J \times \bm B = \grad p
\end{equation}
with $\bm J = \frac{1}{\mu_0}\grad \times \bm B$ for 3D magnetic fields in toroidal geometries.
We'll begin by assuming the existence of closed, nested, toroidal flux surfaces with flux label $\psi$ such that $\bm B \cdot \grad \psi = 0$. These flux surfaces are guaranteed to exist for 2D axisymmetric magnetic fields with non-zero toroidal current \cite{imbert2019introduction}; assuming that they also exist in 3D non-axisymmetric toroidal fields greatly simplifies the solution of \cref{eq:ch3-mhd-equilibrium}.
We can use the non-orthogonal flux coordinates $(\psi, \theta, \zeta)$ to label any point within the nested toroidal flux surfaces; $\psi \in [0, \psi_B)$ is a radial coordinate, $\theta \in [0, 2\pi)$ is a poloidal angle, and $\zeta \in [0, 2\pi)$ is a toroidal angle. For an introduction to flux coordinates, see chapter 6 of \cite{imbert2019introduction}.

\subsubsection{Inverse coordinate representation}

The essential step in solving \cref{eq:ch3-mhd-equilibrium} is to write the flux coordinates in their inverse representation
\begin{subequations}
\begin{equation}
    R = R(\psi, \theta, \zeta)
\end{equation}
\begin{equation}
    \phi = \phi(\psi, \theta, \zeta)
\end{equation}
\begin{equation}
    Z = Z(\psi, \theta, \zeta)
\end{equation}
\end{subequations}
where $(R, \phi, Z)$ are the cylindrical coordinates.
$R$, $\phi$, and $Z$ are periodic functions of $\theta$ and $\zeta$.
For simplicity, $\phi = \zeta$ or $\phi = -\zeta$ is often chosen.

The inverse representation of the flux coordinates allows us to write the contravariant basis vectors in the $(\psi, \theta, \phi)$ coordinate system (choosing $\phi = \zeta$):
\begin{subequations}
\begin{equation}
    \frac{\partial \bm r}{\partial \psi} = \big(\frac{\partial R}{\partial \psi}, 0, \frac{\partial Z}{\partial \psi}\big)
\end{equation}
\begin{equation}
    \frac{\partial \bm r}{\partial \theta} = \big(\frac{\partial R}{\partial \theta}, 0, \frac{\partial Z}{\partial \theta}\big)
\end{equation}
\begin{equation}
    \frac{\partial \bm r}{\partial \phi} = \big(\frac{\partial R}{\partial \phi}, R, \frac{\partial Z}{\partial \phi}\big ).
\end{equation}
\end{subequations}
These derivatives can be computing using (forward mode) AD, though they can also be computed analytically as well.
The contravariant basis vectors allow us to compute the Jacobian determinant
\begin{equation}
    \sqrt{g} = \Big( \frac{\partial \bm r}{\partial \psi} \times \frac{\partial \bm r}{\partial \theta}\Big) \cdot \frac{\partial \bm r}{\partial \phi}.
\end{equation}

\subsubsection{Contravariant form of $\bm B$}

Since $\bm B \cdot \grad \psi = 0$, then
\begin{equation}
    \bm B = B^\theta \frac{\partial \bm r}{\partial \theta} + B^\phi \frac{\partial \bm r}{\partial \phi}.
\end{equation}
$\grad \cdot \bm B = 0$ gives
\begin{equation}
        \bm \nabla \cdot \bm B = \frac{1}{\sqrt{g}} \bigg(\frac{\partial}{\partial \theta}(B^\theta \sqrt{g}) + \frac{\partial}{\partial \phi} (B^\phi \sqrt{g}) \bigg) = 0
\end{equation}
which implies that the magnetic field components can be written as
\begin{subequations}
\begin{equation}
    B^\phi \sqrt{g} = j(\psi)\Big(1 + \frac{\partial \lambda(\psi, \theta, \phi)}{\partial \theta}\Big) 
\end{equation}    
\begin{equation}
    B^\theta \sqrt{g} = h(\psi) - j(\psi)\frac{\partial \lambda(\psi, \theta, \phi)}{\partial \phi}.
\end{equation}
\end{subequations}
A short calculation \cite{imbert2019introduction} shows that $j(\psi) = \Psi_T'(\psi)/2\pi$ and $h(\psi) = \Psi_P'(\psi)/2\pi$, where $\Psi_T(\psi)$ and $\Psi_P(\psi)$ are the toroidal and poloidal fluxes within the flux surface labeled by coordinate $\psi$.
Thus, the contravariant form of the magnetic field can be written as
\begin{subequations}
\begin{equation}
    B^\phi = \frac{\Psi_T(\psi)'}{2\pi \sqrt{g}}(1 + \frac{\partial \lambda}{\partial \theta})
\end{equation}
\begin{equation}
    B^\theta = \frac{\Psi_T'(\psi)}{2\pi \sqrt{g}}(\iota(\psi) - \frac{\partial \lambda}{\partial \phi} ).
\end{equation}
\end{subequations}
$\iota(\psi)$ is the rotational transform, defined to be $\Psi'_P/\Psi'_T$.

\subsubsection{Covariant form of $\bm B$}

The magnetic field can be written in covariant form using
\begin{equation}\label{eq:ch3-covariantB}
    B_i = g_{ij} B^j
\end{equation}
where $g$ is the metric tensor with components
\begin{equation}
    g_{ij} = \frac{\partial \bm r}{\partial x_i} \cdot \frac{\partial \bm r}{\partial x_j}.
\end{equation}
\subsubsection{Contravariant form of $\bm J$}
The current can be calculated using
\begin{equation}\label{eq:ch3-J}
    \bm J = \frac{1}{\mu_0} \grad \times \bm B = \sum_{k=1}^3 \frac{1}{\sqrt{g}} (\frac{\partial B_j}{\partial x_i} - \frac{\partial B_i}{\partial x_j}) \frac{\partial \bm r}{\partial x_k}.
\end{equation}
Once again, these derivatives can be computed analytically or using forward mode AD.
AD is easier, though possibly slower by a small factor.

\subsubsection{Computing MHD equilibrium using optimization}

There are two main strategies for solving \cref{eq:ch3-mhd-equilibrium}. The first is to find a minimum of 
\begin{equation}\label{eq:ch3-energy}
    W = \int \Big(\frac{B^2}{2\mu_0} - p \Big) dV
\end{equation}
subject to certain constraints.
As chapter 11 of \cite{imbert2019introduction} derives, if the first-order variation $\delta W$ is equal to $0$ due to first-order variations in $\bm B$ and $p$, subject to the constraints listed below, then \cref{eq:ch3-mhd-equilibrium} must be satisfied.
These constraints are:
\begin{itemize}
    \item Exists closed nested flux surfaces such that $\psi$ is a flux surface label and $\bm B \cdot \grad \psi = 0$.
    \item $\grad \cdot \bm B = 0$.
    \item Pressure as a function of $\psi$ is fixed, $p(\psi) = $ constant.
    \item Poloidal transform as a function of $\psi$ is fixed, $\iota(\psi) = $ constant.
    \item The toroidal flux enclosed by the outer flux surface is constant, $\Psi_T(\psi = \psi_B) = $ constant.
\end{itemize}
Finding the parameters of the magnetic field representation that minimize $W$ subject to these constraints ensure that $\grad W = 0$, which implies that $\delta W = 0$ due to first-order variations in $\bm B$ (and not $p$, since $p(\psi)$ is constant).
$W$ can be minimized using the techniques discussed in \cref{sec:ch3-optimization-ad}.

The second approach to solving \cref{eq:ch3-mhd-equilibrium} is to compute the force $\bm f = \bm J \times \bm B - \grad p$ at $m$ points, then solve the vector-valued equation $\bm F(\bm x) = 0$
where $\bm F : \mathbb{R}^n \to \mathbb{R}^{3m}$ outputs each component of $\bm f$ at each of the $m$ evaluation points. In practice, since $n\ne m$ in general, a nonlinear least squares algorithm such as Gauss-Newton or Levenberg-Marquardt will be used to find the vector of parameters $\bm x$ describing the magnetic equilibrium.

\subsubsection{Example: axisymmetric DESC}

DESC is a 3D MHD equilibrium code that can find solutions to \cref{eq:ch3-mhd-equilibrium} as well as optimize the geometry and parameters of the equilibrium to improve plasma performance \cite{dudt2020desc,panici2023desc,conlin2023desc,dudt_conlin_panici_kolemen_2023}.
DESC uses AD for both equilibrium solves and optimization.
For the pedagogical purpose of better understanding how AD can be used for stellarator optimization, it is illustrative to consider a simplified version of DESC for axisymmetric magnetic fields.

DESC uses flux coordinates $\rho \in [0, 1)$, $\theta \in [0, 2\pi)$, and $\phi \in [0, 2\pi)$.
$\rho$ is a flux function, $\theta$ is a poloidal angle, and $\phi$ is a toroidal angle.
In this simplified example we assume axisymmetry so that $\frac{\partial}{\partial \phi} = 0$ for all quantities.
DESC makes the choice $\rho = \sqrt{\psi/\psi_b}$ where $\psi =\Psi_T(r)/2\pi$ is the toroidal flux per radian within the flux surface with flux label $r$
\begin{equation}
    \Psi_T(r) = \int_0^{r(\theta)}\int_0^{2\pi} \bm B \cdot \bm{\hat{n}} \mathop{dS}
\end{equation}
with $\bm{\hat{n}} = \frac{\grad \phi}{|\grad \phi|}$ and $dS = \sqrt{g}|\grad{\phi}| d\theta dr$.
$\psi_b$ is the toroidal flux per radian at the outer flux surface.
$\rho$ can be understood as the normalized `minor radius' of the plasma, since the toroidal flux will be proportional to the minor radius squared.

The axisymmetric magnetic field can be written as
\begin{subequations}
\begin{equation}
    B^\phi = \frac{2\psi_B \rho}{ \sqrt{g}}(1 + \frac{\partial \lambda}{\partial \theta})
\end{equation}
\begin{equation}
    B^\theta = \frac{2\psi_B \rho}{\sqrt{g}}\iota(\psi)
\end{equation}
\end{subequations}
using $\psi = \Psi_T/2\pi$ and (from the definition of $\rho$) $\psi = \psi_B \rho^2$ implying $\psi' = 2\psi_B \rho$.
The covariant form of the magnetic field and the current $\bm J$ can be computed using \cref{eq:ch3-covariantB,eq:ch3-J}.

We use a simplified 2D version of DESC's 3D representation for the flux surface. 
We write both the position of the flux surface in cylindrical coordinates $\{R, Z \}$ and the function $\lambda(\rho, \theta)$ as a sum over Zernike polynomials
\begin{subequations}\label{eq:ch3-fourier_series_for_rz}
\begin{equation}
    R(\rho, \theta) = \sum_{m=-M}^M \sum_{l \in L} R_{lm} \mathscr{L}_l^m(\rho, \theta)
\end{equation}
\begin{equation}
    Z(\rho, \theta) = \sum_{m=-M}^M \sum_{l \in L} Z_{lm} \mathscr{L}_l^m(\rho, \theta)
\end{equation}
\begin{equation}
    \lambda(\rho, \theta) = \sum_{m=-M}^M \sum_{l \in L} \lambda_{lm} \mathscr{L}_l^m(\rho, \theta)
\end{equation}
\end{subequations}
for $L = |m|, |m+2|, \dots, 2M-|m|$. Here $\mathscr{L}^m_l$ are the Zernike polynomials
\begin{equation}
    \mathscr{L}_{l}^m(\rho, \theta) = \begin{cases}
    \mathscr{R}_l^{|m|}(\rho)\cos{(|m|\theta)},& \text{if } m \ge 0\\
    \mathscr{R}_l^{|m|}(\rho) \sin{(|m|\theta)}              & \text{if } m < 0.
\end{cases}
\end{equation}
$\mathscr{R}_l^{|m|}(\rho)$ is the shifted Jacobi polynomial
\begin{equation}
    \mathscr{R}_l^{|m|}(\rho) = \sum_{s=0}^{(l-|m|)/2}  \frac{(-1)^s(l-s)!}{s![\frac{1}{2}(l+|m|) - s]![\frac{1}{2}(l-|m| - s)]!}\rho^{l-2s}.
\end{equation}
One of the required inputs to DESC is the shape of the last closed flux surface (LCFS) at $\rho = 1$, given by a Fourier series in $\theta$.
\begin{subequations}
\begin{equation}
    R^b(\theta) = \sum_{m=-M}^M R^b_m \mathscr{F}_m(\theta)
\end{equation}
\begin{equation}
    Z^b(\theta) = \sum_{m=-M}^M Z^b_m \mathscr{F}_m(\theta)
\end{equation}
\end{subequations}
$\mathscr{F}_m(\theta)$ are the Fourier polynomials
\begin{equation}
    \mathscr{F}_m(\theta) = \begin{cases}
        \cos{(|m|\theta)},& \text{if } m \ge 0\\
        \sin{(|m|\theta)}              & \text{if } m < 0.
    \end{cases}
\end{equation}
Using $\mathscr{R}^{|m|}_l(\rho = 1)=1$ for all $m$ and $l$, the boundary condition requires
\begin{subequations}
\begin{equation}
    \sum_{l\in L} R_{lm} = R_m^b,
\end{equation}
\begin{equation}
    \sum_{l\in L} Z_{lm} = Z_m^b.
\end{equation}
\end{subequations}
We initialize the basis coefficients $\{R_{lm}, Z_{lm}\}$ to be
\begin{subequations}
\begin{equation}
    R_{lm} = 
    \begin{cases}
        R^b_m,& \text{if } l = |m|\\
        0 & \text{otherwise.} 
    \end{cases}
\end{equation}
\begin{equation}
    Z_{lm} = 
    \begin{cases}
        Z^b_m,& \text{if } l = |m|\\
        0 & \text{otherwise.} 
    \end{cases}
\end{equation}
\end{subequations}
DESC finds a solution to \cref{eq:ch3-mhd-equilibrium} by finding a solution $\bm x^*$ to the non-linear least-squares problem for $\bm F(\bm x, \bm p) = 0$:
\begin{equation}
    \bm x^* = \argminC_{\bm x} \frac{1}{2} || \bm F(\bm x, \bm c)||^2.
\end{equation}
$\bm x$ represents the Zernike coefficients of $\{R, Z, \lambda\}$
\begin{equation}
    \bm x = \{ R_{lm}, Z_{lm}, \lambda_{lm}\}
\end{equation}
while $\bm p$ represents the parameters given as inputs to DESC
\begin{equation}
    \bm p = \{R^b_m, Z^b_m, p(\rho), \iota(\rho), \psi_B \}.
\end{equation}
There are five parameters in $\bm p$: the two sets of Fourier coefficients describing the boundary $\{R^b_m, Z^b_m\}$, the pressure $p(\rho)$ as a function of $\rho$, the rotational transform $\iota(\rho)$ as a function of $\rho$, and the toroidal flux per unit radian at the last closed flux surface $\psi_B$.
Each component of $\bm F(\bm x, \bm p) : \mathbb{R}^n \times \mathbb{R}^p \to \mathbb{R}^{3m}$ is one of the $3$ components of the force $\bm f = \bm J \times \bm B - \grad p$ evaluated at one of the $m$ evaluation points. 

\subsubsection{Gradient-based optimization with DESC}

Suppose that I would like to choose the parameters $\bm p$ to minimize a scalar cost function $f(\bm x, \bm p)$ subject to the constraint that $\bm x$ satisfies MHD equilibrium. If $\bm x$ is computed using DESC, then
\begin{equation}
    \bm x = \argminC_{\bm x} \frac{1}{2} ||\bm F(\bm x, \bm p)||^2.
\end{equation}
Suppose that we'd like to minimize $f$ using gradient descent. The gradient $\frac{df}{d\bm p}$ can be computed via PDE-constrained optimization discussed in \cref{sec:ch3-optimization-ad}, and the required equations are \cref{eq:ch2-adjoint-variable-overdetermined,eq:ch2-implicit-gradient-overdetermined}.

\subsubsection{Gradient-based optimization with VMEC}

Suppose instead that $\bm x$ is computed using VMEC, which minimizes $W$ (\cref{eq:ch3-energy}) instead of finding a root of $\bm F$. Unlike DESC, VMEC isn't written in JAX and doesn't support AD. Nevertheless, if VMEC were written in JAX, it would be possible to perform gradient-based optimization of VMEC. Since $\bm x$ is a minimum of $W$, then $\bm x$ is given by the stationary point equation
\begin{equation}
    \frac{\partial W}{\partial \bm x}(\bm x, \bm p) = 0.
\end{equation}
Suppose that we'd like to minimize a cost function $f$ using gradient descent. The gradient $\frac{df}{d\bm p}$ can be computed via PDE-constrained optimization (\cref{sec:ch3-optimization-ad}) for stationary points (\cref{sec:ch2-implicit-differentiation}). $\frac{df}{d\bm p}$ is given by \cref{eq:ch3-pde-constrained-optimization,eq:ch3-pde-constrained-adjoint} with $\bm F = \frac{\partial W}{\partial \bm x}$.

\section{Higher-order convergence of the Biot-Savart law}

In the previous two sections, I discussed how AD can be used for stellarator coil optimization (\cref{sec:ch3-coil}) and magnetic field optimization (\cref{sec:ch3-magnetic-field}). In this section, I discuss a related but separate topic: how to more efficiently compute the magnetic field from a current-carrying coil using the Biot-Savart law.


The Biot-Savart formula for the magnetic field produced by a static current-carrying wire loop is well known:
\begin{equation}\label{eq:ch3-filamentary_biot_savart}
    \bm B(\bm r) = \frac{\mu_0 I}{4\pi}{\oint} \frac{ d\bm r' \times (\bm r - \bm r')}{|\bm r - \bm r'|^3}.
\end{equation}
Researchers in plasma physics numerically compute the Biot-Savart law in a variety of applications, especially those related to stellarator optimization. Examples include field line tracing \cite{Pedersen2016}, particle tracking, magnetohydrodynamic equilibrium codes \cite{stellopt}, and stellarator coil optimization codes \cite{Zhu_2017}. 
If the magnetic field cannot be calculated analytically, equation \ref{eq:ch3-filamentary_biot_savart} must be computed numerically.

There are three main approaches to computing the Biot-Savart line integral given by equation \ref{eq:ch3-filamentary_biot_savart}. The first approach involves using a numerical quadrature and is usually highly accurate, but {gives $\bm{\nabla}\times\bm{B}\ne 0$ and} requires storing the coil parameterization $\bm r'(s)$ or a series of position vectors $\bm r'_i$ and tangent vectors $\delta \bm r'_i$. The second approach, described in the next paragraph, is inaccurate {relative to the first approach} but is conceptually simple and is widely used because it only requires storing a series of discrete points $\bm r'_i$ {and because, to machine precision, $\bm\nabla \times \bm B =0$ when the evaluation point does not lie on the curve}. The third approach relies on the Fast Multipole Method \cite{greengard_rokhlin_1997} and is discussed in the related work section (section \ref{sec:ch3-biotsavart-relatedwork}). The purpose of this section is to propose a new approach to evaluating the Biot-Savart line integral which is both highly accurate and conceptually simple {while giving $\bm\nabla \times \bm B =0$},\footnote{{It is well known that the Biot-Savart integral has zero curl in regions where $\bm J = 0$, as predicted by Ampere's law. However, the integrand of the Biot-Savart law has non-zero curl even where $\bm J = 0$. Therefore, in contrast with many other methods (e.g., quadrature, discontinuous piecewise linear), the explicit construction of a closed filamentary curve ensures that the resulting magnetic field is curl-free to machine precision.}} thereby enjoying the benefits of both the first and second approaches.

\begin{figure}
    \centering
	\begin{subfigure}[t]{\textwidth}
		\centering%
		\includegraphics[width=0.8\linewidth]{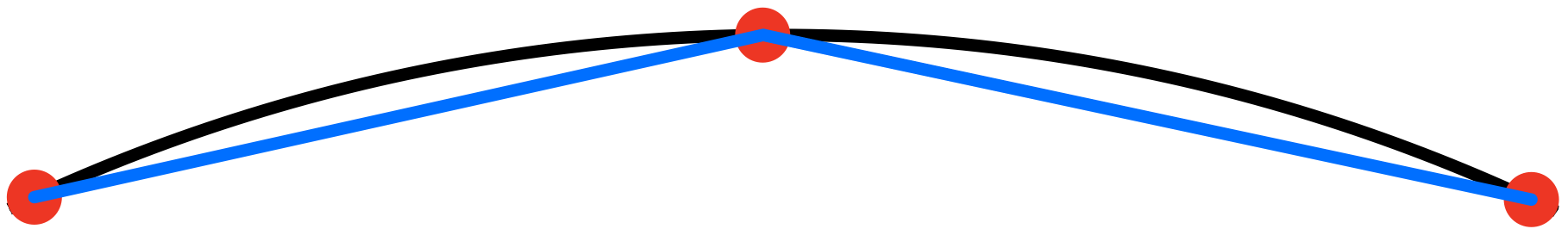}
    \caption{}
    \label{fig:ch3-biotsavart-piecewiselinear}
	\end{subfigure}
	\begin{subfigure}[t]{\textwidth}	
		\centering%
		\includegraphics[width=0.8\linewidth]{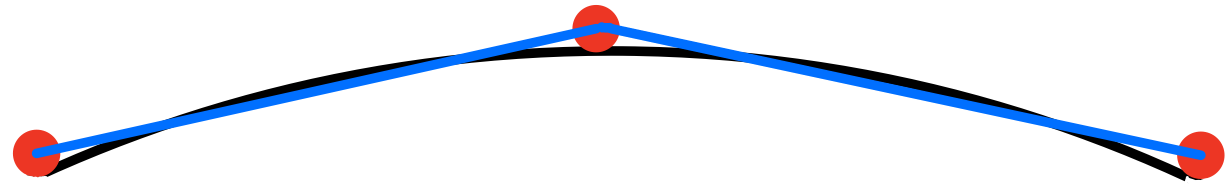}
		\caption{}
		\label{fig:ch3-biotsavart-shiftedpiecewiselinear}
	\end{subfigure}
	\caption{(a) Standard piecewise linear approach. Here the endpoints (red) lie on the filamentary curve (black). (b) Shifted piecewise linear approach. Here the endpoints are shifted in the outwards normal direction relative to the curve.}
	\label{fig:ch3-biotsavart-sketch}
\end{figure}

The second approach to numerically computing equation \ref{eq:ch3-filamentary_biot_savart} is to approximate the coil-carrying filament as a series of straight segments, then sum the magnetic field from each segment. As a convenient analytic expression is known for the magnetic field resulting from a straight segment \cite{Hanson_Hirshman}, a filamentary coil may be simply represented as an array of discrete endpoints. These endpoints lie on the coil and are typically equally spaced in distance or parameterization angle. This approach to computing equation \ref{eq:ch3-filamentary_biot_savart}, which we call the `standard piecewise linear approach', is illustrated in figure \ref{fig:ch3-biotsavart-piecewiselinear}. In figure \ref{fig:ch3-biotsavart-piecewiselinear}, the filamentary curve is shown in black, the discrete endpoints are shown in red, and the piecewise linear segments are shown in blue. 

Although the standard piecewise linear approach is convenient, simple, and flexible, it is not a particularly accurate way of computing the Biot-Savart line integral. One source of numerical inaccuracy can be understood from figure \ref{fig:ch3-biotsavart-piecewiselinear}: because the red endpoints lie on the curve, then the blue straight segments always lie locally inside the black curve. 
Because the straight segment always lies locally inside the curve, the field from the segment is biased and the result is that the accuracy of the Biot-Savart computation is only second-order in the number of endpoints.

We therefore propose a slightly different approach. Instead of choosing endpoints which lie on the filament, we choose endpoints which are slightly shifted in the outwards normal direction. This alternative approach, which we call the `shifted piecewise linear approach', is sketched in figure \ref{fig:ch3-biotsavart-shiftedpiecewiselinear}. Now the blue segments sometimes lie inside and sometimes lie outside the black curve, which eliminates the source of bias mentioned in the previous paragraph. The result is that, if the proper shift is chosen, then the accuracy of the Biot-Savart computation will be fourth order in the number of endpoints and the error will be dramatically reduced.

\begin{figure}
    \centering%
	\includegraphics[width=\linewidth]{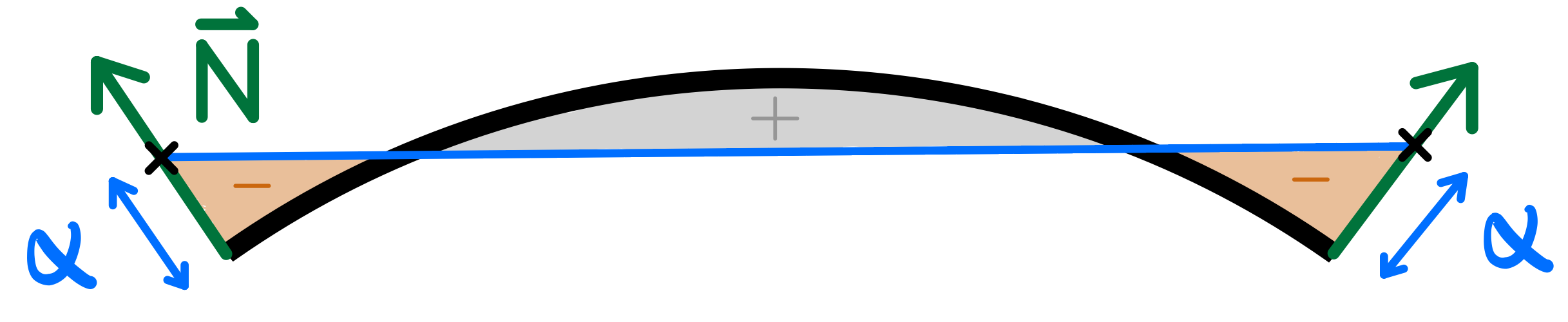}
	
	\caption{A close-up of the geometry of a shifted linear segment. The filamentary coil is shown in black. The normal vector to the curve $\bm N$ is drawn in green, and the straight segment is drawn in blue. The black crosses represent the endpoints of the straight segments, and the magnitude of the shift is given by $\alpha$. To lowest order, the area shaded in grey (positive) is equal to the area shaded in brown (negative) if $\alpha$ is given by equation \ref{eq:ch3-biotsavart-alpha}. }
	\label{fig:ch3-biotsavart-shift}
	
\end{figure}

It remains to determine by what magnitude to shift the endpoints in the outwards normal direction. Figure \ref{fig:ch3-biotsavart-shift} sketches a close-up of the geometry of a shifted linear segment, where the magnitude of the shift is labeled by $\alpha$. In appendix \ref{sec:ch3-biotsavart-appendixA}, we compare the Biot-Savart fields from a curved segment and shifted straight segment, and show that a shift of
\begin{equation}\label{eq:ch3-biotsavart-alpha}
    \alpha = \frac{\kappa |\delta\bm r'|^2}{12}
\end{equation}
in the outwards normal direction cancels the second-order error in the Biot-Savart computation near the center of the coil. Similarly, in appendix \ref{sec:ch3-biotsavart-appendixB} we show that the same shift minimizes the mean squared deviation between the curved segment and the straight segment. Here, the curvature \unstretch{$\kappa \equiv 1/R$}, where $R$ is the radius of curvature, \unstretch{$\delta \bm r' \equiv (d\bm r'/ds) \delta s$}, $s$ is a coordinate parametrizing the coil, $\delta s$ is the coordinate spacing between points, and the normal direction is defined by the Frenet-Serret formulas. 

{Inspired by the calculations in appendices \ref{sec:ch3-biotsavart-appendixA} and \ref{sec:ch3-biotsavart-appendixB}, we propose choosing $\alpha$ using equation \ref{eq:ch3-biotsavart-alpha}. Although we have not discovered an analytic proof for the order of convergence with this shift, we do present careful numerical convergence studies (see section \ref{sec:ch3-biotsavart-tests}) to test our method. Based on these numerical convergence experiments, we conclude that once there are enough endpoints such that the curve between endpoints is locally well-described by a parabola, then the convergence will be fourth-order for a closed curve with continuous curvature. To ensure fourth-order convergence of a piecewise differentiable curve or non-closed curve requires a slightly modified approach, which is described in appendix \ref{sec:ch3-biotsavart-appendixD}.}

In summary, the shifted piecewise linear approach is a highly accurate method of computing the magnetic field produced by a twice-differentiable filamentary coil. The piecewise linear approach results in a {continuous curve and a} piecewise constant representation of the tangent vector and therefore {gives $\bm \nabla \times \bm B = 0$ and} only requires storing a series of discrete points $\bm r'_i$. This is in contrast to the numerical quadrature method, which {gives $\bm \nabla \times \bm B \ne 0$ and which} requires storing a set of discrete points $\bm r'_i$ along with their associated tangent vectors $\delta \bm r'_i$.

\subsubsection{Numerical tests \label{sec:ch3-biotsavart-tests}}

In this section, we perform a few numerical tests of our proposed approach, the so-called `shifted piecewise linear approach'. To do so, we compare the Biot-Savart accuracy of the shifted piecewise linear approach to the accuracy of the standard piecewise linear approach as a function of the number of discretization points $N$ for three separate test problems: (1) an off-axis measurement from a circular coil, (2) an off-axis measurement from a non-planar modular stellarator-like coil, and (3) an on-axis measurement of the magnetic field from a D-shaped, tokamak-like coil. We measure the error as a function of the number of discretization points $N$, $\epsilon_N$, using the normalized $\ell^2$ norm between the produced magnetic field for a given number of discretization points, \unstretch{$\bm B_N(\bm r)$}, and the magnetic field as \unstretch{$N \rightarrow \infty$}, \unstretch{$\bm B_\infty (\bm r)$}:
\begin{equation}\label{eq:error}
    \epsilon_N = \frac{|\bm B_N (\bm r) - \bm B_\infty (\bm r)|}{|\bm B_\infty (\bm r)|}.
\end{equation}
To avoid possible biases related to the particular choice of measurement point $\bm r$, in each test problem we average the error $\epsilon_N$ over a set of ${R=100}$ measurement points $\{\bm r_j\}_{j=1}^R$ where \unstretch{$\bm r_j = \bm r + \bm \delta_j$} and $\bm \delta_j$ is a small random vector. {In} each test {problem}, the {maximum deviation of $\epsilon_N$ relative to the average is less than an order of magnitude}. {For test problem 1, $\bm B_\infty$ is known analytically and involves a computationally efficient elliptic integral \cite{urankar_part1}. }

\begin{figure*}
\centering
\includegraphics[width=\linewidth]{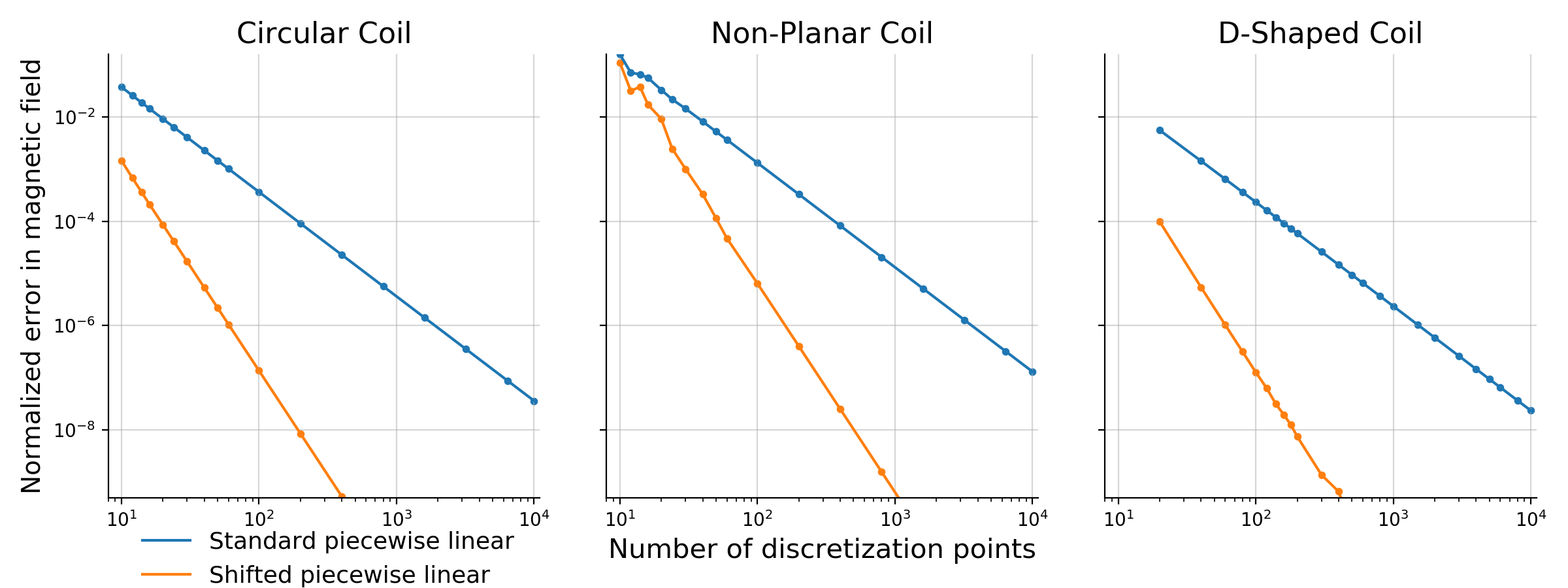}
\caption{The average error $\epsilon_N$ as a function of the number of discretization points $N$ for (left) a circular coil (center) a stellarator-like, non-planar coil, and (right) a tokamak-like, D-shaped coil. The shifted piecewise linear approach is fourth-order in each experiment, while the standard piecewise linear approach is second-order. Code for these experiments, as well as additional experiments, can be found at \url{https://github.com/nickmcgreivy/biot-savart-line-integral}}.
\label{fig:ch3-biotsavart-results}
\end{figure*}

Our results are shown in figure \ref{fig:ch3-biotsavart-results}. In each case, $\epsilon_N$ is smaller using the shifted piecewise linear approach by at least an order of magnitude compared to the standard piecewise linear approach for all \unstretch{$N > 20$}. Additionally, while the accuracy of the standard piecewise linear approach is second-order, the accuracy of the shifted piecewise linear approach is fourth-order for each experiment. The shifted piecewise linear approach is significantly more accurate than the standard piecewise linear approach for a given number of discretization points.

We also {have found} that our results are {qualitatively} robust to the choice of measurement point $\bm r$: it doesn't matter whether $\bm r$ is on-axis, off-axis, near the coil or far from the coil, inside or outside the coil. In each case, the {convergence of the shifted piecewise linear approach remains fourth-order}.

\subsubsection{Related work\label{sec:ch3-biotsavart-relatedwork}}

Early work by \cite{urankar_part1} has developed analytic expressions for the magnetic field produced by filamentary circular \cite{urankar_part1} and elliptic \cite{urankar_elliptic} arc segments, thin conic current sheets \cite{urankar_part2}, rectangular cross-section circular arcs with azimuthal \cite{urankar_part3} and arbitrary \cite{urankar_part4} current densities, and polygonal cross-section circular arcs \cite{urankar_part5}. \cite{Hanson_Hirshman} present new compact expressions for the analytic fields from a straight line segment \cite{Hanson_Hirshman}. \cite{suh_2000_surfaceintegral} develops closed-form expressions for the magnetic field from volumetric elements with planar edges and constant and/or linear current density which involve line integrals over the boundary edges of the element \cite{suh_2000_surfaceintegral}.
\cite{integration_methods} examine a variety of methods for computing the Biot-Savart fields from a volumetric  current-carrying coil, including (i) splitting the domain into tetrahedral regions and performing a quadrature on each domain, and (ii) numerically integrating a semi-analytic one-dimensional integral based on Urankar's analytic expressions \cite{integration_methods}. \cite{rectangular_prism_and_curved_segments} generate analytic expressions for the magnetic field from trapezoidal prisms \cite{rectangular_prism_and_curved_segments}.
\cite{nunes_curvilinear_elements} approximate filamentary, surface current, and volumetric current densities as first-order or second-order finite elements, then use a quadrature to evaluate the current from each finite element and an adaptive method which reduces the required number of Gauss points \cite{nunes_curvilinear_elements}.

\cite{magnetic_field_of_modular_coils} rewrite the Biot-Savart volume integral for a rectangular cross-section finite-build modular stellarator coil as a simpler, tractable integral over finite-build coil coordinates \cite{magnetic_field_of_modular_coils}.
Although the integrals over the coil cross-section have no analytic solution, they can either be evaluated numerically or semi-numerically in a power series. \cite{mcgreivy2020optimized} have approximated the same integral using a multi-filament approximation and a quadrature-based approach to computing the Biot-Savart line integral \cite{mcgreivy2020optimized}. The approach of using a numerical quadrature requires computing the tangent vector to the filamentary coil; this approach is often also significantly more accurate than the standard piecewise linear approach but is not examined here to simplify the presentation.

A different approach to computing the Biot-Savart law uses the Fast Multipole Method (FMM) to improve the runtime scaling of the Biot-Savart law from \unstretch{$\mathcal{O}(NM)$} to \unstretch{$\mathcal{O}(N+M)$} where $N$ is the number of sources and $M$ is the number of evaluation points. This approach to computing the Biot-Savart law will significantly improve the runtime of a Biot-Savart {calculation} which requires a large number of evaluation points $M$, but will be less efficient for small $M$ and $N$. A runtime comparison of these approaches as $N$ and $M$ vary is outside the scope of this presentation. Recently, \cite{ProjectRat} have developed an open-source library called Project Rat which implements this approach \cite{ProjectRat}.

\subsubsection{Conclusion\label{sec:ch3-biotsavart-conclusion}}

We investigated the standard piecewise linear approach to computing the magnetic field from a filamentary coil. We found that this approach is only second-order accurate if the endpoints lie on the coil, because the straight segments are biased to always lie locally inside the coil. We propose a new approach, the shifted piecewise linear approach, where the endpoints are shifted in the outwards normal direction by an amount proportional to the local curvature. We find that a shift of \unstretch{$\kappa |\delta\bm r'|^2/12$} minimizes the squared distance in real space between the filament and the straight segment, while also cancelling the second-order errors in the Biot-Savart field near the radius of curvature of the filament. We conduct simple numerical tests, and find that the shifted piecewise linear approach has higher-order convergence than the standard piecewise linear approach and is dramatically more accurate for a given number of discretization points.

There are two main conclusions of this section. First, the shifted piecewise linear approach is superior to the standard piecewise linear approach, and should be used when possible. Second, when filamentary coils are stored in databases, the stored endpoints should not lie on the coil but rather should be shifted normally outwards.

\chapter{Invariant preservation in machine learned PDE solvers \label{ch:invariant}}
\noindent\rule{\textwidth}{1pt}

Part of the contents of this chapter have been published in the following paper: (i) McGreivy, N., \& Hakim, A. ``Invariant preservation in machine learned PDE solvers via error correction.'' arXiv preprint arXiv:2303.16110 (2023) \cite{mcgreivy2023invariant}.

\noindent\rule{\textwidth}{1pt}

Scientists and engineers are interested in solving partial differential equations (PDEs). Many PDEs cannot be solved analytically, and must be approximated using discrete numerical algorithms. We refer to these algorithms as `PDE solvers.' The fundamental challenge for PDE solvers is to balance between two competing objectives: first, to find an accurate approximation to the solution of the equation, and second, to do so with as few computational resources as possible.

Decades of research into discrete numerical algorithms have resulted in reliable solvers for most PDEs of interest. For time-dependent PDEs, these so-called `standard numerical methods' use hand-crafted rules to update the solution at each timestep. 
Successful hand-crafted update rules have two key properties. First, the property of \textit{convergence}. Convergent methods converge to the exact solution in the limit that the grid spacing $\Delta x$ and the timestep $\Delta t$ approach zero \cite{lax1956survey}. Second, the property of \textit{invariant preservation}. Time-dependent PDEs often have one or more invariants. Examples of such invariants include conservation of mass \cite{eymard2000finite}, conservation of energy \cite{hakim2019discontinuous}, non-decreasing entropy \cite{merriam1989entropy}, and/or non-negative density \cite{perthame1996positivity}. Invariant preserving PDE solvers satisfy discrete analogues of these continuous invariants when $\Delta x$ and $\Delta t$ are positive. As a result, they are numerically stable and do not violate qualitatively important properties.

In recent years, scientists and engineers have attempted to use machine learning (ML) to design new and better PDE solvers  \cite{tompson2017accelerating,bar2019learning,zhuang2021learned,embedding_hard_constraints,hsieh2019learning,rose_yu_paper,les_closure,um2020solver, mishra2018machine}.
The goal of these machine learned PDE solvers is as follows. Suppose there is a PDE we would like to find an approximate solution to for many different initial or boundary conditions (ICs or BCs) or on a very large domain. We first generate training data, usually from a highly accurate standard solver. Next, we design and train a learnable update rule; often this involves the use of neural networks. If the learned update rule is faster than the best-performing standard solvers with equal accuracy, it can then be used to amortize the initial cost of training over many different ICs or BCs or a much larger domain, by finding sufficiently accurate solutions at reduced computational cost \cite{mishra2018machine,xue2020amortized}. 

For time-dependent PDEs, the main strategy for making solvers faster with ML is to solve the equations at coarser resolution, i.e., with increased $\Delta x$ and/or $\Delta t$ relative to standard methods \cite{vinuesa2022enhancing}.\footnote{There are other possibilities for accelerating the solution of PDEs with ML. For stiff PDEs, it may be possible to replace a solver that requires implicit (slow) timestepping with an explicit machine learned surrogate \cite{wang2023long,di2021deeponet}. For other PDEs, it may be possible to accelerate a solver by learning a surrogate for some particularly costly operator at fixed grid resolution. One can, for example, use ML to accelerate a Poisson solve \cite{cheng2021using,greenfeld2019learning} or learn to approximate a costly plasma collision operator \cite{miller2021encoder,dener2020training,holloway2021acceleration}.}
To do so, machine learned solvers use a radically different approach from standard PDE solvers. Instead of designing hand-crafted update rules that converge as $\Delta x \rightarrow 0$ and $\Delta t \rightarrow 0$, machine learned solvers learn an update rule from data that is accurate at some large value(s) of $\Delta x$ and/or $\Delta t$.\footnote{Note that for time-\textit{independent} PDEs, it is possible to guarantee convergence by interfacing ML with a standard iterative solver. See, for example, \cite{illarramendi2022performance}.} Ideally, this update rule would be faster than standard methods at a given level of accuracy.
Machine learned PDE solvers were found, for certain problems, to have successfully achieved high accuracy at low computational cost \cite{kochkov2021machine, stachenfeld2021learned, li2020fourier,greenfeld2019learning, luz2020learning}.
A more recent study \cite{dresdner2022learning}, which makes a comparison between a very efficient standard numerical method and the best ML-based solvers, tacitly implies that earlier studies compared to weak baselines and thus casts doubt on some purportedly impressive claims about improved performance from ML-based solvers. Nevertheless, there is still potential for ML-based solvers to accelerate the solution of PDEs, and this remains an active area of research.

Because the only way for a numerical method to guarantee perfect accuracy (i.e., to output the exact solution) is (except in trivial cases) to use a convergent method in the limit $\Delta x \rightarrow 0$ and $\Delta t \rightarrow 0$, there is no way to guarantee that a solver outputs the exact solution while using large $\Delta x$ and/or $\Delta t$.
Some amount of error is inevitable, so the question becomes: how can we constrain machine learned solvers to give us the sorts of errors that we are willing to tolerate? In other words, how can we build more reliable machine learned PDE solvers?

One approach to building more reliable machine learned PDE solvers is to change the ML model and training procedure. This approach is quite natural to students of ML. Improving the model \cite{brandstetter2022message}, increasing the dataset size \cite{brandstetter2022lie}, changing the loss function \cite{um2020solver}, adding random noise to the training procedure \cite{stachenfeld2021learned,sanchez2020learning}, and adding regularization \cite{Kaptanoglu_2021} are all examples of this approach. To some extent, these techniques have been successful at improving robustness and numerical stability \cite{klimesch2022simulating}. However, none of these ML-based techniques are capable of \textit{guaranteeing} numerical stability. While these solvers may give reliable results for some inputs, on other inputs the solution might blow up or be nonsensical.

The purpose of this paper is to propose and demonstrate a different and mutually compatible approach to building more reliable machine learned PDE solvers: \textit{preserving discrete invariants}. This approach is quite natural to students of computational physics, as so much of the theory and development of standard numerical methods is related to ensuring those methods preserve the right invariants. This approach can be used with any solver that uses an update rule (including standard solvers) and is otherwise agnostic to the details of the solver.

Why do we want our machine learned solvers to preserve invariant quantities? A simple but incomplete explanation is that ML models which use physical knowledge as an inductive bias tend to outperform models that don't \cite{thuerey_guaranteed_momentum,ling2016reynolds,brandstetter2022clifford}. Invariant preservation is, when done correctly, free lunch. We know that for a given PDE our solution should preserve certain invariants, so by enforcing those invariants at each timestep we improve the solution. A second reason has to do with \textit{numerical stability}. By preserving the right combination of invariants, we can design machine learned PDE solvers which are numerically stable by construction. These solvers, like well-designed standard solvers, are guaranteed not to blow up as $t \rightarrow \infty$. A third reason has to do with \textit{trust}.
People are unlikely to use solvers they do not trust.
People are more likely to trust a numerical method if it preserves the correct set of invariants.

In standard numerical methods, most of the theory of invariant preservation has been developed for hyperbolic PDEs in conservation form
\begin{equation}\label{eq:hyperbolic}
    \frac{\partial \bm u}{\partial t} + \bm \nabla \cdot \bm F(\bm u) = 0
\end{equation}
where $\bm u(\bm x, t) \in \mathbb{R}^m$, $\bm x \in \Omega$, the domain $\Omega \subset \mathbb{R}^d$, $t \in \mathbb{R}^+$, and $\bm F(\bm u) \in C^1(\mathbb{R}^{m \times d})$. This is because hyberbolic PDEs are conservation laws which maintain a variety of conserved quantities and other physical invariants. Because hyperbolic PDEs are so important to research and production applications, because their theory is so well-developed, and because they make a useful playground for studying invariants in time-dependent PDEs, this paper is focused on preserving invariants in machine learned solvers for hyberbolic PDEs. However, as we discuss in \cref{sec:mlpdesolvers}, the strategy and techniques we introduce can be applied to other invariant-preserving PDEs as well as PDEs with non-invariant terms.

The key insight of this paper is simple: to preserve discrete invariants in machine learned PDE solvers, at each timestep apply an error-correcting algorithm to the update rule. Let us now sketch how this works. Suppose that we represent the continuous solution $\bm u(\bm x, t)\in \mathbb{R}^m$ to \cref{eq:hyperbolic} with a discrete solution $\hat{\bm u}(\bm x, t) \in \mathbb{R}^m$ which is a linear sum of $N$ basis functions $\bm \phi_k(\bm x) \in \mathbb{R}^m$ and coefficients $\bm c_k(t) \in \mathbb{R}^m$ for $k \in [1, \dots, N]$, such that $\hat{u}_j(\bm x, t) = \sum_{k=1}^N c_{jk}(t) \phi_{jk}(\bm x)$ for $j \in [1, \dots, m]$. 
Suppose also that the update rule predicts that $\hat{\bm u}$ will change at a rate $\frac{\partial \hat{\bm u}}{\partial t}$ at time $t$ (or an amount $\Delta \hat{\bm u}$ in time $\Delta t$). Suppose also that, due to a priori knowledge about the underlying equations, we would like the solution to satisfy $L$ discrete invariants $\mathcal{I}_\ell(\hat{\bm u}, \frac{\partial \hat{\bm u}}{\partial t})$ for $\ell = [1, \dots, L]$ satisfying either equalities ($\mathcal{I}_\ell= 0$) or inequalities ($\mathcal{I}_\ell \ge 0$). 
If $\frac{\partial \hat{\bm u}}{\partial t}$ (or $\Delta \hat{\bm{u}})$ does not already satisfy these discrete invariants, then we use an error correcting algorithm to modify $\frac{\partial \hat{\bm u}}{\partial t}$ (or $\Delta \hat{\bm{u}})$ to ensure that each of the invariants is satisfied. We repeat this process each timestep. This process is meant to used at inference, though it could also be used while training.

Students of computational physics will notice that this strategy is very different from how invariants are preserved in standard solvers. As we discuss in \cref{sec:standardsolvers}, standard solvers typically conserve linear invariants by using some type of finite volume (FV) method where the rate of change of the discrete solution in each grid cell is equal to the flux through cell boundaries. Standard solvers typically conserve non-linear invariants by putting locally-derived constraints on the flux \cite{mishra2019numerical,hakim2019discontinuous}. Examples of such constraints include upwinding, centered fluxes, or limiters. Yet the error-correcting algorithms we introduce in \cref{sec:scalarhyperbolic,sec:systemshyperbolic} involve using \textit{global} constraints to constrain either the flux through cell boundaries or the update $\frac{\partial \hat{\bm u}}{\partial t}$ (or $\Delta \hat{\bm u}$). In these algorithms, instead of the time-derivative in a cell depending only on its nearest neighbors, the time-derivative depends on the value of the solution over the entire domain. 

One might ask why machine learned solvers should use global, rather than local, constraints to preserve invariants? It is \textit{possible} for a machine learned solver to preserve invariants by putting local constraints on the flux. The problem with doing so, as we show in \cref{sec:why_standard_dont_work}, is that for some invariants standard approaches can be too constraining -- in particular, they introduce too much numerical diffusion -- to give accurate results at large $\Delta x$, and are too constrained by the Courant-Friedrichs-Lewy (CFL) condition to work at large $\Delta t$. Global constraints, in contrast, allow machine learned solvers the flexibility to make accurate predictions at large $\Delta x$ and/or $\Delta t$, while ensuring that whatever errors are made at least do not violate invariants. 

The error-correcting algorithms we introduce have two key properties. First, by preserving the right invariants these algorithms guarantee numerical stability. Numerical stability is about whether the solution blows up; formally, this can be defined as $||\hat{\bm{u}}(\bm x, T)|| \le C_T ||\hat{\bm{u}}(\bm x, 0)||$ for time $T>0$ for some norm and some constant $C_T$ which can depend on $T$ but not on $\Delta x$ or $\Delta t$. For invariant-preserving PDEs, the inequality $||\hat{\bm{u}}(\bm x, T)|| \le ||\hat{\bm{u}}(\bm x, 0)||$ is often a better definition of stability \cite{durran2013numerical}. For scalar hyperbolic PDEs and for many systems of hyperbolic PDEs, it can be shown that preserving discrete analogues of a subset of the continuous invariants is sufficient to guarantee stability \cite{mishra2019numerical,merriam1989entropy,juno2018discontinuous}.

The second key property is that, as we show in \cref{sec:verification}, in closed or periodic systems this strategy preserves invariants without degrading the accuracy of an already-accurate solver. In closed or periodic systems, we know a priori the discrete invariants ($\mathcal{I}_\ell = 0$ or $\mathcal{I}_\ell \ge 0$). If the solver doesn't satisfy those invariants at each timestep, then we know that our solver has made an error. In open systems, we may not know the invariants due to uncertainty in the fluxes through the boundary. We can estimate these fluxes, but this estimate can be inaccurate. Thus, as we demonstrate in \cref{sec:verification_compressible_euler}, in open systems the error-correction process can degrade accuracy if the rate of change of the invariants cannot be estimated accurately.



Be careful not to confuse the property of invariance or equivariance under a transformation with the preservation of an invariant. Noether's theorem says that for certain physical systems, a continuous symmetry (invariance) leads to the existence of a conserved (invariant) quantity. However, for discrete systems no such theorem exists. While equivariant neural networks may be used to enforce discrete symmetries and thereby improve the generalization capabilities of machine learned PDE solvers \cite{wang2020incorporating}, whether or not a solver is equivariant is unrelated to whether or not that solver preserves an invariant.

The rest of this paper is structured as follows. 
In \cref{sec:mlpdesolvers}, we define what a machine learned PDE solver is and for what types of PDEs this strategy can be applied. 
In \cref{sec:standardsolvers}, we review the theory of invariant preservation in scalar hyperbolic PDEs and in systems of hyperbolic PDEs.
In \cref{sec:scalarhyperbolic}, we design invariant-preserving error-correcting algorithms for a variety of popular solver types for scalar hyperbolic PDEs.
In \cref{sec:systemshyperbolic}, we design invariant-preserving error-correcting algorithms for systems of hyperbolic PDEs, focusing on FV-like solvers. In \cref{sec:verification}, we turn our attention to machine learned solvers. We computationally verify that, at least in periodic systems, these error-correcting invariant-preserving algorithms do not degrade the accuracy of an already-accurate machine learned solver.   
Code and instructions for reproducing the figures in \cref{sec:scalarhyperbolic,sec:systemshyperbolic,sec:verification} can be found at \href{https://github.com/nickmcgreivy/InvariantPreservingMLSolvers}{https://github.com/nickmcgreivy/InvariantPreservingMLSolvers}.
\Cref{sec:relatedwork} is related work. We conclude in \cref{sec:limitations} by discussing limitations and trade-offs associated with invariant preservation.

\section{Machine learned PDE solvers}\label{sec:mlpdesolvers}

\noindent We are interested in finding numerical solutions to PDEs of the form
\begin{equation}\label{eq:generalpde}
    \frac{\partial \bm u}{\partial t} + \mathcal{N}[\bm u] = 0
\end{equation}
where the solution $\bm u(\bm x, t) \in \mathbb{R}^m$, $\bm x \in \Omega$, $\Omega \subset \mathbb{R}^d$, $t \in \mathbb{R}^+$, and $\mathcal{N} : \mathbb{R}^m \rightarrow \mathbb{R}^m$ is an operator which preserves some invariant quantities across time. Examples of PDEs in the form of \cref{eq:generalpde} include the parabolic diffusion equation
\begin{equation}
    \frac{\partial \bm u}{\partial t} = D{\nabla}^2 \bm u,
\end{equation}
scalar hyperbolic PDEs in the form
\begin{equation}
    \frac{\partial u}{\partial t} + \bm \nabla \cdot \bm f(u) = 0,
\end{equation}
Hamiltonian systems evolving under the influence of a Hamiltonian $H$ and Poisson bracket $\{\cdot,\cdot\}$ 
\begin{equation}\label{eq:hamiltoniansystem}
    \frac{\partial f}{\partial t} + \{ f, H\} = 0,
\end{equation}
systems of hyperbolic PDEs
\begin{equation}
    \frac{\partial \bm u}{\partial t} + \bm \nabla \cdot \bm F(\bm u) = 0,
\end{equation}
and the Boltzmann equation of kinetic physics which both advects particles in phase space and evolves them according to a mass, energy, and momentum-conserving collision operator
\begin{equation}
    \frac{\partial f}{\partial t} + \bm v \cdot \frac{\partial f}{\partial \bm x} + \bm a \cdot \frac{\partial f}{\partial \bm v} = \bigg(\frac{\partial f}{\partial t}\bigg)_{\textnormal{coll}}.
\end{equation}
To numerically approximate the solution to an equation in the form \cref{eq:generalpde}, we begin by representing the continuous solution $\bm u(\bm x, t)$ with a discrete solution $\hat{\bm u}(\bm x, t)\in \mathbb{R}^m$. As we discussed in the introduction, the discrete solution is represented as the linear sum of $N$ basis functions $\bm \phi_k(\bm x) \in \mathbb{R}^m$ such that the $j$th dimension of $\hat{\bm u}$ is given by $\hat{u}_j = \sum_{k=1}^N c_{jk}(t) \phi_{jk}(\bm x)$. Examples of this discretization process include the finite volume (FV) method, the discontinuous Galerkin (DG) method, spectral methods, and the finite element method (FEM). 

In the FV method, the domain $\Omega$ is partitioned into $N$ cells. The solution in each cell is a piecewise constant function; thus the $k$th basis function $\bm \phi_k$ is a vector of ones $\bm 1 \in \mathbb{R}^m$ inside the cell and zero outside the cell. The matrix of coefficients $\bm c_{jk} \in \mathbb{R}^{m \times N}$ thus represents, for each of the $m$ components of $\bm u$, the constant scalar value within each cell. The DG method is similar to the FV method, except that the solution within each cell is represented by a piecewise polynomial of degree $p \in \mathbb{N}$. With DG methods, the domain is  partitioned into cells. Each cell contains a discrete set of basis functions which are usually orthogonal polynomials. A FV method is equivalent to a DG method with $p=0$. Spectral methods represent the solution with global basis functions; often these basis functions are Fourier modes. FEM represents the solution as a sum of polynomial basis functions, though unlike DG methods the solution is piecewise continuous instead of piecewise discontinuous. In this paper we mostly consider FV solvers, though we also consider DG and spectral solvers. We don't consider FEM solvers, nor do we consider other possible basis functions.

Any solver which uses basis functions to represent the solution to \cref{eq:generalpde} will need to find some way of updating the basis function coefficients $\bm c_{jk} \in \mathbb{R}^{m\times N}$ in time. We define a machine learned PDE solver, for the purposes of this paper, as any solver which uses an update rule for the coefficients $\bm c_{jk}$.
This can also be called an `autoregressive' solver. The error-correcting strategy we introduce in \cref{sec:scalarhyperbolic,sec:systemshyperbolic} can be applied to any such solver. Note that the use of ML is not actually part of the above definition.

There are two main types of update rule. One is to use a discrete-time update. The discrete-time update iteratively updates time from $t$ to $t + \Delta t$ and the solution from $\hat{\bm u}$ to $\hat{\bm u} + \Delta t \hat{\mathcal{N}}[\bm \hat{\bm u}]$. The second is to use a continuous-time update. The continuous-time update combined with discretization in space is often called the method of lines (MOL). MOL involves approximating $\hat{\mathcal{N}}$, setting $\frac{\partial \hat{\bm u}}{\partial t} = \hat{\mathcal{N}}$, and using an ODE integrator to advance $\hat{\bm u}$ in time. Using a strong stability preserving Runge Kutta (SSPRK) \cite{ssprk,ssp_gottlieb} ODE integration method and choosing the timestep to satisfy a CFL condition is usually sufficient to ensure that, for hyperbolic PDEs, invariants which are preserved in the continuous-time limit are preserved as time advances. 

Note that some methods which use ML to find a solution to a PDE do not satisfy the above definition of a machine learned PDE solver. For example, the physics-informed neural network (PINN) approach \cite{karniadakis2021physics} does not iteratively update the solution in time, and so the results in this paper do not apply to PINNs. The same is true of the Fourier Neural Operator (FNO) approach \cite{li2020fourier} if the FNO convolves in both space and time. However, the strategy introduced in this paper could be applied to a recurrent FNO which iteratively updates Fourier coefficients in time.

In practical applications, we usually don't care about solving an invariant-preserving equation $\frac{\partial \bm u}{\partial t} + \mathcal{N}[\bm u]=0$, but rather a more complicated equation $\frac{\partial \bm u}{\partial t} + \mathcal{N}[\bm u]=\mathcal{F}[\bm u]$ where $\mathcal{F}[\bm u]$ is some operator that breaks one or more of the invariants preserved by $\mathcal{N}$. In these cases, we can model $\mathcal{N}$ using machine learning, apply an error-correcting algorithm to $\mathcal{N}$, and model $\mathcal{F}$ using some other technique. Doing so ensures that invariants are violated only due to the presence of $\mathcal{F}$, not because of faulty numerics in the calculation of $\mathcal{N}$. Consider a concrete example. The Navier-Stokes equations can be written as a sum of the Euler equations and viscous and forcing terms. The viscous and forcing terms break the invariants of the Euler equations. In this case, we can model the conservative terms in the Euler equations using ML, apply an error-correcting algorithm to these terms, and use some other technique to model the viscous and forcing terms. For an example of how this can be done, see \cite{kochkov2021machine}.

\section{Invariants of hyperbolic PDEs }\label{sec:standardsolvers}

\noindent PDEs have certain invariants. Computational scientists usually try to design numerical solvers which preserve some, or all, of those invariants. By preserving the right set of invariants, it becomes possible to design solvers which are numerically stable
and which give physically reasonable results when $\Delta x$ and $\Delta t$ are positive. 
In this section, we review the theory of invariant preservation in hyperbolic PDEs. We examine generic scalar hyperbolic PDEs, generic systems of hyperbolic PDEs, and a few example PDEs in each class. We briefly discuss how standard FV methods preserve invariants.


\subsection{Scalar hyperbolic PDEs}\label{sec:standardsolversscalar}

\noindent Scalar hyperbolic PDEs can be written as
\begin{equation}\label{eq:scalarhyperbolic}
\frac{\partial u}{\partial t} + \bm \nabla \cdot \bm f(u) = 0.
\end{equation}
The solution $u(\bm x, t) \in \mathbb{R}$, $\bm x \in \Omega$, $\Omega \in \mathbb{R}^d$, and the flux $\bm f \in C^1(\mathbb{R}^d)$. 

\subsubsection{Continuous invariants}
Scalar hyperbolic PDEs have one linear invariant which is constant in time and three non-linear invariants which are non-increasing in time \cite{mishra2019numerical}:
\begin{itemize}
    \item Total mass $\int_\Omega u \mathop{d\bm x}$, which is conserved in time.
    \item The $\ell_p$-norm $\int_\Omega |u|^p \mathop{d\bm x}$ for $p > 1$, which is non-increasing in time.
    \item The $\ell_\infty$-norm $\textnormal{max}_{\bm x} u(\bm x, t)$, which is non-increasing in time.
    \item The total variation, which for continuous $u$ in 1D is $\int_0^L \big|\frac{\partial u}{\partial x}\big| \mathop{dx}$. The total variation is non-increasing in time. This is usually called the total variation diminishing (TVD) property.
\end{itemize}
We can prove that hyperbolic PDEs conserve mass by integrating \cref{eq:scalarhyperbolic} over the domain $\Omega$ and using the divergence theorem:
\begin{equation}
    \int_{\Omega} \Big[\frac{\partial u}{\partial t} + \bm \nabla \cdot \bm f(u) \Big]d\bm x = \frac{d}{dt}\int_{\Omega}u \mathop{d\bm x} + \int_{\partial \Omega} \bm f \cdot d\bm n = 0.
\end{equation}
In words: the rate of change of $\int_\Omega u \mathop{d\bm x}$ with respect to time is equal to the negative flux of $u$ through the domain boundary $\partial \Omega$. In an infinite or periodic system $\int_\Omega u \mathop{d\bm x}$ is constant. 

To prove that the $\ell_p$-norm and $\ell_\infty$-norm are non-increasing in time, we use the entropy inequality for scalar hyperbolic PDEs. The entropy inequality can be derived as follows. First multiply \cref{eq:scalarhyperbolic} by a scalar function $\eta'(u)$. $\eta(u)$ is an entropy function. This gives
\begin{equation}\label{eq:entropy_deriv_1}
    \frac{\partial \eta(u)}{\partial t} + \frac{\partial \eta}{\partial u} \Big(\bm \nabla \cdot \bm f(u)\Big) = 0.
\end{equation}
If $u$ is smooth, we can use the chain rule to rearrange \cref{eq:entropy_deriv_1} as
\begin{equation}\label{eq:entropy_deriv_2}
    \frac{\partial \eta}{\partial t} + \frac{\partial \eta}{\partial u} \frac{\partial \bm f}{\partial u} \cdot \bm \nabla u = \frac{\partial \eta}{\partial t} + \frac{\partial \bm \psi}{\partial u} \cdot \bm \nabla u = \frac{\partial \eta}{\partial t} + \bm \nabla \cdot \bm \psi = 0
\end{equation}
where the entropy flux $\bm \psi$ is defined by $\bm \psi'(u) \coloneqq \eta' \bm f'$. \Cref{eq:entropy_deriv_2} states that, for smooth solutions, the entropy $\eta$ satisfies a conservation law. 
For discontinuous $u$, a longer derivation \cite{mishra2019numerical} reveals that the equality in \cref{eq:entropy_deriv_2} is replaced with an inequality for any convex entropy function $\eta(u)$ with corresponding entropy flux $\bm \psi$:
\begin{equation}\label{eq:entropy_inequality}
\frac{\partial \eta(u)}{\partial t} + \bm \nabla \cdot \bm \psi(u) \ge 0.
\end{equation}
Integrating \cref{eq:entropy_inequality} over $\Omega$ shows that the rate of change of total entropy $\int_\Omega \eta \mathop{d\bm x}$ is equal to the negative entropy flux through the domain boundary $\partial \Omega$; in an infinite or periodic system total entropy is non-decreasing.
In brief: for continuous $u$ entropy is conserved, while for discontinuous $u$ entropy increases.

By choosing $\eta(u) = -|u|^p$ and integrating \cref{eq:entropy_inequality} over $\Omega$, we have the non-increasing $\ell_p$-norm property. Taking the limit as $p \rightarrow \infty$ gives the non-increasing $\ell_\infty$-norm property. The TVD property is derived in \cite{mishra2019numerical}.

\subsubsection{Finite volume (FV) method} 
A common approach for solving hyperbolic PDEs is by using a finite volume (FV) method. FV methods divide the spatial domain  $\Omega$ into a number of discrete cells $\Omega_j$, then use a scalar value to represent the solution average within each cell. For example, on the 1D domain $x \in [0,L]$ with uniform cell width, a FV method divides the domain into $N$ cells of width $\Delta x = \nicefrac{L}{N}$ where the left and right boundaries of the $j$th cell for $j = 1, \dots, N$ are $x_{j-\nicefrac{1}{2}} = (j-1)\Delta x$ and $x_{j+\nicefrac{1}{2}}=j\Delta x$ respectively. FV methods use a scalar value $u_j(t)$ to represent the solution average within each cell where $u_j(t) \coloneqq \int_{x_{j-\nicefrac{1}{2}}}^{x_{j+\nicefrac{1}{2}}} u(x, t)  \mathop{dx}$. The standard FV equations for the time-derivative of $u_j$ in 1D and $u_{i,j}$ in 2D are simply discrete versions of \cref{eq:scalarhyperbolic}:
\begin{subequations}
\begin{equation}\label{eq:fv_a}
    \frac{\partial u_j}{\partial t} + \frac{f_{j+\frac{1}{2}} - f_{j-\frac{1}{2}}}{\Delta x} = 0
\end{equation}
\begin{equation}\label{eq:fv_b}
    \frac{\partial u_{i,j}}{\partial t} + \frac{f^x_{i+\frac{1}{2},j} - f^x_{i-\frac{1}{2},j}}{\Delta x} + \frac{f^y_{i,j+\frac{1}{2}} - f^y_{i,j-\frac{1}{2}}}{\Delta y} = 0.
\end{equation}
$f_{j+\nicefrac{1}{2}}$ is the flux at the cell boundary $x_{j+\nicefrac{1}{2}}$. $f^x_{i+\nicefrac{1}{2},j}$ and $f^y_{i,j+\nicefrac{1}{2}}$ are the average x-directed and y-directed fluxes through the right and top cell boundaries, e.g., $f^x_{i+\nicefrac{1}{2},j} \coloneqq \frac{1}{\Delta y}\int_{y=y_{j-\nicefrac{1}{2}}}^{y=y_{j+\nicefrac{1}{2}}} \bm{\hat{x}} \cdot \bm f(x_{i+\nicefrac{1}{2}},y) \mathop{dy}$. 
In higher dimensions, the FV update equation in the $j$th cell can be written as
\begin{equation}\label{eq:fv_c}
    \frac{\partial u_j}{\partial t} + \frac{1}{|\Omega_j|}\oint_{\partial \Omega_j} \bm{f}\cdot \mathop{d\bm s} = 0
\end{equation}
\end{subequations}
where $|\Omega_j| \coloneqq \int_{\Omega_j} \mathop{d\bm x}$ is the volume in cell $\Omega_j$ and $\mathop{d\bm s}$ is the outward normal vector at cell boundary $\partial \Omega_j$.
In 1D, \cref{eq:fv_a} can be derived by applying the integral $\int_{x_{j-\nicefrac{1}{2}}}^{x_{j+\nicefrac{1}{2}}} (...) \mathop{dx}$ to \cref{eq:scalarhyperbolic} for all $j \in 1, \dots, N$; a similar calculation gives \cref{eq:fv_b} and \cref{eq:fv_c}.
So long as $f_{j+\nicefrac{1}{2}}$ or $f_{i+\nicefrac{1}{2},j}^x$ and $f_{i,j+\nicefrac{1}{2}}^y$ are exact for all $t$, then $u_j$ or $u_{ij}$ will be exact for all $t$. Thus, the key challenge for a FV scheme is to accurately reconstruct the flux at cell boundaries.

\subsubsection{Discrete invariants}
FV schemes conserve a discrete analogue of the continuous linear invariant $\int_\Omega u \mathop{d \bm x}$ by construction. In 1D, we can see this with a short proof:  
$\nicefrac{d}{d t} \sum_{j=1}^N u_j \Delta x = \Delta x \sum_{j=1}^N \nicefrac{\partial u_j}{\partial t} = - \sum_{j=1}^N (f_{j+\nicefrac{1}{2}} - f_{j-\nicefrac{1}{2}}) = f_{\nicefrac{1}{2}} - f_{N+\nicefrac{1}{2}}$. The rate of change of the discrete mass is equal to the flux of $u$ through the boundaries; in a periodic system this equals 0.

Although FV schemes preserve a discrete analogue of conservation of mass by construction, they do not automatically preserve discrete analogues of any of the non-linear invariants of the continuous PDE. Instead, FV methods preserve non-linear invariants through careful choice of flux.

The only known way of inheriting discrete analogues of all three non-linear invariants of \cref{eq:scalarhyperbolic} (non-increasing $\ell_p$-norm, non-increasing $\ell_\infty$-norm, and TVD) is to use a consistent monotone flux function while satisfying a CFL condition \cite{mishra2019numerical}. See \cite{durran2013numerical} for definitions of consistency and monotonicity. An example of a monotone flux function for the linear advection equation $f=cu$ is the upwind flux
\begin{equation}
    f_{j+\frac{1}{2}} = \begin{cases}
        c u_j &\textnormal{ if } c \ge 0 \\
        c u_{j+1} &\textnormal{ if } c < 0.
    \end{cases}
\end{equation}
For non-linear $f(u)$, a Reimann solver or approximate Reimann solver results in a monotone flux function.  Examples of monotone flux functions include the Godunov flux 
\begin{equation}
    f_{j+\frac{1}{2}} = \begin{cases}
        \textnormal{min}_{u_j \le u \le u_{j+1}} f(u) &\textnormal{ if } u_j \le u_{j+1} \\
        \textnormal{max}_{u_j \le u \le u_{j+1}} f(u) &\textnormal{ if } u_j > u_{j+1}
    \end{cases}
\end{equation}
and the Lax-Friedrichs flux
\begin{equation}
    f_{j+\frac{1}{2}} = \frac{f(u_j) + f(u_{j+1})}{2} - \frac{\Delta x}{2\Delta t} (u_{j+1} - u_j).
\end{equation}
Unfortunately, Godunov's famous theorem from 1959 implies that monotone schemes can be at most first-order accurate \cite{godunov1959finite}. This means that while monotone schemes preserve all the invariants of the underlying PDE, they are usually not very accurate.

Godunov's theorem reveals a more general lesson. For some PDEs, it is impossible to design highly accurate numerical methods that preserve discrete analogues of \textit{every} invariant of the continuous PDE. Designers of numerical methods must therefore determine which invariants of the continuous system should be preserved by the discrete system and which invariants either cannot be preserved or degrade the accuracy of the discrete system.

For scalar hyperbolic PDEs, it turns out that it is possible to design accurate and stable numerical solvers by preserving just \textit{one} of the three non-linear invariants of \cref{eq:scalarhyperbolic} \cite{durran2013numerical}.
One such scheme is the MUSCL scheme, introduced in a seminal paper by Van Leer \cite{muscl}. MUSCL uses limiters to reconstruct the solution at cell boundaries. In 1D, this preserves a discrete analogue of the TVD property. Doing so guarantees numerical stability and prevents spurious oscillations\footnote{Spurious oscillations are unphysical oscillations which develop in numerical methods that do not have enough numerical diffusion to damp high-$k$ modes that develop near steep gradients \cite{john2007spurious,shocks_artificial_viscosity}.},  while retaining second-order accuracy.

It is also possible to design stable numerical methods by preserving a discrete analogue of an $\ell_p$-norm non-increasing property of the solution. For the linear advection equation $f=cu$, the centered flux 
\begin{equation}
    f_{j+\frac{1}{2}} = \frac{c}{2}(u_j + u_{j+1})
\end{equation}
conserves the discrete $\ell_2$-norm in the continuous-time limit, though if used with a forward Euler update the centered flux increases the discrete $\ell_2$-norm leading to numerical instability. For non-linear $f(u)$ the flux formula
\begin{equation}
    f_{j+\frac{1}{2}} = \int_0^1 f(\hat{u})d\theta \textnormal{\hspace{0.1cm} where \hspace{0.1cm}} \hat{u}(\theta) = u_j + \theta (u_{j+1}-u_j)
\end{equation}
conserves the discrete $\ell_2$-norm \cite{jameson2008construction} in the continuous-time limit.

\subsubsection{2D incompressible Euler equations\label{sec:standardsolversincompressible}}

The 2D incompressible Euler equations in vorticity form are a scalar hyperbolic PDE coupled to an elliptic PDE:
\begin{align}\label{eq:euler}
    \frac{\partial \chi}{\partial t} + \bm \nabla \cdot (\bm u \chi) = 0\textnormal{,} && \bm u = \bm \nabla \psi \times \hat{e}_z\textnormal{,} && -\bm\nabla^2 \psi = \chi.
\end{align}
These equations can be written as a Hamiltonian system (\cref{eq:hamiltoniansystem}) for the vorticity $\chi(x,y,t)$ evolving under a Hamiltonian given by the streamfunction $\psi(x,y,t)$ \cite{hamiltonian_structure}. 

\subsubsection{Continuous invariants} 
\Cref{eq:euler} has the same invariants as the generic scalar hyperbolic \cref{eq:scalarhyperbolic}. In particular, \cref{eq:euler} conserves the mass $\int_\Omega \chi \mathop{dx}\mathop{dy}$ and the $\ell_2$-norm $\int_{\Omega} \chi^2 \mathop{dx}\mathop{dy}$, also called the Enstrophy. \Cref{eq:euler} has an additional conserved invariant, the energy $\frac{1}{2} \int_{\Omega} \bm u^2 \mathop{dx}\mathop{dy}$.
We introduce the notation $\mathop{dz} = \mathop{dx}\mathop{dy}$. We prove conservation of mass using integration by parts, and canceling the boundary term using periodic BCs:
\begin{equation}
    \frac{d}{dt} \int_\Omega \chi \mathop{d z}= \int_\Omega \frac{\partial \chi}{\partial t} \mathop{d z} = - \int_\Omega \bm \nabla \cdot (\bm u \chi) \mathop{d z} = - \int_{\partial \Omega} \bm u\chi \cdot \mathop{d\bm s} = 0.
\end{equation}
We prove Enstrophy conservation as follows. From incompressibility, $\bm \nabla \cdot \bm u = 0$, which implies $\bm \nabla \cdot (\bm u \nicefrac{\chi^2}{2}) =\chi\bm u\cdot \bm \nabla \chi = \chi \bm \nabla \cdot (\bm u \chi)$. Using Gauss's theorem and periodicity,
\begin{equation}
    \frac{d}{dt} \int_\Omega \frac{1}{2}\chi^2 \mathop{d z} = - \int_\Omega \chi \bm \nabla \cdot (\bm u \chi) \mathop{d z} = -\int_{\partial \Omega} \frac{1}{2} {\bm u \chi^2} \cdot \mathop{d\bm s} = 0.
\end{equation}
We prove energy conservation using integration by parts and $\bm u \cdot \bm \nabla \psi = 0$:
\begin{equation}
\begin{split}
    \frac{d}{dt}\int_{\Omega} \frac{1}{2}\bm u^2 \mathop{d z} &= \frac{d}{dt}\int_{\Omega} \frac{1}{2}(\bm \nabla \psi)^2 \mathop{d z} = \int_{\Omega} \bm \nabla \psi \cdot \bm \nabla \frac{\partial \psi}{\partial t} \mathop{d z}\\ &=- \int_\Omega \psi \nabla^2 \frac{\partial \psi}{\partial t} \mathop{d z} = \int_\Omega \psi \frac{\partial \chi}{\partial t} \mathop{d z} \\&= -\int_\Omega \psi \bm \nabla \cdot (\bm u \chi) = \int_\Omega \chi \bm u \cdot \bm \nabla \psi \mathop{d z} = 0.
\end{split}
\end{equation}

\subsection{Systems of hyberbolic PDEs\label{sec:standard_systems}}

\noindent We consider systems of hyperbolic PDEs in 1D written in conservation form, given by
\begin{equation}\label{eq:system_conservation_form}
    \frac{\partial \bm u}{\partial t} + \frac{\partial}{\partial x} \bm F(\bm u) = 0
\end{equation}
where $\bm u \in \mathbb{R}^m$ and $\bm F(u) \in C^1(\mathbb{R}^m)$.
\Cref{eq:system_conservation_form} is hyperbolic if the Jacobian matrix $\frac{\partial \bm F}{\partial \bm u}$ has real eigenvalues and a complete set of linearly independent eigenvectors \cite{leveque1992numerical}.

\subsubsection{Continuous invariants}
\Cref{eq:system_conservation_form} implies that each component of $\int \bm u(x, t) \mathop{dx}$ is conserved. In a 1D periodic system with $x \in [0, L]$, an integral over $x$ makes this apparent: $\frac{d}{dt}\int_{0}^L \bm u \mathop{dx} = \int \frac{\partial \bm u}{\partial t} \mathop{dx} = - \int_0^L \frac{\partial \bm F}{\partial x} \mathop{dx} = \bm F(0) - \bm F(L)$. 
The total rate of change of $\bm u$ equals the flux $\bm F$ through the boundaries; in a periodic or infinite system this equals zero.

For certain systems of hyperbolic PDEs, it is possible to define a generalized scalar convex entropy function $\eta(\bm u)$ and entropy flux $\psi(\bm u)$ such that the entropy satisfies an inequality \cite{leveque1992numerical}
\begin{equation}\label{eq:system_entropy_inequality}
    \frac{\partial \eta(\bm u)}{\partial t} + \frac{\partial\psi(\bm u)}{\partial x}  \ge 0.
\end{equation} 
Integrating \cref{eq:system_entropy_inequality} over $x$ for a 1D periodic system where $x \in [0, L]$ shows that the total entropy is non-decreasing: $\frac{d}{dt} \int \eta(\bm u) \mathop{dx} \ge \psi(0) - \psi(L) = 0$.
It can be shown that \cref{eq:system_conservation_form} satisfies the entropy inequality \cref{eq:system_entropy_inequality} if there exists a change of variables from $\bm u$ to $\bm w$ such that $\nicefrac{\partial \bm u}{\partial \bm w}$ and $\nicefrac{\partial \bm F}{\partial \bm w}$ are both symmetric \cite{jameson2008construction,harten1983symmetric}. 
If this change of variables is made, it turns out that the entropy variable $\bm w = \nicefrac{\partial \eta}{\partial \bm u}$, $\nicefrac{\partial \eta}{\partial t} = \bm w^T \nicefrac{\partial \bm u}{\partial t}$, and $\nicefrac{\partial \psi}{\partial \bm u} = (\nicefrac{\partial \eta}{\partial \bm u})^T \nicefrac{\partial \bm F}{\partial \bm u}$.

Notice that while generic scalar hyperbolic PDEs all conserve the same non-linear invariants, generic systems of hyperbolic PDEs are not guaranteed to have any non-linear invariants. Many physically relevant systems of hyperbolic PDEs do have non-linear invariants, but these invariants differ between systems of PDEs.

\subsubsection{Discrete invariants}
FV schemes conserve a discrete analogue of the continuous linear invariant $\int_\Omega \bm u \mathop{d x}$ by construction. In 1D, we can see this with a short proof:  
$\nicefrac{d}{d t} \sum_{j=1}^N \bm u_j \Delta x = \Delta x \sum_{j=1}^N \nicefrac{\partial \bm u_j}{\partial t} = - \sum_{j=1}^N (\bm F_{j+\nicefrac{1}{2}} - \bm F_{j-\nicefrac{1}{2}}) = \bm F_{N+\nicefrac{1}{2}}- \bm F_{\nicefrac{1}{2}}$. The rate of change of the discrete mass is equal to the flux of $\bm u$ through the boundaries; in a periodic system this equals 0.

When a generalized entropy function $\eta(\bm u)$ exists, FV schemes do not guarantee that the discrete entropy $\sum_{j=1}^N \eta_j(\bm u_j) \Delta x$ is non-decreasing in time. Instead, FV schemes inherit a discrete analogue of the non-decreasing total entropy property by satisfying a discrete entropy inequality in each grid cell:
\begin{equation}\label{eq:discrete_entropy_inequality}
    \frac{\partial \eta_j}{\partial t} + \frac{\psi_{j+\frac{1}{2}} - \psi_{j - \frac{1}{2}}}{\Delta x} \ge 0.
\end{equation}
One way of satisfying \cref{eq:discrete_entropy_inequality} is to use a Reimann solver or approximate Reimann solver to compute $\bm F_{j+\nicefrac{1}{2}}$ \cite{leveque1992numerical}. Doing so preserves the non-decreasing total entropy invariant.

\subsubsection{Compressible Euler equations}

\noindent The compressible Euler equations of gas dynamics in 1D are given by 
\begin{equation}\label{eq:compressibleeuler}
    \frac{\partial }{\partial t} \begin{bmatrix}
           \rho \\
           \rho v \\
           E
         \end{bmatrix} + \frac{\partial}{\partial x} \begin{bmatrix}
             \rho v \\
             \rho v^2 + p \\
            v(E+p)
         \end{bmatrix} = 0.
\end{equation}
where $\rho$ is the density, $u$ is the velocity, $p$ is the pressure and $E$ is the energy. The equation of state for an ideal gas is
\begin{equation}\label{eq:equationofstate}
    E = \frac{p}{\gamma - 1} + \frac{1}{2}\rho v^2
\end{equation}
where $\gamma$ is the ratio of specific heat at constant pressure to the specific heat at constant volume. In the notation of \cref{eq:system_conservation_form}, $\bm u = \begin{bmatrix}
    \rho, \rho v, E
\end{bmatrix}$ and $\bm F = \begin{bmatrix}
    \rho v, \rho v^2 + p, v(E+p)
\end{bmatrix}$.

\subsubsection{Continuous invariants}
As with all hyperbolic PDEs, the time rate of change of the linear invariant $\int \bm u \mathop{dx}$ is equal to the negative flux $\bm F$ of $\bm u$ through the domain boundaries. In an infinite or periodic system $\int \bm u \mathop{dx}$ is constant in time. 

The Euler equations have two positivity invariants: the density $\rho \in \mathbb{R}$ and the pressure $p \in \mathbb{R}$ are both everywhere non-negative. From the equation of state \cref{eq:equationofstate}, non-negativity of density and pressure imply that the energy $E \in \mathbb{R}$ is everywhere non-negative.

The Euler equations also satisfy an entropy inequality \cref{eq:system_entropy_inequality}. This implies that the total entropy $\int \eta(\bm u) \mathop{dx}$ is non-decreasing in time. It has been shown \cite{harten1983symmetric} that the generalized entropy function $\eta(s) = \rho g(s)$ is convex for any $g(s)$ with specific entropy $s = \log{(\nicefrac{p}{\rho^\gamma})}$ for which $\nicefrac{g''}{g'} < \gamma^{-1}$. In this case the entropy flux $\psi = \rho g(s) v$. 
If we choose $g(s) = e^{\nicefrac{s}{\gamma + 1}}$, some tedious algebra shows that the entropy variable $\bm w = \nicefrac{\partial \eta}{\partial \bm u}$ equals
\begin{equation}\label{eq:compressibleeuler_entropy_variable}
    \bm w = \frac{p^*}{p}\begin{bmatrix}
        E \\
        -\rho v \\
        \rho
    \end{bmatrix}
\end{equation}
where
\begin{equation}
    p^* = \frac{\gamma - 1}{\gamma + 1} \bigg(\frac{p}{\rho^\gamma}\bigg)^{\frac{1}{\gamma + 1}}.
\end{equation}

\section{Invariant-preserving algorithms for scalar hyperbolic PDEs}\label{sec:scalarhyperbolic}

\noindent We now turn to the main purpose of this paper: designing machine learned solvers that preserve discrete invariants.
As we discussed in \cref{sec:standardsolversscalar}, scalar hyperbolic PDEs conserve the linear invariant $\int_\Omega u \mathop{d\bm x}$, have non-increasing $\ell_p$-norm, have non-increasing $\ell_\infty$-norm, and are TVD. The only known way of inheriting discrete analogues of all three non-linear invariants of \cref{eq:scalarhyperbolic} is to use a consistent monotone flux function. Godunov's theorem implies that such schemes can be at most first-order accurate. Thus, we will not try to design monotone machine learned solvers. Instead, we will design machine learned solvers that preserve two invariants: mass conservation and one of the non-linear invariants. This ensures that the machine learned solver is numerically stable, while being flexible enough to accurately predict the rate of change of the solution. 

In this section, we introduce error-correcting algorithms for scalar hyperbolic PDEs in periodic domains which modify an update rule to enforce mass conservation and change the time-derivative of the discrete $\ell_2$-norm from $\nicefrac{d\ell_2^{\textnormal{old}}}{dt}$ to $\nicefrac{d\ell_2^{\textnormal{new}}}{dt}$.
These error-correcting algorithms make it easy to design invariant-preserving machine learned solvers: first, at each timestep use the error-correcting algorithm to modify the output of the update rule to ensure that the discrete mass is conserved. Second, if $\nicefrac{d\ell_2^{\textnormal{old}}}{dt} > 0$, set $\nicefrac{d\ell_2^{\textnormal{new}}}{dt} =0$.
If $\nicefrac{d\ell_2^{\textnormal{old}}}{dt} \le 0$, set $\nicefrac{d\ell_2^{\textnormal{new}}}{dt} = \nicefrac{d\ell_2^{\textnormal{old}}}{dt}$. If we have a priori information about the expected rate of change of the discrete $\ell_2$-norm, we can set $\nicefrac{d\ell_2^{\textnormal{new}}}{dt}$ to that value.

We consider a variety of different solver types and update rules. \Cref{sec:continuoustimeFV,sec:fluxpredictingFV,sec:DG,sec:fourier} show how to design invariant-preserving FV solvers, DG solvers, and spectral solvers with a continuous-time update. \Cref{sec:continuoustimeFV} is derived for arbitrary mesh shapes. \Cref{sec:discretetimeFV} shows how to design invariant-preserving FV solvers when using a discrete-time update.
In \cref{sec:energyconservation} we consider a different scalar hyperbolic PDE, the 2D incompressible Euler equations. For these equations we show how to preserve an additional invariant, conservation of energy.
For each type of solver, the general strategy is the same: at each timestep, apply an error-correcting algorithm to the update rule.
We only consider periodic BCs in this section. With the exception of the spectral solver in \cref{sec:fourier}, we can derive invariant-preserving algorithms for non-periodic BCs by estimating the flux through the boundaries.

\subsection{Flux-predicting FV solvers \label{sec:fluxpredictingFV}}

\noindent Suppose we are interested in designing a machine learned solver for \cref{eq:scalarhyperbolic} which, like the FV method, divides the domain $\Omega$ into a number of grid cells $\Omega_j$ and represents the solution average in the $j$th cell as a scalar $u_j$. Suppose also we use the continuous-time FV update \cref{eq:fv_a,eq:fv_b,eq:fv_c} to compute $\nicefrac{\partial u_j}{\partial t}$. Because \cref{eq:fv_a,eq:fv_b,eq:fv_c} are exact, to evolve the discrete solution accurately a machine learned solver has to predict the (average) flux across cell boundaries accurately. We refer to this type of machine learned solver as a `flux-predicting finite volume' solver. 

Flux-predicting FV solvers conserve the discrete mass by construction. We now show how to modify the predicted flux to ensure that the discrete $\ell_2$-norm is non-increasing in the continuous-time limit. In 1D, the discrete $\ell_2$-norm is non-increasing if 
\begin{equation}
    \frac{d}{dt}\sum_{j=1}^N \frac{1}{2}u_j^2 \Delta x_j \le 0
\end{equation}
for all $t$. $\Delta x_j$ is the width of cell $\Omega_j$. Some simple algebra and \cref{eq:fv_a} gives
\begin{equation}
    \frac{d}{dt} \sum_{j=1}^N \frac{\Delta x_j}{2} u_j^2 = \sum_{j=1}^N u_j \frac{\partial u_j} {\partial t} \Delta x_j =  -\sum_{j=1}^N u_j (f_{j+\frac{1}{2}} - f_{j-\frac{1}{2}}).
\end{equation}
Performing summation by parts gives
\begin{equation}
-\sum_{j=1}^N u_j (f_{j+\frac{1}{2}} - f_{j-\frac{1}{2}}) = \sum_{j=1}^{N-1} f_{j+\frac{1}{2}}\big(u_{j+1} - u_j \big) + f_{\frac{1}{2}} u_1 - f_{N+\frac{1}{2}} u_N \le 0.
\end{equation}
In a periodic domain, this is simply
\begin{equation}\label{eq:stability_energy_method}
    \frac{d}{dt}\sum_{j=1}^N  \frac{\Delta x_j }{2}(u_j)^2 = \sum_{j=1}^{N} f_{j+\frac{1}{2}}\big(u_{j+1} - u_j \big) \le 0.
\end{equation}
For the rest of the section we will assume a periodic domain, though our results can easily be generalized to non-periodic domains. 
Let us now define 
\begin{equation}
    \frac{d\ell_{2}^{\textnormal{old}}}{dt} \coloneqq \sum_{j=1}^N f_{j+\frac{1}{2}}(u_{j+1} - u_j) 
\end{equation}
as the original rate of change of the discrete $\ell_2$-norm, and $\nicefrac{d\ell_{2}^{\textnormal{new}}}{dt}$ as the desired rate of change of the discrete $\ell_2$-norm. To ensure non-increasing $\ell_2$-norm, we want $\nicefrac{d\ell_2^{\textnormal{new}}}{dt}\le 0$.
We also define $\bm u_j \coloneqq \{u_j\}_{j=1}^N$ as a vector representation of the discrete solution. 
We can change the time-derivative of the discrete $\ell_2$-norm from $\nicefrac{d\ell_2^{\textnormal{old}}}{dt}$ to $\nicefrac{d\ell_2^{\textnormal{new}}}{dt}$ by making the following transformation to $f_{j+\nicefrac{1}{2}}$:
\begin{equation}\label{eq:1d_stability}
	f_{j+\frac{1}{2}} \Rightarrow 
	f_{j+\frac{1}{2}} + 
	\frac{(\nicefrac{d\ell_2^{\textnormal{new}}}{dt} - \nicefrac{d\ell_2^{\textnormal{old}}}{dt}) G_{j+\nicefrac{1}{2}}(\bm u_j)}
		{\sum_{k=1}^N G_{k+\nicefrac{1}{2}}(\bm u_k)(u_{k+1} - u_k) }
\end{equation}
for any scalar $\nicefrac{d\ell_2^{\textnormal{new}}}{dt}$ and any non-constant, finite function $G_{j+\nicefrac{1}{2}}(\bm u_j)$ for which 
\begin{equation}
    \sum_{k=1}^N G_{k+\nicefrac{1}{2}}(\bm u_k)(u_{k+1} - u_k) \ne 0.
\end{equation}
As the reader can verify by plugging \cref{eq:1d_stability} into \cref{eq:stability_energy_method}, \cref{eq:1d_stability} modifies $f_{j+\nicefrac{1}{2}}$ in a way that adds a constant $(\nicefrac{d\ell_2^{\textnormal{new}}}{dt} - \nicefrac{d\ell_2^{\textnormal{old}}}{dt})$ to \cref{eq:stability_energy_method} via cancellation of the denominator.
Note that $G_{j+\nicefrac{1}{2}}(\bm u_j)$ is a hyperparameter that determines how each $f_{j+\nicefrac{1}{2}}$ is modified and $\nicefrac{d\ell_2^{\textnormal{new}}}{dt}$ is a user-defined quantity which sets the rate of change of the discrete $\ell_2$-norm. 
A similar calculation in a 2D periodic rectangular domain with uniform grid spacing reveals that the rate of change of the discrete $\ell_2$-norm is given by
\begin{equation}\label{eq:l2cons2d}
\begin{split}
    \frac{d}{dt}\sum_{i,j} \frac{u_{i,j}^2}{2} \Delta x \Delta y = &\Delta y\sum_{i,j} f^x_{i+\frac{1}{2},j}(u_{i+1,j} - u_{i,j}) \\ + &\Delta x \sum_{i,j} f^y_{i,j+\frac{1}{2}}(u_{i,j+1} - u_{i,j}) \le 0.
\end{split}
\end{equation}
We define 
\begin{subequations}
\begin{equation}
    \frac{d\ell^{\textnormal{old},x}_{2}}{dt} \coloneqq \Delta y \sum_{i,j} f^x_{i+\frac{1}{2},j}(u_{i+1,j} - u_{i,j}),
\end{equation}
\begin{equation}
    \frac{d\ell^{\textnormal{old},y}_{2}}{dt} \coloneqq \Delta x \sum_{i,j} f^y_{i,j+\frac{1}{2}} (u_{i,j+1} - u_{i,j}.
\end{equation}
\end{subequations}
\Cref{eq:l2cons2d} will be satisfied if the following transformations are made to $f^x_{i+\frac{1}{2},j}$ and $f^y_{i,j+\frac{1}{2}}$:
\begin{subequations}
\begin{equation} \label{eq:2d_stability_a}
	f^x_{i+\frac{1}{2},j} \Rightarrow 
	f^x_{i+\frac{1}{2},j} + \frac{(\nicefrac{d\ell^{\textnormal{new},x}_{2}}{dt} - \nicefrac{d\ell^{\textnormal{old},x}_{2}}{dt}) G^x_{i+\nicefrac{1}{2},j}(\bm u_{ij})}{\Delta y\sum_{k,l} G^x_{k+\nicefrac{1}{2},l}(\bm u_{kl})(u_{k+1,l} - u_{k,l})}
\end{equation}
\begin{equation} \label{eq:2d_stability_b} 
	f^y_{i,j+\frac{1}{2}} \Rightarrow 
	f^y_{i,j+\frac{1}{2}} + \frac{(\nicefrac{d\ell^{\textnormal{new},y}_{2}}{dt} - \nicefrac{d\ell^{\textnormal{old},y}_{2}}{dt}) G^y_{i,j+\nicefrac{1}{2}}(\bm u_{ij})}{\Delta x\sum_{k,l} G^y_{k,l+\nicefrac{1}{2}}(\bm u_{kl}) (u_{k,l+1} - u_{k,l})}
\end{equation}
\end{subequations}
for any scalars $\nicefrac{d\ell^{\textnormal{new},x}_{2}}{dt}$ and $\nicefrac{d\ell^{\textnormal{new},y}_{2}}{dt}$ where $\nicefrac{d\ell^{\textnormal{new},x}_{2}}{dt} + \nicefrac{d\ell^{\textnormal{new},y}_{2}}{dt} \le 0$ and any non-constant, finite functions $G^x_{i+\nicefrac{1}{2},j}(\bm u_{ij})$ and $G^y_{i,j+\nicefrac{1}{2}}(\bm u_{ij})$ for which 
\begin{subequations}
    \begin{equation}
        \sum_{k,l} G^x_{k+\nicefrac{1}{2},l}(\bm u_{kl})(u_{k+1,l} - u_{k,l}) \ne 0
    \end{equation}
    \begin{equation}
        \sum_{k,l} G^y_{k,l+\nicefrac{1}{2}}(\bm u_{kl}) (u_{k,l+1} - u_{k,l}) \ne 0
    \end{equation}
\end{subequations}
\Cref{eq:1d_stability,eq:2d_stability_a,eq:2d_stability_b} are the main results of section \ref{sec:fluxpredictingFV}; for scalar hyperbolic PDEs in 1D and 2D periodic domains they ensure that the discrete $\ell_2$-norm will be non-increasing in the continuous-time limit.

How should the hyperparameters $G_{j+\nicefrac{1}{2}}(\bm u_j)$, $G^x_{i+\nicefrac{1}{2},j}(\bm u_{ij})$, and $G^y_{i,j+\nicefrac{1}{2}}(\bm u_{ij})$ be set? In our experiments, we set $G_{j+\nicefrac{1}{2}}(\bm u_j) = (u_{j+1}-u_j)$, $G^x_{i+\nicefrac{1}{2},j}(\bm u_{ij}) = (u_{i+1,j} - u_{i,j})$, and $G^y_{i,j+\nicefrac{1}{2}}(\bm u_{ij}) = (u_{i,j+1} - u_{i,j})$. These choices have a simple physical interpretation: they correspond to the addition of a spatially constant diffusion coefficient everywhere in space \cite{artificial_viscosity}. Possible alternatives include setting $G_{j+\nicefrac{1}{2}}(\bm u_j) = (u_{j+1}-u_j)^\beta$ for $\beta>1$ or $G_{j+\nicefrac{1}{2}}(\bm u_j) = \alpha_{j+\nicefrac{1}{2}} (u_{j+1}-u_j)$ for $\alpha_{j+\nicefrac{1}{2}} \in \mathbb{R}$. Choosing large $\beta$ increases the amount of numerical diffusion added at discontinuities and decreases the amount of diffusion added in smooth regions, while $\alpha_{j+\nicefrac{1}{2}}$ is a spatially dependent scalar which determines a spatially varying distribution of added numerical diffusion.

\begin{figure}
  \centering
  \includegraphics[width=\textwidth]{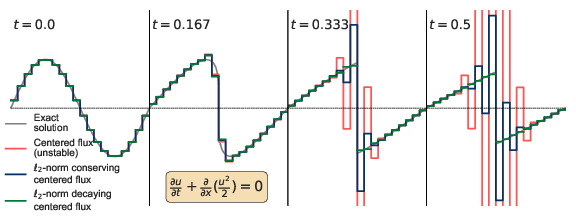}
  \caption{We modify the flux at cell boundaries to ensure non-increasing $\ell_2$-norm, thereby turning an unstable solver into a stable solver. While the centered flux $f_{j+\nicefrac{1}{2}}=\nicefrac{(u_{j}^2 + u_{j+1}^2)}{4}$ (red) is an unstable choice of flux on the inviscid Burgers equation and blows up by $t=0.5$, \cref{eq:1d_stability} with $\nicefrac{d\ell_2^{\textnormal{new}}}{dt}=0$ (blue) ensures that the discrete $\ell_2$-norm is conserved and is thus stable. Setting $\nicefrac{d\ell_2^{\textnormal{new}}}{dt} = \nicefrac{d\ell_2^{\textnormal{exact}}}{dt}$ (green) results in a more accurate solver.}
  \label{fig:fluxFV}
\end{figure}

We now illustrate the effect of modifying an unstable solver in 1D using \cref{eq:1d_stability}.
In \cref{fig:fluxFV}, we solve the inviscid Burgers' equation using a 3rd-order SSPRK ODE integrator \cite{ssprk} with the initial condition $u_0(x) = \sin{x}$.
It turns out that the centered flux $f_{j+\nicefrac{1}{2}}=\nicefrac{(u_j^2 + u_{j+1}^2)}{4}$, shown in red in \cref{fig:fluxFV}, gives $\nicefrac{d\ell_2^{\textnormal{old}}}{dt} > 0$ and is thus unstable and inaccurate.
If we transform $f_{j+\nicefrac{1}{2}}$ according to \cref{eq:1d_stability} with $\nicefrac{d\ell_2^{\textnormal{new}}}{dt}=0$, shown in blue, the solver becomes $\ell_2$-norm conserving and is stable.
Because the $\ell_2$-norm of the exact solution (grey) is decreasing in time, we can further improve the accuracy of the unstable centered flux by setting $\nicefrac{d\ell_2^{\textnormal{new}}}{dt} < 0$.
If we instead transform $f_{j+\nicefrac{1}{2}}$ according to \cref{eq:1d_stability} with $\nicefrac{d\ell_2^{\textnormal{new}}}{dt}=\nicefrac{d\ell_2^{\textnormal{exact}}}{dt}$, shown in green, the solver becomes much more accurate.

Note that the blue solution in \cref{fig:fluxFV} conserves both the discrete mass and the discrete $\ell_2$-norm, but does not maintain a discrete analogue of the total variation diminishing (TVD) property \cite{durran2013numerical} of the scalar Burgers equation. As a result, the blue solution does not fully eliminate the high-$k$ oscillations that develop in the red solution.
Note also that, in practice, we will usually not have a priori knowledge of $\nicefrac{d\ell_2^{\textnormal{exact}}}{dt}$.
Even if we did, it would be unrealistic to expect that error-correcting invariant preservation algorithms will in general be able to turn a very inaccurate solver (red in \cref{fig:fluxFV}) into an accurate solver (green in \cref{fig:fluxFV}). The takeaway from \cref{fig:fluxFV} is simply that solvers which preserve the right set of invariants tend to be more accurate than solvers which do not preserve those invariants.

\subsection{Continuous-time FV solvers with arbitrary time-derivative \label{sec:continuoustimeFV}}

\noindent In section \ref{sec:fluxpredictingFV}, we considered schemes that use ML to predict the flux $f$ across cell boundaries. Using integration by parts, we found that we could modify the fluxes to control the rate of change of the discrete $\ell_2$-norm and thereby preserve the non-increasing $\ell_2$-norm invariant. However, some machine learned PDE solvers may use an alternative form for the time-derivative which does not involve predicting the flux across cell boundaries. Thus, we now consider the more general problem of how to design invariant-preserving solvers for \cref{eq:scalarhyperbolic} with arbitrary time-derivative in arbitrary number of dimensions with arbitrary cell shapes. Once again we assume periodic boundary conditions and consider continuous-time solvers that predict $\nicefrac{\partial u_j}{\partial t}$ and use an ODE integration algorithm to advance $u_j$ in time. 
We again use a FV representation where the domain $\Omega$ is divided into $N$ grid cells $\Omega_j$ with volume $|\Omega_j|$ and the solution average in the $j$th grid cell is a scalar $u_j$.
We introduce bracket notation where $\langle \bm a \rangle \coloneqq \nicefrac{1}{N|\Omega|}\sum_{j=1}^N a_j |\Omega_j|$ denotes an average over the domain while the inner product notation $\langle \bm a | \bm b \rangle \coloneqq \sum_{j=1}^N a_j b_j |\Omega_j|$.

Suppose that the rate of change of $\bm u_j$ is given by
\begin{equation}\label{eq:blackbox}
 \frac{d \bm u_j}{d t}  = \bm N_j(\bm u_j)
\end{equation}
where $\bm N_j(\bm u_j) \in \mathbb{R}^N$ is an arbitrary update function. A machine learned solver would use ML to predict $\bm N_j$. Note that \cref{eq:blackbox} does not guarantee mass conservation by construction. Ensuring conservation of mass and the non-increasing $\ell_2$-norm property therefore requires modifying $\bm N_j$. Assuming periodic BCs, conservation of mass requires
\begin{equation}\label{eq:fv_arb_mass}
    \frac{d}{dt} \sum_{j=1}^N u_j |\Omega_j| = \sum_{j=1}^N \frac{d u_j}{d t} |\Omega_j| = \sum_{j=1}^N N_j  |\Omega_j|=0
\end{equation}
and non-increasing discrete $\ell_2$-norm requires
\begin{equation}\label{eq:fv_arb_l2}
    \frac{d}{dt} \sum_{j=1}^N \frac{1}{2} u_j^2 |\Omega_j| = \sum_{j=1}^N u_j \frac{d u_j}{d t} |\Omega_j| =  \sum_j u_j N_j |\Omega_j| \le 0.
\end{equation}
In vector-bracket notation, \cref{eq:fv_arb_mass,eq:fv_arb_l2} can be written as
$\langle \bm N_j \rangle = 0$
and
$\langle \bm u_j | \bm N_j \rangle \le 0$.
 These conditions will be satisfied if the following transformation is applied to $\bm N_j$:
\begin{equation}\label{eq:black_box_stability}
\begin{split}
    \bm U_j \coloneqq \bm u_j - &\langle \bm u_j\rangle \hspace{1.0cm} \bm M_j \coloneqq \bm N_j - \langle \bm N_j\rangle \hspace{1.0cm} \frac{d\ell_2^{\textnormal{old}}}{dt} = \langle \bm U_j | \bm M_j\rangle \\
     &\bm N_j \Rightarrow \bm M_j + \bigg(\frac{d\ell_2^{\textnormal{new}}}{dt} - \frac{d\ell_2^{\textnormal{old}}}{dt}\bigg)\frac{\bm G_j(\bm u_j)}{\langle \bm U_j | \bm G_j(\bm u_j)\rangle}
\end{split}
\end{equation}
for any $\nicefrac{d\ell_2^{\textnormal{new}}}{dt} \le 0$ and any finite function $\bm G_j(\bm u_j)$ where $\langle \bm G_j(\bm u_j) \rangle = 0$ and $\langle \bm G_j(\bm u_j) | \bm U_j \rangle \ne 0$. The choice $G_j(\bm u_j) = (\nabla^2 u)_j$ adds a spatially constant diffusion coefficient. In 1D with a spatially uniform grid, $(\nabla^2 u)_j = u_{j+1} - 2 u_j + u_{j-1}$.

\begin{figure}
  \centering
  \includegraphics[width=\textwidth]{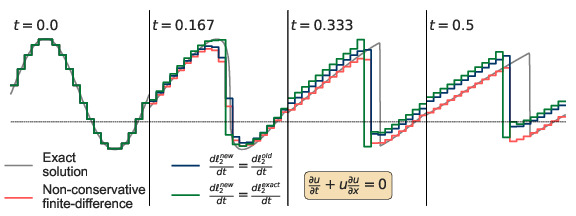}
  \caption{Using \cref{eq:black_box_stability}, we modify the time-derivative of a non-conservative finite-difference scheme. While the original finite-difference scheme (red) does not conserve the discrete mass $\sum_{j=1}^N u_j \Delta x$, the modified schemes (blue and green) conserve the discrete mass. All of these schemes decay the discrete $\ell_2$-norm. None of these finite-difference schemes result in a shock front traveling at the correct speed.} 
  \label{fig:burgers_nonconservative}
\end{figure}

We now demonstrate the effect of modifying a solver using \cref{eq:black_box_stability} to ensure the solver inherits discrete analogues of conservation of mass and non-increasing $\ell_2$-norm. We again solve the inviscid Burgers' equation $\frac{\partial u}{\partial t} + u \frac{\partial u}{\partial x } = 0$. The original scheme is the finite-difference scheme with
\begin{equation}
    N_j = -u_j (\delta u)_j\textnormal{, \hspace{0.5cm} } (\delta u)_j = \begin{cases}
        \nicefrac{u_j - u_{j-1}}{\Delta x}, & \text{if } u_j \ge 0\\
        \nicefrac{u_{j+1} - u_j}{\Delta x}, & \text{if } u_j < 0
        \end{cases}.
\end{equation}
We again use a 3rd-order SSPRK ODE integrator. The initial condition is $u_0(x) = 0.5 + \sin{x}$. The original finite-difference scheme, shown in red in \cref{fig:burgers_nonconservative}, does not conserve the discrete mass but preserves the non-increasing $\ell_2$-norm invariant. Using \cref{eq:black_box_stability} to modify $\bm N_j$, shown in blue and green in \cref{fig:burgers_nonconservative}, ensures that the discrete mass is conserved and maintains the property of non-increasing discrete $\ell_2$-norm. As $\nicefrac{d\ell_2^{\textnormal
old}}{dt}$ and $\nicefrac{d\ell_2^{\textnormal
exact}}{dt}$ are both negative and similar in magnitude, setting $\nicefrac{d\ell_2^{\textnormal
new}}{dt} = \nicefrac{d\ell_2^{\textnormal
exact}}{dt}$ (green) does not significantly improve accuracy over $\nicefrac{d\ell_2^{\textnormal
new}}{dt} = \nicefrac{d\ell_2^{\textnormal
old}}{dt}$. For all three of the schemes in \cref{fig:burgers_nonconservative}, the shock travels at the wrong speed.

\subsection{Discrete-time FV solvers with arbitrary time-derivative \label{sec:discretetimeFV}}

\noindent We now show how to design invariant-preserving solvers when using the discrete-time update 
\begin{equation}\label{eq:discretetimeupdaterule}
    \bm u_{j}^{n+1} = \bm u_{j}^n + \Delta \bm u_j^n
\end{equation}
rather than the continuous-time update $\nicefrac{d \bm u_j}{d t} = \bm N_j(\bm u_j)$. The same error-correction strategy can be used to modify $\Delta \bm u_j^n$, though preserving the non-linear invariant requires solving a quadratic equation.
A discrete-time machine learned solver would use ML to predict $\Delta \bm u_j^n$. 
To ensure conservation of mass, we want
$\langle \bm u^{n+1}_j \rangle = \langle \bm u^{n}_j \rangle$
which requires that
\begin{equation}\label{eq:discretetime_masscondition}
    \langle \Delta \bm u^n_j \rangle = 0.
\end{equation}
To ensure that the discrete $\ell_2$-norm is non-increasing, we want
$\langle (\bm u_{j}^{n+1})^2 \rangle \le  \langle (\bm u_j^{n})^2 \rangle$. Suppose that we want the change of the discrete $\ell_2$ norm to be some scalar $\Delta \ell_2$ where $\Delta \ell_2 \le 0$. Thus, we want $\frac{1}{2}\langle (\bm u_{j}^{n+1})^2 \rangle  =  \frac{1}{2}\langle (\bm u_j^{n})^2 \rangle  + \Delta \ell_2$. Some simple algebra gives
\begin{equation}\label{eq:discretetime_l2condition}
     2 \langle \bm u_j^n |  \Delta \bm u_j^n \rangle +  \langle ( \Delta \bm u_j^n)^2 \rangle = 2 \Delta \ell_2 .
\end{equation}
\Cref{eq:discretetime_masscondition,eq:discretetime_l2condition} will be satisfied if the following transformation is made to $\Delta \bm u_j^n$:
\begin{equation}\label{eq:discretetime_updatedelta}
     \Delta \bm u_j^n \rightarrow \overline{\Delta \bm u_j^n}+ \epsilon \bm G_j(\bm u_j^n)
\end{equation}
where $\overline{\Delta \bm u_j^n} \coloneqq \bm{{\Delta \bm u}}_j^n - \langle \bm{{\Delta \bm u}}_j^n \rangle$, for any function $\bm G_j(\bm u_j)$ for which $\langle \bm G_j(\bm u_j^n) \rangle =0$ and $\epsilon$ is some yet-to-be-determined scalar. Plugging \cref{eq:discretetime_updatedelta} into \cref{eq:discretetime_l2condition} gives
\begin{equation}
    2\langle \bm u_j^n | \overline{\Delta \bm u_j^n} \rangle + \langle (\overline{\Delta \bm u_j^n})^2 \rangle + 2\epsilon\Big( \langle \bm u_j^n + \overline{\Delta \bm u_j^n} | \bm G_j \rangle \Big) + \epsilon^2 \langle  (\bm G_j)^2 \rangle = 2 \Delta \ell_2
\end{equation}
which is a quadratic equation for epsilon. \Cref{eq:discretetime_l2condition} will thus be satisfied if
\begin{equation}\label{eq:discretetime_epsilon}
    \epsilon = \frac{\langle \bm u_j^n + \overline{\Delta \bm u_j^n} | \bm G_j \rangle}{\langle  (\bm G_j)^2 \rangle}\Bigg[ -1 \pm \sqrt{1 - \frac{\langle  (\bm G_j)^2 \rangle \Big(2\langle \bm u_j^n | \overline{\Delta \bm u_j^n} \rangle + \langle (\overline{\Delta \bm u_j^n})^2 \rangle -  2 \Delta \ell_2\Big)}{\Big( \langle \bm u_j^n + \overline{\Delta \bm u_j^n} | \bm G_j \rangle \Big)^2}}\Bigg]
\end{equation}
To ensure that $\epsilon$ is small when ${{{\Delta \bm u}}_j^n}$ is small, we choose the plus sign for $\epsilon$ in \cref{eq:discretetime_epsilon}.
Modifying the discrete update $\Delta \bm u_j^n$ according to  \cref{eq:discretetime_updatedelta} where $\epsilon$ is given by the plus sign in \cref{eq:discretetime_epsilon} ensures that mass is conserved and the discrete $\ell_2$-norm changes by an amount $\Delta \ell_2$. 

Notice that \cref{eq:discretetime_epsilon} can have no solution. Depending on the values of $\bm u_j$, $\Delta \bm u_j^n$, and the hyperparameter $\bm G_j(\bm u_j)$, there is a minimum allowed value of $\Delta \ell_2$. We discuss the implications of this in \cref{sec:limitations}.

\begin{figure}
  \centering
  \includegraphics[width=\textwidth]{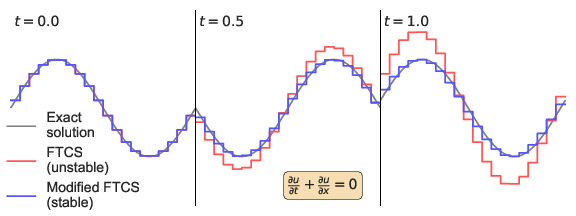}
  \caption{
    Solving the advection equation $\nicefrac{\partial u}{\partial t} + \nicefrac{\partial u}{\partial x} = 0$ with the forward-time centered-space (FTCS) update \cref{eq:ftcs} results in an $\ell_2$-norm increasing solution which is unconditionally unstable. Modifying the FTCS update using \cref{eq:discretetime_updatedelta,eq:discretetime_epsilon} allows us to control $\Delta \ell_2$, the change in the discrete $\ell_2$ norm. Setting $\Delta \ell_2 = 0$ (blue) results in an $\ell_2$-norm conserving and numerically stable solver.
  } 
  \label{fig:advection_ftcs}
\end{figure}

We now demonstrate the effect of using \cref{eq:discretetime_updatedelta,eq:discretetime_epsilon} to ensure that a discrete-time solver conserves mass and does not increase the discrete $\ell_2$-norm. We solve the advection equation $\frac{\partial u}{\partial t} + c \frac{\partial u}{\partial x} = 0$ with the forward-time, centered-space (FTCS) update 
\begin{equation}\label{eq:ftcs}
    u_{j}^{n+1} = u_{j}^n - \frac{ c\Delta t}{2\Delta x} (u_{j+1}^n - u_{j-1}^n)
\end{equation}
where $n$ is an index representing the values at the $n$th timestep. Our initial condition is $u_0(x) = \sin(x)$. We set $c=1$ and use a CFL number of $0.5$. We use periodic BCs in a domain of width $1$.
The exact solution to the advection equation is $u(x,t) = u_0(x - ct)$, which simply means that the solution translates to the right with speed $c$. The exact solution is shown in grey in \cref{fig:advection_ftcs}. 
FTCS conserves the discrete mass but increases the discrete $\ell_2$-norm for any $\Delta t$ and $\Delta x$ and is thus unconditionally unstable. The unstable FTCS solution is shown in red in \cref{fig:advection_ftcs}.
We can modify the FTCS update to control the change in the discrete $\ell_2$ norm using \cref{eq:discretetime_updatedelta,eq:discretetime_epsilon} with $G_j = u_{j+1} - 2 u_j + u_{j-1}$. For the 1D advection equation with smooth initial conditions we can set $\Delta \ell_2 = 0$, though for other PDEs we might want $\Delta \ell_2 < 0$. The $\ell_2$-norm conserving modified FTCS update is shown in blue in \cref{fig:advection_ftcs}. The modified update adds numerical diffusion to the FTCS update and results in a stable solver.

\subsection{Discontinuous Galerkin solvers \label{sec:DG}}

\noindent In \cref{sec:continuoustimeFV}, we developed a technique for preserving invariants of any solver for \cref{eq:scalarhyperbolic} which uses a FV-like solution representation in the continuous-time limit. We now show how to do the same for any solver which uses a discontinuous Galerkin (DG) solution representation. A machine learned DG solver would use ML to predict the time-derivative of the DG coefficients. 

\subsubsection{Discontinuous Galerkin}
We now give a brief introduction to DG methods \cite{durran2013numerical}. Intuitively, the main difference between FV methods and DG methods is the solution representation: FV methods represent the solution as piecewise constant within each cell, while DG methods represent the solution as a polynomial within each cell.
With DG, we partition our domain $\Omega \in \mathbb{R}^d$ into $N$ cells $\{I_j\}_{j=1}^N$. In 1D, the solution representation within cell $I_j \in [x_{j-\nicefrac{1}{2}}, x_{j+\nicefrac{1}{2}}]$ is
\begin{equation}
    u_{j}(x,t) = \sum_{k=0}^p a_{jk}(t) \psi_k(x)
\end{equation}
where $\psi_k(x)$ are $p+1$ polynomial basis functions and $a_{jk}(t)$ are time-dependent coefficients. Notice that $u_j(x,t)$ is continuous within a cell but discontinuous across cell boundaries. The equation for the time-evolution of $a_{jk}(t)$ is found by minimizing the $\ell_2$-norm of the PDE residual within the subspace of basis functions spanned by $\psi_k(x)$. For \cref{eq:scalarhyperbolic}, this residual $E_j$ is given by
\begin{equation}
E_j = \int_{I_j}\bigg(\sum_{k=0}^p \dot{a}_{jk} \psi_k(x) + \frac{\partial f(u_j)}{\partial x}\bigg)^2\mathop{dx}.   
\end{equation}
The minimum of $E_j$ can be found from
\begin{equation}\label{eq:dg_weak_form}
    0 = \frac{\partial E_j}{\partial \dot{a}_{jk}} = 2 \int_{I_j} \psi_k \bigg(\sum_{k'=0}^p \dot{a}_{jk'} \psi_{k'} + \frac{\partial f}{\partial x}\bigg) \mathop{dx}.
\end{equation}
For general $\psi_k(x)$, \cref{eq:dg_weak_form} is a matrix equation for $\dot{a}_{jk}$. Typically, the $p+1$ basis functions are chosen to span the vector space of polynomials of degree $p$ and are orthogonal polynomials such that
\begin{equation}
    \int_{I_j}\psi_k(x)\psi_{k'}(x) dx = \Delta x_j \langle \psi_k | \psi_k \rangle {\delta_{kk'}}
\end{equation}
where $\langle \psi_k | \psi_k \rangle$ is a scalar and $\delta_{kk'}$ is the kronecker delta. In 1D, Legendre polynomials are usually chosen as basis functions so that $\langle \psi_k | \psi_k \rangle = 1/(2k+1)$. If orthogonal polynomials are chosen, then \cref{eq:dg_weak_form} can be inverted to solve for $\dot{a}_{jk}$:
\begin{equation}\label{eq:dg_before_ibp}
    \dot{a}_{jk} = -\frac{1}{\Delta x_j \langle \psi_k | \psi_k \rangle } \int_{I_j} \psi_k(x) \frac{\partial f}{\partial x} \mathop{dx}.
\end{equation}
The final step is to integrate by parts, giving
\begin{equation}\label{eq:dg_time_evolution}
     \dot{a}_{jk} =\frac{1}{\Delta x_j \langle \psi_k | \psi_k \rangle }\Big(- f_{j+\frac{1}{2}} \psi_k^+ + f_{j-\frac{1}{2}} \psi_k^- + \int_{I_j} f \frac{\partial \psi_k}{\partial x} \mathop{dx}\Big).
\end{equation}
where $\psi_k^+$ and $\psi_k^-$ are the values of $\psi_k$ at the right ($+$) and left ($-$) cell boundaries respectively. \Cref{eq:dg_time_evolution} has two terms: a volume term $\int_{I_j} f \frac{\partial \psi_k}{\partial x}dx$, and a boundary term $-f_{j+\nicefrac{1}{2}} \psi_k^+ + f_{j-\frac{1}{2}} \psi_k^-$. Depending on the form of \cref{eq:scalarhyperbolic}, the volume term can either be computed analytically or with a high-order quadrature. Because the discrete solution is discontinuous, the boundary term $f_{j+\frac{1}{2}}$ cannot be computed exactly and requires reconstructing the flux $f$ at cell boundaries. Note that if $k=0$ and $\psi_0(x)$ is chosen to be the zeroth Legendre polynomial $P_0(x)=1$, then \cref{eq:dg_time_evolution} reduces to the FV time-evolution equation \ref{eq:fv_a}. Note also that \cref{eq:dg_time_evolution} can be written as
\begin{equation}\label{eq:dg_general_update}
    \frac{d\bm a_{jk}}{dt} = \frac{1}{\Delta x \langle \psi_k | \psi_k \rangle} \bm N_{jk}
\end{equation}
where $\bm a_{jk}$ is the vector representation of the $N \times (p+1)$ solution coefficients. 
A machine learned DG solver would use ML to predict $\bm N_{jk}$.

\subsubsection{Discrete invariants}
DG schemes conserve a discrete analogue $\sum_j \int_{I_j} u_j(x,t)\mathop{dx}$ of the continuous invariant $\int_\Omega u \mathop{dx}$ by construction. In a 1D periodic system, assuming the DG basis functions $\psi_k(x)$ are given by Legendre polynomials $P_k(x)$, we can see this with a short proof: 
\begin{equation}
\begin{split}
    \frac{d}{dt} \sum_{j=1}^N \int_{I_j} u_j(x,t) \mathop{dx} = &\sum_{j=1}^N \sum_{k=0}^p \dot{a}_{jk} \int_{I_j} P_k(x) \mathop{dx} \\= &\sum_j \dot{a}_{j0} \Delta x_j = -\sum_j(f_{j+\frac{1}{2}} - f_{j-\frac{1}{2}}) = 0.
\end{split}
\end{equation}
DG schemes do not automatically preserve discrete analogues of any of the non-linear invariants of the continuous PDE \cref{eq:scalarhyperbolic}. We now show how to modify a DG solver for \cref{eq:scalarhyperbolic} to ensure that the discrete $\ell_2$-norm is non-increasing in the continuous-time limit. We want
\begin{equation}
    \frac{d}{dt} \sum_j \int_{I_j} u_j(x,t)^2 \mathop{dx} \le 0.
\end{equation}
Using orthogonality of Legendre polynomials,
\begin{equation}\label{eq:DG_dl2dt}
    \begin{split}
        \frac{1}{2} \frac{d}{dt} \sum_j \int_{I_j} u_j(x,t)^2 dx = \frac{1}{2} \frac{d}{dt} \sum_j \int_{I_j} (\sum_k a_{jk} \psi_k)(\sum_{k'} a_{jk'} \psi_k') \mathop{dx} \\= \frac{1}{2}\frac{d}{dt}\sum_j \sum_k a_{jk}^2(t) \langle \psi_k | \psi_k \rangle \Delta x_j = \sum_j \sum_k a_{jk} \dot{a}_{jk} \langle \psi_k | \psi_k \rangle \Delta x_j.
    \end{split}
\end{equation}
Using \cref{eq:dg_general_update},
\begin{equation}
        \frac{1}{2} \frac{d}{dt} \sum_j \int_{I_j} u_j(x,t)^2 dx = \sum_j \sum_k a_{jk} N_{jk} = \langle \bm a_{jk} | \bm N_{jk} \rangle.
\end{equation}
Let us now define 
\begin{equation}
    \frac{d\ell_2^{\textnormal{old}}}{dt} \coloneqq \langle \bm a_{jk} | \bm N_{jk} \rangle
\end{equation}
as the original rate of change of the discrete $\ell_2$-norm, and $\nicefrac{d\ell_2^{\textnormal{new}}}{dt}$ as the desired rate of change of the discrete $\ell_2$-norm. To ensure non-increasing $\ell_2$-norm, we want $\nicefrac{d\ell_2^{\textnormal{new}}}{dt} \le 0$. To modify the rate of change of the discrete $\ell_2$-norm, we can add numerical diffusion (or anti-diffusion) to the time-derivative.
Unlike FV methods, where diffusion can be written as the sum of fluxes $f_{j+\nicefrac{1}{2}} \propto (u_{j+1} - u_j)$, DG diffusion requires computing both boundary terms and volume terms. The details of DG diffusion can be found in \cite{dgdiffusion}. For our purposes, it is sufficient to know that the time-derivative of the DG coefficients due to a diffusion term $\nu \nabla^2 u$ with diffusion coefficient $\nu$ can be written as
\begin{equation}\label{eq:dg_diffusion_term}
    \frac{d \bm a_{jk}}{dt} = \frac{\nu}{\Delta x \langle \psi_k | \psi_k \rangle} \bm N_{jk}^{\textnormal{diffusion}}.
\end{equation}
Using \cref{eq:DG_dl2dt}, the rate of change of the discrete $\ell_2$-norm due to the diffusion term is
\begin{equation}
    \frac{1}{2} \frac{d}{dt} \sum_j \int_{I_j} u_j(x,t)^2 dx = \nu \sum_{j}\sum_k a_{jk} N_{jk}^{\textnormal{diffusion}} = \nu \langle \bm a_{jk} | \bm N_{jk}^{\textnormal{diffusion}} \rangle.
\end{equation}
If we set
\begin{equation}\label{eq:dg_nu}
    \nu = \bigg(\frac{d\ell_2^{\textnormal{new}}}{dt}  - \frac{d\ell_2^{\textnormal{old}}}{dt}  \bigg) \frac{1}{\langle \bm a_{jk} | \bm N_{jk}^{\textnormal{diffusion}} \rangle}
\end{equation}
and add the diffusion term \cref{eq:dg_diffusion_term} to the original update equation \cref{eq:dg_general_update}, then the total rate of change of the discrete $\ell_2$-norm will be $\nicefrac{d\ell_2^{\textnormal{new}}}{dt}$.
Although the addition of a diffusion term is not the most general way of updating $\nicefrac{d\ell_2}{dt}$, it is straightforward to compute and has a clear physical interpretation.

\begin{figure}
    \centering
    \begin{subfigure}[b]{\textwidth}
        \centering
        \includegraphics[width=\textwidth]{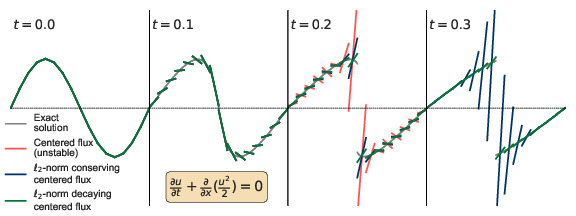}
        \caption{}
        \label{fig:dg_a}
    \end{subfigure}
    \begin{subfigure}[b]{\textwidth}
         \centering
         \includegraphics[width=\textwidth]{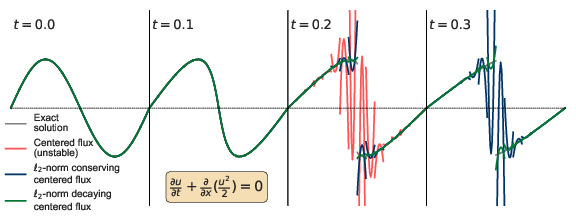}
         \caption{}
         \label{fig:dg_b}
     \end{subfigure}
    \caption{We add a diffusion term to two unstable DG solvers to control the rate of change of the discrete $\ell_2$-norm, thereby turning an unstable solver into a stable solver. Shown in red are standard DG solvers for the inviscid Burgers' equation with centered flux $f_{j+\nicefrac{1}{2}} = \frac{(u_{j+\nicefrac{1}{2}}^- + u_{j+\nicefrac{1}{2}}^+)^2}{8}$. This choice of flux increases the discrete $\ell_2$-norm, resulting in an unstable solver which blows up by $t=0.3$. (a) uses a degree-1 polynomial representation within each cell, while (b) uses a degree-2 polynomial representation within each cell. Adding a diffusion term with diffusion coefficient $\nu$ given by \cref{eq:dg_nu} controls the rate of change of the $\ell_2$-norm and results in a stable solver so long as $\nicefrac{d\ell_2^{\textnormal{new}}}{dt} \le 0$. Setting $\nicefrac{d\ell_2^{\textnormal{new}}}{dt} = 0$ (blue) gives large spurious oscillations and a stable but inaccurate solver, while setting $\nicefrac{d\ell_2^{\textnormal{new}}}{dt} = \nicefrac{d\ell_2^{\textnormal{exact}}}{dt}$ (green) adds additional diffusion and results in a more accurate solver.}
        \label{fig:dg_demo}
\end{figure}

Let us now demonstrate the effect of adding a diffusion term $\nu \nabla^2 u$ with $\nu$ given by \cref{eq:dg_nu} to an $\ell_2$-norm increasing DG solver. We use the same setup as in \cref{fig:fluxFV}. In \cref{fig:dg_demo} we show the time-evolution of DG methods with degree-1 polynomials ($p=1$, \cref{fig:dg_a}) and degree-2 polynomials ($p=2$, \cref{fig:dg_b}). The original DG solver uses the update \cref{eq:dg_time_evolution} with centered flux $f_{j+\nicefrac{1}{2}} = \frac{(u_{j+\nicefrac{1}{2}}^- + u_{j+\nicefrac{1}{2}}^+)^2}{8}$ where $u_{j+\nicefrac{1}{2}}^-$ is the solution just to the left of the ($j+\frac{1}{2}$)th cell boundary and $u_{j+\nicefrac{1}{2}}^+$ is the solution just to the right.
With this choice of flux, the original DG solvers are unstable and blow up by $t = 0.3$. By adding a diffusion term to the DG solver with diffusion coefficient given by \cref{eq:dg_nu}, we can control the rate of change of the $\ell_2$-norm and improve the accuracy of the solver. Like in \cref{fig:fluxFV}, setting $\nicefrac{d\ell_2^{\textnormal{new}}}{dt} = 0$ (blue) results in a highly oscillatory solution, but setting $\nicefrac{d\ell_2^{\textnormal{new}}}{dt} = \nicefrac{d\ell_2^{\textnormal{exact}}}{dt}$ damps many of the oscillations and results in a more accurate solution.


\subsection{Spectral solvers \label{sec:fourier}}

\noindent In \cref{sec:fluxpredictingFV,sec:continuoustimeFV,sec:discretetimeFV} we considered solvers for \cref{sec:scalarhyperbolic} which represent the solution in a FV basis. In \cref{sec:DG}, we considered solvers which represent the solution in a DG basis. We now consider spectral solvers which represent the solution in a Fourier basis. In a 1D periodic domain where $x \in [0,L]$, the Fourier representation $\hat{u}$ of the solution $u$ is
\begin{equation}
    \hat{u}(x, t) = \sum_{m=-N}^{N} \tilde{u}_{m}(t) e^{\frac{2\pi i m x} {L}}.
\end{equation}
The coefficients $\tilde{u}_m = u_m^r + i u_m^i \in \mathbb{C}$. To ensure that $\hat{u}(x,t)\in \mathbb{R}$, we require $\tilde{u}_{-m} = \tilde{u}_{m}^*$ which gives $2N+1$ degrees of freedom in the solution representation. We consider the update equation
\begin{equation}
    \frac{d \tilde{\bm u}_m}{d t} = \tilde{\bm N}_m
\end{equation}
where $\tilde{\bm u}_m \in \mathbb{R}^{2N+1}$ is a vector representation of the $2N+1$ degrees of freedom in the complex solution coefficients.

\subsubsection{Discrete invariants}
To ensure conservation of mass, we require that
\begin{equation}
    \frac{d}{dt} \int_0^L \hat{u}(x, t) \mathop{dx} = \int_0^L \sum_{m=-N}^N\frac{d\tilde{u}_m}{dt} e^{\frac{2\pi i m x} {L}} \mathop{dx} = L\tilde{N}_0 = 0.
\end{equation}
The rate of change of the $0$th Fourier coefficient must be zero. To ensure that the $\ell_2$-norm is non-increasing, we require that
\begin{equation}
    \frac{d}{dt}\int_0^L\frac{1}{2}|\hat{u}(x,t)|^2 dx 
    \le 0.
\end{equation}
We then use the Plancherel theorem
\begin{equation}
    \int_0^L |\hat{u}(x,t)|^2 dx = L\sum_{m=-N}^N |\tilde{u}_m|^2.
\end{equation}
Using $|\tilde{u}_m|^2=|\tilde{u}_{-m}|^2$ and $\nicefrac{d\tilde{u}_0}{dt}=0$, we have
\begin{equation}
   \frac{L}{2}\frac{d}{dt}\sum_{m=-N}^N |\tilde{u}_m|^2 = L\frac{d}{dt}\sum_{m=1}^N |\tilde{u}_m|^2 = 2L \sum_{m=1}^N u_m^r \frac{d{u}_m^r}{dt} + {u}_m^i \frac{d{u}_m^i}{dt} \le 0.
\end{equation}
In vector notation, this can be written as
\begin{equation}
    2L \langle \tilde{\bm u}_m | \tilde{\bm N}_m \rangle \le 0.
\end{equation}
These conditions will be satisfied if the following transformations are applied to $\tilde{\bm N}_m$:
\begin{equation}\label{eq:fourier_stability}
\begin{split}
    &\tilde{N}_0 \Rightarrow 0 \hspace{2.0cm} \frac{d\ell_2^{\textnormal{old}}}{dt} = 2L \langle \tilde{\bm{u}}_m | \tilde{\bm N}_m \rangle
    \\
    &\tilde{\bm{N}}_m \Rightarrow \tilde{\bm N}_m + \bigg(\frac{d\ell_2^{\textnormal{new}}}{dt} - \frac{d\ell_2^{\textnormal{old}}}{dt}\bigg)\frac{\bm G_m(\tilde{\bm u}_m)}{2L \langle \tilde{\bm u}_m | \bm G_m(\tilde{\bm u}_m)\rangle}
\end{split}
\end{equation}
for any $\nicefrac{d\ell_2^{\textnormal{new}}}{dt} \le 0$ and any finite function $\bm G_m(\tilde{\bm u}_m)$ where $G_0 = 0$ and $\langle \bm G_m(\tilde{\bm u}_m) | \tilde{\bm u}_m \rangle \ne 0$.

\subsection{2D incompressible Euler equations \label{sec:energyconservation}}

\noindent In \cref{sec:fluxpredictingFV,sec:continuoustimeFV,sec:discretetimeFV,sec:DG,sec:fourier}, we designed mass-conserving and $\ell_2$-norm non-increasing solvers for the generic scalar hyperbolic PDE, \cref{eq:scalarhyperbolic}. In this section, we design invariant-preserving solvers for a specific scalar hyperbolic PDE, the 2D incompressible Euler equations \cref{eq:euler}.
As we learned in \cref{sec:standardsolversincompressible}, the incompressible Euler equations exactly conserve the mass $\int \chi \mathop{dx}\mathop{dy}$, the energy $\frac{1}{2}\int \bm u^2 \mathop{dx}\mathop{dy}$, and the enstrophy $\int \chi^2 \mathop{dx}\mathop{dy}$. We will now design an error-correcting algorithm which preserves discrete analogues of these three invariants.

Suppose we want to solve \cref{eq:euler} on a 2D periodic rectangular domain. We divide the domain $\Omega$ into $N_x \times N_y$ cells with indices $i \in [1, \dots, N_x]$ and $j \in [1, \dots, N_y]$, with average vorticity $\chi_{ij}$, and with uniform grid spacing $\Delta x = \nicefrac{L_x}{N_x}$ and $\Delta y = \nicefrac{L_y}{N_y}$. Each cell has volume $|\Omega_{ij}| = \Delta x \Delta y$. We again use the notation $\mathop{dz} = \mathop{dx}\mathop{dy}$. Suppose also that the rate of change of $\bm \chi_{i,j}$ is given by
\begin{equation}\label{eq:update_incompressible}
    \frac{d\bm \chi_{i,j}}{dt} = \bm N_{i,j}(\bm \chi_{i,j})
\end{equation}
where $\bm N_{i,j} \in \mathbb{R}^{N_x \times N_y}$ is an arbitrary update function. A machine learned solver would use ML to predict $\bm N_{i,j}$ and would solve the elliptic equation $\nabla^2 \psi = - \chi$ using a standard FEM Poisson solver. Notice that \cref{eq:update_incompressible} is a subset of \cref{eq:blackbox} and that while $\bm \chi_{i,j}$ is represented in a discontinuous FV basis, $\bm \psi_{i,j}$ is represented in a continuous FEM basis.

\subsubsection{Discrete invariants}

In \cref{sec:continuoustimeFV}, we showed that discrete conservation of mass requires $\langle \bm N_{i,j} \rangle = 0$ and non-increasing $\ell_2$-norm requires $\langle \bm u_{i,j} | \bm N_{i,j} \rangle \le 0$. The rate of change of the discrete energy can be computed using integration by parts and continuity of $\psi_{i,j}$ across cell boundaries:
\begin{equation}
\begin{split}
        \frac{d}{dt} \sum_{i,j} \int_{\Omega_{i,j}} \frac{1}{2} \bm \nabla \psi_{i,j} \cdot \bm \nabla \psi_{i,j} \mathop{dz} = 
    \sum_{i,j} \int_{\Omega_{i,j}}\bm \nabla \psi_{i,j} \cdot \bm \nabla \frac{\partial \psi_{i,j}}{\partial t} \mathop{dz}
    \\
    = -\sum_{i,j}\int_{\Omega_{i,j}} \psi_{i,j} \frac{\partial}{\partial t} \nabla^2 \psi_{i,j} + \int_{\partial \Omega_{i,j}} \psi_{i,j} \frac{\partial}{\partial t} \bm \nabla \psi_{i,j} \cdot \mathop{d\bm s} \\= \sum_{i,j} \int_{\Omega_{i,j}} \psi_{i,j} \frac{\partial \chi_{i,j}}{\partial t}\mathop{dz} = \sum_{i,j}  \frac{\partial \chi_{i,j}}{\partial t} \int_{\Omega_{i,j}} \psi_{i,j} \mathop{dz}.
\end{split}
\end{equation}
To ensure that the discrete energy is conserved, we require that
\begin{equation}
    \frac{d}{dt} \sum_{i,j} \int_{\Omega_{i,j}} \frac{1}{2} \bm \nabla \psi_{i,j} \cdot \bm \nabla \psi_{i,j} \mathop{dz} = 
    \sum_{i,j} \frac{\partial \chi_{i,j}}{\partial t} \int_{\Omega_{i,j}} \psi_{ij} \mathop{dz} = \langle \overline{\bm \psi}_{i,j} | \bm N_{i,j} \rangle = 0
\end{equation}
where $\overline{\psi}_{ij}$ is the average value of $\psi$ within cell $\Omega_{ij}$. Conservation of mass, conservation of energy, and the non-increasing $\ell_2$-norm property will therefore all be guaranteed for \cref{eq:euler} if the following transformation is applied to $\bm N_{ij}$:
\begin{align*}
    \bm U_{i,j} = \bm \chi_{i,j} - \langle \bm \chi_{i,j} \rangle && \bm M_{i,j} = \bm N_{i,j} - \langle \bm N_{i,j} \rangle && \bm{\Bar{\phi}}_{i,j} = \bm{\Bar{\psi}}_{i,j} - \langle \bm{\Bar{\psi}}_{i,j} \rangle \\
    \bm W_{i,j} = \bm U_{i,j} - \frac{\langle \bm U_{i,j} | \bm{\Bar{\phi}}_{i,j}  \rangle}{\langle \bm{\Bar{\phi}}_{i,j}  | \bm{\Bar{\phi}}_{i,j}  \rangle} \bm{\Bar{\phi}}_{i,j} && \bm P_{i,j} = \bm M_{i,j} - \frac{\langle \bm M_{i,j} | \bm{\Bar{\phi}}_{i,j}  \rangle}{\langle \bm{\Bar{\phi}}_{i,j}  | \bm{\Bar{\phi}} _{i,j} \rangle} \bm{\Bar{\phi}}_{i,j} && \frac{d\ell_2^{\textnormal{old}}}{dt} = \langle \bm W_{i,j} | \bm P_{i,j}  \rangle
\end{align*}
\begin{equation}\label{eq:eulerenergycons}
    \bm N_{i,j} \Rightarrow \bm P_{i,j} + \bigg(\frac{d\ell_2^{\textnormal{new}}}{dt} - \frac{d\ell_2^{\textnormal{old}}}{dt}\bigg)\frac{\bm G(\bm \chi_{i,j})}{\langle \bm W_{i,j} | \bm G( \bm \chi_{i,j}) \rangle}
\end{equation}
for any $\nicefrac{d\ell_2^{\textnormal{new}}}{dt}\le 0$ and any non-constant scalar function $\bm G_{i,j}(\bm \chi_{i,j})$ for which $\langle \bm G_{i,j}(\bm \chi_{i,j}) \rangle = 0$, $\langle \bm G_{i,j}(\bm \chi_{i,j}) | \bm {\bar{\psi}}_{i,j} \rangle = 0$ and $\langle \bm W_{i,j} | \bm G_{i,j}(\bm \chi_{i,j}) \rangle \ne 0$. The physically motivated choice 
\begin{equation}\label{eq:g_euler}
    \bm G_{i,j}(\bm \chi_{i,j}) = (\nabla^2 \bm W)_{i,j} - \frac{\langle (\nabla^2 \bm W)_{i,j} | \bm{\Bar{\phi}}_{i,j}  \rangle}{\langle \bm{\Bar{\phi}}_{i,j}  | \bm{\Bar{\phi}}_{i,j}  \rangle} \bm{\Bar{\phi}}_{i,j}.
\end{equation}
corresponds to the addition of a spatially constant diffusion coefficient, projected into an energy-conserving subspace.

We now illustrate the effect of modifying a stable standard solver for the 2D incompressible Euler equations using \cref{eq:2d_stability_a,eq:2d_stability_b,eq:eulerenergycons}. This standard solver is the second-order MUSCL scheme with monotonized central (MC) flux limiters \cite{sweby,muscl}. We use a linear finite element (FE) solver for the poisson equation \cite{fe_solver} and an SSP-RK3 ODE integrator \cite{ssprk}. The MUSCL scheme does not have any provable guarantees of $\ell_2$-norm conservation, but in 1D the MUSCL scheme is provably TVD \cite{osher1985convergence} and in practice the MUSCL scheme tends to decay the discrete $\ell_2$-norm as well as the discrete energy.

\begin{figure}
    \centering
    \begin{subfigure}[b]{0.98\textwidth}
        \centering
        \includegraphics[width=\textwidth]{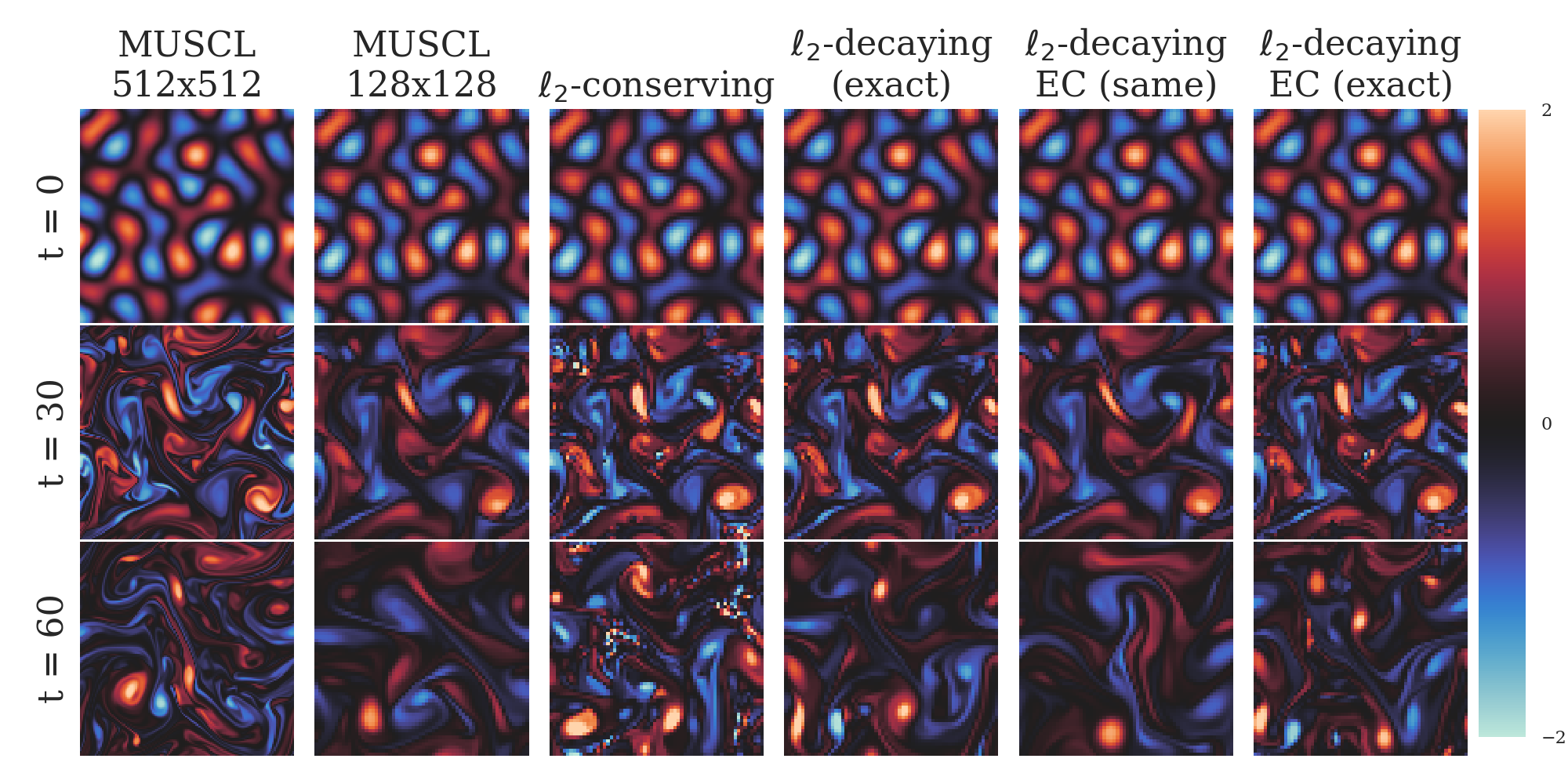}
        \caption{}
        \label{fig:euler_data}
    \end{subfigure}
    \begin{minipage}[c]{0.54\textwidth}
    \begin{subfigure}[b]{\textwidth}
         \centering
         \includegraphics[width=\textwidth]{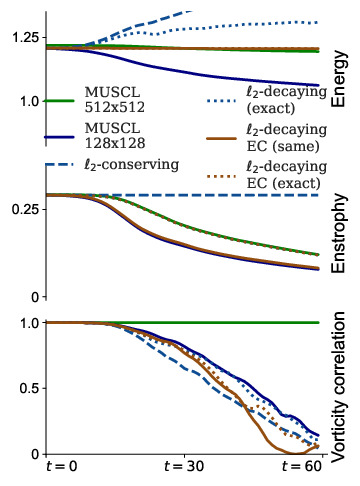}
         \caption{}
         \label{fig:euler_diag_demo}
     \end{subfigure}
     \end{minipage}
     \begin{minipage}[c]{0.44\textwidth}
    \caption{By modifying the time-derivative using \cref{eq:eulerenergycons}, we can design mass-conserving, energy-conserving, and enstrophy non-increasing solvers for the 2D incompressible Euler equations. (a) Images of the vorticity $\chi$ evolving under different numerical schemes for the incompressible Euler equations \cref{eq:euler}. The first and second columns show the standard MUSCL scheme at high and low resolution. The third and fourth columns show the low resolution MUSCL scheme, modified using \cref{eq:2d_stability_a,eq:2d_stability_b} to control the rate of change of the $\ell_2$-norm. The fifth and six columns use \cref{eq:eulerenergycons} to modify the low resolution MUSCL time-derivative to ensure energy conservation (EC) and again control the rate of change of the discrete $\ell_2$-norm. (b) Energy, enstrophy, and vorticity correlation over time. We use vorticity correlation as a benchmark measure of accuracy.}
    \label{fig:euler_2d}
    \end{minipage}
\end{figure}

In each of the six columns of \cref{fig:euler_data}, we see snapshots of the vorticity $\chi$ evolved using different numerical schemes. The first column is at high resolution ($512 \times 512$), while the other five columns are at low resolution ($128 \times 128$). The first and second columns use the unmodified MUSCL scheme. The third and fourth columns use \cref{eq:2d_stability_a,eq:2d_stability_b} to modify the MUSCL fluxes and set $\nicefrac{d\ell_2^{\textnormal{new}}}{dt}$. We use $\bm G_{i+\nicefrac{1}{2},j}^x = u_{i+1,j} - u_{ij}$ and $\bm G_{i,j+\nicefrac{1}{2}}^y = u_{i,j+1} - u_{ij}$. In the third column, we set $\nicefrac{d\ell_2^{\textnormal{new}}}{dt} = 0$ with $\nicefrac{d\ell_2^{\textnormal{new},x}}{dt} = \nicefrac{d\ell_2^{\textnormal{new},y}}{dt} = 0$. We find, similarly to \cref{fig:fluxFV}, that ensuring $\ell_2$-norm conservation introduces spurious high-$k$ oscillations. In the fourth column, we set 
$\nicefrac{d\ell_2^{\textnormal{new}}}{dt} = \nicefrac{d\ell_2^{\textnormal{exact}}}{dt}$, the rate of change of the discrete $\ell_2$-norm of the high resolution `exact' simulation.
We set $\nicefrac{d\ell_2^{\textnormal{new},x}}{dt} = \frac{1}{2} \nicefrac{d\ell_2^{\textnormal{exact}}}{dt}$, and $\nicefrac{d\ell_2^{\textnormal{new},y}}{dt} = \frac{1}{2}\nicefrac{d\ell_2^{\textnormal{exact}}}{dt}$. 
This allows for spurious oscillations to form, but much fewer than with $\nicefrac{d\ell_2^{\textnormal{new}}}{dt} = 0$. In the fifth and sixth columns, we modify the MUSCL time-derivative to enforce energy conservation (EC) using \cref{eq:eulerenergycons} with $\bm G_{i,j}$ set according to \cref{eq:g_euler}. In the fifth column we set $\nicefrac{d\ell_2^{\textnormal{new}}}{dt} =\nicefrac{d\ell_2^{\textnormal{old}}}{dt}$. In the sixth column we set $\nicefrac{d\ell_2^{\textnormal{new}}}{dt} = \nicefrac{d\ell_2^{\textnormal{exact}}}{dt}$. The energy-conserving schemes tend not to form spurious oscillations.

In \cref{fig:euler_diag_demo}, bottom row, we plot the vorticity correlation between the high resolution baseline and each of the five other schemes. Vorticity correlation has been used previously as a benchmark measure of accuracy for \cref{eq:euler} \cite{kochkov2021machine}. We find that setting $\nicefrac{d\ell_2^{\textnormal{new}}}{dt} = 0$ worsens accuracy relative to the unmodified MUSCL scheme at the same resolution, while setting $\nicefrac{d\ell_2^{\textnormal{new}}}{dt} = \nicefrac{d\ell_2^{\textnormal{exact}}}{dt}$ neither helps nor harms accuracy. In \cref{fig:euler_diag_demo}, middle and top rows, we plot the discrete enstrophy $\frac{1}{2}\sum_{i,j}\int \chi_{i,j}^2 \Delta x \Delta y$ and discrete energy $\frac{1}{2}\sum_{i,j} \int (\bm u_{ij})^2\Delta x \Delta y$. The unmodified MUSCL schemes in the first and second columns of \cref{fig:euler_data} decay energy and enstrophy. 
The modified MUSCL schemes in the third and fourth columns of \cref{fig:euler_data} monotonically increase the discrete energy.
The energy-conserving schemes in the fifth and sixth columns of \cref{fig:euler_data} conserve the discrete energy.

\subsubsection{Enstrophy and coarse graining \label{sec:coarsegraining}}

\noindent The incompressible Euler equations in vorticity form exactly conserve the L2 norm of the solution, called the enstrophy. Naively, we might expect that a good PDE solver would conserve enstrophy as well. Yet algorithms that exactly conserve enstrophy on \cref{eq:euler}, such as the centered flux (not shown) or the modified MUSCL scheme with $\nicefrac{d\ell_2^{\textnormal{new}}}{dt} = 0$ (shown in \cref{fig:euler_2d}), perform significantly worse than schemes that allow enstrophy to decay.

To understand this puzzling result, we consider the relationship between the continuous enstrophy $\int \chi^2 \mathop{dx}\mathop{dy}$ and the discrete enstrophy $\sum_{i,j} \chi_{i,j}^2 \Delta x \Delta y$. Suppose that $\chi^{\textnormal{exact}} = \chi(x,y,t)$ is the exact solution to \cref{eq:euler}. As we know, $\chi^{\textnormal{exact}}$ has constant enstrophy. Now suppose that we coarse grain $\chi^\textnormal{exact}$, such that
$\chi^\textnormal{exact}_{i,j} = \int_{i,j} \chi^\textnormal{exact} \mathop{dx} \mathop{dy}$.
It turns out that, with very high probability, the discrete enstrophy of $\chi^{\textnormal{exact}}_{i,j}$ will decay in time \cite{basictypesofcoarsegraining}. This happens because $\chi^{\textnormal{exact}}$ tends to develop structures on a scale smaller than the grid size. These structures cannot be represented by $\chi_{i,j}$ and are replaced via coarse graining by a low-dimensional representation of the solution with lower enstrophy. 

Although the continuous equations for $\chi^\textnormal{exact}$ conserve enstrophy, the discrete equations for $\chi^\textnormal{exact}_{i,j}$ decay enstrophy.
Because machine learned PDE solvers solve discrete equations that are designed to approximate $\chi^{\textnormal{exact}}_{i,j}$, then machine learned PDE solvers should preserve the invariants of the discrete equations for $\chi^{\textnormal{exact}}_{i,j}$, not the invariants of the continuous equations for $\chi^{\textnormal{exact}}$.
As a result, solvers of \cref{eq:euler} should guarantee that enstrophy is non-increasing even though the continuous equations conserve enstrophy.

\section{Invariant-preserving algorithms for systems of hyperbolic PDEs}
\label{sec:systemshyperbolic}

\noindent Unlike scalar hyperbolic PDEs, systems of hyperbolic PDEs are not guaranteed to all have the same non-linear invariants. While it is possible to design invariant-preserving algorithms for the generic scalar hyperbolic PDE \cref{eq:scalarhyperbolic}, invariant-preserving algorithms for systems of hyperbolic PDEs need to be tailored to the specific equation.

In this section, we design invariant-preserving error-correcting algorithms for an important example system: the compressible Euler equations of gas dynamics. We consider continuous-time, flux-predicting FV update rules in 1D. To produce invariant-preserving machine learned solvers, the procedure is the same as in \cref{sec:scalarhyperbolic}: at each timestep, if the invariants are not preserved apply an error-correcting algorithm to the update rule.

\subsection{Compressible Euler equations\label{sec:compressible_euler}}

\noindent The 1D compressible Euler equations are given by \cref{eq:compressibleeuler}. They are in the form \cref{eq:system_conservation_form}. Recall from \cref{sec:standardsolvers} that some equations in the form \cref{eq:system_conservation_form} satisfy an entropy inequality \cref{eq:system_entropy_inequality} and that the compressible Euler equations satisfy an entropy inequality with generalized entropy function $\eta(s) = \rho g(s)$ with specific entropy $s = \log(\nicefrac{p}{\rho^\gamma})$ for which $\nicefrac{g''}{g'}\le \gamma^{-1}$. Recall that the choice $g(s) = e^{\nicefrac{s}{\gamma+1}}$ has the entropy variable $\bm w = (\nicefrac{\partial \eta}{\partial \bm u})^T$ given by \cref{eq:compressibleeuler_entropy_variable}.

Suppose we want to solve \cref{eq:compressibleeuler} for $\bm u = \begin{bmatrix}
    \rho, \rho v, E
\end{bmatrix}$
on a 1D domain with $x \in [0, L]$. We consider non-periodic BCs, though our results can easily be extended to periodic BCs. We use a FV discretization and divide the domain into $N$ cells of width $\Delta x = \nicefrac{L}{N}$ where the left and right boundaries of the $j$th cell for $j = 1, \dots, N$ are $x_{j-\nicefrac{1}{2}} = (j-1)\Delta x$ and $x_{j+\nicefrac{1}{2}}=j\Delta x$ respectively. We use a vector $\bm u_j = \begin{bmatrix}
        \rho_j,
        (\rho v)_j,
        E_j
    \end{bmatrix}$ to represent the solution average within each cell where $\bm u_j(t) \coloneqq \int_{x_{j-\nicefrac{1}{2}}}^{x_{j+\nicefrac{1}{2}}} \bm u(x, t)  \mathop{dx}$. We apply $\int_{x_{j-\nicefrac{1}{2}}}^{x_{j+\nicefrac{1}{2}}} (\dots) \mathop{dx}$ to \cref{eq:compressibleeuler} to derive the continuous-time FV update equation:
\begin{equation}\label{eq:compressible_euler_update}
    \frac{d \bm u_j}{d t} + \frac{1}{\Delta x} \bigg( \bm F_{j+\frac{1}{2}} - \bm F_{j-\frac{1}{2}}\bigg) = 0, \hspace{0.5cm} \bm F_{j+\frac{1}{2}} =  
    \begin{bmatrix}
        (\rho v)_{j+\frac{1}{2}} \\
        (\rho v^2 + p)_{j+\frac{1}{2}} \\
        (v(E + p))_{j+\frac{1}{2}}
    \end{bmatrix}.
\end{equation}
A machine learned solver would output the flux $\bm F_{j+\nicefrac{1}{2}}$ at $N-1$ cell boundaries for $j = 1, \dots, N-1$. $\bm F_{\nicefrac{1}{2}}$ and $\bm F_{N+\nicefrac{1}{2}}$ can be computed either from the boundary conditions or by using ML.

\subsubsection{Continuous invariants}
As we saw in \cref{sec:standard_systems}, with non-periodic BCs the rate of change of the discrete mass is equal to the negative flux through the domain boundaries: $\frac{d}{dt}\int_{0}^L \bm u \mathop{dx} = \bm F(0) - \bm F(L)$. We also saw that the rate of change of the total entropy is greater than or equal to the entropy flux through the domain boundaries: $\frac{d}{dt} \int_{0}^L \eta(\bm u) \mathop{dx} \ge \psi(0) - \psi(L)$. The compressible Euler equations also maintain the positivity invariants $\rho \ge 0$ and $p = (\gamma - 1)(E - \frac{1}{2}\rho v^2) \ge 0$.

\subsubsection{Discrete invariants}
With FV solvers, the rate of change of $\sum_{j=1}^N \bm u_j(t) \Delta x$ is equal to the negative flux through the boundaries $\bm F_{\nicefrac{1}{2}} - \bm F_{N+\nicefrac{1}{2}}$. Thus, the discrete analogue of the linear invariant $\int_\Omega u \mathop{dx}$ is preserved. We now show how to modify the predicted flux to ensure that the other two invariants are preserved. We want the discrete positivity invariants $\rho_j \ge 0$ and $p_j \ge 0$ to be maintained for all $j \in 1, \dots, N$. We also want the rate of change of a discrete analogue of the entropy to be greater than or equal to the entropy flux through the domain boundaries.

To ensure that the positivity invariants are preserved, we first limit $\bm F_{j+\nicefrac{1}{2}}$. One possible limiter introduced in \cite{Hu_2013} transforms $\bm F_{j+\nicefrac{1}{2}}$ for $j = 0, \dots, N$ according to the update
\begin{equation}\label{eq:compressible_euler_positivity_transformation}
    \bm F_{j+\nicefrac{1}{2}} \Rightarrow \theta_{j+\nicefrac{1}{2}} \bm F_{j+\nicefrac{1}{2}} + (1-\theta_{j+\nicefrac{1}{2}}) \bm F^{LF}_{j+\nicefrac{1}{2}}
\end{equation}
where $\bm F^{LF}_{j+\nicefrac{1}{2}}$ is the first-order Lax-Friedrichs flux and $0 \leq \theta_{j+\nicefrac{1}{2}} \leq 1$ is chosen to ensure positivity of $\rho_j$ and $p_j$. For details of how $\theta_{j+\nicefrac{1}{2}}$ is chosen, see \cite{Hu_2013}.

We now show how to further modify $\bm F_{j+\nicefrac{1}{2}}$ to ensure that the discrete entropy is greater than or equal to the entropy flux through the domain boundaries.
The discrete cell entropy $\eta_j = \rho_j g_j(s_j)$. We can choose $g_j(s_j) = e^{\nicefrac{s_j}{\gamma + 1}}$ with $s_j = \log{\nicefrac{p_j}{\rho_j^\gamma}}$ which results in the discrete entropy variable $\bm w_j = \frac{p^*_j}{p_j} \begin{bmatrix}
    E_j, -(\rho v)_j, \rho_j
\end{bmatrix}$ where $p^*_j = \frac{\gamma - 1}{\gamma + 1} (\nicefrac{p_j}{\rho_j^\gamma})^{\nicefrac{1}{\gamma + 1}}$ and $p_j = (\gamma - 1) (E_j - \nicefrac{(\rho v)_j^2}{2\rho_j} )$. Using $\nicefrac{d \eta_j}{d \bm u_j} = \bm w_j^T$, the rate of change of the discrete entropy is
\begin{equation}
    \frac{d}{dt}\sum_{j=1}^N \eta_j(\bm u_j) \Delta x = \sum_{j=1}^N \bm w_j^T \frac{d\bm u_j}{dt} \Delta x = -\sum_{j=1}^N \bm w_j^T (\bm F_{j+\frac{1}{2}} - \bm F_{j - \frac{1}{2}}).
\end{equation}
Using summation by parts, the rate of change of the discrete entropy is
\begin{equation}
        \frac{d}{dt} \sum_{j=1}^N \eta_j \Delta x = \bm F_{\frac{1}{2}}^T \bm w_1 - \bm F_{N+\frac{1}{2}}^T \bm w_N + \sum_{j=1}^{N-1} \bm F_{j+\frac{1}{2}}^T (\bm w_{j+1} - \bm w_j).
\end{equation}
To ensure that rate of change of the discrete entropy is greater than or equal to the entropy flux through the domain boundaries, we want
\begin{equation}\label{eq:compressibleeuler_entropyincreasing}
    \bm F_{\frac{1}{2}}^T \bm w_1 - \bm F_{N+\frac{1}{2}}^T \bm w_N + \sum_{j=1}^{N-1} \bm F_{j+\nicefrac{1}{2}}^T (\bm w_{j+1} - \bm w_j) \ge \psi(0) - \psi(L).
\end{equation}
In open systems, the entropy flux through the domain boundaries $\psi(0) - \psi(L)$ may or may not be known a priori. If it is not known a priori, we can make an estimate of the entropy flux using the boundary conditions.

Next, we transform $\bm F_{j+\nicefrac{1}{2}}$ to ensure that \cref{eq:compressibleeuler_entropyincreasing} is satisfied. We define scalars $\nicefrac{d\eta^{\textnormal{old}}}{dt} = \bm F_{\frac{1}{2}}^T \bm w_1 - \bm F_{N+\frac{1}{2}}^T \bm w_N + \sum_{j=1}^{N-1} \bm F_{j+\nicefrac{1}{2}}^T (\bm w_{j+1} - \bm w_j)$ and $\nicefrac{d\eta^{\textnormal{new}}}{dt} \ge \psi(0) - \psi(L)$. We then transform $\bm F_{j+\nicefrac{1}{2}}$ for $j = 1, \dots, N-1$ as follows:
\begin{equation}\label{eq:compressible_euler_entropy_transformation}
    \bm F_{j+\frac{1}{2}} \Rightarrow \bm F_{j+\frac{1}{2}} + \frac{ (\nicefrac{d\eta^{\textnormal{new}}}{dt} - \nicefrac{d\eta^{\textnormal{old}}}{dt})\bm G_{j+\nicefrac{1}{2}}}{\sum_{k=1}^{N-1} \bm G_{k+\nicefrac{1}{2}}^T(\bm w_{k+1} - \bm w_k)}
\end{equation}
for any finite non-constant function $\bm G_{j+\nicefrac{1}{2}}(\bm u)$ for which $\sum_{k=1}^{N-1} \bm G_{k+\nicefrac{1}{2}}^T(\bm w_{k+1} - \bm w_k) \ne 0$ and for which the addition of $\bm G_{j+\nicefrac{1}{2}}$ does not violate the positivity invariants.
The choice $\bm G_{j+\nicefrac{1}{2}}(\bm u) = \bm w_{j+1} - \bm w_j$ ensures that $\sum_{k=1}^{N-1} \bm G_{k+\nicefrac{1}{2}}^T(\bm w_{k+1} - \bm w_k) \ne 0$ so long as the discrete solution varies in space, but does not guarantee positivity of $\rho_j$ and $p_j$.
If we instead make the choice $\bm G_{j+\nicefrac{1}{2}}(\bm u) = \begin{bmatrix} 0, v_{j+1} -  v_j, p_{j+1} - p_j \end{bmatrix}^T$ corresponding to the physical diffusion terms $\nabla^2 u$ and $\nabla^2 p$ in the Navier-Stokes momentum and energy equations, we can guarantee positivity of $\rho_j$ and $p_j$ so long as $\nicefrac{d\eta^{\textnormal{new}}}{dt} > \nicefrac{d\eta^{\textnormal{old}}}{dt}$ (otherwise $\bm G_j$ is adding anti-diffusion) and a timestep restriction is satisfied. In our experiments, we empirically find that this choice also satisfies $\sum_{k=1}^{N-1} \bm G_{k+\nicefrac{1}{2}}^T(\bm w_{k+1} - \bm w_k) \ne 0$.

We now illustrate the effects of applying \cref{eq:compressible_euler_entropy_transformation} to an entropy-increasing, positivity-preserving numerical method. We use the Sod shock tube setup \cite{sod1978survey} with open (Dirichlet) BCs.
We apply the entropy-modifying algorithm \cref{eq:compressible_euler_entropy_transformation} to the MUSCL scheme \cite{muscl} with reconstruction in characteristic variables \cite{miyoshi2020short}.
We estimate the entropy flux through the boundaries ($\psi(0) - \psi(L)$) using the BCs and the formula $\psi = \rho g(s) v$. 
The MUSCL scheme is an entropy-increasing scheme, implying that $\nicefrac{\partial \eta^{\textnormal{old}}}{dt} > \psi(0) - \psi(L)$.
The entropy-modifying algorithm updates the rate of change of the entropy from $\nicefrac{\partial \eta^{\textnormal{old}}}{dt}$ to $\nicefrac{\partial \eta^{\textnormal{new}}}{dt}$, where $\nicefrac{\partial \eta^{\textnormal{new}}}{dt} = (\psi(0) - \psi(L)) + R (\nicefrac{\partial \eta^{\textnormal{old}}}{dt} - (\psi(0) - \psi(L)))$. We set $\bm G_{j+\nicefrac{1}{2}}(\bm u) = \begin{bmatrix} 0, v_{j+1} -  v_j, p_{j+1} - p_j \end{bmatrix}^T$.

\begin{figure}
    \centering
    \begin{subfigure}[b]{0.49\textwidth}
        \centering
        \includegraphics[width=0.9\textwidth]{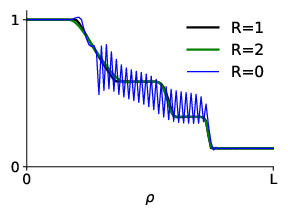}
        \caption{}
        \label{fig:3a}
    \end{subfigure}
    \begin{subfigure}[b]{0.49\textwidth}
         \centering
         \includegraphics[width=0.9\textwidth]{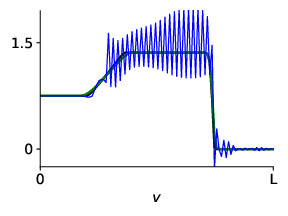}
         \caption{}
         \label{fig:3b}
     \end{subfigure}
    \begin{subfigure}[b]{0.49
    \textwidth}
         \centering
         \includegraphics[width=0.9\textwidth]{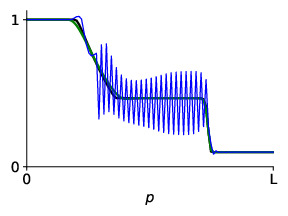}
         \caption{}
         \label{fig:3c}
     \end{subfigure}
    \caption{By adding diffusion and limiters to machine learned solvers using \cref{eq:compressible_euler_entropy_transformation,eq:compressible_euler_positivity_transformation} with $\nicefrac{\partial \eta^{\textnormal{new}}}{dt} > \psi(0) - \psi(L)$, we can design entropy-preserving and positivity-preserving solvers for the 1D compressible Euler equations. The discrete density $\rho$ in (a), velocity $v$ in (b), and pressure $p$ in (c) for three numerical methods for the 1D compressible Euler equations using a Sod shock tube setup with Dirichlet BCs. We modify the flux $\bm F_{j+\nicefrac{1}{2}}$ at cell boundaries of a standard MUSCL scheme using \cref{eq:compressible_euler_entropy_transformation}. We set $\nicefrac{\partial \eta^{\textnormal{new}}}{dt} = (\psi(0) - \psi(L)) + R (\nicefrac{\partial \eta^{\textnormal{old}}}{dt} - (\psi(0) - \psi(L)))$. The original MUSCL scheme ($R=1$) is shown in black. Increasing the rate of change of the discrete entropy ($R=2$), shown in green, adds diffusion to the original scheme. Decreasing the rate of change of the discrete entropy ($R=0$), shown in blue, adds anti-diffusion to the original scheme.}
        \label{fig:compressible_euler_demo}
\end{figure}

In \cref{fig:compressible_euler_demo} we plot the density $\rho$, velocity $v$, and pressure $p$ for three values of $R$. The original scheme ($R=1$) is plotted in black in \cref{fig:compressible_euler_demo}. Plotted in green is a scheme which increases entropy at a faster rate than the original scheme ($R=2$). Plotted in blue $(R=0)$ is a scheme with an entropy that changes only due to the entropy flux through the domain boundaries.
Notice that the scheme in blue adds anti-diffusion and is not guaranteed to preserve positivity, as $\nicefrac{d\eta^{\textnormal{new}}}{dt} < \nicefrac{d\eta^{\textnormal{old}}}{dt}$.
We do not show the result of setting $\nicefrac{d\eta^{\textnormal{new}}}{dt} < \nicefrac{d\eta^{\textnormal{BC}}}{dt}$ as the spurious oscillations result in $\rho_j < 0$ leading to NaNs in the reconstruction of the characteristic variables.

In summary: applying the transformation \cref{eq:compressible_euler_positivity_transformation} followed by the transformation \cref{eq:compressible_euler_entropy_transformation} with  $\nicefrac{d\eta^{\textnormal{new}}}{dt} \ge \psi(0) - \psi(L)$ to $\bm F_{j+\nicefrac{1}{2}}$ ensures that the discrete invariant $\sum_{j=1}^N \bm u_j(t) \Delta x$ is conserved, the positivity invariants $\rho_j\ge 0$ and $p_j \ge 0$ are maintained, and the discrete entropy $\sum_{j=1}^N \eta_j \Delta x$ is greater than or equal to the entropy flux through the domain boundaries.
These error-correcting transformations give us a procedure for designing invariant-preserving machine learned solvers. At each timestep or at each stage of an Runge-Kutta ODE integration, use \cref{eq:compressible_euler_positivity_transformation} to modify $\bm F_{j+\nicefrac{1}{2}}$ to ensure positivity of $\rho_j$ and $p_j$. If the resulting $\nicefrac{d\eta^{\textnormal{old}}}{dt} < \psi(0) - \psi(L)$, use \cref{eq:compressible_euler_entropy_transformation} with $\bm G_{j+\nicefrac{1}{2}}(\bm u) = \begin{bmatrix} 0, v_{j+1} -  v_j, p_{j+1} - p_j \end{bmatrix}^T$ to set $\nicefrac{d\eta^{\textnormal{new}}}{dt} = \psi(0) - \psi(L)$.
Because this transformation adds physically motivated diffusion terms to a positivity-preserving scheme, in the continuous-time limit the resulting machine learned solvers maintain positivity while ensuring that the discrete entropy is non-decreasing.

\subsection{Vlasov-Maxwell}

We now consider a system of hyperbolic PDEs, the Vlasov-Maxwell system of equations:
\begin{subequations}
\begin{align}
    \frac{\partial f_s}{\partial t} &+ \bm \nabla_{\bm z} \cdot (\bm \alpha_s f_s) = 0 \label{eq:vm_dfdt}\\
    \mu_0 \epsilon_0 \frac{\partial \bm E}{\partial t} &= \bm \nabla_{\bm x} \times \bm B - \mu_0 \bm J \label{eq:vm_dEdt}\\
    \frac{\partial \bm B}{\partial t} &= -\bm \nabla_{\bm x} \times \bm E \label{eq:vm_dBdt}
\end{align}
\end{subequations}
where $t$ is time, $\bm x$ is a 3D spatial coordinate, $\bm v$ is a 3D velocity coordinate, $\bm z = \{\bm x, \bm v\}$ is a 6D phase space coordinate, $f_s = f_s(\bm z, t)$ is the particle distribution function for species of particle $s$ with charge $q_s$ and mass $m_s$, $\bm E = \bm E(\bm x, t)$ is the electric field, $\bm B = \bm B(\bm x, t)$ is the magnetic field, $\bm J = \bm J(\bm x, t) = \sum_s q_s\int_{-\infty}^{\infty} f_s \bm v \mathop{d\bm v}$ is the current density from all species, and $\bm \alpha_s = \{\bm v, \frac{q_s}{m_s}( \bm E + \bm v \times \bm B)\}$ is the phase space velocity for species $s$. The initial conditions for $\bm E$ and $\bm B$ are given by Gauss's law
\begin{subequations}
\begin{align}
    \bm \nabla_{\bm x} \cdot \bm E &= \epsilon_0 \rho \\
    \bm \nabla_{\bm x} \cdot \bm B &= 0
\end{align}
\end{subequations}
where $\rho(\bm x, t) = \sum_s q_s \int_{-\infty}^{\infty} f_s d\bm v$ is the charge density. The initial condition for $f_s$ is that $f_s \ge 0$ everywhere in phase space and $f_s$ decays towards $0$ as $|\bm v| \rightarrow \infty$ faster than $|\bm v|^w, \forall w \in \mathbb{N}$.
The Vlasov-Maxwell system is system of equations in 6 dimensions plus time for $6 +\sum_s(1)$ variables which describes the self-consistent interaction between electromagnetic fields and multiple species of charged particles in the continuum limit. \Cref{eq:vm_dfdt} is a collisionless model of a plasma.

Our objective is to demonstrate how to design a machine learned solver for the Vlasov-Maxwell system which preserves the same invariants and stability properties as the underlying system of equations.

We consider a FV scheme for the Vlasov-Maxwell system of equations. We consider a spatially periodic domain $\Omega$ in physical space $\bm{{x}}$ and an infinitely large domain $\mathcal{V}$ in velocity space $\bm{{v}}$. Our 6-dimensional phase space domain is $\mathcal{K} = \Omega \cup \mathcal{V}$. To partition $\mathcal{K}$ into a cartesian mesh, we begin by choosing some maximum velocity $v_{\textnormal{max}}$ where we know a priori that $f_s \approx 0$ and restricting to the domain $\mathcal{K}'$ where $\mathcal{K}' \in \mathcal{K}$ and $|v_i| < v_{\textnormal{max}} \forall v_i \in \mathcal{K}'$ for $i=1,2,3$. We then divide $\mathcal{K}'$ into uniform discrete rectangular cells $\mathcal{K}_{jk}$ of volume $|\Delta \bm z|=|\Delta \bm x\Delta \bm v|$. The index `$j$' defines a cell's index in real space, while the index `$k$' refers to a cell's index in velocity space. The restriction of cell $\mathcal{K}_{jk}$ to physical space is cell $\Omega_j$. The boundary of $\mathcal{K}_{jk}$ is $\partial \mathcal{K}_{jk}$, while the boundary of $\Omega_j$ is $\partial \Omega_j$.

The discrete representation for each of the $6 + \sum_s(1)$ variables is a piecewise constant function within each cell. We denote the cell average with the subscript $h$. Thus, $f_{sh}$ is the discrete representation of $f_s$ while $\bm{E}_h$ and $\bm{B}_h$ are the discrete representations of $\bm E$ and $\bm B$.

Integrating \cref{eq:vm_dfdt,eq:vm_dEdt,eq:vm_dBdt} over each cell and using the divergence and Stokes' theorems gives the FV update equations. For phase space cell $\mathcal{K}_{jk}$ and real space cell $\Omega_j$, these equations are
\begin{subequations}
\begin{align}
    \int_{\mathcal{K}_{jk}} \frac{\partial f_{sh}}{\partial t} \mathop{d\bm z} + \oint_{\partial \mathcal{K}_{jk}} {\bm{\hat{F}}}_s\cdot \mathop{d\bm s} &= 0 \\
    \int_{\Omega_j} \frac{\partial \bm{B}_h}{\partial t} \mathop{d\bm x} - \oint_{\partial \Omega_j} \bm{\hat{E}}_h \times \mathop{d\bm{s}} &= 0 \\
    \mu_0\epsilon_0 \int_{\Omega_j} \frac{\partial \bm{E}_h}{\partial t} \mathop{d\bm{x}} + \oint_{\partial \Omega_j}\bm{\hat{B}}_h{\times}\mathop{d \bm s} &= -\mu_0 \int_{\Omega_j}\bm{J}_h \mathop{d\bm x}.
\end{align}
\end{subequations}
$\bm{\hat{F}}_s$ is a reconstruction of the flux $\bm{\alpha}_s f_s$ at cell boundary $\partial \mathcal{K}_{jk}$. $d\bm s$ is the outward normal vector at each cell boundary. $\bm{\hat{E}}_h$ and $\bm{\hat{B}}_h$ are reconstructions of $\bm E$ and $\bm B$ at cell boundary $\partial \Omega_j$. 

\textbf{Conservation properties}: The Vlasov-Maxwell system of equations conserves the following invariants. For proofs of these conservation properties, see \cite{juno2018discontinuous}.
\begin{itemize}
    \item Particles: $\frac{d}{dt}\int_\mathcal{K} f_s \mathop{d \bm z}=0$
    \item $\ell_2$-norm of the distribution function: $\frac{d}{dt}\int_\mathcal{K} f_s^2 \mathop{d\bm z}=0$
    \item Entropy: $\frac{d}{dt} \int_\mathcal{K} -f_s \log{f_s} \mathop{d\bm z} = 0$
    \item Momentum: $\frac{d}{dt}\big(\sum_s \int_\mathcal{K} f_s \bm v \mathop{d\bm z} + \int_{\Omega} \epsilon_0 \bm{E}{\times}\bm{{B}} \mathop{d\bm x} \big)=0$
    \item Energy: $\frac{d}{dt}\big( \sum_s \frac{1}{2} \int_\mathcal{K} m_s |\bm v|^2 f_s \mathop{d\bm z} + \int_\Omega \frac{\epsilon_0}{2} |\bm E|^2 + \frac{1}{2\mu_0}|\bm B|^2 \mathop{d\bm x}\big) = 0$
\end{itemize}
The Vlasov-Maxwell system also preserves non-negativity of the distribution function, $f_s\ge0$.

\textbf{Invariant-Preserving Flux-Predicting Schemes:} We want to show how to design a flux-predicting FV scheme that preserves some combination of the above invariants. 

\textbf{Conservation of particles}: FV schemes conserve a discrete analogue $\sum_j \sum_k \int_{\mathcal{K}_{jk}} f_{sh} \mathop{d \bm z}$ of the continuous invariant $\int_\mathcal{K} f_s \mathop{d\bm z}$. \textbf{Proof}:
$\frac{d}{dt}\sum_j \sum_k \int_{\mathcal{K}_{jk}} f_{sh} \mathop{d\bm z} = 
\sum_j \sum_k \int_{\mathcal{K}_{jk}} \frac{\partial f_{sh}}{\partial t} \mathop{d\bm z} = -\sum_j \sum_k \oint_{\partial \mathcal{K}_{jk}} {\bm{\hat{F}}}_s\cdot \mathop{d\bm s}
 = 0$
because $\bm{\hat{F}}_s\cdot d\bm s$ is equal and opposite at adjacent cell boundaries. Here we have relied on spatially periodic boundary conditions and the assumption that $f_{sh}$ and $\bm{\hat{F}}_s$ are zero at the boundaries of velocity space.

\textbf{Energy stability}: Although the electromagnetic energy of the continuous equations is conserved, to prevent spurious oscillations of $\bm E_h$ and $\bm B_h$ and to ensure stability of the electromagnetic fields we demand that the electromagnetic energy by non-increasing. In particular, we demand that
\begin{equation}\label{eq:vm_particle_energy}
    \frac{d}{dt} \sum_j \sum_k \int_{\mathcal{K}_{jk}} \frac{1}{2} m_s |\bm v|^2 f_{sh} \mathop{d\bm{z}} = \sum_{j} \int_{\Omega_j} \bm{E}_h \cdot \bm{J}_{sh} \mathop{d\bm{x}}
\end{equation}
where $\bm J_{sh}$ is the current due to species $s$ and
\begin{equation}\label{eq:vm_electromagnetic_energy}
\frac{d}{dt} \sum_j \int_{\Omega_j} \frac{\epsilon_0}{2} |\bm E_h|^2 + \frac{1}{2\mu_0}|\bm B_h|^2 \mathop{d\bm x} \le -\sum_j \int_{\Omega_j} \bm{E}_h \cdot \bm{J}_{h} \mathop{d\bm{x}}
\end{equation}
so that the total energy is non-increasing:
\begin{equation}\label{eq:vm_total_energy_nonincreasing}
    \frac{d}{dt} \sum_s \sum_j \sum_k \int_{\mathcal{K}_{jk}} \frac{1}{2} m_s |\bm v|^2 f_{sh} \mathop{d\bm{z}} + \frac{d}{dt} \sum_j \int_{\Omega_j} \frac{\epsilon_0}{2} |\bm E_h|^2 + \frac{1}{2\mu_0}|\bm B_h|^2 \mathop{d\bm x} \le 0.
\end{equation}
To ensure that \cref{eq:vm_electromagnetic_energy,eq:vm_particle_energy,eq:vm_total_energy_nonincreasing} are satisfied, we will modify the reconstructed values of $\bm{\hat{F}}_s$, $\bm{\hat{E}}_h$, and $\bm{\hat{B}}_h$. To start, we expand the LHS of \cref{eq:vm_particle_energy}:
\begin{equation}
    \sum_j \sum_k \int_{\mathcal{K}_{jk}} \frac{1}{2} m_s |\bm v|^2 \frac{\partial f_{sh}}{\partial t} \mathop{d\bm{z}} = -\sum_j \sum_k \frac{m_s}{2}\frac{\int_{\mathcal{K}_{jk}} |\bm v|^2 \mathop{d\bm z }}{\int_{\mathcal{K}_{jk}} \mathop{d\bm z}} \oint_{\partial \mathcal{K}_{jk}} {\bm{\hat{F}}}_s\cdot \mathop{d\bm s}.
\end{equation}
For notational simplicity, we define $\langle |\bm v|^2 \rangle \coloneqq \nicefrac{\int_{\mathcal{K}_{jk}} |\bm v|^2 \mathop{d\bm z }}{\int_{\mathcal{K}_{jk}}\mathop{d\bm z}}$ and use summation by parts.
This turns the sum over cell boundaries $\sum_j \sum_k \oint_{\partial \mathcal{K}_{jk}}$ into a sum over boundary segments $\sum_{b \in \mathcal{S}_x \cup \mathcal{S}_v} \int_{\partial \mathcal{B}_b}$ where $\mathcal{S}_x$ is a set of all boundary segments with a real-space directed normal vector and $\mathcal{S}_v$ is a set of all boundary segments with a velocity-space directed normal vector. For each boundary segment $\partial \mathcal{B}_b$, we choose a positively-directed normal vector $\bm{n}_b$ which defines a positive ($+$) and negative ($-$) orientation relative to the segment. 
Summation by parts thus gives
\begin{equation}\label{eq:vm_energy_intermediate}
    \frac{d}{dt}\sum_j \sum_k \int_{\mathcal{K}_{jk}} \frac{1}{2} m_s |\bm v|^2 f_{sh} \mathop{d\bm{z}} = \frac{m_s}{2} \sum_{b\in \mathcal{S}_v} \int_{\mathcal{B}_b} (\langle |\bm v|^2 \rangle^+ - \langle |\bm v|^2 \rangle^-) \bm{\hat{F}}_s \cdot d\bm{n}_b
\end{equation}
where we have used the fact that $\langle |\bm v|^2 \rangle^+ = \langle |\bm v|^2 \rangle^-$ for all $b \in \mathcal{S}_x$. 

From \cref{eq:vm_particle_energy}, we want the rate of change of the particle energy to equal $\sum_j \int_{\Omega_j} \bm{E}_h \cdot \bm{J}_{sh} \mathop{d\bm{x}}$. To ensure that \cref{eq:vm_particle_energy} is satisfied, we can make the following transformation to $\bm{\hat{F}}_s \cdot d{\bm{n}}_b$ for each $b \in \mathcal{S}_v$. We define $P_{\textnormal{new}} \coloneqq \frac{2}{m_s}\int_{\Omega_j} \bm E_h \cdot \bm J_{sh} \mathop{d\bm x}$ where $\Omega_j$ has the same spatial coordinate as cell boundary $\mathcal{B}_b$. We define $P_{\textnormal{old}} \coloneqq \sum_{b\in \mathcal{S}_{jv}} \int_{\mathcal{B}_b} (\langle |\bm v|^2 \rangle^+ - \langle |\bm v|^2 \rangle^-) \bm{\hat{F}}_s \cdot d\bm{n}_b$ where $\mathcal{S}_{jv}$ is the set of all velocity-space directed segments at spatial index $j$. We then transform each integral in \cref{eq:vm_energy_intermediate} according to
\begin{equation}\label{eq:vm_mod_f}
    \int_{\mathcal{B}_b}\bm{\hat{F}}_s \cdot d\bm{n}_b \rightarrow \int_{\mathcal{B}_b} \bm{\hat{F}}_s \cdot d\bm{n}_b + \frac{(P_{\textnormal{new}}-P_{\textnormal{old}}) G_{jb}}{ \sum_{b' \in \mathcal{S}_{jv}} (\langle |\bm v|^2 \rangle^+ - \langle |\bm v|^2 \rangle^-)G_{jb'}}.
\end{equation}
$G_{jb}$ is a non-constant finite function for which $\sum_{b' \in \mathcal{S}_{jv}} (\langle |\bm v|^2 \rangle^+ - \langle |\bm v|^2 \rangle^-)G_{jb'} \ne 0$. $G_{jb}$ is a hyperparameter which determines the distribution of velocity-directed flux of $f_s$ at spatial index $j$.
\Cref{eq:vm_mod_f}, at each spatial index $j$, modifies the velocity-space directed fluxes to ensure that the local rate of change of particle energy is equal to the expected rate of change due to electromagnetic work.

Next, we expand the LHS of \cref{eq:vm_electromagnetic_energy}:
\begin{subequations}
\begin{align}
&\frac{d}{dt} \sum_j \int_{\Omega_j} \frac{\epsilon_0}{2} |\bm E_h|^2 + \frac{1}{2\mu_0}|\bm B_h|^2 \mathop{d\bm x} = \sum_j \int_{\Omega_j} {\epsilon_0} \bm E_h \cdot \frac{\partial \bm E_h}{\partial t} + \frac{1}{\mu_0}\bm B_h\cdot \frac{\partial \bm B_h}{\partial t} \mathop{d\bm x}\\
&=
\sum_j \frac{1}{\mu_0} \oint_{\partial \Omega_j}\Big(-\bm E_h \cdot (\bm{\hat{B}}_h{\times}\mathop{d \bm s})  +   \bm B_h\cdot (\bm{\hat{E}}_h \times \mathop{d\bm{s}})\Big) - \sum_j\int_{\Omega_j}\bm E_h \cdot \bm{J}_h \mathop{d\bm x}\\
&= -\sum_j \frac{1}{\mu_0} \oint_{\partial \Omega_j} (\bm E_h \times \bm{\hat{B}}_h + \bm{\hat{E}}_h \times \bm{B}_h) \cdot d\bm s - \sum_j \int_{\Omega_j}\bm E_h \cdot \bm{J}_h \mathop{d\bm x}.
\end{align}
\end{subequations}
Thus, for \cref{eq:vm_electromagnetic_energy} to be satisfied, we want 
\begin{equation}\label{eq:vm_em_intermediate}
    \sum_j \oint_{\partial \Omega_j} (\bm E_h \times \bm{\hat{B}}_h + \bm{\hat{E}}_h \times \bm{B}_h) \cdot d\bm s \ge 0.
\end{equation}
Performing summation by parts on \cref{eq:vm_em_intermediate} gives
\begin{equation}
    \sum_{b \in \mathcal{S}_x} \int_{\partial \mathcal{B}_b} \Big[ \bm{\hat{B}}_h \times (\bm{E}_h^+ - \bm{E}_h^-) - \bm{\hat{E}}_h \times (\bm{B}_h^+ - \bm{B}_h^-) \Big] \cdot d\bm n_b \ge 0
\end{equation}

\begin{equation}\label{eq:vm_em_inequality}
    \sum_{b \in \mathcal{S}_x} \int_{\partial \mathcal{B}_b} \big((\bm{E}_h^+ - \bm{E}_h^-) \times d\bm{n}_b\big) \cdot \bm{\hat{B}}_h - \big((\bm{B}_h^+ - \bm{B}_h^-) \times d\bm{n}_b \big) \cdot \bm{\hat{E}}_h  \ge 0
\end{equation}
We then define $Q^{\textnormal{old}} = \sum_{b \in \mathcal{S}_x} \int_{\partial \mathcal{B}_b} \big((\bm{E}_h^+ - \bm{E}_h^-) \times d\bm{n}_b\big) \cdot \bm{\hat{B}}_h - \big((\bm{B}_h^+ - \bm{B}_h^-) \times d\bm{n}_b \big) \cdot \bm{\hat{E}}_h$. Then \cref{eq:vm_em_inequality} will be satisfied if the following transformations are made to $\bm{\hat{E}}_h$ and $\bm{\hat{B}}_h$:
\begin{equation}\label{eq:vm_mod_B}
\begin{split}
    &\sum_{b \in \mathcal{S}_x} \int_{\partial \mathcal{B}_b} \big((\bm{E}_h^+ - \bm{E}_h^-) \times d\bm{n}_b\big) \cdot \bm{\hat{B}}_h \rightarrow \\ &\sum_{b \in \mathcal{S}_x} \int_{\partial \mathcal{B}_b} \big((\bm{E}_h^+ - \bm{E}_h^-) \times d\bm{n}_b\big) \cdot \bm{\hat{B}}_h + \frac{(Q^{\textnormal{new}}_B - Q_B^{\textnormal{old}}) G_b^B}{\sum_{b\in \mathcal{S}_x} G_{b}^B |\int_{\partial \mathcal{B}_b}(\bm{E}_h^+ - \bm{E}_h^-) \times d\bm{n}_b|}
\end{split}
\end{equation}
\begin{equation}\label{eq:vm_mod_E}
\begin{split}
    &\sum_{b \in \mathcal{S}_x} \int_{\partial \mathcal{B}_b} \big((\bm{B}_h^+ - \bm{B}_h^-) \times d\bm{n}_b \big) \cdot \bm{\hat{E}}_h \rightarrow \\
    &\sum_{b \in \mathcal{S}_x} \int_{\partial \mathcal{B}_b} \big((\bm{B}_h^+ - \bm{B}_h^-) \times d\bm{n}_b \big) \cdot \bm{\hat{E}}_h - \frac{(Q_E^{\textnormal{new}} - Q_E^{\textnormal{old}}) G_{b}^E}{\sum_{b \in \mathcal{S}_x} G_b^E |\int_{\partial \mathcal{B}_b} (\bm B_h^+ - \bm B_h^-) \times d\bm{n}_b|}
\end{split}
\end{equation}
for any scalars $Q_E^{\textnormal{new}}$, $Q_E^{\textnormal{old}}$, $Q_B^{\textnormal{new}}$, $Q_B^{\textnormal{old}}$ where $Q_B^{\textnormal{old}} + Q_E^{\textnormal{old}} = Q^{\textnormal{old}}$ and $Q_E^{\textnormal{new}} + Q_B^{\textnormal{new}} \ge 0$ and for any scalar functions $G_b^E$ and $G_b^B$ for which the denominators are non-zero. \Cref{eq:vm_mod_f,eq:vm_mod_E,eq:vm_mod_B} ensure that the total energy of the Vlasov-Maxwell system is non-increasing in time.

\textbf{$\ell_2$-norm stability}: Although the $\ell_2$-norm of the continuous distribution function $f_s$ is constant, enforcing conservation of the $\ell_2$-norm introduces spurious oscillations into the distribution function. Instead of conserving the discrete $\ell_2$-norm, we demand that the $\ell_2$-norm to be non-increasing. Using \cref{eq:vm_dfdt} and the fact that $f_{sh}$ is constant within a cell, we have
\begin{equation}\label{eq:l2-norm-a}
    \frac{d}{dt} \sum_j \sum_k \int_{\mathcal{K}_{jk}} \frac{1}{2}f_{sh}^2 \mathop{d\bm z} = \sum_j \sum_k \int_{\mathcal{K}_{jk}} f_{sh} \frac{\partial f_{sh}}{\partial t} \mathop{d\bm z} = -\sum_j \sum_k \oint_{\partial \mathcal{K}_{jk}} f_{sh} {\bm{\hat{F}}}_s\cdot \mathop{d\bm s} \le 0
\end{equation}
Next, we use summation by parts which gives
\begin{equation}\label{eq:vm_l2_norm}
    -\sum_j \sum_k \oint_{\partial \mathcal{K}_{jk}} f_{sh} {\bm{\hat{F}}}_s\cdot \mathop{d\bm s} = \sum_{b \in \mathcal{S}_x \cup \mathcal{S}_v} \int_{\partial \mathcal{B}_b} (f_{sh}^+ - f_{sh}^-)\bm{\hat{F}}_s \cdot \mathop{d \bm{n}_b} \le 0.
\end{equation}
\Cref{eq:vm_l2_norm} is a straightforward extension of \cref{eq:stability_energy_method,eq:l2cons2d} to six dimensional phase space. Since we have already modified $\bm{\hat{F}}_s$ for all $b \in \mathcal{S}_v$ to ensure energy is non-increasing, we will only modify $\bm{\hat{F}}_s$ for $b \in \mathcal{S}_x$ to ensure that the $\ell_2$-norm decays. We define $\nicefrac{d\ell_2^{\textnormal{old}}}{dt} \coloneqq \sum_{b \in \mathcal{S}_x \cup \mathcal{S}_v}\int_{\partial \mathcal{B}_b} (f_{sh}^+ - f_{sh}^-)\bm{\hat{F}}_s \cdot \mathop{d \bm{n}_b}$. To ensure that \cref{eq:vm_l2_norm} is satisfied, we apply the following transformation to $\bm{\hat{F}}_s$ for all $b \in \mathcal{S}_x$:
\begin{equation}
    \int_{\partial \mathcal{B}_b} \bm{\hat{F}}_s \cdot d\bm{n}_s \rightarrow \int_{\partial \mathcal{B}_b} \bm{\hat{F}}_s \cdot d\bm{n}_s + \frac{\nicefrac{(d\ell_2^{\textnormal{new}}}{dt} - \nicefrac{d\ell_2^{\textnormal{old}}}{dt}) G_b}{\sum_{b^{'} \in \mathcal{S}_x} (f_{sh}^+ - f_{sh}^-) G_{b'}} 
\end{equation}
for any scalar $d\ell_2^{\textnormal{new}}{dt} \le 0$ and any finite function $G_b$ for which the denominator is non-zero.




\section{Computational verification}\label{sec:verification}

\noindent In \cref{sec:scalarhyperbolic,sec:systemshyperbolic} we algebraically derived invariant-preserving error-correcting algorithms. For most of these algorithms, we computationally verified that they do in fact preserve the correct invariants by applying them to standard numerical methods as shown in \cref{fig:fluxFV,fig:burgers_nonconservative,fig:advection_ftcs,fig:dg_demo,fig:euler_2d,fig:compressible_euler_demo}. In this section, we computationally verify two additional claims about these error-correcting algorithms.

First, in \cref{sec:why_standard_dont_work} we argue that, for certain invariants, standard approaches to preserving invariants will not work with machine learned solvers.
Intuitively, this is because standard approaches are either too restrictive or add too much numerical diffusion, degrading the accuracy of the solution.
Second, we claim that in closed or periodic systems the error-correcting algorithms we introduce do not degrade the accuracy of an already-accurate machine learned solver.
Intuitively, this is because an already-accurate solver will either satisfy or nearly satisfy the desired invariants, so the invariant-preserving correction will be small and only applied if necessary.
We verify this claim for the 1D advection equation in \cref{sec:why_standard_dont_work}, for the 1D Burgers' equation in \cref{sec:verification_burgers}, for the 2D incompressible Euler equation in \cref{sec:verification_2d_euler}, and for the 1D compressible Euler equations in \cref{sec:verification_compressible_euler}. We also demonstrate, using the 1D compressible Euler equations, how in an open system the accuracy of a machine learned solver can be degraded due to errors in the estimates of the rate of change of the invariant(s).

We emphasize that in this section, the purpose is not to design high-performing machine learned solvers but rather to use simple machine learned solvers trained to solve simple tasks as tools to illustrate the claims in the previous paragraph. Except for the 1D Burgers' solver in \cref{sec:verification_burgers}, we train extremely simple models. In particular, we don't use any ML-based strategies to improve performance, as the focus is not on the performance of the machine learned solvers. Instead, the focus is on the relative performance of the machine learned solvers with and without the invariant-preserving corrections.

\subsection{Why standard invariant-preserving methods don't work with ML\label{sec:why_standard_dont_work}}

\noindent Some invariants \textit{can} be preserved in machine learned solvers using standard approaches to invariant preservation without degrading accuracy. For example, the linear invariant  $\frac{d}{dt} \int_{\Omega} \bm u \mathop{dx} = 0$ can be preserved by predicting the flux across cell boundaries. Likewise, limiters can be used to ensure positivity of the solution. 
Yet for other invariants, standard approaches will not work with machine learned solvers. Why not? 

To be useful, machine learned PDE solvers must outperform standard numerical methods. To do so, invariant-preserving algorithms are needed which do not degrade accuracy at large $\Delta x$ and/or $\Delta t$. Standard approaches to preserving non-linear invariants are unstable at large $\Delta t$ due to the CFL condition and add local numerical diffusion proportional to $(\Delta x)^2$ \cite{artificial_viscosity}, thereby degrading accuracy at large $\Delta x$. As a result, machine learned solvers cannot simultaneously preserve non-linear invariants and outperform standard solvers using standard approaches.
The whole point of the invariant-preserving algorithms introduced in \cref{sec:scalarhyperbolic,sec:systemshyperbolic} is they guarantee invariant preservation at large $\Delta x$ and/or $\Delta t$ while adding the minimum correction needed to preserve the invariant.

We use an example to illustrate. We consider the simplest hyperbolic PDE, the 1D advection equation with periodic BCs.
As we will see, standard approaches can be used to design invariant-preserving machine learned PDE solvers, but those solvers cannot outperform standard methods. In contrast, the error-correcting algorithms we introduce preserve invariants without degrading accuracy. As a result, it becomes possible to design invariant-preserving machine learned solvers that outperform standard solvers.

\begin{figure}
  \centering
  \includegraphics[width=0.9\textwidth]{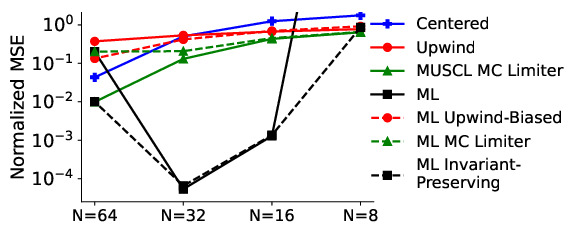}
  \caption{While there are various ways of designing invariant preserving machine learned PDE solvers, only the error-correction strategy we propose (black dotted line) preserves invariants without degrading the accuracy of an already-accurate machine learned solver (black line, $N=16$ and $N=32$ where $N$ is the number of spatial grid cells). Here we compare seven different methods for solving the 1D advection equation. Three are standard numerical methods (centered flux, upwind flux, and MUSCL flux) while four are machine learned solvers. After training the ML models with the same setup (see \ref{sec:appendix_verification_details}), we compute the normalized mean squared error (MSE) from $t=0$ to $t=1$ over 25 samples drawn from the training distribution.}
  \label{fig:compare_advection}
\end{figure}

We use the continuous-time FV update function \cref{eq:fv_a} and compare seven different choices for the flux at cell boundaries $f_{j+\nicefrac{1}{2}}$. Because they are flux-predicting methods, all seven solvers guarantee that the discrete mass is conserved. Three are standard numerical methods:
\begin{enumerate}
    \item The centered flux $f_{j+\nicefrac{1}{2}} = \frac{u_j + u_{j+1}}{2}$. This flux conserves the discrete $\ell_2$-norm.
    \item The upwind flux $f_{j+\nicefrac{1}{2}} = u_j$. This flux is TVD and decays the discrete $\ell_2$-norm. 
    \item The MUSCL flux with a Monotonized Central (MC) limiter. This flux is TVD.
\end{enumerate}
All three standard numerical methods (solvers 1, 2, 3) preserve one or more of the non-linear invariants and are thus numerically stable. The MUSCL scheme (solver 3) is the most accurate of the three standard numerical methods we consider. The other four methods are flux-predicting machine learned PDE solvers:
\begin{enumerate}
    \setcounter{enumi}{3}
    \item A machine learned solver which outputs $f_{j+\nicefrac{1}{2}}$. This solver is not guaranteed to conserve any non-linear invariants.
    \item An upwind-biased flux-predicting solver which outputs $\alpha_{j+\nicefrac{1}{2}} \ge 0$ where $f_{j+\nicefrac{1}{2}} = \alpha_{j+\nicefrac{1}{2}} f_{j+\nicefrac{1}{2}}^{\textnormal{Upwind}} + (1 - \alpha_{j+\nicefrac{1}{2}}) f_{j+\nicefrac{1}{2}}^{\textnormal{Centered}}$. This solver decays the $\ell_2$-norm.
    \item The same as solver 4, except with an MC flux limiter. This solver is TVD.
    \item The same as solver 4, but using the error-correcting algorithm \cref{eq:1d_stability} with $\nicefrac{d\ell_2^{\textnormal{new}}}{dt} \le 0$ to ensure that the discrete $\ell_2$-norm is non-increasing. 
\end{enumerate}
The upwind-biased solver (solver 5) and the flux-limited solver (solver 6) are examples of how standard approaches can be used to preserve invariants in machine learned PDE solvers. The error-corrected solver (solver 7) is an example of the strategy proposed in this paper. The details of the initial conditions, loss function, training data, and ML models are included in \ref{sec:appendix_verification_details}. In \cref{fig:compare_advection}, we plot the normalized mean squared error (MSE) for all seven solvers averaged over time from $t=0$ to $t=1$ averaged over 25 samples drawn from the training distribution. We compare solvers with $N$ grid cells, where $N=8$, 16, 32, and 64. 

The important takeaways from \cref{fig:compare_advection} are the following. The machine learned solver (solver 4) is highly accurate for $N=16$ and $N=32$. The upwind-biased machine learned solver (solver 5) decays the $\ell_2$-norm but is too constrained and too diffusive to outperform the standard solvers. The flux-limited machine learned solver (solver 6) is TVD but adds numerical diffusion proportional to $(\Delta x)^2$ to extremum and sharp gradients. At coarse resolution a high proportion of grid cells are either extremum or have sharp gradients. Thus flux-limited TVD-stable machine learned numerical methods operating at large $\Delta x$ will add large amounts of numerical diffusion to many of the grid cells which will degrade the accuracy of the solution.
In contrast, the invariant-preserving error-corrected machine learned solver (solver 7) adds the minimum amount of numerical diffusion necessary to ensure that $\nicefrac{d\ell_2^\textnormal{new}}{dt} \le 0$ and only does so when $\nicefrac{d\ell_2^\textnormal{old}}{dt} > 0$. Because the invariant-preserving correction adds the minimum amount of numerical diffusion necessary to preserve the invariant and only does so when the invariant is violated, the error-correcting algorithm preserves a discrete analogue of the non-increasing $\ell_2$-norm invariant without degrading the accuracy of an already-accurate solver ($N=16$ and $N=32$). 
For $N=8$ and $N=64$ the machine learned solver (solver 4) is unstable and increases the $\ell_2$-norm of the solution without bound, decreasing accuracy. There are a variety of possible causes of this poor performance, but the key takeaway is that the error-correcting solver (solver 7) improves the reliability and accuracy of a numerically unstable machine learned solver which has failed to preserve the desired invariants.

\subsection{Burgers' equation \label{sec:verification_burgers}}

\noindent In this section, we verify for the 1D Burgers' equation that the error-correcting algorithms we introduce do not degrade the accuracy of an already-accurate machine learned solver. To do so, we need to train a machine learned solver to solve the 1D Burgers' equation.
Instead of designing our own solver, we attempt to replicate the solvers in fig 3c of \cite{bar2019learning}; fig 3c compares the accuracy of a highly-accurate flux-predicting `data-driven discretization' 1D Burgers' solver with the accuracy of standard solvers. The solver, model, training, and evaluation are nearly identical to \cite{bar2019learning}; details are included in \ref{sec:appendix_verification_details}. 

\begin{figure}
    \centering
    \includegraphics[width=0.86\textwidth]{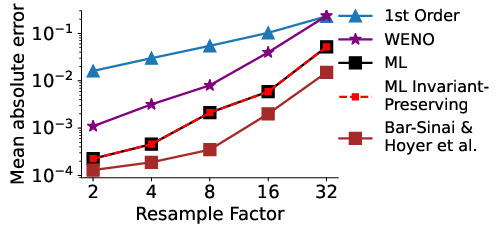}
    \caption{We demonstrate that the invariant-preserving error-correcting algorithm we introduce in \cref{sec:fluxpredictingFV} (red dotted line) doesn't degrade the accuracy of an already-accurate machine learned solver (black line).
    We plot the mean absolute error (MAE) averaged over 100 draws from the training distribution and over $t \le 15$.
    Instead of designing and training our own solver, we attempt to replicate the highly accurate 1D Burgers' machine learned solver in fig. 3 of \cite{bar2019learning}.
    Our goal is for the accuracy of our replicated solver (black line) to roughly match the accuracy of the original solver (brown line). While we match the overall trend, our error is higher. Thus, our replication attempt is not fully successful.}
    \label{fig:ml_burgers}
\end{figure}

Our attempt to replicate fig 3c of \cite{bar2019learning} is shown in \cref{fig:ml_burgers}. We plot the mean absolute error (MAE) for various solvers averaged over 100 samples drawn from the training distribution and over time $t$ less than 15. 
Our attempt at replicating the machine learned solver in \cite{bar2019learning} is plotted in black.
In the dotted red line of \cref{fig:ml_burgers}, we apply the invariant-preserving error-correcting algorithm introduced in \cref{sec:fluxpredictingFV} to that machine learned solver. The accuracy of the invariant-preserving machine learned solver (red dotted line) is practically identical to that of the already-accurate machine learned solver (black line). This is because the already-accurate machine learned solver guarantees mass conservation and tends not to increase the discrete $\ell_2$-norm within its training distribution, so the correction is usually not needed. When the correction is needed, the error-correcting algorithm applies the minimum amount of numerical diffusion necessary to preserve the non-increasing $\ell_2$-norm invariant.

The MAE as originally reported in fig. 3 and fig. S8 of \cite{bar2019learning} is plotted in brown. While are able to match the same trend as \cite{bar2019learning}, we achieve worse accuracy. Unfortunately, our attempt at replicating fig. 3 of \cite{bar2019learning} was only partially successful. We do not know the cause of the discrepancy. In any case, this failed replication attempt does suggest that it can be tricky to train machine learned solvers to achieve very high accuracy.

\subsection{2D incompressible Euler equation \label{sec:verification_2d_euler}}

\begin{figure}
    \centering
    \includegraphics[width=0.85\textwidth]{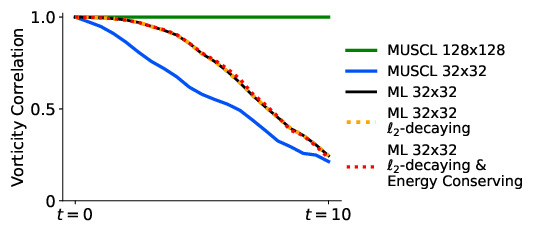}
    \caption{We demonstrate that the invariant-preserving error-correcting algorithms we introduce in \cref{sec:continuoustimeFV,sec:energyconservation} don't degrade the accuracy of an already-accurate machine learned solver (black line) for the 2D incompressible Euler equations. For each solver, we plot the correlation with the high-resolution `exact' solution; the correlation serves as a measure of accuracy. The low-resolution machine learned solver (black line) is more accurate than the low-resolution MUSCL scheme, but doesn't guarantee preservation of any of the invariants of the incompressible Euler equations. Applying error-correcting algorithms to either conserve mass and ensure non-increasing $\ell_2$-norm (\cref{eq:black_box_stability}, orange dotted line) or to conserve mass, conserve energy, and ensure non-increasing $\ell_2$-norm (\cref{eq:eulerenergycons}, red dotted line) do not degrade the accuracy of the machine learned solver. }
    \label{fig:corr_2d_euler}
\end{figure}

\noindent In this section, we verify for the 2D incompressible Euler equations that the invariant-preserving error-correcting algorithms we introduce do not degrade the accuracy of an already-accurate machine learned solver. 
The 2D incompressible Euler equations conserve mass, have non-increasing $\ell_2$-norm, and conserve energy. We consider the effect of two different invariant-preserving algorithms applied to a machine learned solver which is accurate at coarse resolution. First, we apply the mass conserving, $\ell_2$-norm non-increasing error-correcting algorithm in \cref{eq:black_box_stability} of \cref{sec:continuoustimeFV}. Second, we apply the mass conserving, energy-conserving, and $\ell_2$-norm non-increasing algorithm in \cref{eq:eulerenergycons} of \cref{sec:energyconservation}. We solve the 2D incompressible Euler equations with non-invariant forcing and diffusion terms, and use ML to approximate the invariant-preserving terms. We use standard methods to approximate the forcing and diffusion terms. The details of the solver, data generation, training and evaluation are in \ref{sec:appendix_verification_details}. \Cref{fig:corr_2d_euler} demonstrates that these invariant-preserving algorithms don't degrade the accuracy of an already-accurate machine learned solver.

\subsection{Compressible Euler equations \label{sec:verification_compressible_euler}}

\noindent In this section, we verify for the 1D compressible Euler equations that in a periodic domain the invariant-preserving error-correcting algorithm we introduce in \cref{sec:compressible_euler} does not degrade the accuracy of an already-accurate machine learned solver, but that in a open system errors in the estimate of the rate of change of the invariant can degrade the accuracy of an already-accurate machine learned solver. The 1D compressible Euler equations conserve density, momentum, and energy, preserve positivity of density $\rho$ and pressure $p$,
and have non-decreasing entropy $\eta$. We train simple machine learned solvers in domains with periodic boundary conditions as well as domains with Dirichlet (open) boundary conditions. The initial conditions are given by a random draw from a (relatively simple) distribution of possible initial conditions; details of the data generation process, training procedure, and evaluation are given in \ref{sec:appendix_verification_details}.

\begin{figure}
    \centering
    \begin{subfigure}[b]{\textwidth}
        \centering
        \includegraphics[width=0.65\textwidth]{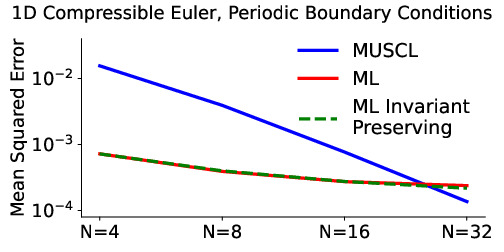}
        \caption{}
        \label{fig:1d_euler_periodic}
    \end{subfigure}
    \begin{subfigure}[b]{\textwidth}
         \centering
         \includegraphics[width=0.65\textwidth]{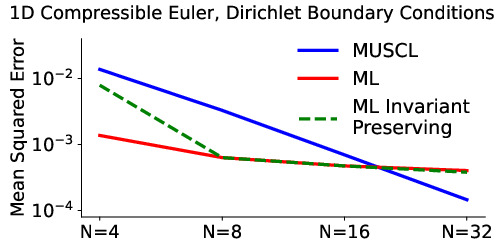}
         \caption{}
         \label{fig:1d_euler_open}
     \end{subfigure}
    \caption{We demonstrate that (a) the invariant-preserving algorithm for the 1D compressible Euler equations introduced in \cref{sec:compressible_euler} does not degrade the accuracy of a machine learned solver in periodic boundary conditions, and (b) that this algorithm can degrade the accuracy of a machine learned solver with open boundary conditions if the rate of change of the invariant is not estimated correctly. This error-correcting algorithm uses the transformation in \cref{eq:compressible_euler_positivity_transformation} to ensure positivity of $\rho_j$ and $p_j$, followed by the transformation in \cref{eq:compressible_euler_entropy_transformation} to ensure that entropy either is non-decreasing or is no less than the entropy flux through the domain boundaries.
    }
    \label{fig:1d_euler_verification}
\end{figure}

In \cref{fig:1d_euler_periodic}, we compare the performance of two machine learned solvers with a baseline MUSCL scheme (blue line) at various grid resolutions. The original machine learned solver (red line) learns a correction to the flux term of the MUSCL scheme. This solver conserves density, momentum, and energy, but doesn't guarantee positivity of $\rho_j$ or $p_j$ and doesn't guarantee that the discrete entropy will be non-decreasing. We apply the error-correcting algorithm introduced in \cref{eq:compressible_euler_positivity_transformation,eq:compressible_euler_entropy_transformation}, modified to support periodic BCs. This invariant-preserving machine learned solver (dotted green line) preserves the desired invariants without degrading the accuracy of the original solver.

In \cref{fig:1d_euler_open}, we do the same comparison except with Dirichlet boundary conditions. Due to the open boundary conditions, we do not have an exact estimate of the entropy flux through the boundaries. Thus, to use the invariant-preserving error-correcting algorithm \cref{eq:compressible_euler_positivity_transformation,eq:compressible_euler_entropy_transformation}, we need to estimate the entropy flux through the domain boundaries. Two possible estimates are to use the value from the Dirichlet boundary condition $\psi(0) = (\rho v)(0) g(s(0))$ or to use the value within the grid cell closest to the boundary $\psi(0) = (\rho v)_{1} g(s_1)$; we find that both estimates can add too much numerical diffusion and thereby degrade performance. We find better performance if we estimate the entropy flux to be the minimum of the two values, i.e., $\psi(0) = \textnormal{min}\{(\rho v)(0) g(s(0)), (\rho v)_{1} g(s_1)\}$. Applying that estimate to the error-correcting algorithm given by \cref{eq:compressible_euler_positivity_transformation}, in \cref{fig:1d_euler_open} we see that performance is degraded for the solver with $N=4$ grid cells but that the already-accurate machine learned solvers for $N=8$, $N=16$, and $N=32$ grid cells are able to guarantee positivity and satisfy the entropy-increasing invariant of \cref{eq:compressibleeuler_entropyincreasing} without degrading accuracy.

\section{Related work}\label{sec:relatedwork}

\subsubsection{Invariant preservation in machine learned numerical methods} \cite{bar2019learning,zhuang2021learned,kochkov2021machine,stevens2020enhancement,stevens2020finitenet,stevens2022applications} use a finite volume representation of the solution to solve a variety of 1D and 2D PDEs. These so-called `hybrid' solvers predict the fluxes through cell boundaries and thus conserve mass by construction. They do not, however, preserve the non-increasing $\ell_2$-norm invariant and therefore do not guarantee stability. \cite{zhuang2021learned,kochkov2021machine} attempt to promote stability by unrolling the loss function over multiple timesteps. \cite{kochkov2021machine} trains a hybrid solver for the 2D incompressible Euler equations that ``remains stable during long simulations.'' This result is likely facilitated by the addition of physical diffusion to the PDE, which decays the $\ell_2$-norm at each timestep. \cite{thuerey_guaranteed_momentum} approximates a fluid with a system of particles and learns to predict the forces between particles using antisymmetric continuous convolutional layers; this ensures that the forces between particles are equal and opposite so that momentum is conserved. In such a system, we have found an error correction strategy could be used to conserve momentum \textit{and} energy. \cite{brandstetter2022clifford} solve Maxwell's equations in 3D and use Clifford Algebra and specially designed neural networks to ensure that geometric invariants relating the electric and magnetic fields are preserved. \cite{holloway2021acceleration} use ML to approximate the Boltzmann collision operator and apply the error-correcting algorithm introduced in \cite{zhang2018conservative}.
\cite{zhang2018conservative} introduce an error-correcting algorithm (intended for standard solvers) for the Boltzmann collision operator that enforces conservation of mass, momentum, and energy.
We have found an iterative error-correcting algorithm for the Boltzmann collision operator that preserves an additional invariant, non-decreasing entropy. We intend to discuss this algorithm in a future paper. \cite{richter2022neural} learn continuous divergence-free vector fields which satisfy incompressibility. \cite{alguacil2021predicting} applies an \textit{a posteriori} correction (i.e., error correction) to conserve energy in a deep learning-based solver for propagating acoustic waves; this correction results in improved performance.

\subsubsection{Other machine learned PDE solvers} \cite{list2022learned, um2020solver, pathak2020using, kochkov2021machine} use convolutional neural networks to correct errors in low-resolution simulations; these hybrid solvers promote stability and improve accuracy by unrolling the loss function over multiple timesteps. \cite{stachenfeld2021learned} solves 2D and 3D hyperbolic PDEs using the `fully learned' update equation $u_{i,j}(t+\Delta t) = u_{i,j}(t) + N_{i,j}(u_{i,j}(t), \Delta t))$ where $N_{i,j}$ is a the output of a convolutional neural network; this update equation is identical to \cref{eq:discretetimeupdaterule}.
\cite{stachenfeld2021learned} attempts to ensure stability by adding noise to the training distribution and by using very large timesteps. 
\cite{brandstetter2022message} argues that instability in machine learned iterative numerical algorithms arises due to a \textit{distribution shift} where the distribution of training data differs from the outputs of the solver during inference due to small errors that accumulate over time. \cite{brandstetter2022message} uses the update equation $u_{i,j}(t+\ell\Delta t) = u_{i,j}(t) + \ell \Delta t N^\ell_{i,j}$ for $1 \le \ell \le K$ where $N^\ell_{i,j}$ is the output of a message passing graph neural network that predicts the next $K$ timesteps.
\cite{brandstetter2022message} attempts to ensure stability by modifying the loss function, adding random noise, and by predicting multiple timesteps into the future. A variety of papers have attempted to promote stability of dynamical systems that result from data-driven reduced order models, including by adding sparsity-promoting priors to a loss function \cite{Kaptanoglu_2021,erichson} and by constraining the eigenvalues of a learned Koopman operator \cite{koopman_stability}. 

\subsubsection{LES models and backscattering} The objective of large eddy simulation (LES) is identical to that of many machine learned solvers: both attempt to find an accurate approximation to the solution of the PDE with fewer computational resources than classical numerical methods. Both also attempt to do so without resolving the smallest scales of the problem, relying on either an explicit or implicit subgrid model to do so  \cite{bar2019learning,zhuang2021learned,kochkov2021machine,stachenfeld2021learned,um2020solver,list2022learned,pathak2020using,Guan_2022,small_data_les, subgrid_modeling_2d_turbulence, les_closure, rose_yu_paper, ml_flux_limiters}.
Of particular relevance to the stability of subgrid models (both in LES and ML) are the concepts of `forward-scatter' and `backscatter'. In 2D LES turbulence, forward-scattering involves the transfer of enstrophy from resolved to unresolved scales, while backscattering involves the transfer of enstrophy from unresolved to resolved scales. Analysis across a wide range of flows demonstrates two important facts \cite{backscatter}. First, to be accurate a subgrid model must allow both forward-scatter and backscatter. This means that to be accurate a subgrid model must allow a discrete analogue of the entropy inequality \cref{eq:entropy_inequality} to be locally violated. Second, averaged over the entire domain there is always more forward-scatter than backscatter. If on average there were more backscatter than forward-scatter, then the subgrid model would be unstable \cite{Guan_2022}. In 2D turbulence, invariant-preserving error-correcting algorithms can be interpreted as ways of constraining a subgrid model to ensure that on average there is always at least as much forward-scatter as backscatter.

\section{Limitations \& trade-offs \label{sec:limitations}}

\noindent Invariant preservation in machine learned solvers is, when applicable, free lunch. We know a priori our solution should preserve certain invariants, so by enforcing these invariants at each timestep we can improve the reliability of our solver without degrading accuracy. Nevertheless, there are some limitations and trade-offs to consider when designing and deploying these algorithms; we discuss eight.

First, the invariant-preserving updates in the error-correcting algorithms we introduce can all be derived analytically and applied in a single correction step. As a result, these algorithms are extremely efficient. However, in some situations it may be impossible to derive an analytic correction that preserves all the desired invariants. In such cases, iterative (and possibly less efficient) algorithms would need to be derived. In kinetic physics, for example, ensuring that the entropy $-\sum f \log{f}$ is non-decreasing seems to require an iterative error-correcting algorithm.

Second, many of the error-correcting algorithms we introduce add numerical diffusion in the continuous-time limit. If the machine learned solver is inaccurate and the algorithm needs to add large amounts of numerical diffusion, this can restrict the allowable timestep. Thus, inaccurate machine learned solvers might still be numerically unstable if the timestep is not modified to account for the additional diffusion.

Third, only one of the error-correcting algorithms we introduce (\cref{eq:discretetime_updatedelta,eq:discretetime_epsilon} in \cref{sec:discretetimeFV}) uses a discrete-time update rule. We have observed that for some invariants it seems to be more difficult (perhaps impossible) to derive analytic error-correcting algorithms with a discrete-time update than with a continuous-time update. 
Notice also that \cref{eq:discretetime_epsilon} has no solution if the change in the $\ell_2$-norm $\Delta \ell_2$ is below a solution- and solver-dependent minimum value. In practice, this means that inaccurate machine learned solvers that use a discrete-time update can still be numerically unstable. Specifically, since $\Delta \ell_2$ and $\Delta \hat{\bm u}$ both scale with the timestep $\Delta t$, more accurate machine learned solvers can guarantee numerical stability while using larger $\Delta t$, while less accurate solvers need to use smaller $\Delta t$ to do so.

Fourth, although designers of numerical algorithms usually would prefer to inherit discrete analogues of all of the properties of the continuous equation, it is usually impossible to design highly accurate numerical methods that inherit all of these properties.\footnote{Note also that the invariants of the continuous PDE are not necessarily the same as the invariants of the discrete equations. This was discussed in \cref{sec:coarsegraining}. }
This idea is formalized in Godunov's theorem \cite{godunov1959finite} for scalar hyperbolic PDEs, but is a pattern seen more generally in computational physics. For some PDEs there may be a trade-off between invariant preservation (i.e., reliability or robustness) and accuracy. Designers of numerical methods must choose which subset of invariants to preserve; being constrained by too many invariants can prevent a solver from making accurate predictions. In brief, it is possible to have \textit{too much} inductive bias.

Fifth, the invariant-preserving algorithms we propose are error-correcting algorithms which use global instead of local constraints. There is a trade-off associated with this decision: our algorithms violate the property that in hyperbolic PDEs information propagates at finite speed. In practice, machine learned PDE solvers violate the finite speed property anyway when outputting the update rule, so applying an additional non-local correction is a small price to pay for the benefits of numerical stability and improved reliability.

Sixth, while this error-correction strategy can still be applied to invariant-preserving PDEs containing non-invariant terms, the error-correcting algorithms should only be applied to the invariant-preserving terms. Thus, when solving non-invariant-preserving PDEs, machine learned solvers will need to separate out the contribution due to the invariant-preserving terms from the contribution due to the non-invariant-preserving terms, and apply an error-correcting algorithm only to the invariant-preserving contribution.

Seventh, in open systems it can be difficult to estimate the fluxes through the boundary. Thus, while it is possible to design invariant-preserving error-correcting algorithms in open systems, if the rate of change of the invariant(s) is not estimated correctly then these algorithms can degrade accuracy. We suspect that this problem could be alleviated by using a higher density of grid cells near boundaries.

Eighth, while these error-correcting algorithms are designed to solve the problem of preserving invariants in machine learned solvers without degrading the accuracy of an already-accurate solver, they do not solve the problem of \textit{finding} accurate machine learned solvers. These algorithms adjust the update at each timestep if the solver violates invariants, but a solver which frequently commits large violations is likely to perform poorly. Alternatively, a solver could preserve the correct invariants but give inaccurate results. Building accurate, fast, and robust machine learned PDE solvers will require not only well-designed numerical methods but also well-engineered learning systems which consistently make accurate predictions about the time evolution of the solution.

\chapter{Reproducibility issues in ML-for-PDE research \label{ch:reproducibility}}
\noindent\rule{\textwidth}{1pt}

Part of the contents of this chapter are currently under review in the following paper: (i) McGreivy, N., \& Hakim, A. ``Weak baselines and reporting biases lead to overoptimism in machine learning for fluid-related partial differential equations." Nature machine intelligence (under review).

\noindent\rule{\textwidth}{1pt}

\subsubsection{Summary}

One of the most promising applications of machine learning (ML) in computational physics is to accelerate the solution of partial differential equations (PDEs). The key objective of ML-based PDE solvers is to output a sufficiently accurate solution faster than standard numerical methods, which are used as a baseline comparison.
We first perform a systematic review of the ML-for-PDE solving literature.
Of articles that use ML to solve a fluid-related PDE and claim to outperform a standard numerical method, we determine that 79\% (60/76) compare to a weak baseline.
Second, we find evidence that reporting biases, especially outcome reporting bias and publication bias, are widespread.
We conclude that ML-for-PDE solving research is overoptimistic:
weak baselines lead to overly positive results, while reporting biases lead to underreporting of negative results.
To a large extent, these issues appear to be caused by factors similar to those of past reproducibility crises: researcher degrees of freedom and a bias towards positive results.
We call for bottom-up cultural changes to minimize biased reporting as well as top-down structural reforms intended to reduce perverse incentives for doing so.

\newcommand{\cmark}{\textcolor{green}{\ding{51}}}%
\newcommand{\xmark}{\textcolor{red}{\ding{54}}}%
\newcommand\dangersign[1][1.8ex]{%
  \scaleto{\stackengine{0.3pt}{\scalebox{1.6}[1.4]{%
  \color{orange}$\blacktriangle$}}{\small\bfseries !}{O}{c}{F}{F}{L}}{#1}%
}

\newcommand{\appendixentry}[4]{
\vspace{0.1cm}
\noindent \textbf{Article #1}: #2
\\ \textit{Title}: #3
\\ \textit{Citations}: #4
}
\newcommand{\appendixdata}[8]{
\\ \textit{Fluids-relevant PDE(s)}: #1
\\ \textit{Primary outcome(s)}: #2
\\ \textit{Baseline}: #4
\\ \textit{Rule 1}: #5
\\ \textit{Rule 2}: #6
\\ \textit{Fair comparison}: #8
}

\newcommand{\fnoli}{1}
\newcommand{\fnolicitations}{941}
\newcommand{\citefnoli}{\cite{li2020fourier}}

\newcommand{\deeponetlulu}{2}
\newcommand{\deeponetlulucitations}{911}
\newcommand{\citedeeponetlulu}{\cite{lu2021learning}}

\newcommand{\tompson}{3}
\newcommand{\tompsoncitations}{558}
\newcommand{\citetompson}{\cite{tompson2017accelerating}}

\newcommand{\mlacceleratedcfd}{4}
\newcommand{\mlacceleratedcfdcitations}{429}
\newcommand{\citemlacceleratedcfd}{\cite{kochkov2021machine}}

\newcommand{\meshbasedpfaff}{5}
\newcommand{\meshbasedpfaffcitations}{390}
\newcommand{\citemeshbasedpfaff}{\cite{pfaff2020learning}}

\newcommand{\barsinai}{6}
\newcommand{\barsinaicitations}{382}
\newcommand{\citebarsinai}{\cite{bar2019learning}}

\newcommand{\deepfluids}{7}
\newcommand{\deepfluidscitations}{382}
\newcommand{\citedeepfluids}{\cite{kim2019deep}}

\newcommand{\deeponetwang}{8}
\newcommand{\deeponetwangcitations}{230}
\newcommand{\citedeeponetwang}{\cite{wang2021learning}}

\newcommand{\solverinthe}{9}
\newcommand{\solverinthecitations}{143}
\newcommand{\citesolverinthe}{\cite{um2020solver}}

\newcommand{\deepmmnetcvt}{10}
\newcommand{\deepmmnetcvtcitations}{128}
\newcommand{\citedeepmmnetcvt}{\cite{cai2021deepm}}

\newcommand{\belbuteperez}{11}
\newcommand{\belbuteperezcitations}{124}
\newcommand{\citebelbuteperez}{\cite{belbute2020combining}}

\newcommand{\pinoli}{12}
\newcommand{\pinolicitations}{124}
\newcommand{\citepinoli}{\cite{li2021physics}}

\newcommand{\projectionyang}{13}
\newcommand{\projectionyangcitations}{122}
\newcommand{\citeprojectionyang}{\cite{yang2016data}}

\newcommand{\neuralconverge}{14}
\newcommand{\neuralconvergecitations}{101}
\newcommand{\citeneuralconverge}{\cite{hsieh2019learning}}

\newcommand{\messagepassing}{15}
\newcommand{\messagepassingcitations}{87}
\newcommand{\citemessagepassing}{\cite{brandstetter2022message}}

\newcommand{\deepmmnet}{16}
\newcommand{\deepmmnetcitations}{84}
\newcommand{\citedeepmmnet}{\cite{mao2021deepm}}

\newcommand{\frameworkmishra}{17}
\newcommand{\frameworkmishracitations}{80}
\newcommand{\citeframeworkmishra}{\cite{mishra2018machine}}

\newcommand{\optimizemultigrid}{18}
\newcommand{\optimizemultigridcitations}{79}
\newcommand{\citeoptimizemultigrid}{\cite{greenfeld2019learning}}

\newcommand{\donga}{19}
\newcommand{\dongacitations}{55}
\newcommand{\citedonga}{\cite{dong2021local}}

\newcommand{\rayb}{20}
\newcommand{\raybcitations}{52}
\newcommand{\citerayb}{\cite{ray2019detecting}}

\newcommand{\novelcnn}{21}
\newcommand{\novelcnncitations}{47}
\newcommand{\citenovelcnn}{\cite{xiao2018novel}}

\newcommand{\wandel}{22}
\newcommand{\wandelcitations}{46}
\newcommand{\citewandel}{\cite{wandel2020learning}}

\newcommand{\shan}{23}
\newcommand{\shancitations}{46}
\newcommand{\citeshan}{\cite{shan2020study}}

\newcommand{\algebraic}{24}
\newcommand{\algebraiccitations}{46}
\newcommand{\citealgebraic}{\cite{luz2020learning}}

\newcommand{\zhuang}{25}
\newcommand{\zhuangcitations}{40}
\newcommand{\citezhuang}{\cite{zhuang2021learned}}

\newcommand{\pathak}{26}
\newcommand{\pathakcitations}{36}
\newcommand{\citepathak}{\cite{pathak2020using}}

\newcommand{\leoni}{27}
\newcommand{\leonicitations}{34}
\newcommand{\citeleoni}{\cite{di2021deeponet}}

\newcommand{\fnodeformation}{28}
\newcommand{\fnodeformationcitations}{33}
\newcommand{\citefnodeformation}{\cite{li2022fourier}}

\newcommand{\stevensa}{29}
\newcommand{\stevensacitations}{27}
\newcommand{\citestevensa}{\cite{stevens2020enhancement}}

\newcommand{\illarramendi}{30}
\newcommand{\illarramendicitations}{27}
\newcommand{\citeillarramendi}{\cite{ajuria2020towards}}

\newcommand{\stachenfeld}{31}
\newcommand{\stachenfeldcitations}{27}
\newcommand{\citestachenfeld}{\cite{stachenfeld2021learned}}

\newcommand{\han}{32}
\newcommand{\hancitations}{27}
\newcommand{\citehan}{\cite{han2022predicting}}

\newcommand{\stevensb}{33}
\newcommand{\stevensbcitations}{26}
\newcommand{\citestevensb}{\cite{stevens2020finitenet}}

\newcommand{\ozbay}{34}
\newcommand{\ozbaycitations}{25}
\newcommand{\citeozbay}{\cite{ozbay2021poisson}}

\newcommand{\zili}{35}
\newcommand{\zilicitations}{24}
\newcommand{\citezili}{\cite{li2022graph}}

\newcommand{\peng}{36}
\newcommand{\pengcitations}{23}
\newcommand{\citepeng}{\cite{peng2022attention}}

\newcommand{\chen}{37}
\newcommand{\chencitations}{21}
\newcommand{\citechen}{\cite{chen2021numerical}}

\newcommand{\alguacilb}{38}
\newcommand{\alguacilbcitations}{21}
\newcommand{\citealguacilb}{\cite{alguacil2021predicting}}

\newcommand{\wandelb}{39}
\newcommand{\wandelbcitations}{21}
\newcommand{\citewandelb}{\cite{wandel2021teaching}}

\newcommand{\blist}{40}
\newcommand{\blistcitations}{16}
\newcommand{\citeblist}{\cite{list2022learned}}

\newcommand{\cheng}{41}
\newcommand{\chengcitations}{15}
\newcommand{\citecheng}{\cite{cheng2021using}}

\newcommand{\wen}{42}
\newcommand{\wencitations}{13}
\newcommand{\citewen}{\cite{wen2020edge}}

\newcommand{\delaraa}{43}
\newcommand{\delaraacitations}{12}
\newcommand{\citedelaraa}{\cite{de2022accelerating}}

\newcommand{\zhao}{44}
\newcommand{\zhaocitations}{10}
\newcommand{\citezhao}{\cite{zhao2022learning}}

\newcommand{\assessments}{45}
\newcommand{\assessmentscitations}{8}
\newcommand{\citeassessments}{\cite{illarramendi2022performance}}

\newcommand{\holloway}{46}
\newcommand{\hollowaycitations}{7}
\newcommand{\citeholloway}{\cite{holloway2021acceleration}}

\newcommand{\azulay}{47}
\newcommand{\azulaycitations}{7}
\newcommand{\citeazulay}{\cite{azulay2022multigrid}}

\newcommand{\wulatent}{48}
\newcommand{\wulatentcitations}{7}
\newcommand{\citewulatent}{\cite{wu2022learning}}

\newcommand{\liu}{49}
\newcommand{\liucitations}{6}
\newcommand{\citeliu}{\cite{liu2022predicting}}

\newcommand{\zhang}{50}
\newcommand{\zhangcitations}{5}
\newcommand{\citezhang}{\cite{zhang2022hybrid}}

\newcommand{\duarte}{51}
\newcommand{\duartecitations}{4}
\newcommand{\citeduarte}{\cite{duarte2022black}}

\newcommand{\alguacil}{52}
\newcommand{\alguacilcitations}{4}
\newcommand{\citealguacil}{\cite{alguacil2022deep}}

\newcommand{\bezginb}{53}
\newcommand{\bezginbcitations}{4}
\newcommand{\citebezginb}{\cite{bezgin2022weno3}}

\newcommand{\shang}{54}
\newcommand{\shangcitations}{4}
\newcommand{\citeshang}{\cite{shang2022deep}}

\newcommand{\kube}{55}
\newcommand{\kubecitations}{3}
\newcommand{\citekube}{\cite{kube2021machine}}

\newcommand{\shi}{56}
\newcommand{\shicitations}{3}
\newcommand{\citeshi}{\cite{shi2022lordnet}}

\newcommand{\ranadea}{57}
\newcommand{\ranadeacitations}{3}
\newcommand{\citeranadea}{\cite{ranade2021latent}}

\newcommand{\chenb}{58}
\newcommand{\chenbcitations}{3}
\newcommand{\citechenb}{\cite{chen2022machine}}

\newcommand{\ranadeb}{59}
\newcommand{\ranadebcitations}{3}
\newcommand{\citeranadeb}{\cite{ranade2021composable}}

\newcommand{\pengb}{60}
\newcommand{\pengbcitations}{3}
\newcommand{\citepengb}{\cite{peng2023linear}}

\newcommand{\delarab}{61}
\newcommand{\delarabcitations}{2}
\newcommand{\citedelarab}{\cite{de2023accelerating}}

\newcommand{\ranadec}{62}
\newcommand{\ranadeccitations}{2}
\newcommand{\citeranadec}{\cite{ranade2022composable}}

\newcommand{\fang}{63}
\newcommand{\fangcitations}{2}
\newcommand{\citefang}{\cite{fang2022immersed}}

\newcommand{\shukla}{64}
\newcommand{\shuklacitations}{2}
\newcommand{\citeshukla}{\cite{shukla2023deep}}

\newcommand{\zhangb}{65}
\newcommand{\zhangbcitations}{2}
\newcommand{\citezhangb}{\cite{zhang2022learning}}

\newcommand{\bezgin}{66}
\newcommand{\bezgincitations}{2}
\newcommand{\citebezgin}{\cite{bezgin2021fully}}

\newcommand{\yang}{67}
\newcommand{\yangcitations}{2}
\newcommand{\citeyang}{\cite{yang2023rapid}}

\newcommand{\tang}{68}
\newcommand{\tangcitations}{2}
\newcommand{\citetang}{\cite{tang2022neural}}

\newcommand{\nastorg}{69}
\newcommand{\nastorgcitations}{1}
\newcommand{\citenastorg}{\cite{nastorg2022ds}}

\newcommand{\gopakumar}{70}
\newcommand{\gopakumarcitations}{1}
\newcommand{\citegopakumar}{\cite{gopakumar2023fourier}}

\newcommand{\shit}{71}
\newcommand{\shitcitations}{0}
\newcommand{\citeshit}{\cite{shit2021semi}}

\newcommand{\suforecast}{72}
\newcommand{\suforecastcitations}{0}
\newcommand{\citesuforecast}{\cite{suforecasting}}

\newcommand{\jeon}{73}
\newcommand{\jeoncitations}{0}
\newcommand{\citejeon}{\cite{jeon2022physics}}

\newcommand{\dai}{74}
\newcommand{\daicitations}{0}
\newcommand{\citedai}{\cite{dai2022fournetflows}}

\newcommand{\sun}{75}
\newcommand{\suncitations}{0}
\newcommand{\citesun}{\cite{sun2022local}}

\newcommand{\shao}{76}
\newcommand{\shaocitations}{0}
\newcommand{\citeshao}{\cite{shao2022poisson}}

\newcommand{\discacciati}{77}
\newcommand{\discacciaticitations}{54}
\newcommand{\citediscacciati}{\cite{discacciati2020controlling}}

\newcommand{\magiera}{78}
\newcommand{\magieracitations}{44}
\newcommand{\citemagiera}{\cite{magiera2020constraint}}

\newcommand{\bezginc}{79}
\newcommand{\bezginccitations}{16}
\newcommand{\citebezginc}{\cite{bezgin2021data}}

\newcommand{\dongb}{80}
\newcommand{\dongbcitations}{13}
\newcommand{\citedongb}{\cite{dong2022computing}}

\newcommand{\dresdner}{81}
\newcommand{\dresdnercitations}{8}
\newcommand{\citedresdner}{\cite{dresdner2022learning}}

\newcommand{\toshev}{82}
\newcommand{\toshevcitations}{0}
\newcommand{\citetoshev}{\cite{toshev20233}}

\subsubsection{Reproducibility issues in science}

Many fields of science have experienced reproducibility issues \cite{randall2018irreproducibility,ritchie2020science,munafo2017manifesto}.
Often, these reproducibility issues impact a significant percentage of published research \cite{ioannidis2005most,prinz2011believe}.
For example, a large-scale attempt to replicate results from 100 studies in three psychology journals found that between a half and a third of studies did not successfully replicate \cite{open2015estimating}. 
In a different example, the biotechnology company Amgen tried to reproduce 53 `landmark' studies in haematology and oncology, and confirmed the findings in only 6 cases (11\%) \cite{begley2012raise}.
Because these issues can undermine the credibility and authority of an entire field, they are often referred to as a `reproducibility crisis' or `replication crisis' \cite{baker2016reproducibility}.

\begin{figure}
    \centering
    \includegraphics[width=\textwidth]{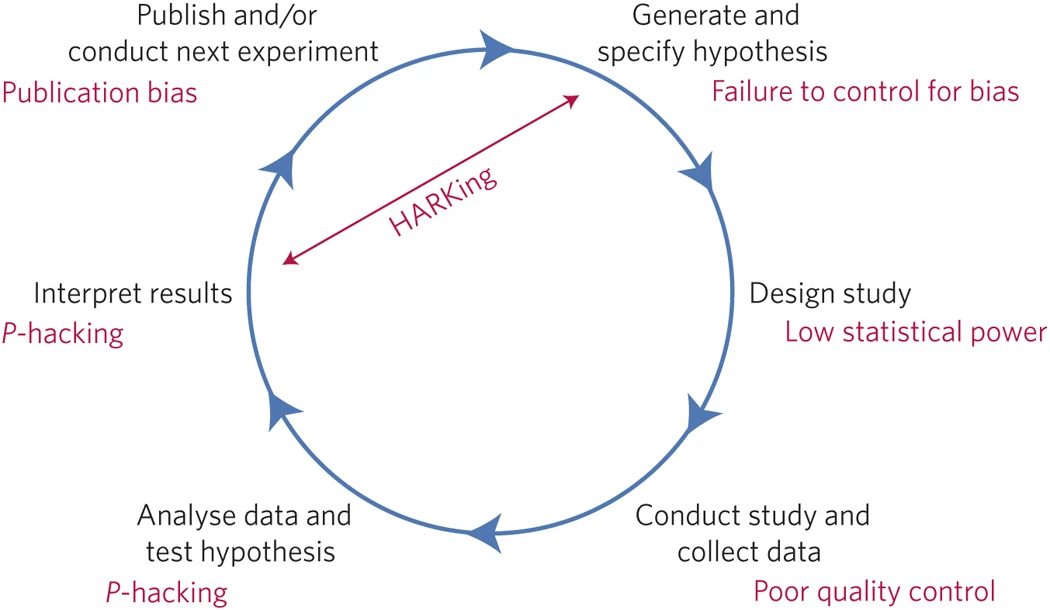}
    \caption{Threats to reproducible science. From \cite{munafo2017manifesto}.}
    \label{fig:reproducibilitythreats}
\end{figure}

Reproducibility issues have a range of causes.
In some extraordinary cases, researchers have simply faked data.
However, scientific fraud is uncommon (and not something I've encountered in scientific ML research).
Much more common are pitfalls with data analysis and statistical techniques, as well as a systematic bias towards publishing negative results \cite{randall2018irreproducibility,ritchie2020science,gelman2013garden}.
Figure \ref{fig:reproducibilitythreats} (from \cite{munafo2017manifesto}) presents a sketch of the scientific process and many of the factors that lead to irreproducible science. 
Many of these factors can be understood as methodological failures related to the use of statistical techniques to draw inferences from data. Studies with low statistical power (i.e., low sample sizes) are not only more likely to give false negatives, but also more likely to give false positives \cite{button2013power}. Multiple hypothesis testing, through p-hacking, HARKing (hypothesizing after results are known), and outcome switching, also tends to increase the likelihood of false positive findings. Multiple hypothesis testing can happen in subtle ways, even when researchers are not consciously probing the data in multiple ways \cite{gelman2013garden}. 

\begin{figure}
    \centering
    \includegraphics[width=0.7\textwidth]{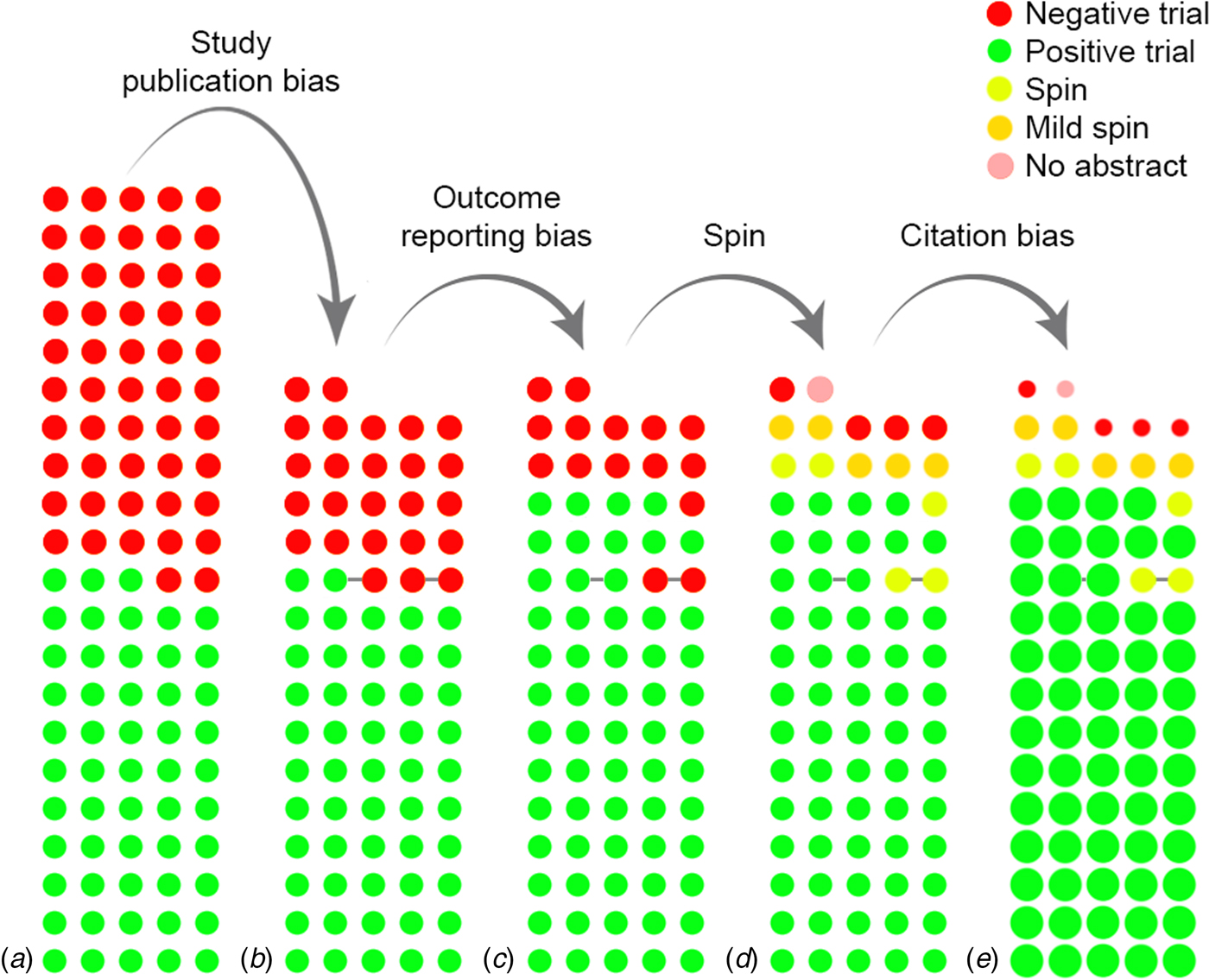}
    \caption{The cumulative effect of reporting and citation biases on 105 pre-registered trials of anti-depressant drugs. From \cite{de2018cumulative}.}
    \label{fig:devriescumulative}
\end{figure}

Figure \ref{fig:devriescumulative}, from \cite{de2018cumulative}), illustrates nicely how these factors can cause reproducibility crises.
The left side of figure \ref{fig:devriescumulative} shows 105 trials of anti-depressant drugs that had, in accordance with U.S. law, been preregistered in an FDA database. About half (53/105) of these trials had statistically significant results and half (52/105) did not.
Even though 50\% of trials had negative results, only 5\% of published articles unambiguously reported a negative result:
\begin{quote}
    While all but one of the positive trials (98\%) were published, only 25 (48\%) of the negative trials were published [due to publication bias]\dots Ten negative trials, however, became `positive' in the published literature [due to outcome reporting bias], by omitting unfavorable outcomes or switching the status of the primary and secondary outcomes. \dots Among the remaining 15 (19\%) negative trials, five were published with spin in the abstract (i.e. concluding that the treatment was effective) \dots [and f]ive additional articles contained mild spin (e.g. suggesting the treatment is at least numerically better than placebo). \cite{de2018cumulative}
\end{quote}
To make matters worse, articles with positive results were cited three times as often as those with negative results (citation bias).
In each step, a bias towards reporting positive outcomes skews the accuracy and reliability of the scientific literature.

\subsubsection{Reproducibility issues in ML}

ML has also experienced reproducibility issues \cite{hutson2018artificial}.
For example, \cite{sculley2018winner} performs an informal meta-analysis of five articles in ML research which concluded that prior work had come to incorrect conclusions, either overestimating the performance of newly proposed models due or misunderstood the source of empirical improvements, due to insufficient hyperparameter tuning or a lack of ablation studies.
They argue that a major cause is a research culture which emphasizes wins -- introducing a new method that beats previous methods on an established benchmark problem -- rather than empirical rigor.
\cite{lin2019neural} agrees, describing how research in neural information retrieval tends to compare ``to leaderboards selectively populated by entries of the authors’ choosing, where better results are often ignored.''
In addition to the cultural origins of these issues, many methodological decisions can lead to irreproducibility in ML research. 
\cite{gundersen2022sources} comprehensively reviews these factors, dividing the causes of irreproducibility into 41 categories. 

There are also reproducibility issues in research that uses ML for science.
Compiling evidence from 22 articles across 17 fields analyzing reproducibility issues in 294 articles, \cite{kapoor2023leakage} argues that there is a `reproducibility crisis' in ML-based science. 
\cite{demasi2017meaningless} performs a meta-analysis of articles that use ML to predict mental wellbeing, and finds that 77\% (30/39) perform ``meaningless comparisons'' to weak baselines, producing ``baseless optimism''. 
\cite{roberts2021common} performs a systematic review of articles that use ML for diagnosis or prognosis of COVID-19 from CT scans or CXR images, and finds that ``none of the models identified are of potential clinical use due to methodological flaws and/or underlying biases.''
\cite{wynants2020prediction} performs a systematic review of data-driven prediction models for COVID-19 prognosis, finding that 90\% (545/606) had high risk of bias (mostly due to small sample sizes or incomplete evaluation).
Common pitfalls in ML-based science include data leakage \cite{kapoor2023leakage,whalen2022navigating}, poor data quality \cite{kapoor2023leakage,roberts2021common,artrith2021best}, weak baselines \cite{demasi2017meaningless}, and insufficient external validation \cite{roberts2021common,wynants2020prediction}. In each case, pitfalls result in overoptimistic assessments about the performance of ML.

\subsubsection{ML for solving PDEs}

In recent years, there has been interest in using ML to advance research into partial differential equations (PDEs) \cite{thuerey2021physics,brunton2023machine,vinuesa2022enhancing,karniadakis2021physics,cuomo2022scientific,duraisamy2019turbulence}. Scientists and engineers study PDEs because they accurately model the behavior of many physical systems.
PDEs relate the output of a function to partial derivatives with respect to input variables (usually space and/or time).
ML-for-PDE research has mostly focused on either solving `forward' well-posed boundary value problems, `inverse' ill-posed problems that use data to infer equation parameters or missing data \cite{karnakov2024solving}, or `reduced order models' that learn low-dimensional representations from data \cite{brunton2023machine}.

Computational scientists have spent decades developing numerical algorithms to approximate the solution of forward PDE problems \cite{durran2013numerical,leveque1992numerical} -- we call these algorithms `standard numerical methods' or `standard solvers' -- but there is great interest in using ML to do so more efficiently \cite{brunton2023machine,vinuesa2022enhancing}.
Standard numerical methods have a basic trade-off between accuracy and efficiency; computing a more accurate approximation takes more time.
In principle, ML could be used to learn new algorithms or surrogate models that reduce the time required to output an approximate solution compared to standard solvers.
Faster ML-based solvers could be useful for downstream applications such as optimization, inverse problems, or uncertainty quantification \cite{mishra2018machine}, and to improve or even replace the standard numerical methods used in simulation codes for research and commercial applications \cite{kochkov2021machine}.
Indeed, many articles claim to have used ML to accelerate the solution of PDEs. These articles compare to standard numerical methods which serve as baselines.

To be useful as a surrogate model, an ML-based solver must reduce the total computational cost in downstream applications.
This includes the cost of generating data and training models, both of which are unaccounted for when comparing speed to standard solvers \cite{kadapa2021machine}.
There are also serious concerns about the accuracy and numerical stability of ML-based solvers when generalizing to a new parameter space, areas where standard solvers excel \cite{ross2023benchmarking}.
Increased speed seems to be the one area where ML-based solvers might have an advantage over standard solvers.
Speed is thus necessary, but not sufficient, to be useful for forward problems.

The results in this Analysis call into question whether ML has actually been as successful at solving PDEs from fluid mechanics and related fields as the scientific literature would suggest. We identify two issues, weak baselines and reporting biases, that lead to overoptimism and affect the interpretation reproducibility \cite{gundersen2021fundamental} of ML-for-PDE solving research. To determine the frequency of weak baselines, we conduct a systematic review of research that uses ML to accelerate the solution of fluid-related PDEs. We observe two common pitfalls with baseline comparisons and identify them in a majority (79\%) of published articles. 
We then use anecdotal and statistical evidence from the systematic review to argue that reporting biases are causing negative results to be underreported.
Due to these reproducibility issues, we caution that at present the scientific literature is not a reliable source for evaluating the success of ML at solving PDEs.

\section{Weak baselines \label{sec:weakbaselines}}

ML-based PDE solvers use neural networks, deep learning, and other techniques from ML to output approximate solutions to forward PDE problems.
To determine whether ML can improve the efficiency of PDE solvers, ML-based solvers must be compared to standard numerical methods which serve as a baseline.
For these comparisons to reach accurate conclusions, they must be fair.
A fair comparison should not overestimate or underestimate the performance of either method.
Underestimating the performance of a baseline is called using a weak baseline. In ML-for-PDE solving research, we've observed two common pitfalls that lead to weak baselines. As a result, we introduce two rules (rule 1 and rule 2) that must be followed for comparisons between ML-based solvers and standard numerical methods to be fair. 
We also introduce three additional recommendations for fair comparisons (see \cref{sec:methodsreview}) but do not require that they be satisfied to ensure a fair comparison.

The first pitfall is to compare the efficiency of the (highly accurate) standard solver used to generate the training data to that of the (less accurate) ML-based solver.
The problem is that the standard solver can trade off accuracy for runtime, making such comparisons meaningless. 
Comparing two methods with different accuracies could, for example, lead to the nonsensical conclusion that a method is orders of magnitude faster than itself.
Thus, rule 1 is to make comparisons at either equal accuracy or equal runtime.
To satisfy rule 1, either (a) reduce the resolution and/or the number of iterations of the standard solver until the two methods have either equal accuracy or equal runtime (or a proxy for runtime), or (b) demonstrate that reducing the resolution of the standard solver any further would give worse accuracy than the ML-based solver.
This must be done even when the two solutions look qualitatively similar, because a lower resolution baseline will often look qualitatively similar as well.

The second pitfall is to compare to a numerical method which is much less efficient than a state-of-the-art method for that problem.
State-of-the-art numerical methods can be orders of magnitude faster than less efficient numerical methods.
Choosing the right algorithm for a given PDE can be difficult and requires a combination of expertise, background knowledge, and effort.
Thus, rule 2 is to compare to an efficient numerical method.
The methods that satisfy rule 2 depend on the particular PDE being solved; we discuss the criteria we used to evaluate rule 2 later in this section.

How frequently does the ML-for-PDE solving literature compare to weak baselines? To answer this question, we perform a systematic review \cite{aromataris2014systematic}. This systematic review attempts to find every article that (i) uses ML to solve a fluid-related PDE and (ii) compares speed, or some proxy for speed, to a standard numerical method.
We define additional inclusion and exclusion criteria in \cref{sec:methodsreview}.
We restrict ourselves to PDEs that are related to fluid mechanics both because these PDEs have received the most attention but also because this is our area of expertise. 

We find 82 articles (see Supplementary Information) matching the inclusion criteria. 76 articles claim to outperform and 4 claim to underperform relative to a standard numerical method \cite{magiera2020constraint,bezgin2021data,dresdner2022learning,toshev20233}, while 2 claim to have similar or varied performance \cite{discacciati2020controlling,dong2022computing}.
For each article that claims to outperform a standard numerical method, we ask whether that article's `primary outcome' (defined in \cref{sec:methodsreview}) followed rules 1 and 2.
We evaluate each rule using a three-point scale. We give a check ({\cmark}) if the rule was satisfied, or if we are unsure. We give an `X' ({\xmark}) if the rule was not satisfied. 
We give a warning sign ({\dangersign}) if the rule was only partially satisfied (rule 1) or if we believe the rule was likely not satisfied but we don't have enough evidence to say for sure (rule 2).
For rule 2, we gave an ({\xmark}) if
\begin{itemize}
    \item we could replicate the article’s primary outcome and achieve significantly improved performance with a different baseline (six articles).
    \item the article reported performance relative to a weaker baseline in the abstract but a stronger baseline in the results section or the appendix (three articles).
    \item the article used a 2D code as a baseline for a 1D problem (one article).
    \item the computational implementation of a state-of-the-art numerical method was orders of magnitude slower than our implementation of that method (one article).
    \item the baseline used implicit timestepping when explicit timestepping would have been faster (one article).
    \item for elliptic PDEs, the baseline method was much less efficient than a state-of-the-art numerical method (eight articles).
    \item for advection-dominated PDEs, the baseline method was much less efficient than a state-of-the-art numerical method (five articles).
\end{itemize}
For elliptic and advection-dominated PDEs, which have been the focus of our replication efforts, we explain which numerical methods we consider state-of-the-art in \cref{sec:methodsreview}.
Eight articles received a ({\dangersign}) for rule 2, either because (a) we believe the numerical method used is much less efficient than a state-of-the-art method, but we haven't performed a direct comparison for that PDE so we can't say for sure, or (b) the general-purpose software package being used as a baseline has been shown to be computationally slow relative to other solvers for that PDE, but we haven't performed direct comparisons ourselves and so we can't say for sure. One article received a ({\dangersign}) for rule 1 for reducing the resolution of the highly accurate numerical method, but not by enough to compare at equal accuracy.

We list the 76 articles that claim to outperform a standard numerical method in Table \ref{tab:baselines}, ordered from highest to lowest number of citations.
60/76 (79\%) receive an ({\xmark}) for rule 1 and/or rule 2 and thus compare to a weak baseline. 2/76 (2.6\%) receive a ({\dangersign}), indicating that they may be comparing to a weak baseline and the performance claims should be treated with caution. 14/76 (18.4\%) receive a ({\cmark}), indicating that we believe they compare to a strong baseline. See Supplementary Information for detailed explanations of each entry in Table \ref{tab:baselines}.
Articles which receive a ({\cmark}) tend to have quantitatively smaller relative improvements than articles which receive an ({\xmark}), suggesting that the more impressive the result, the more likely the article used a weak baseline. 

\afterpage{
\setlength{\tabcolsep}{2pt}
\renewcommand{\arraystretch}{.75}
\begin{longtable}{@{\extracolsep\fill}llllccc}
\caption{Weak baselines in ML-for-PDE solving research.}\label{tab:baselines}\\
\endfirsthead
\toprule
\endhead
\bottomrule
\endfoot
\endlastfoot
\toprule
Article & Cited & PDE & Primary outcome & Rule 1 & Rule 2 & Fair? \\
\midrule
    \fnoli{ }\citefnoli& \fnolicitations & a,b &up to $10^3 \times$ faster& \xmark & \cmark & \xmark \\
    \deeponetlulu{ }\citedeeponetlulu& \deeponetlulucitations & c,d& substantially lower cost & \xmark & \cmark & \xmark \\
    \tompson{ }\citetompson& \tompsoncitations & e &$4.1\times$ faster& \cmark & \cmark & \cmark \\
    \mlacceleratedcfd{ }\citemlacceleratedcfd& \mlacceleratedcfdcitations & b & 8-10$\times$ coarser, 40-80$\times$ speedup & \cmark & \xmark & \xmark \\
    \meshbasedpfaff{ }\citemeshbasedpfaff& \meshbasedpfaffcitations & f,g & $10^1$-$10^2\times$ faster& \xmark & \cmark& \xmark \\
    \barsinai{ }\citebarsinai& \barsinaicitations & a,h,i &4-8$\times$ coarser& \cmark & \xmark & \xmark \\
    \deepfluids{ }\citedeepfluids& \deepfluidscitations & j &700$\times$ faster& \xmark & \cmark & \xmark \\
    \deeponetwang{ }\citedeeponetwang& \deeponetwangcitations & a &up to $10^3\times$ faster& \xmark & \cmark & \xmark \\
    \solverinthe{ }\citesolverinthe& \solverinthecitations &a,b,e,f&68$\times$ speedup& \xmark & \cmark & \xmark  \\
    \deepmmnetcvt{ }\citedeepmmnetcvt & \deepmmnetcvtcitations & k &$10^4\times$ speedup& \xmark & \cmark& \xmark \\
    \belbuteperez{ }\citebelbuteperez & \belbuteperezcitations & g & substantial speedup & \xmark & \cmark & \xmark \\
    \pinoli{ }\citepinoli& \pinolicitations & a,b & orders of magnitude speedup & \cmark & \xmark & \xmark \\
    \projectionyang{ }\citeprojectionyang& \projectionyangcitations & e & drastic speedup & \xmark & \xmark & \xmark \\
    \neuralconverge{ }\citeneuralconverge& \neuralconvergecitations & e &2-3$\times$ speedup& \cmark & \xmark & \xmark \\
    \messagepassing{ }\citemessagepassing& \messagepassingcitations & a,j,l &outperforms state-of-the-art & \cmark & \xmark & \xmark \\
    \deepmmnet{ }\citedeepmmnet&\deepmmnetcitations & m & over $10^5\times$ faster & \xmark & \xmark & \xmark \\
    \frameworkmishra{ }\citeframeworkmishra& \frameworkmishracitations & a,c,n &significant gain in efficiency& \cmark & \xmark & \xmark \\
    \optimizemultigrid{ }\citeoptimizemultigrid& \optimizemultigridcitations & o & improved convergence rates & \cmark& \cmark & \cmark \\
    \donga{ }\citedonga& \dongacitations & a,c,e,t & often exceeds performance & \cmark & \xmark & \xmark \\
    \rayb{ }\citerayb& \raybcitations & a,c,n &outperforms TVB limiter& \cmark & \cmark & \cmark \\
    \novelcnn{ }\citenovelcnn& \novelcnncitations & e & $10^2\times$ faster projection step & \xmark & \xmark& \xmark \\
    \wandel{ }\citewandel& \wandelcitations & f & $11$/$40\times$ faster on CPU/GPU & \xmark & \cmark & \xmark \\
    \shan{ }\citeshan& \shancitations & e & significant speedup & \xmark & \xmark & \xmark \\
    \algebraic{ }\citealgebraic& \algebraiccitations & o & improved convergence rates & \cmark & \cmark & \cmark \\
    \zhuang{ }\citezhuang& \zhuangcitations & c & 4$\times$ lower resolution & \cmark & \xmark & \xmark \\
    \pathak{ }\citepathak& \pathakcitations & b & lower resolution & \cmark & \cmark & \cmark \\
    \leoni{ }\citeleoni& \leonicitations & p & very small cost & \xmark & \cmark & \xmark \\
    \fnodeformation{ }\citefnodeformation& \fnodeformationcitations & b,g & $10^5\times$ faster & \xmark & \cmark & \xmark \\
    \stevensa{ }\citestevensa& \stevensacitations & a,c,n &outperforms WENO& \cmark & \cmark & \cmark \\
    \illarramendi{ }\citeillarramendi& \illarramendicitations & e & 3.2$\times$ faster & \cmark & \xmark & \xmark \\
    \stachenfeld{ }\citestachenfeld& \stachenfeldcitations & b,q & outperforms state-of-the-art & \cmark & \dangersign & \dangersign \\
    \han{ }\citehan& \hancitations & f,q & 100-800$\times$ speedup & \xmark & \dangersign & \xmark \\
    \stevensb{ }\citestevensb& \stevensbcitations & a,c,i & 2-3$\times$ lower error & \cmark & \cmark & \cmark \\
    \ozbay{ }\citeozbay& \ozbaycitations & e & improved preconditioner & \cmark & \cmark & \cmark \\
    \zili{ }\citezili& \zilicitations & j & 5-8$\times$ acceleration & \cmark & \xmark & \xmark \\
    \peng{ }\citepeng& \pengcitations & b & 8000$\times$ speedup & \xmark & \xmark & \xmark \\
    \chen{ }\citechen& \chencitations & b & 300-600$\times$ speedup & \xmark & \dangersign & \xmark \\
    \alguacilb{ }\citealguacilb& \alguacilbcitations & l & 15.5$\times$ speedup & \xmark & \dangersign & \xmark \\
    \wandelb{ }\citewandelb& \wandelbcitations & f &  considerably faster & \xmark &  \cmark & \xmark \\
    \blist{ }\citeblist& \blistcitations & b & 14.4$\times$ speedup & \cmark & \xmark & \xmark \\
    \cheng{ }\citecheng& \chengcitations & e & 2$\times$ faster & \cmark & \cmark & \cmark \\
    \wen{ }\citewen& \wencitations & n,r & fewer grid points & \cmark & \cmark & \cmark \\
    \delaraa{ }\citedelaraa& \delaraacitations & a & significant cost savings & \xmark & \cmark & \xmark \\
    \zhao{ }\citezhao& \zhaocitations & f,l & 8$\times$ or $35\times$ faster & \dangersign & \cmark & \dangersign \\
    \assessments{ }\citeassessments& \assessmentscitations & e & 10-25$\times$ faster & \cmark & \xmark & \xmark \\
    \holloway{ }\citeholloway& \hollowaycitations & s & $270\times$ speedup & \xmark& \cmark& \xmark \\
    \azulay{ }\citeazulay& \azulaycitations & t & favorable runtime on GPU & \cmark & \cmark & \cmark \\
    \wulatent{ }\citewulatent& \wulatentcitations & a,b & 840$\times$ speedup & \xmark & \cmark & \xmark \\
    \liu{ }\citeliu& \liucitations & a,b & 10-60$\times$ speedup & \xmark & \xmark & \xmark \\
    \zhang{ }\citezhang& \zhangcitations & e,t & up to $\mathcal{O}(10^2)$ more efficient & \cmark & \xmark & \xmark \\
    \duarte{ }\citeduarte& \duartecitations & u & $10^4\times$ faster & \xmark & \cmark & \xmark \\
    \alguacil{ }\citealguacil& \alguacilcitations & l & 141$\times$ acceleration & \xmark & \dangersign & \xmark \\
    \bezginb{ }\citebezginb& \bezginbcitations & c,n & similar or better performance & \cmark & \xmark & \xmark \\
    \shang{ }\citeshang& \shangcitations & e,l & much more accurate & \cmark & \cmark & \cmark \\
    \kube{ }\citekube& \kubecitations & v & 25\% fewer iterations & \xmark & \cmark & \xmark \\
    \shi{ }\citeshi& \shicitations & b,e & over $50\times$ faster & \xmark & \cmark & \xmark \\
    \ranadea{ }\citeranadea& \ranadeacitations & w & over 200$\times$ speedup & \xmark & \cmark & \xmark \\
    \chenb{ }\citechenb& \chenbcitations & e & fewer iterations & \cmark & \xmark & \xmark \\
    \ranadeb{ }\citeranadeb& \ranadebcitations & b,w,x & 40-100$\times$ faster & \xmark & \cmark & \xmark \\
    \pengb{ }\citepengb& \pengbcitations & b & 20$\times$ speedup & \xmark & \cmark & \xmark \\
    \delarab{ }\citedelarab& \delarabcitations & q & 4-5$\times$ faster & \cmark & \cmark & \cmark \\
    \ranadec{ }\citeranadec& \ranadeccitations & e,w,x & 40-50$\times$ faster & \xmark & \cmark & \xmark \\
    \fang{ }\citefang& \fangcitations & f & 38.5\% faster & \xmark & \cmark & \xmark \\
    \shukla{ }\citeshukla& \shuklacitations & g & 32,253$\times$ speedup & \xmark & \cmark & \xmark \\
    \zhangb{ }\citezhangb& \zhangbcitations & l & around $10\times$ faster & \xmark & \cmark & \xmark \\
    \bezgin{ }\citebezgin& \bezgincitations & q & outperforms Rusanov flux & \cmark & \xmark & \xmark \\
    \yang{ }\citeyang& \yangcitations & l & nearly $10^2\times$ faster & \xmark & \cmark & \xmark \\
    \tang{ }\citetang& \tangcitations & e & up to $12\times$ speedup & \cmark & \xmark & \xmark \\
    \nastorg{ }\citenastorg& \nastorgcitations & e & $10\times$ faster & \xmark & \cmark & \xmark \\
    \gopakumar{ }\citegopakumar& \gopakumarcitations & y & $10^6\times$ faster & \xmark & \dangersign & \xmark \\
    \shit{ }\citeshit& \shitcitations & d & 19.2\% faster & \cmark & \xmark & \xmark \\
    \suforecast{ }\citesuforecast& \suforecastcitations & q & over $10^3\times$ faster & \xmark & \cmark & \xmark \\
    \jeon{ }\citejeon& \jeoncitations & b & 1.8$\times$ acceleration & \xmark & \dangersign & \xmark \\
    \dai{ }\citedai& \daicitations & f & $10^4$-$10^5\times$ faster & \xmark & \dangersign & \xmark \\
    \sun{ }\citesun& \suncitations & e,t & better accuracy & \cmark & \cmark & \cmark \\
    \shao{ }\citeshao& \shaocitations & e & improves convergence & \cmark & \xmark & \xmark \\
\bottomrule
\caption*{\footnotesize{PDEs: (a) Burgers' (b) incompressible Navier-Stokes (INS) (c) advection (d) advection-diffusion (e) Poisson (f) INS wake dynamics (g) compressible Navier-Stokes airfoil wing (h) Korteweg-de Vries (i) Kuramoto-Sivashinsky (j) INS graphics/particle-based (k) electroconvection (l) wave (m) reacting Navier-Stokes (n) Euler (o) elliptic diffusion (p) parabolized stability equations (q) compressible Navier-Stokes (r) shallow water (s) Boltzmann collision operator (t) Helmholtz (u) black hole hydrodynamics (v) particle-in-cell (w) convective heat transfer (x) Laplace (y) magnetohydrodynamics}}
\end{longtable}
\clearpage
}

We reproduce results from ten articles using a stronger baseline; these articles are listed in Table \ref{tab:reproducebaselines}.
We primarily focus on reproducing results from highly cited articles solving 1D and 2D PDEs in regular geometries.
4/10 articles violate rule 1, while at least 6/10 violate rule 2 because they use an inefficient numerical method and/or an inefficient implementation.
In 9/10 cases the stronger baseline is at least two orders of magnitude faster than the slower baseline; the exception is article 6 \cite{bar2019learning}.
In 7/10 cases the stronger baseline outperforms the ML-based solver; the exceptions are articles 1 \cite{li2020fourier}, 6 \cite{bar2019learning}, and 12 \cite{li2021physics}. In \cref{sec:methodsreview}, we give additional details about each of the stronger baselines listed in Table \ref{tab:reproducebaselines}. In Appendix A, we give additional details about the results of each reproducibility experiment. The code, instructions for computationally reproducing our experiments, and explanations for interpreting the computational results can be found at \url{https://github.com/nickmcgreivy/WeakBaselinesMLPDE/}.

\afterpage{
\begin{table}
\setlength{\tabcolsep}{2pt}
\begin{tabular}{@{}lllp{22mm}p{22mm}p{30mm}p{30mm}@{}}
\toprule
Article & Cited  & PDE & Weaker \newline baseline & Stronger \newline baseline & Old \newline outcome & New \newline outcome \\
\midrule
\fnoli{ }\citefnoli & \fnolicitations & b & PS $64\times64$ & DG2 $7\times7$ & $10^3\times$ faster & $7\times$ faster \\
\deeponetlulu{ }\citedeeponetlulu & \deeponetlulucitations & c  & FD $n_x$=100 & DG2 $n_x$=13 & $24\times$ faster & $10\times$ slower \\
\mlacceleratedcfd{ }\citemlacceleratedcfd & \mlacceleratedcfdcitations & b & FV & PS & 80$\times$ faster & slightly slower$^\dagger$ \\
\barsinai{ }\citebarsinai& \barsinaicitations & a & WENO & DG2/DG3 & 4-8$\times$ fewer DOF & 2-4$\times$ fewer DOF\\
\deeponetwang{ }\citedeeponetwang& \deeponetwangcitations & a & SP $n_x$=100 & FV $n_x$=100 & $10^3\times$ faster & 10$\times$ slower \\
\pinoli{ }\citepinoli& \pinolicitations & b & PS $64\times64$ & DG2 $3\times3$ & $10^3\times$ faster & $7\times$ faster \\
\neuralconverge{ }\citeneuralconverge& \neuralconvergecitations & e  & MG & LU & faster & $10^3\times$ slower \\
\messagepassing{ }\citemessagepassing& \messagepassingcitations & a,l & WENO, PS & WENO, FV & much faster & $10^3\times$ slower \\
\delaraa{ }\citedelaraa& \delaraacitations & a & DG28 $n_x$=1 & DG9 $n_x=1$ & 22-75$\times$ faster & 4-10$\times$ slower \\
\tang{ }\citetang& \tangcitations & e & CG \& MG & LU & $12\times$ faster & 35-500$\times$ slower \\
\bottomrule
\end{tabular}%
\caption{Reproducing results in ML-for-PDE solving research using stronger baselines. \newline PDEs: (a) 1D Burgers' (b) 2D incompressible Navier-Stokes (c) 1D advection (e) 2D Poisson (l) 1D wave. Numerical methods: (PS) pseudo-spectral (FD) second-order finite-difference (DG2) Discontinuous galerkin, polynomial basis functions of order 2 (FV) finite volume (SP) spectral (MG) multigrid (LU) LU decomposition (CG) conjugate gradient. Abbreviations: (DOF) degrees of freedom ($n_x$) number of cells in the $x$ direction ($64\times64$) 64 cells in both the $x$ and $y$ directions. \newline 
{$\dagger$} This result on GPU is consistent with \cite{dresdner2022learning} who reproduce the result on TPU.
}\label{tab:reproducebaselines}
\end{table}
\clearpage
}

\subsection{Details of systematic review \label{sec:methodsreview}}
A systematic review attempts to answer a predefined research question by collecting and analyzing evidence from all available research studies on the topic. Our research question is: how frequently does the ML-for-PDE solving literature compare to weak baselines?

\subsubsection{Inclusion criteria}

To restrict the scope of the systematic review to our area of expertise, we only consider articles that use ML to output an approximate solution to one or more fluid-related PDEs.
The PDEs we include in the review are: advection, advection-diffusion, Burgers', Euler, Navier-Stokes, reacting Navier-stokes, advection-diffusion-reaction, Korteweg–de Vries (KdV), Kuramoto–Sivashinsky (KS), shallow water, parabolized stability equations, Poisson, wave, elliptic diffusion, Helmholtz, Laplace, convective heat transfer, plasma models including MHD, PIC, \& Hasegawa-Wakatani, particle-based fluid dynamics, Boltzmann or plasma collision operators, \& black hole hydrodynamics. 

We only include articles that compare the speed, computational cost, or some proxy for speed, of an ML-based solver to that of a standard numerical method used to solve that PDE. Examples of proxies for speed or cost include number of iterations or resolution in space or time. The comparison must be made in a figure, in a table, or in a quantitative statement in the text. A qualitative statement (e.g., ``our method is more efficient'') counts as a valid comparison if it is supported by quantitative or visual supporting evidence.

We define the `primary outcome' as follows. First, if the article has a quantitative comparison (e.g., ``$56\times$ faster'' or ``$4\times$ coarser'') in the abstract, we use that comparison. If no quantitative comparisons are made in the abstract, we look for quantitative comparisons in the conclusion, followed by the introduction, followed by the main text. If no quantitative comparison is made in the entire text, then we look for a qualitative comparison (e.g., ``significantly faster'' or ``outperforms'') beginning in the abstract, followed by the conclusion, followed by the introduction. If there are multiple quantitative or qualitative comparisons, we use our best judgement to determine which should count as the `primary outcome.'

We ended the search process on April 1st, 2023 and thus only include articles available on or before that date. We didn't find any articles published before 2016 that matched our inclusion criteria. Tables \ref{tab:baselines} and \ref{tab:reproducebaselines} show the number of citations each article has according to Google Scholar as of July 3rd, 2023.


\subsubsection{Exclusion criteria}

We exclude from the review articles that only consider PDEs related to solid mechanics, quantum mechanics, multiscale modeling, or other non-fluid-related topics. We exclude Reynolds-averaged Navier-Stokes (RANS) and large-eddy simulation (LES).
We also exclude the following PDEs and problems: weather, climate, Schrodinger, fractional, multiphase flows including gas-particle flow, Darcy flow, reaction-diffusion, Eikonal, parabolic diffusion, very high-dimensional PDEs, Compton scattering, meta-materials, hyper-elasticity, ice flow, vessel dynamics, and CO2 injection. 
We also exclude review articles, theses, presentations, technical reports, articles published in languages other than English, ill-posed \& inverse problems, backstepping \& control problems, surrogates for macroscopic quantities, stochastic differential equations, and ordinary differential equations (ODEs).
We exclude model order reduction (MOR) methods, including SVD-based methods such as proper orthogonal decomposition (POD), Sparse Identification of Non-linear Dynamics (SINDy) and Dynamic Mode Decomposition (DMD). 
We don't exclude kernel-based methods, though we didn't find any kernel-based solvers matching our inclusion criteria. 
We exclude physics-informed neural network (PINN)-based methods, because (a) standard numerical methods are known to outperform PINNs for solving forward problems, (b) the PINN literature is too vast to comprehensively review (e.g., \cite{raissi2019physics} has over 6,900 citations), and (c) we only know of a few articles that have ever reported superior performance with PINNs compared to standard numerical methods, and to the best of our knowledge all of these articles either compare to weak baselines or fail to account for the PINN optimization time.
Of the articles included in our systematic review, it turns out that ``machine learning'' invariably involved the use of neural networks and/or deep learning. 

We exclude articles that compare to no baselines or articles that compare to ML baselines but not a standard numerical method as a baseline.
We exclude articles that compare the accuracy of ML-based solvers with standard numerical methods but not the speed or computational cost.
We exclude six articles \cite{ray2018artificial,wang2023long,ovadia2021beyond,li2021learning,ni2023numerical,dong2021modified} that make a qualitative statement of comparison (e.g., ``our method is more efficient'') that are not supported by quantitative or visual supporting evidence about the relative computational cost.
We exclude five articles \cite{mueller2022leveraging,wang2021fast,schwander2021controlling,donon2020deep,wan2023evolve} that might implicitly be suggesting that their proposed method is more efficient on a fluid-related PDE but never make an explicit statement or comparison about the relative speed (or a proxy for speed).
We excluded four articles \cite{di2023neural,kovachki2021neural,holl2020phiflow,nemmen2020first,wandel2020unsupervised} for having duplicate results with other articles.
We excluded one article \cite{haridas2022deep} that uses neural networks to correct floating-point errors in a 16-bit simulation.

\subsubsection{Search process}

The process of systematically searching for every article matching our search criteria was informal at first, but eventually turned into a formal process that happened in two stages. In the first stage, we compiled in list A the names of every author we knew of who worked on ML and PDEs. For each author in list A, we used their Google Scholar profile to look at every title of every article published since 2016. If the title seemed potentially relevant to ML and PDE solving, we read the abstract. If the abstract suggested that the article might possibly satisfy our search criteria, we added it to list B. 
In the second stage, we used Google Scholar to find every article that cites one of two key articles \cite{kochkov2021machine,brandstetter2022message}. We read the title and abstract of each article. If the abstract suggested that the article might satisfy our search criteria, we added it to list B. 
We also tried using Google Scholar to search for key words such as ``machine learning physics", ``machine learning partial differential equations'', ``machine learning fluids'', ``machine learning accelerate pde'', etc. This third approach did not discovery any new articles that were not already added to list B.

For every article added to list B, we read the introduction and conclusion. We also searched the text for key words such as ``fast'', ``speed'', ``improve'', ``pde'', ``equation'', ``compare'', etc., to determine whether the article might have matched the inclusion criteria. Articles that were once again deemed to potentially match the inclusion criteria were read fully to determine whether they should be included in the systematic review. If we found a citation to a new article that might match the inclusion criteria, we added it to list B as well. We also added every co-author of every article that matched the inclusion criteria to list A.

We didn't count the exact number of titles or abstracts we read in total. We added 258 authors to list A and 358 articles to list B. 82 of the articles in list B matched our inclusion criteria.

While we did our best to find every article matching our search criteria, it is possible we missed some articles. If we missed any articles, they are likely articles with fewer citations and/or articles that didn't cite a few key articles.

\subsubsection{Criteria for evaluating baselines \label{sec:methods_evaluation_rules}}

We introduce two necessary but not sufficient conditions (rules 1 and 2) which must be satisfied to ensure a fair comparison between a ML-based PDE solver and a standard numerical method. Rule 1 is to make comparisons at either equal accuracy or equal runtime. Rule 2 is to compare to an efficient numerical method. These rules are discussed in detail in section \ref{sec:weakbaselines}.

We also introduce three recommendations that we recommend following, but do not require that they be satisfied to ensure a fair comparison.
Recommendation 1 is to be cautious of comparing between general-purpose tools and specialized algorithms.
In order to solve a wide class of PDEs, general-purpose libraries are forced to make choices that trade off efficiency for robustness, making them suboptimal for many PDEs.
In contrast, ML-based solvers are specialized algorithms trained to be optimal for a specific PDE or a narrow class of problems. 
Comparisons between a specialized ML-based solver and a general-purpose library are likely to be unfairly biased in favor of the specialized solver.
None of the articles we found explicitly mentioned any reasons to be wary of comparisons between general-purpose tools and the highly specialized ML-based solver.
Moving forward, we encourage articles to be cautious about making these comparisons and to warn readers of the potential for an unfair comparison.

Recommendation 2 is to justify why the choice of hardware (CPU/GPU/TPU) used for comparison is fair.
Some methods, including neural networks, achieve significant reductions in runtime using graphics processing units (GPU) or tensor processing units (TPU) rather than central processing units (CPU).
Other methods achieve only minimal speedups, or no speedup at all, using GPU/TPU compared to CPU. 
Some methods are not implemented on GPU/TPU, only CPU.
In practice, what type of hardware to use for a fair comparison can be context-dependent and to some extent subjective.
Usually, GPU-GPU or TPU-TPU comparisons will be most fair, but in some contexts CPU-CPU or CPU-GPU comparisons can be considered fair.
Most of the articles we found made reasonable choices for the hardware used when comparing different methods.
Moving forward, rather than making definitive rules regarding the choice of hardware, we encourage articles to explain why they chose the hardware they did and to justify why that choice is fair.

In order to account for the cost of generating data and training models, recommendation 3 is to report the number of surrogate evaluations $N$ needed to reduce the total computational cost in downstream applications. $N$ is only defined if the ML-based solver is faster than the numerical baseline. $N$ can be computed using the formula
\begin{equation}
    C_{\textnormal{data}} + C_{\textnormal{train}} + N \frac{t_B}{s} = N t_B
\end{equation}
where $C_{\textnormal{data}}$ is the time required to generate the training data, $C_{\textnormal{train}}$ is the time required to train the model(s), $t_B$ is the time required for the standard numerical method baseline to compute one surrogate evaluation at equivalent accuracy to the ML model, and $s$ is the speedup of the ML-based solver relative to the numerical baseline.

We now explain which standard numerical methods we consider state-of-the-art for elliptic and advection-dominated PDEs.
For elliptic PDEs (such as Poisson’s, Laplace’s, or Helmholtz equations) finite element methods (FEM) are standard; direct solvers such as LU decomposition are most efficient for small problems, while iterative solvers are more efficient for large problems. For elliptic PDEs, multigrid solvers are typically state-of-the-art for large problems. We suggest using Eigen \cite{guennebaud2010eigen} for LU decomposition and HYPRE \cite{falgout2002hypre} for multigrid methods, though other libraries can also be extremely efficient. 
For advection-dominated PDEs, we recommend using second or higher-order shock-capturing finite volume (FV) methods for problems with shocks (such as compressible Navier-Stokes and Burgers’ equations), while using higher-order methods for problems with smooth solutions (such as the advection equation with smooth solutions, the incompressible Navier-Stokes equations or the compressible Navier-Stokes equations in the weakly compressible limit). For advection-dominated problems with smooth solutions, pseudo-spectral methods are usually state-of-the-art when applicable, though discontinuous Galerkin (DG) methods are also extremely efficient. We have found that higher-order DG methods (polynomial order 2 or higher) work better than lower-order DG methods (order 0 or 1), though there are diminishing returns for using very high-order (polynomial order 3 or higher) DG methods. Moving forward, for the Navier-Stokes equations we recommend comparing the performance of FV, DG, and (if applicable) spectral methods and choosing the strongest baseline.
In general, first-order methods should not be used as baselines for fluid-related PDEs. First-order methods tend to be extremely diffusive and require high grid resolution.
Explicit time-stepping schemes are usually preferred for advection-dominated PDEs; for these problems, using implicit time-stepping typically leads to inefficient numerical methods. The time-step restriction from the Courant–Friedrichs–Lewy (CFL) condition is usually sufficiently small that the dominant error is from spatial discretization, and so the choice of explicit time-stepping scheme is less important than the spatial discretization.
However, in some cases (such as fluid-structure interaction \cite{mayr2018adaptive}) the temporal discretization errors can dominate, in which case specialized time-stepping schemes that compute error estimates and perform adaptive time-stepping perform best \cite{reynolds2014arkode}.

To help ensure that we applied these rules fairly, we emailed the authors of each article to give them an opportunity to point out any errors we might have made in applying rules 1 and 2. We received 15 responses about 23 articles. Seven responses expressed agreement and gave suggestions for improvement, two provided additional information, and six expressed disagreement. Based on the responses, we modified six entries in Table \ref{tab:baselines} and one entry in Table \ref{tab:reproducebaselines}.

\subsubsection{Details of stronger baselines in Table \ref{tab:reproducebaselines}}

For articles 1, 4, and 12 we use a Runge-Kutta discontinuous galerkin (RKDG) method to solve the 2D incompressible Navier-Stokes equations on a periodic domain. We use a third-order strong-stability preserving (SSPRK3) ODE integration \cite{gottlieb2001strong}. We use second-order discontinuous polynomial basis functions with a serendipity basis function which uses 8 basis functions per grid cell. We use LU decomposition and a continuous Galerkin formulation to solve the Poisson equation at each Runge-Kutta stage. The full scheme is explained in \cite{hakim2019discontinuous}. \cite{dresdner2022learning} also reproduces article 4 using a pseudo-spectral implementation. The details can be found at \url{https://github.com/google/jax-cfd}.
For articles 2 and 6 we use again a RKDG scheme, except this time to solve the 1D advection and 1D Burgers' equations with periodic boundary conditions. We again use SSPRK3 ODE integration and second-order Legendre polynomial basis functions. The full scheme is explained in \cite{cockburn1989tvb}. For article 43 we again use an RKDG scheme to solve the 1D Burgers' equation, except with Dirichlet instead of periodic boundary conditions.
For articles 8 and 15 we use a finite volume scheme with Godunov flux \cite{durran2013numerical} and SSPRK3 ODE integration to solve the 1D Burgers' equation with periodic boundary conditions.
For articles 14 and 68 we solve the Poisson equation on a square periodic domain using a continuous Galerkin formulation with linear basis functions. We use an LU decomposition to solve the resulting linear system.
For article 15 we solve the 1D wave equation with Dirichlet boundary conditions, using SSPRK3 timestepping and a finite-volume method with irregular grid spacing.

\section{Reporting biases \label{sec:reportingbiases}}

Reporting biases is an umbrella term for a set of biases that can arise when the analysis, reporting, and/or interpretation of research findings are influenced by the nature and direction of results \cite{reportingbiascatalogue}.
Types of reporting biases include publication bias \cite{thornton2000publication}, spin bias \cite{boutron2018misrepresentation}, and different flavors of outcome reporting bias \cite{outcomereportingbiascatalogue} such as p-hacking \cite{head2015extent}, selective reporting \cite{saini2014selective}, outcome switching \cite{altman2017outcomeswitching}, and data-dredging \cite{erasmus2022dredging}.

Because reporting biases cause negative results to get suppressed \cite{de2018cumulative}, the percentage of positive results is believed to correlate with the frequency of reporting biases \cite{ritchie2020science,fanelli2010positive}.
To estimate the percentage of positive results in ML-for-PDE solving research, we analyzed a random sample of articles (see \cref{sec:methods-random-sample-ml-for-pde}). Of articles whose abstracts mention positive and/or negative experimental results, 94.8\% (220/232) mention only positive results, 5.2\% (12/232) mention both positive and negative results, and 0\% (0/232) mention only negative results.
This is an unusually high percentage of positive results compared to other fields of science \cite{fanelli2010positive} and motivates us to investigate whether reporting biases are causing negative results to be underreported in ML-for-PDE research.

During our systematic review, we found anecdotal and statistical evidence of publication bias and outcome reporting bias.
Of the 82 articles matching the inclusion criteria, 76 (93\%) claimed to outperform a standard numerical method baseline, while only 4 (5\%) claim to underperform relative to a baseline.
This suggests that many negative results are either not being published (publication bias) or being reported as positive due to outcome reporting bias.
Sure enough, a close reading of articles in the systematic review reveals evidence of outcome reporting bias, especially selective reporting and outcome switching: reporting the runtime of some PDEs but not others \cite{li2020fourier,wang2021learning,um2020solver,brandstetter2022message,dong2021local,shang2022deep}, only highlighting the results from the most successful PDE \cite{li2021physics,li2022fourier}, reporting performance in non-standard ways to seem more successful or to conceal a negative result \cite{bar2019learning,zhuang2021learned,kube2021machine,wang2021learning,stevens2020finitenet,dresdner2022learning,alguacil2021predicting,alguacil2022deep,shang2022deep}, or comparing to a stronger baseline in the main text but a weaker baseline in the abstract \cite{dong2021local,bezgin2022weno3,xiao2018novel}.
By cross-referencing with other articles, we also find evidence consistent with publication bias: some methods, which are successful on one PDE \cite{sanchez2020learning,wang2021learning,li2020fourier,bar2019learning}, either have worse performance when tested with different parameters or on different PDEs \cite{klimesch2022simulating,wang2022improved,gupta2022towards} or don't reproduce under nearly identical conditions \cite{mcgreivy2023invariant}.

We can more directly observe the collective effects of outcome reporting bias using a natural experiment in the ML-for-PDE solving literature.  
We gather two groups of articles, which we call sample A and sample B. 
Sample A includes all 76 articles in Table \ref{tab:baselines}.
Sample B is a random sample of 60 articles which use the so-called `physics-informed neural networks' (PINN) method \cite{raissi2019physics} to solve a fluid-related PDE.
There is one key difference between sample A and sample B: while every article in sample A claims to outperform a standard numerical method in speed or computational cost, the authors of every article in sample B believe that their ML-based solver underperforms relative to standard numerical methods
(see \cref{sec:methods-random-sample-pinn} for an explanation of why we can assume that they believe this).
If outcome reporting bias were not present, we would expect -- given that both samples try to solve fluid-related PDEs using ML and both report the accuracy of their proposed method relative to standard solvers -- that both samples would report the efficiency relative to standard solvers at similar rates.
Yet the percentage of articles that report this in the abstract is 80\% (61/76) in sample A and 8\% (5/60) in sample B. Only 12\% (7/60) of articles in sample B report the efficiency relative to standard solvers in the entire article.
In other words, when articles have a positive result they almost always highlight it, but when they have a negative result they rarely report it. 

\begin{figure}[ht]
\centering
\includegraphics[width=0.6\textwidth]{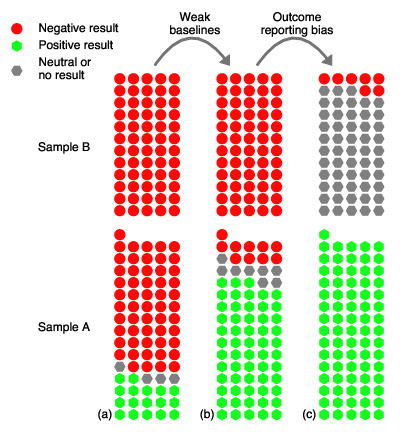}
\caption{The cumulative effects of weak baselines and reporting biases on samples A and B. Each circle or hexagon represents an article, while each color represents the result of comparing the relative speed and accuracy to a standard numerical method. In (a) we estimate what the results would be with strong baselines and without outcome reporting bias. (b) shows what the results would likely be without outcome reporting bias. (c) shows the results in the published literature.}
\label{fig:cumulativeeffects}
\end{figure}

Figure \ref{fig:cumulativeeffects} shows the cumulative effects of weak baselines and outcome reporting bias on samples A and B.
Weak baselines lead to overly positive results, while reporting biases lead to underreporting of negative results. The result is overoptimism about ML.

\subsection{Details of random sample of ML-for-PDE articles \label{sec:methods-random-sample-ml-for-pde}}

To approximate a random sample of articles in ML-for-PDE solving, we use Google Scholar to find all 400 articles that cite \cite{kochkov2021machine} as of June 18th, 2023. We include articles whose abstracts mention positive and/or negative experimental results. We define a result as: proposing a method to tackle a problem and using quantitative metrics or qualitative descriptions to describe the performance of the method on the problem.
We classify each abstract based on whether it mentions positive and/or negative results.
Many articles have the pattern `method A has negative aspects, we introduce method B which solves those negative aspects'; we don't consider this pattern a negative result unless the article discusses negative results associated with method B. 
Three articles comment that a method has limitations or is limited in scope; we do not count those comments as negative results. We exclude review articles, theses, duplicates, articles written in languages other than English, articles that don't mention experimental results anywhere in the abstract, and articles that are unrelated to ML or statistical learning. 
We excluded ten articles which mention experimental results, but don't give any indication as to whether those results are positive or negative.
232 out of 400 articles (58\%) in the sample match our inclusion criteria.
We report our classification for each article, including additional explanations for some articles, at \url{https://osf.io/gq5b3/}.

\subsection{Details of random sample of PINN articles \label{sec:methods-random-sample-pinn}}

Physics-informed neural networks (PINNs) \cite{raissi2019physics} are a popular method which can solve PDEs and inverse problems associated with PDEs. We look for a random sample of articles which (a) use PINNs to solve a fluid-related PDE, (b) which focus on solving ``forward'' problems (and are not focused on solving inverse or ill-posed problems), (c) which report the accuracy of the PINN-based solver, and (d) which generate their own data using a standard numerical method or general-purpose solver to measure accuracy. We define ``fluid-related'' using the same inclusion and exclusion criteria defined earlier. Once again, we exclude review articles, theses, duplicates, technical reports, and articles not written in English. 
We also exclude articles that use an analytic solution to measure accuracy.
If an article uses PINNs to solve both forward and inverse (i.e., ill-posed) problems, we only include the article if a majority of the problems it solves are forward problems.

To obtain a random sample of articles matching these criteria, we use Google Scholar to search within all 5,640 articles which cite the original PINN article \cite{raissi2019physics} as of June 21st, 2023. Using the ``search within citing articles'' function of Google Scholar, we type into the search bar

\begin{quote}
    ``PINN" AND ``Burgers" OR ``navier" OR ``stokes" OR ``fluid" OR ``advection" OR ``KdV" OR ``Kuramoto" OR ``Sivashinsky" OR ``Euler" OR ``convection" OR ``Laplace" OR ``poisson" OR ``parabolized stability" OR ``plasma" OR ``collision" OR ``MHD" OR ``Helmholtz"
\end{quote}
and sort by relevance. This returns 1,000 articles, which is the most articles Google Scholar will return in a single search. We focus only on the first 250 articles, sorting by relevance rather than by date. If the title and abstract potentially matches the inclusion criteria, we add it to a list of potentially relevant articles. We added 155 articles (62\%) to the list of potentially relevant articles. We then read each article in the list closely to see if it matches our inclusion criteria. 
We exclude eight articles which claim to outperform standard numerical methods with PINNs, though each of these articles compares to a weak baseline or, more often, doesn't account for the PINN optimization time.
This search process ultimately returns 60 articles, which we use as sample B.
We report our classification for each of the 155 articles at \url{https://osf.io/gq5b3/}.

It is well known that PINN-based solvers are slower than standard numerical methods (except possibly when meta-learning is used, e.g., in \cite{qin2022meta}). See, for example, the review article \cite{karniadakis2021physics}, which states that ``collectively, the results from these works demonstrate that \dots for forward, well-posed problems that do not require any data assimilation the existing numerical grid-based solvers currently outperform PINNs.'' 
Importantly, we can assume that the authors of each article in sample B know this. We can assume this because (a) is it common knowledge, (b) we only include articles that generate their own data from standard numerical methods to measure accuracy, so they must have known the runtime of that method, and (c) most of the articles in sample B are focused on developing strategies to improve the speed of PINNs, and thus these authors recognize that speed is a limitation of PINNs. Using PINNs to get good accuracy on most PDEs takes (in most cases) hours to days, while doing so with a standard numerical method takes between a fraction of a second to minutes \cite{chuang2022experience,chuang2023predictive}, depending on the PDE and the solver used. The authors doubtlessly noticed the difference in runtime between the PINN and the standard solver, and must have known (at least implicitly) that the PINN-based solver was less efficient than the standard numerical method.

\section{Causes and recommendations \label{sec:causes}}

To some extent, the issue of weak baselines (especially violations of rule 2) appears to be caused by three factors specific to this subfield: a lack of domain expertise in the ML community, insufficient benchmarking by the numerical analysis community, and little awareness about the difficulty and importance of choosing a strong baseline.
While benchmarking is a critical way of evaluating model performance in ML research, numerical analysis research tends to focus more on the theoretical properties of algorithms.
Furthermore, the relative performance of different numerical methods depends heavily on PDE parameters and implementation details.
It can be quite difficult, even for researchers with years of experience developing algorithms to solve PDEs, to predict the relative performance of different numerical methods on a given PDE.
Although a few researchers have begun to informally discuss issues with baseline comparisons in ML-for-PDE research \cite{hoyertweet,scientificmltofoolmasses}, widespread awareness about the extent of the problem appears to be missing.

To reduce the frequency of baseline-related reproducibility issues in ML-for-PDE solving research, we make a few recommendations for best practices.
While failing to beat a baseline should not cause an article to be rejected, failing to follow best practices when evaluating models can and should be treated as grounds for rejection.
First, we recommend that all articles using ML to solve PDEs compare to two types of baselines: standard numerical methods and other ML-based solvers.
This allows readers to better evaluate model performance and reduces selective reporting.
A good example is \cite{stachenfeld2021learned}.
If other ML-based solvers cannot be implemented as baselines, articles should explain why.
Second, articles should follow rule 1 when comparing to standard numerical methods.
To satisfy rule 1, reduce the spatial resolution and/or the number of iterations of the standard solver until the two methods have equal accuracy or equal runtime.
Ideally, articles would also make plots of cost versus accuracy.
A good example is figure 1a of \cite{kochkov2021machine}.
Third, articles should discuss in a separate paragraph or subsection how the baselines were chosen and justify why the comparisons are unbiased. In particular, articles should explain why the standard numerical method being used as a baseline is highly efficient (or state-of-the-art) for that PDE.
Ideally, articles would compare multiple numerical methods for each PDE and use the most efficient method as a baseline.
A good example is Appendix C of \cite{cheng2021using}.
If authors are unsure which baseline is state-of-the-art for a given PDE, they should (a) talk to and/or collaborate with domain scientists or other experts, and (b) clearly acknowledge their uncertainty. A good example of (b) is the speed comparison in appendix D.1 of \cite{lippe2023pde}.
Fourth, besides rules 1 and 2 we make three additional recommendations for fair comparisons (see \cref{sec:methodsreview}).

To a large extent, however, weak baselines (especially violations of rule 1) and reporting biases appear be caused by factors similar to those that have led to reproducibility crises in other fields: researcher degrees of freedom, combined with a bias towards positive results.
In the process of writing an article about ML for PDEs, researchers make many choices.
Researchers choose not only PDEs, boundary conditions, hyperparameters, evaluation metrics, etc., but also which hypotheses to test, experiments to report, and results to emphasize.
Choices about experimental design, analysis, and reporting are called researcher degrees of freedom \cite{simmons2011false,wicherts2016degrees}.
Researcher freedom is valuable, but it becomes a problem when decisions about analysis and reporting are made or altered after results are known \cite{gelman2013garden}.
The conditional probability distribution of each decision given the experimental results tends to be biased in favor of positive outcomes.
The cumulative sum of these many biased decisions can significantly affect the reproducibility of published research \cite{de2018cumulative}.

We emphasize both good intentions and perverse incentives as explanations for the apparent bias towards positive results.
The culture of scientific ML is one in which well-intentioned researchers try to figure out ways that ML might be useful for science.
In the process of doing so, they tend to be less interested in reporting ways that ML isn't useful. 
Perverse incentives also contribute.
Because ML research rewards novel ideas and positive experimental results \cite{sculley2018winner}, all else being equal articles with weak baselines and/or reporting biases are more likely to get accepted to prestigious venues and more likely to be widely cited \cite{serra2021nonreplicable}.
Incentives against negative results are particularly strong in scientific ML, because career advancement (in academia) and lucrative jobs (in industry) depend on the presumption that ML will be a useful tool for scientific problems.
Negative results could cast doubt on that presumption, thereby undermining justification for one's research area.

Regardless of whether negative results are valuable in ML research \cite{borji2018negative}, the fact is that ML is now being used in science. In science, negative results matter. Without negative results, scientists cannot accurately determine whether and how ML is useful for advancing knowledge in their field. Unchecked, overoptimism can lead to misunderstanding of applicability, misallocation of research priorities, and slowdown in scientific progress.

Because the causes of biased reporting in ML-for-PDE research appear to be similar to those of past reproducibility crises, we recommend two types of reforms similar to those implemented by other fields: bottom-up cultural changes intended to minimize biased reporting, as well as top-down structural reforms intended to reduce perverse incentives for doing so. Most of these reforms will not benefit ML researchers – research projects will require more work and report more negative results -  but they will benefit science.

Cultural changes start at the level of the individual, research group, and/or department. ML researchers should have widespread awareness about and understanding of reproducibility issues. \cite{ritchie2020science} is a good place to start, while \cite{gundersen2022sources} discusses many issues unique to ML. Research groups should develop internal controls to ensure that reporting is complete and unbiased. Departments should teach about reproducibility pitfalls in ML classes. Individuals should commit to eliminating biased reporting, using strong baselines, discussing limitations honestly and transparently \cite{smith2022real}, and publishing negative results. To increase confidence in their conclusions, researchers can preregister their experiments or use registered reports.

We recommend two structural reforms.
First, ML journals and conferences could allow for registered reports \cite{gundersen2021case}.
These would be peer-reviewed before experiments are performed and evaluated based on whether the proposed analysis answers an interesting research question and is methodologically sound.
Accepted proposals would be guaranteed publication in the journal or a future conference, so long as the final paper conforms to the registered report.
Second, funding agencies could fund domain scientists to propose and setup challenge problems for the ML community to tackle.
A desirable challenge problem for scientific ML -- for example the CASP protein folding challenge \cite{jumper2021highly} -- would have three qualities.
First, the problem must be unsolved or extremely tedious to solve using standard methods.
Second, there must be a way of verifying whether or how well the problem was solved.
Third, the scientific community must agree that solving the problem would be a valuable contribution to science.
A challenge problem with these qualities would have clear evaluation metrics and either a standard baseline or no baseline at all, thereby eliminating the potential for weak baselines, reducing opportunities for outcome reporting bias, while also directing ML research away from toy problems and towards those of greatest scientific importance.

\chapter{Machine learning for solving PDEs: lessons learned and future outlook \label{ch:conclusion}}
This thesis has introduced the theory of automatic differentiation (AD) and explored applications of differentiable programming in computational plasma physics. We've identified and studied two main applications of differentiable programming in computational plasma physics: (a) stellarator optimization and (b) solving partial differential equations (PDEs) with machine learning (ML). For stellarator optimization, differentiable programming can be used to compute derivatives needed for derivative-based optimization algorithms. For solving PDEs with ML, differentiable programming is used to optimize parameters in ML models using first-order gradient-based optimization.

One of the main conclusions of this dissertation is that there are clear benefits to using AD in stellarator optimization, with very few drawbacks. The main benefits of using AD are that it is accurate (to machine precision), efficient (in most applications equally as efficient as an optimal hand-coded algorithm), and easier to implement than alternatives. The ease of computing derivatives is probably the main benefit to using AD: researchers simply program the result they want to differentiate, and AD computes the derivatives with no additional effort. This allows for much more rapid prototyping of ideas and objective functions. The main drawback of using AD is that it restricts which programming languages and packages can be used; only certain libraries support AD. The best AD libraries at the moment appear to be written in Python and, to some extent, Julia. Fortunately, it is possible to add custom routines that call code written in another library or other languages, but the user needs to also program the derivates of that routine. In practice, this means that old stellarator optimization packages (written in languages such as Fortran) will not benefit from AD, but new stellarator optimization tools should consider using AD.
As evidenced by the DESC stellarator optimization code \cite{dudt2020desc,conlin2023desc,panici2023desc}, this shift towards AD tools appears to already be happening.

One of the goals of this dissertation was to use ML to develop faster algorithms to solve PDEs in plasma physics, which could then be used to tackle research problems in fusion energy and/or astrophysics that would otherwise be intractable. We were unsuccessful at achieving this goal. Instead, we learned that ML has shown much less potential for efficiently solving PDEs than we initially thought, for the reasons discussed in \cref{ch:reproducibility}. As a result, we believe that ML is unlikely to be useful for developing improved algorithms for solving PDEs in plasma physics.

Unlike in stellarator optimization, where the role of AD is clear, in PDE solving the role of ML remains unclear. Many researchers continue to study ML as a tool for solving PDEs, and different researchers tend to have different perspectives on how ML can be useful in computational physics and even disagree on whether it can be useful at all. In the rest of this chapter, we give our perspective on the role of ML in PDE solving, focusing on the lessons learned during this dissertation as well as the most important questions for which we don't yet have solid answers.

\section*{Introduction}

Partial differential equations (PDEs) are mathematical models of physical systems. PDEs describe the relationship between variables in terms of their partial derivatives. PDEs are of great interest to engineers, physicists, and mathematicians, because they accurately describe the behavior of many physical systems of interest. Solving a PDE or system of PDEs requires formulating a well-posed initial boundary value problem. The solution variable(s), such as density, temperature, velocity, etc., satisfy initial and/or boundary conditions on the domain boundary, and satisfy the PDE within the domain.

The standard approach to solving PDEs uses numerical algorithms \cite{durran2013numerical}. While PDEs describe continuous variables, standard numerical algorithms use a discrete representation of solution variables and/or the domain. These algorithms are typically derived analytically using discrete formulations of the PDE and boundary conditions. In practice, standard numerical algorithms are almost always solved using computers. Every numerical algorithm has some error relative to the exact solution of the PDE. The error (in most cases) can be decreased by discretizing the domain more finely. Decreased error via finer discretization (i.e., smaller grid spacing $\Delta x$ and/or $\Delta t$, more degrees of freedom) leads to increased computational cost. Because standard numerical methods are derived analytically from the PDE, their theoretical properties (such as consistency, order of accuracy, convergence, stability, etc) are often known analytically. As a result, scientists and engineers are capable of having confidence in the accuracy of standard numerical methods, at least in certain regimes and if enough computational resources are used.

Machine learning (ML) is a set of techniques for using data to perform a desired task, usually by approximating input-output relationships. ML methods, especially deep learning methods such as neural networks, form the basis of modern artificial intelligence (AI) technology including facial recognition and language models such as ChatGPT. The ML paradigm of learning input-output relationships from data can be applied in many other disciplines, including science. Many scientists are interested in understanding whether and how ML can be useful for scientific research.

ML is capable of learning to solve PDEs. There are many examples of this being done in the scientific literature, using a wide variety of techniques \cite{raissi2019physics,li2020fourier,kochkov2021machine,luz2020learning,brandstetter2022message}. However, just because ML can solve PDEs does not mean it is useful. To be useful, ML-based PDE solvers must both (i) outperform standard numerical methods and (ii) either facilitate scientific discoveries in research applications or improve efficiency and/or accuracy in engineering applications. It is unclear whether, and how, ML can be useful in scientific research and engineering applications involving PDEs.

In this essay, we give our perspective on whether and how ML will be useful for solving PDEs. To do so, we pose three questions. We believe the answers to these questions will be critical for determining whether or not ML can be useful for solving PDEs. The three questions are:
\begin{enumerate}
    \item Do ML-based PDE solvers actually outperform standard numerical methods?
    \item Can the predictions of ML-based PDE solvers ever be trusted?
    \item What applications could ML-based PDE solvers be useful for?
\end{enumerate}
In the rest of this essay, we will attempt to answer each question. While there is much we have learned, there is still some uncertainty in our answers, and ultimately, some uncertainty about the future role of ML in PDE solving.

Our answers to each question are mostly pessimistic. We are therefore pessimistic about the possibility that ML will be a useful tool in PDE solving. This pessimism stems from comparisons with standard numerical methods for solving PDEs and theoretical arguments about the reliability of PDE solvers. As a result, we conclude that ML-based PDE solvers can only be useful in applications where the same system must be simulated repeatedly under similar conditions. However, we have not been able to identify (nor have we read about or talked with anyone who has identified) specific applications which have that property. Unless an important application with the right properties can be identified, or a theoretical breakthrough can be made guaranteeing (or dramatically improving) the reliability of ML-based PDE solvers, we anticipate that any benefits from using ML to solve PDEs in computational science and engineering will remain only a hypothetical possibility.

In all of the examples where ML has been used to solve PDEs, to the best of our knowledge ML has not helped create any new scientific knowledge, nor has it yet served as a useful tool in any research applications. For the reasons discussed in the previous paragraph, we expect this is likely to continue, at least when it comes to solving PDEs with ML.

We believe that ML could possibly provide some value in other areas of computational physics and related research. Indeed, there are a few examples where it has already done so. ML (specifically, reinforcement learning) has been used to control the motion of fish in schooling formations by coupling to 3D simulations of Navier-Stokes equations, ultimately demonstrating how and why fish schooling can be an energetically favorable mode of swimming \cite{verma2018efficient}. This helps answer a longstanding and apparently previously unsolved question in fluid mechanics. Gaussian process regression coupled to gyrokinetic simulations has been used within a Bayesian optimization framework to predict core temperature and density profiles for the SPARC tokamak using fewer calls to an expensive gyrokinetic solver compared to Newton's method \cite{rodriguez2022nonlinear}; this demonstrates some modest benefit to using ML in a research-level problem in fusion energy research. However, we are only aware of a few examples of ML creating value in computational physics, and these are in relatively niche applications or engineering problems. None of these examples involve using ML to output an approximate solution to a PDE.

One area of computational science where ML appears to be unusually well-suited is weather prediction. Weather prediction shares certain aspects of PDE solving, but is a different topic. Because the equations describing the evolution of the weather are not known exactly (at least on the macroscopic scales for which weather simulation is possible), the best alternative methods for weather prediction are not as accurate and efficient as when the PDE is known exactly \cite{gneiting2005weather}. Furthermore, this is an application where the same system (the Earth) must be simulated repeatedly under similar conditions; for applications with these characteristics, we argue that ML-based simulators can potentially be useful. While there are certainly reasons to be cautious about whether ML can eventually replace numerical weather prediction (NWP) models \cite{schultz2021can}, in one recent example ML has apparently been used ML to outperform state-of-the-art NWP models on many metrics \cite{lam2023learning}. 

We now turn to the three questions posed earlier, each of which we believe is critical for determining the future of ML in PDE solving.

	\section{Do ML-based PDE solvers actually outperform standard numerical methods?}

Why use ML to solve PDEs? The justification for doing so is stated concisely in the first sentences of \cite{lippe2023pde}:

\begin{displayquote}
	Recently, mostly due to the high computational cost of traditional solution techniques, \dots deep neural network based PDE surrogates have gained significant momentum as a more computationally efficient solution methodology. 
\end{displayquote}

This quote reflects what has, in recent years, become a consensus view in the ML community: that standard numerical methods for solving PDEs are slow, while ML-based methods are fast.\footnote{The view that ML-based solvers are fast excludes ``physics-informed neural networks'' (PINNs), which are known to be much less efficient than standard numerical methods. The term `PINN' was branded in 2019 in \cite{raissi2019physics}. \cite{raissi2019physics} reintroduced the idea of representing the solution to a PDE with a neural network, rather than by discrete points on a grid or polynomials on a mesh; this idea had originally been proposed in the 1990s \cite{dissanayake1994neural,van1995neural,lagaris1998artificial} but regained in popularity recently. \cite{raissi2019physics} has now been cited over 8,500 times, making it one of the most highly-cited papers in the history of computational physics. Despite the massive interest in using PINNs, all evidence points to the conclusion that PINNs simply do not work very well \cite{grossmann2023can,krishnapriyan2021characterizing,chuang2022experience,chuang2023predictive,karnakov2024solving}. For forward problems (i.e., solving PDEs), PINNs often do not converge to a qualitatively or quantitatively reasonable solution \cite{krishnapriyan2021characterizing,wang2022and}, and when they do converge they do so extremely slowly \cite{chuang2022experience}. This is because solving PDEs with PINNs requires optimizing a neural network loss function, which requires a very large number of floating point operations compared to the runtime of standard numerical methods. PINNs can also be used for inverse problems (i.e., using data to estimate the equations that created them or imputing missing data). While PINNs are often said to perform well at inverse problems \cite{karniadakis2021physics,lippe2023pde}, recent evidence suggests that optimization of standard discretizations rather than a neural network (called `optimizing a discrete loss' or `ODIL') performs much better, at least on the benchmark problems where the ODIL method has been tested \cite{karnakov2022optimizing,karnakov2024solving}. PINNs may work well at solving very high-dimensional PDEs, a regime in which standard numerical methods struggle, though there is not yet enough evidence to draw this conclusion \cite{grossmann2023can}. The strongest advantage of PINNs appears to be their simplicity: they require almost no knowledge about PDEs or numerical methods, and can be used by anyone who understands neural networks and automatic differentiation. The simplicity of PINNs most likely explains the massive interest in their use, in spite of their poor performance.} 
While ML-based solvers are known to have certain limitations (such as lack of convergence guarantees or the development of numerical instabilities) \cite{mcgreivy2023invariant,kadapa2021machine}, their performance is evaluated using quantitative comparisons with standard numerical methods or other ML-based solvers. On the quantitative metric of speed (i.e. computational cost, time to solution, etc), ML-based PDE solvers are thought to perform better than standard numerical methods.

Is this consensus view correct? If so, then there is potential for ML to prove useful in computational physics research. Faster solvers could be useful for a variety of tasks, for example optimization or uncertainty quantification \cite{mishra2018machine}. They might even one day replace the standard solvers used in research and engineering applications \cite{kochkov2021machine}. However, if the consensus view is wrong, then there is no potential for ML to be useful in computational physics research. This is because speed, as we will see, is the only advantage that ML-based PDE solvers have over standard numerical methods. If ML-based PDE solvers are not faster, there is no reason to use them.

	\subsection{How ML-based PDE solvers are evaluated}

When ML-based PDE solvers are compared to other ML-based PDE solvers, they are typically compared based on their \textit{accuracy}. If a method is more accurate, this is seen as evidence of better performance.
However, when ML-based PDE solvers are compared to standard numerical methods, they are typically compared based on their \textit{speed}. If an ML-based method is faster, this is seen (either explicitly or implicitly) as evidence of better performance than the standard numerical method(s). Consider an example which illustrates a pattern we've seen repeated across many papers. \cite{li2020fourier} introduces an ML-based solver which compares to both ``traditional PDE solvers" (i.e., standard numerical methods) and ``previous learning-based solvers'' (i.e., other ML-based solvers). When comparing to traditional PDE solvers, the speed is compared, and the ML-based solver is ``up to three orders of magnitude faster.'' When comparing to previous learning-based solvers, the accuracy is compared, and the ML-based solver ``achieves superior accuracy compared to previous learning-based solvers under fixed resolution.'' In short, accuracy is used to compare within-class, while speed is used to compare across-class.

There are certainly exceptions. In some cases, the speed, efficiency, or some other quantitative metric besides accuracy is used to compare multiple ML-based PDE solvers \cite{huang2023neuralstagger}. In other cases, accuracy, or some other quantitative metric besides speed is used to compare an ML-based PDE solver to a standard numerical method \cite{shang2022deep,stevens2020finitenet}. However, most papers we've seen follow this pattern.

For standard numerical methods, speed and accuracy are related via a speed-accuracy tradeoff. Standard numerical methods can be made faster using coarser discretization (larger $\Delta x$ and/or $\Delta t$, fewer degrees of freedom) at the cost of decreased accuracy, or more accurate using finer discretization (smaller $\Delta x$ and/or $\Delta t$, more degrees of freedom) at the cost of decreased speed. Increased speed over standard numerical methods is only meaningful if it takes into account this speed-accuracy tradeoff. In particular, this means that speed comparisons between two methods should be performed at equal accuracy.  As discussed in \cref{sec:weakbaselines}, comparisons between a highly accurate (but slow) standard numerical method and a less accurate (but fast) ML-based solver will be unfairly biased in favor of the ML-based solver. In addition, comparisons to standard numerical methods methods much slower than state-of-the-art will also be unfairly biased in favor of the ML-based solver. This is because some standard numerical methods are much faster than others. The performance of a standard numerical method can depend significantly on the PDE being solved. Since ML-based PDE solvers are trained to perform well at solving a particular PDE, they should be compared to standard numerical methods which work well on the PDE being solved, preferably a state-of-the-art method. 

	\subsection{Weak baselines challenge consensus}

The consensus view that ML-based surrogates for solving PDEs (excluding PINNs) are generally faster than standard numerical methods is supported by a large number of published research papers. In \cref{ch:reproducibility}, we studied the papers supporting this consensus using a systematic review. The systematic review searched for papers that (i) learn to solve one or more fluid-related PDEs using ML, (ii) compare the speed (i.e., time to solution), some proxy for speed, or accuracy, to a standard numerical method, (iii) were available online before April 1st, 2023, and (iv) do not use the PINN method. We found 82 papers matching these criteria. Of these, 76 reported improved performance using ML. Many papers report speedups by multiple orders of magnitude, suggesting that using ML has led to massive improvements in speed. 

We then evaluated whether each paper compared to a weak baseline. In particular, we focused on whether papers were comparing at constant accuracy and whether they were comparing a state-of-the-art (or near state-of-the-art) numerical method. Of the 76 papers reporting improved performance, 59 (79\%) used a weak baseline (see \cref{tab:baselines}). We also found that the larger the quantitative improvement from using ML (i.e., the larger the speedup), the more likely the article used a weak baseline. Our results suggest that using a strong baseline is critical for evaluating model performance. We replicated the results in 10 articles (see \cref{tab:reproducebaselines}) using stronger baselines, and found that in 9 of them the stronger baseline was at least two orders of magnitude faster than the weaker baseline. In most problems, using a strong baseline seems to make a very big difference in terms of speed. 

It is unclear whether, and by how much, ML-based PDE solvers would be faster than standard numerical methods if strong baselines were used. In \cref{fig:cumulativeeffects}, we estimated that about 17 out of the 76 papers (22\%) would report speedups if strong baselines were used, 4 out of the 76 (5\%) would report similar speed to state-of-the-art numerical methods, and 55 out of the 76 (72\%) would report being slower than state-of-the-art standard numerical methods. However, these numbers are estimates, and the true results are not known. In any case, it is clear that the consensus is only partially correct: there are some instances where ML-based PDE solvers have outperformed standard numerical methods, but in many instances (in our estimate, most instances) standard numerical methods are actually faster than ML-based PDE solvers.

The results from the systematic review in \cref{ch:reproducibility} present a significant challenge to the consensus that ML-based PDE solvers are fast while standard numerical methods are slow. In many (probably most) instances, the reverse is true. Nevertheless, there are no doubt some instances where ML-based methods have been able to learn from data to solve PDEs more efficiently than standard numerical methods. Thus, the consensus is at least partially wrong, but it is not entirely wrong. Better comparisons with strong baselines are needed to fully evaluate the potential of ML for accelerating PDE solving.

A few examples where ML has accelerated the solution of PDEs relative to a strong baseline are worth mentioning. For time-independent PDEs such as Poisson's equation, a handful of papers have found faster performance, in most cases very small speedups between 1-$2\times$ \cite{greenfeld2019learning,luz2020learning,ozbay2021poisson,cheng2021using}, and at most up to $4\times$ faster \cite{tompson2017accelerating}. The only instance where ML has accelerated the solution of a time-dependent turbulent PDE relative to a strong baseline is \cite{de2023accelerating}, which found 4-$5\times$ speedups compared to high-order discontinuous Galerkin (DG) methods for 3D time-dependent turbulence by using ML as a forcing function to mimic the influence of higher-order polynomial elements on a lower-order simulation. Another paper, published after the systematic review was finished, found 5-$10\times$ speedups for a time-independent radiative transfer PDE by replacing elements of a hybridizable DG (HDG) solver \cite{du2023element}. Nevertheless, these are relatively modest speedups, and it is unclear to what extent these methods can be extended more broadly to other applications.

\subsection{Efficiency is meaningless without generalization}

As we have seen, ML-based PDE solvers can sometimes be faster than standard numerical methods but are often slower. Thus, we might conclude that ML-based PDE solvers sometimes outperform standard numerical methods. Doing so, however, would rely on a narrow definition of `performance'. Here we argue that the performance of a PDE solver is about much more than just quantitative metrics like speed or accuracy. Instead, performance should also account for other factors like robustness and generalization. In \cref{sec:conclusion-is-speed-right-metric} we argue that, taking a holistic view of performance, ML-based PDE solvers clearly perform worse than standard numerical methods in practical applications.

ML relies on training data to learn how to solve a desired task and testing data to evaluate performance on that task. ML only works (or at least, only works well) under the assumption that the training data, testing data, and real-world inputs are drawn from the same (or at least, a sufficiently similar) probability distribution. ML thus assumes some randomness or variation in the data, and attempts to generalize over the variation by achieving good performance over the entire probability distribution.

Good performance (i.e., speed, accuracy) on solving PDEs is meaningless without generalization. To illustrate why, consider the hypothetical task of repeatedly solving the same PDE with identical initial and boundary conditions. Because there would be no variation in the inputs, the output would not change. In this case, the exact PDE solution could simply be stored in a database and retrieved when necessary. While this would have perfect accuracy and speed, nothing meaningful would have been learned. Therefore, when evaluating the success of an ML-based solver, it is essential to consider not just its speed and accuracy, but also what is being generalized over. If a method cannot generalize successfully to the inputs for which it will be used in applications, it will have no value (at best) or be actively harmful (at worst).

Unfortunately, while generalization capabilities are critical for evaluating the success of ML-based PDE solvers, generalization is often overlooked or ignored when evaluating performance. While papers introducing ML-based PDE solvers tend to state clearly a model's accuracy and/or efficiency, they tend to be much less clear about what is being generalized over. There are exceptions, of course, but in our experience generalization capabilities usually cannot be determined by reading a paper's abstract, introduction, and/or conclusion. Usually, a closer reading is required. The papers we are most familiar with usually attempt to generalize over a random distribution of initial conditions (for a time-dependent PDE) \cite{li2020fourier,bar2019learning,kochkov2021machine} or a random source term (for a time-independent PDE) \cite{cheng2021using}. Some papers also attempt to generalize over a random distribution of PDE coefficients (such as diffusion coefficients) \cite{bar2019learning,luz2020learning}, over the domain size \cite{lippe2023pde}, or over the domain shape and boundary conditions \cite{qin2022meta,hsieh2019learning}. We emphasize that we have not studied the topic of generalization systematically, and no other review papers exist on the topic of generalization in ML-based PDE solvers, so the examples from this paragraph should not necessarily be seen as representative of the entire literature in this area.

The evidence we do have suggests that ML-based PDE solvers generalize well in some areas but poorly in others. For example, ML-based solvers seem to be able to generalize well if the PDE and boundary conditions are held fixed but the initial conditions (for time-dependent PDEs) or source terms (for time-independent PDEs) are drawn randomly from a probability distribution \cite{stachenfeld2021learned,cheng2021using}. These solvers may still develop numerical instabilities or rapidly accumulating errors as the solution is unrolled \cite{lippe2023pde,mcgreivy2023invariant,zhuang2021learned}, limiting their applicability at large $t$; at small enough $t$, however, they can successfully generalize over varying initial conditions within the probability distribution of the training data.

The generalization capability of most interest to scientists -- whether or not ML-based PDE solvers can generalize to qualitatively new physical regimes -- appears to be where ML-based PDE solvers generalize worst.\footnote{This reflects an important philosophical mismatch between ML and science: while ML focuses on generalization within a probability distribution where inputs and outputs are known, science is most interested in extrapolation to areas where the data is unknown.} To the best of our knowledge, only a handful of papers have reported testing generalization to new physical regimes. One example is \cite{stachenfeld2021learned}, which uses ML to solve a 3D compressible turbulent mixing problem due to the Kelvin-Helmholtz instability. They generalize over the size of the 3D box; at larger box sizes, the mixing transitions from fast (i.e., high cooling velocity) to slow (i.e., low cooling velocity). In this case, the ML-based solver is not able to generalize accurately to different cooling velocities as the box size is varied, even when trained on varying box sizes. Similarly, the ML-based solver is not able to generalize accurately to initial conditions drawn from distributions more compressive or more solenoidal than the training data. \cite{de2023accelerating} finds that a trained DG forcing function is not able to generalize to different flow conditions without retraining. Another example is \cite{kochkov2021machine}, which develops an ML-based solver able to generalize accurately over different forcing functions and diffusion coefficients. To do so, however, requires modeling the forcing and diffusion terms in the same way that standard numerical methods do. Therefore, while the generalization results in \cite{kochkov2021machine} are impressive -- more impressive than any other generalization result for time-dependent PDEs -- they reinforce the conclusion that any component of an algorithm involving deep neural networks is unlikely to reliably generalize outside of the distribution of the training data.

\subsection{Is speed even the right metric to use? \label{sec:conclusion-is-speed-right-metric}}

ML is a quantitative field: ML models are typically evaluated based on their performance on quantitative metrics. For example, in December 2023 Google released a new AI model called Gemini. In the press release (\url{https://blog.google/technology/ai/google-gemini-ai/}), immediately after the introduction Google reports the performance of Gemini on ``32 widely-used academic benchmarks used in large language modeling (LLM) research and development." On 30 out of 32 benchmarks, Gemini reports state-of-the-art performance compared to ChatGPT-4 or ChatGPT-4V. While the press release eventually links to videos showing what Gemini can do and how it might be useful, the focus is on how it performs quantitatively on benchmark problems. This focus is typical of AI and ML research. 

The focus on quantitative performance on benchmark problems likely arises from the history of AI research. Historically, quantitative performance on benchmark datasets has been used to determine which models and architectures work better than others. In 1994, the MNIST digit classification benchmark was created, which allowed AI researchers to easily evaluate the success of different architectures at a simple image recognition problem. While the lowest achieved error rate on MNIST did not improve much during the 2000s, in 2010 graphics processing units (GPUs) allowed for a deep multilayer perceptron (MLP) to achieve state-of-the-art performance on MNIST \cite{cirecsan2010deep}. In 2011, a convolutional neural network (CNN) trained on GPUs again beat the state-of-the-art on MNIST \cite{ciresan2011convolutional}. In 2012 researchers at Toronto famously combined the three innovations of GPUs, CNNs, and deep networks to train deep CNNs on GPUs to achieve dramatically improved state-of-the-art performance on a different and harder benchmark problem called ImageNet \cite{krizhevsky2012imagenet}. Quantitative benchmarks provided an objective, fair, and straightforward way of determining which architectures worked best at image recognition problems. They also created a competitive marketplace in which the best ideas `won'. Ultimately, the use of quantitative benchmark problems almost certainly accelerated progress in AI.

As we've discussed, the most common way of comparing the performance of ML-based PDE solvers to standard numerical methods is using the quantitative metric of speed. Speed usually measures the computational time required to output an approximate solution, though it can sometimes be measured in other ways (floating-point operations, degrees of freedom, etc). This use of a quantitative metric to measure performance is consistent with the history of AI/ML research as well as the present-day practices in the ML research community.

However, the use of quantitative metrics to benchmark the performance of different algorithms for solving PDEs is \textit{different} from how the numerical analysis community benchmarks algorithms. Instead of focusing on quantitative metrics like accuracy or speed, the numerical analysis community typically uses qualitative observations and focuses on understanding the theoretical properties of algorithms. The numerical analysis community is often interested in properties such as: the order of convergence, consistency, stability, monotonicity, conservation, positivity-preserving, and/or whether the algorithm is symplectic or total variation diminishing (TVD). In some cases, these properties correlate with quantitative metrics such as $\mathcal{L}_2$-norm error or speed. However, often the numerical analysis community is interested in these properties because they relate to the \textit{generalization} capabilities of the algorithm. For the numerical analysis community, speed is important, but most important is confidence in the correctness and extrapolation capabilities of the algorithm.

For example, consider the simulation journal of A.H., which benchmarks the performance of two standard numerical algorithms for solving the 1D Euler equations. The two algorithms are a wave-propagation scheme with a positivity fix, and a second-order MUSCL-Hancock scheme without positivity-preserving limiters. The simulation journal can be found at \url{https://ammar-hakim.org/sj/je/je2/je2-euler-shock.html}. Notably, quantitative metrics are not used. Instead, the performance of each algorithm is compared qualitatively on eight challenging 1D benchmark problems. These problems are designed to test the generalization capabilities of each algorithm, and also evaluate the accuracy of each model compared to the exact (analytic) solution. On four of the eight benchmark problems, the second-order MUSCL-Hancock scheme fails due to NaNs and/or non-physical negative values. On the four problems where MUSCL-Hancock does not fail, the wave-propagation scheme has less numerical diffusion and a more accurate solution. This approach to benchmarking serves two functions. First, it helps develop theoretical insight into the behavior of each algorithm in different physical scenarios. This helps users understand where an algorithm is likely to work well and where it is not. In the previous example, it was learned that the MUSCL-Hancock scheme was too diffusive and fails to ensure positivity of density and energy. Using this information, in later (unpublished) experiments we modified the MUSCL-Hancock scheme to perform flux reconstruction in characteristic variables and have positivity-preserving limiters, which ultimately led to a much more accurate and robust algorithm. Second, it gives practical insight into both the accuracy and generalization capabilities of each algorithm. While the ML community allows readers to quickly evaluate performance using a single (or a few) quantitative number(s), the numerical analysis community tends to eschew such comparisons and forces readers to develop qualitative understanding of performance based on observation.

Which is the best way of benchmarking the performance of an algorithm for solving a PDE? Is it the ML community's approach of using quantitative metrics like accuracy and speed? Or is it the numerical analysis community's approach of developing a theoretical understanding of expected generalization capabilities and qualitative understanding of accuracy? Different readers may make different arguments, and thus come to different conclusions. In our view, the best argument in favor of the ML community's approach is that it creates a competitive marketplace where the best ideas `win'. As happened for AI problems such as image recognition, this incentivizes researchers to develop state-of-the-art methods and can accelerate progress. However, we believe there is much wisdom in the approach of the numerical analysis community. While image recognition and other problems in AI didn't require much theoretical insight to be solved successfully, progress in algorithm development for solving PDEs has come almost entirely from theoretical insights. Focusing on quantitative metrics disincentivizes the development of theoretical understanding. Most importantly, the performance of a numerical method cannot be summarized in a single number. As the numerical analysis community knows well, methods which work well on one problem may fail entirely on another problem. Knowing the situations in which a method can be expected to output an accurate solution, and the contexts in which it cannot, is crucial for understanding the performance of an algorithm.

We believe that the performance of numerical algorithms should be viewed holistically. Speed has some value, but it should not be the primary method by which a PDE solver is evaluated. Equal, if not greater, importance should be given to the generalization and robustness capabilities of an algorithm. Thus, even if ML-based PDE solvers are faster than standard numerical methods -- and they often aren't -- given their apparent failure to generalize outside the training data, it is difficult to conclude that they outperform standard numerical methods. Because the vast majority of applications where numerical methods are used involve extrapolation to new physical regimes, for almost all applications standard numerical methods outperform ML-based PDE solvers.

\section{Can the predictions of ML-based PDE solvers ever be trusted? \label{sec:conclusion-trusted}}

As we've discussed, ML-based PDE solvers do not seem to generalize (i.e., give accurate results) outside of the distribution of the training data. Even within the distribution of the training data, the only evidence of reliability is usually low error on the testing dataset, which may not be enough to give confidence in the predictions of ML-based PDE solvers. The lack of trust in ML-based PDE solvers is a serious concern that is likely to severely limit the applicability of ML-based PDE solvers in practical applications. Thus, one of the most important topics for this research area is to determine whether the predictions of ML-based PDE solvers can ever be trusted and if so, how.

We know that it is possible to design PDE solvers that are robust and generalize reliably: well-designed standard numerical methods do so. If it is possible to make standard numerical methods that can be trusted, then it is also possible to make ML-based PDE solvers that are trusted as well. However, simply building ML-based PDE solvers that are equally as reliable as well-designed standard numerical methods isn't sufficient: ML-based PDE solvers must also be fast. If they are not faster than standard numerical methods, there is no point in using them. Thus, the great challenge of ML-for-PDE solving research is to develop ML-based PDE solvers that do both simultaneously: they should be faster than standard numerical methods, and equally as reliable.  In practice, the goal would likely be to have superior speed and/or accuracy relative to state-of-the-art numerical methods within the distribution of the training data, while having theoretical guarantees of robustness and accuracy outside the distribution of the training data.

As we discuss in \cref{sec:conclusion-linear-solve}, this can be done when using ML to accelerate the solution of linear systems of equations within standard numerical methods for PDEs. However, for other types of ML-based PDE solvers, it is not clear how to (or whether it is even possible to) combine increased speed with reliable predictions. In our opinion, this is the most important unsolved problem in this research area: figuring out how to combine increased speed with robust predictions. In \cref{sec:conclusion-inductive-bias}  we explain why physically motivated inductive biases are not sufficient to ensure reliability of ML-based PDE solvers, and in \cref{sec:conclusion-challenges-to-trust} we outline some of the challenges to designing ML-based PDE solvers that are both faster than standard numerical methods and have guarantees of reliability.

\subsection{Solving linear systems: speed and convergence mutually compatible \label{sec:conclusion-linear-solve}}

Using the finite-element method (FEM), some time-independent PDEs can be written in discrete form as a linear system of equations. For some time-dependent PDEs, standard numerical methods with implicit timestepping require the solution to a linear system of equations at each timestep. Systems of linear equations can be solved using either direct solvers, which are more efficient for small problem sizes, or iterative numerical algorithms, which are more efficient for large problem sizes. Iterative numerical algorithms require an initial guess for the first iteration. Preconditioners are often used to obtain an accurate initial guess for these iterative algorithms, thereby reducing the number of iterations required.

Two strategies of using ML to accelerate the solution of linear systems of equations have been proposed, both of which provide theoretical guarantees of convergence to the solution of the linear system. As a result, for PDEs requiring the solution of linear systems of equations, it is possible to build ML-based PDE solvers that are faster than standard numerical methods, while also having theoretical guarantees of robustness which ensure that an accurate solution will be found even outside the training data. 
Thus, when it comes to accelerating the solution of linear systems of equations within standard numerical methods, the great challenge of ML-for-PDE solving research has been solved. For these solvers, in the limit that the error of the iterative solver goes to zero, the only error will be the discretization error of the FEM or whatever standard numerical method is being used. 

The two strategies are to either (a) use ML to learn a preconditioner for linear systems of equations arising during the solution of PDEs \cite{ozbay2021poisson,ackmann2020machine,markidis2021old}, or (b) to learn the internal elements of an iterative solver while maintaining convergence guarantees \cite{hsieh2019learning,luz2020learning,zhang2022hybrid}. If the resulting preconditioner is more accurate than state-of-the-art preconditioners, or the learned iterative algorithm converges more quickly than state-of-the-art algorithms, then ML can be used to solve these systems of equations with fewer iterations (and potentially reduced computational cost) while maintaining guarantees of convergence. 

This dissertation did not explicitly focus on using ML to learn preconditioners for linear equations. Therefore, we are not experts on the strengths and limitations of ML for linear systems of equations. However, based on our understanding of the literature in this area, we find reasons to be both cautiously optimistic and somewhat skeptical about the potential of ML-based preconditioners to be useful in PDE solving applications. We are optimistic because (a) at least a few papers have reported modest speed improvements relative to state-of-the-art baselines, up to $2\times$ for Poissons' equation \cite{ozbay2021poisson} and up to $3\times$-$10\times$ for the shallow water equations \cite{ackmann2020machine}, and (b) because the convergence guarantees will ensure the robustness of the solver outside the distribution of the training data. On the other hand, we are skeptical about whether the ML-based preconditioners (a) will be able to generalize to different geometry sizes and shapes, necessary in most practical applications, and (b) will remain faster than state-of-the-art numerical methods outside the training distribution, possibly resulting in a slower solver in practical applications. Ultimately, it remains to be seen whether and how learned preconditioners and other ways of using ML accelerate the solution of linear systems will be useful in PDE solving research and applications.

	\subsection{Inductive biases are not enough \label{sec:conclusion-inductive-bias}}

One common strategy for producing more reliable ML-based algorithms is to embed inductive biases into the model. Inductive bias is a term that refers to the assumptions made by the learning algorithm. Typically, these assumptions involve constraints on the model outputs or a preference for favoring certain input-output relationships. These assumptions are usually based on prior knowledge about the application domain; in physics applications, inductive bias usually refers to ensuring that the predictions of ML algorithms are consistent with known physical laws.

Many PDEs are symmetric under certain transformations. For example, PDEs may have translational symmetry, rotational symmetry, Galilinean invariance, etc. These symmetries can be embedded into neural networks as inductive biases via constraints on the network weights \cite{cohen2016group}. These so-called symmetry group equivariant neural networks are expected to have improved generalization and to learn from data more efficiently \cite{bogatskiy2022symmetry}. However, some authors have claimed, incorrectly, that due to Noether's theorem using symmetry group equivariant neural networks for PDEs will result in ML-based PDE solvers that have corresponding conserved quantities. This is an incorrect understanding of Noether's law. Symmetry group equivariant neural networks may improve the generalization of ML-based PDE solvers, but they will not result in conserved quantities nor are they sufficient to guarantee reliability or robustness.

For time-dependent autoregressive ML-based PDE solvers, it is possible to guarantee numerical stability and conservation laws using error-correcting algorithms at each timestep. This approach was introduced and discussed in detail in \cite{mcgreivy2023invariant} and in \cref{ch:invariant}. While guarantees of numerical stability and conservation laws can improve the generalization and reliability of ML-based PDE solvers, they too are not sufficient to guarantee reliability. PDE solvers can satisfy conservation laws and be numerically stable, but still output qualitatively incorrect predictions. Thus, both equivariances (i.e., symmetries) and invariants (i.e., conservation laws) can improve reliability, but they are insufficient to guarantee robustness.

The only known way of solving a PDE with zero error is to use a convergent algorithm in the limit that $\Delta x$ and/or $\Delta t$ approach zero. A convergent algorithm of order $p$ is one in which the global error is proportional to $(\Delta x)^p$ for $p > 0$; for time-dependent PDEs, a convergent algorithm of order $(p, q)$ has a global error proportional to $c_x(\Delta x)^p + c_t(\Delta t)^q$ for constants $c_x \ge 0$, $c_t \ge 0$, and $p, q > 0$. Taking the limit that $\Delta x$ and/or $\Delta t$ go to zero implies that the global error approaches zero as well. This ensures that the solution to the discrete numerical algorithm is identical to the exact solution to the PDE.

However, standard numerical algorithms are not actually solved in the limit that $\Delta x$ and/or $\Delta t$ go to zero. Instead, they attempt to give qualitative reasonable results at finite $\Delta x$ and/or $\Delta t$. To do so, standard numerical methods combine guarantees of stability -- that the norm of the solution remains bounded as $t \rightarrow \infty$ -- with a known order of convergence -- that the (local) error is proportional to $\Delta x$ and/or $\Delta t$ to some power $p$ and/or $q$, where $p$ and $q$ are usually integers. So long as $\Delta x$ and/or $\Delta t$ are sufficiently small, then the local error will remain small. In practice, sufficiently small typically means that $\Delta x < L_{\textnormal{min}}$, where $ L_{\textnormal{min}}^{-1} = \max |\frac{\grad f }{f}|$ is defined as the smallest scale length in the solution. 

While inductive biases based on physical laws can improve the reliability of ML-based PDE solvers, they are not sufficient to guarantee robustness. Designing ML-based PDE solvers that are robust under all possible inputs likely will require putting analytic constraints on the truncation error, perhaps with a known order of convergence. However, as we now discuss, there are a number of challenges to doing so while also being faster than standard numerical methods.

\subsection{Challenges to designing robust ML-based PDE solvers \label{sec:conclusion-challenges-to-trust}}

Perhaps the main reason that designing robust ML-based PDE solvers is so difficult is that the reliability of standard numerical methods requires using sufficiently small $\Delta x$ and $\Delta t$ (i.e., a large number of degrees of freedom), but ML-based PDE solvers need to use sufficiently large $\Delta x$ and/or $\Delta t$ (i.e., a small number of degrees of freedom) to be significantly faster than standard numerical methods. Even if ML-based PDE solvers have theoretical guarantees of convergence as $\Delta x$ and/or $\Delta t$ approach zero, these solvers must operate in the regime that $\Delta x$ and/or $\Delta t$ are large if they are to be useful. While there is no fundamental limit to how fast a PDE can be solved, there may be a fundamental limit to how few degrees of freedom can be used to solve a PDE while maintaining sufficiently small truncation error under all possible inputs.

Another reason that designing robust ML-based PDE solvers is so difficult is that is is unclear what is even required to guarantee reliability; the problem is not well-formulated. Standard numerical methods for time-dependent PDEs guarantee reliability using a combination of theoretical guarantees: consistency, numerical stability, order of convergence, invariants such as conservation laws and positivity, Courant-Friedrichs-Levy (CFL) conditions or other stability conditions on $\Delta t$, and small enough $\Delta x$ and/or $\Delta t$ to ensure sufficiently small truncation error. However, standard numerical methods do not necessarily need to satisfy every one of these theoretical guarantees to guarantee robustness on a given PDE; sometimes only a subset of properties is sufficient. For ML-based PDE solvers, it is not clear what theoretical properties are necessary to guarantee reliability for all possible inputs.

A third reason that designing robust ML-based PDE solvers is so difficult is that achieving very high accuracy (i.e., zero error) at finite $\Delta x$ and/or $\Delta t$ requires having flexibility in the rate at which the solution is updated; however, achieving robustness requires putting constraints on the rate at which the solution is updated. Standard numerical methods use hand-coded update rules, and thus have no flexibility but are typically robust and reliable except at large $\Delta x$ and/or $\Delta t$. ML-based PDE solvers use learned update rules, which typically have completely flexible predictions that allow for accuracy even at large $\Delta x$ and/or $\Delta t$, but which cannot guarantee robustness. If it is possible to design algorithms which allow for flexible predictions but are still robust, then the predictions of ML-based PDE solvers can be simultaneously fast and reliable. If not -- if the flexibility required for accuracy and the constraints required for robustness are fundamentally incompatible -- then the predictions of ML-based PDE solvers cannot be simultaneously fast and reliable. It is unclear whether the flexibility required for an accurate solver is compatible with the constraints required for robustness. For standard numerical methods, accuracy and robustness are often incompatible: to prevent non-physical numerical oscillations, standard numerical methods must add numerical diffusion at extremum, decreasing the order of accuracy to first- or second-order. Preventing non-physical oscillations in ML-based PDE solvers also requires reverting to second-order accuracy at extremum (see \cref{sec:why_standard_dont_work}), revealing that ML-based PDE solvers also have tradeoffs between accuracy and robustness. As we've discussed, inductive biases motivated by physical laws are not sufficient to guarantee the reliability of ML-based PDE solvers; guaranteeing reliability would most likely require some sort of analytic constraints on the truncation error. While \cite{bar2019learning} invented a method of ensuring that an ML-based PDE solver has a specified order of convergence, \cite{bar2019learning} also found that the best performance (i.e., lowest accuracy at large $\Delta x$) required first-order convergence. Constraining the ML-based solver to have higher-order convergence degraded accuracy, suggesting that the flexibility required for improvements in speed and/or accuracy over standard numerical methods might be incompatible with the constraints required for robustness.

	\section{What applications could ML-based PDE solvers be useful for?}

Thus far, we've made four main observations about ML-based PDE solvers. First, while they are often slower than standard numerical methods, in some cases they've been faster than state-of-the-art numerical methods. Second, empirical evidence suggests that they do not generalize well outside the distribution of the training data. Third, work on designing ML-based PDE solvers that satisfy inductive biases (such as symmetries \cite{wang2020incorporating} and conservation laws \cite{mcgreivy2023invariant}) has not been sufficient to ensure the robustness of ML-based PDE solvers outside the training data. Fourth, there are serious challenges to designing ML-based PDE solvers that are significantly faster than standard numerical methods while simultaneously having theoretical guarantees of robustness; doing so may not be possible. The exception is accelerating the solution of linear systems of equations with ML, for which convergence guarantees are possible while achieving modest speedups relative to state-of-the-art standard numerical methods.

ML-based PDE solvers have an additional limitation relative to standard numerical methods: they require additional computational costs, both to train the ML model and to produce the training data.\footnote{PINNs are an exception to this, as they do not require training data. } Typically, producing the training data requires using standard numerical methods; this can be extremely time-consuming if high-quality training data is required, because high-quality training data can only be obtained from high-resolution (and thus slow) numerical methods.

Because ML-based PDE solvers struggle to generalize outside of the training data, their predictions can only be trusted within the distribution of the training data and on qualitatively similar dynamics. This severely limits the applicability of ML-based PDE solvers, since engineers and scientists are mostly interested in predicting the behavior of PDEs under novel conditions and potentially with qualitatively different dynamics.

Due to these limitations and the effectiveness of standard numerical methods, ML-based PDE solvers can only be useful under three conditions. First, either the ML-based PDE solver must have guarantees of robustness (such as with accelerating linear solves), or the application must involve solving the same PDE repeatedly under similar conditions without any qualitatively new dynamics that aren't found in the training data. Unfortunately, there is no way of rigorously verifying that an application is within the distribution of the training data. Second, the ML-based PDE solver must reduce the total computational cost relative to using standard numerical methods. Otherwise, there is no benefit to using ML-based PDE solvers. The total computational cost will be reduced relative to using standard numerical methods if the following formula is satisfied:
\begin{equation} \label{eq:conclusion-computational-cost}
	C_{\textnormal{data}} + C_{\textnormal{train}} + N \frac{t_B}{s} \le N t_B.
\end{equation}
$N$ is the number of evaluations of the ML-based PDE solver, $C_{\textnormal{data}}$ is the time required to generate the training data, $C_{\textnormal{train}}$ is the time required to train the model(s), $t_B$ is the time required for the standard numerical method baseline to compute one surrogate evaluation at equivalent accuracy to the ML model, and $s$ is the speedup of the ML-based solver relative to the numerical baseline. \Cref{eq:conclusion-computational-cost} only has a solution if $s > 0$. If $C_{\textnormal{data}} \gg t_B$, which will typically be the case, then $N$ must be very large to reduce the total computational cost. Third, the application must not be a toy problem, but rather a problem that standard numerical methods are unable to solve, or at least are unable to solve except with very large computational cost. Solving toy problems can demonstrate the potential for a method to be useful, but is not in itself useful.

Many papers have studied the use of ML for solving PDEs and attempted to develop new and improved PDE solvers. However, to the best of our knowledge, all of these papers have only studied toy problems which can already be solved using standard numerical methods at a reasonable computational cost. As a result, ML has not yet been useful in research or engineering applications involving solving PDEs.

There are specific applications involving the solution to systems of linear equations which satisfy at least two, and potentially all three, of the conditions listed above. For example, the 2D incompressible Navier-Stokes equations or the 2D Hasegawa-Wakatani equations are hyperbolic PDEs that require the solution to a linear system of equations at each timestep. Some ML-based methods for accelerating linear solves, such as learned preconditioners, retain theoretical guarantees of convergence, satisfying the first conditions. Because linear solves must be performed each timestep, then $N$ will be very high for many practical applications, satisfying the second condition.
It is unclear whether any specific applications relevant to research applications are limited by the speed of solving linear systems of equations; however, solving linear systems is such a fundamental aspect of numerical linear algebra and such a common task in standard numerical algorithms that we expect many research-relevant applications would benefit from faster linear solvers. Thus, for developing faster linear solvers with ML, the third condition is likely satisfied as well.


With the possible exception of accelerating linear solves, we are pessimistic about the possibility of ML being useful as a tool in PDE solving. The reason for this is that we have not been able to identify any specific research or engineering applications which satisfy all three conditions mentioned above.
In theory, faster ML-based PDE solvers could be useful for applications such as uncertainty quantification (UQ), optimization, inverse problems, or optimal control \cite{mishra2018machine}. However, we have not been able to identify any \textit{specific} research applications involving UQ, optimization, inverse problems, that satisfy the three conditions listed above, nor have we seen any other researchers identify specific applications in the hundreds or even thousands of papers we've read over the past few years.
Unless specific applications which satisfy the necessary criteria can be identified, or some major breakthrough can be made which improves robustness, ML-based PDE solvers are unlikely to ever be useful.


	\section{Recommendations}

Many papers in ML-for-PDE solving research are overoptimistic. We've identified two main reasons for this overoptimism. First, comparing ML-based PDE solvers to weak baselines. As a result, many papers achieve overly positive results. Second, reporting biases that cause underreporting of negative results. We discussed the reasons for these two issues in \cref{ch:reproducibility}, as well as recommendations to prevent their reoccurrence in \cref{sec:causes}.

In addition to the recommendations in \cref{sec:causes}, we make three recommendations and propose one challenge problem.
First, we recommend that research into ML-based PDE solvers put increased attention and effort into understanding and measuring generalization performance.
While papers often focus on quantitative metrics like speed and accuracy, understanding the generalization capabilities of ML-based PDE solvers is essential to understanding their performance.
The accuracy of ML-based PDE solvers is typically measured based on a testing set drawn from the same distribution as the training set. While this measures the best-case performance of ML-based PDE solvers, it doesn't give much information about the generalization capabilities of the solver. Standard numerical methods, however, use a variety of benchmark problems to test robustness and accuracy. For example, on the 1D advection equation, benchmark problems include advection of a square wave, advection of a Gaussian pulse, and advection of both a square wave and a Gaussian pulse simultaneously. ML-based PDE solvers should include these benchmark robustness tests; preferably, this would include both when these tests are in the training dataset and when they are excluded from the training dataset. These sorts of tests would allow readers to understand the generalization performance of ML-based PDE solvers in the same way that they help test the robustness of standard numerical methods. Additionally, a systematic review of generalization capabilities in ML-for-PDE solving research -- evaluating both successful generalization capabilities and failed generalization tests -- would be helpful to better understand the present generalization capabilities of ML-based PDE solvers.

Second, we recommend that researchers attempt to develop a better theoretical understanding of what is necessary to guarantee reliability of ML-based PDE solvers. We introduced (\cref{ch:invariant}) a method of improving robustness in autoregressive ML-based PDE solvers by preserving invariants (such as numerical stability and conservation laws), but these improvements are not sufficient to guarantee robustness. Researchers might ask questions such as: what properties are required for ensuring robustness of ML-based PDE solvers? Is there a fundamental tradeoff between speed and robustness, and if so, how can that tradeoff be formalized? Most importantly, is there a solution to the great challenge of developing ML-based PDE solvers that are both faster than standard numerical methods and equally as reliable?

Third, we recommend that researchers working in this area put less effort into publishing papers about using ML to solve toy problems and more effort into understanding what specific downstream applications those ML-based PDE solvers might one day be useful for.
This could be difficult to achieve, as researchers typically have strong incentives to publish papers frequently.
Funding agencies can help by allocating funding towards the creation of challenge problems or benchmarks currently unsolvable by standard numerical methods, with prizes given to those who can solve them. Hopefully this would shift the incentives away from publishing papers on toy problems, which have thus far had minimal impact on any research application (except perhaps weather modeling), and towards tackling the unsolved problems of greatest scientific importance.
On the other hand, if scientists are unable to think of any challenge problems which ML might be able to solve, or if ML researchers are unable to solve any of the challenge problems that are proposed, that should be seen as clear evidence that ML is unlikely to be useful for solving PDEs.

We propose a challenge problem for researchers working on ML-based PDE solvers: to develop an ML-based preconditioner suitable for integration into LAPACK, Eigen, BLAS, or other standard packages for solving linear systems of equations. Since these packages are used in many, even most standard numerical algorithms -- solving linear equations is a very fundamental task -- a faster, reliable ML-based preconditioner would be of great benefit to PDE solving research and applications. However, this challenge problem is not an easy task. The preconditioner for the linear equation $\bm A \bm x = \bm b$ would have to work for arbitrary matrix sizes $\bm A$, a range of sparsity patterns, and remain efficient for a variety of possible values of $\bm b$. It would have to be tested extensively to confirm that it is both reliable and faster than standard numerical methods in a known class of problems. Furthermore, it would require sufficiently small memory (at most a few megabytes) so as not to dramatically increase the memory footprint of these commonly used packages. While satisfying these constraints would likely be extremely challenging, if they can all be satisfied and successfully integrated into mature software packages such as LAPACK, Eigen, and/or BLAS, ML would finally and unambiguously have shown the ability to be useful in PDE solving.

	\section{Conclusion}

We are mostly pessimistic about the possibility of ML being useful in PDE solving. While many papers have reported impressive speedups relative to standard numerical methods, most of these comparisons are relative to weak baselines. ML-based PDE solvers are sometimes faster than state-of-the-art numerical methods, but in many cases they seem to be much slower. Furthermore, in most applications speed is not as important as robustness and reliability. Unlike standard numerical methods, ML-based PDE solvers do not have theoretical guarantees of reliability and empirically struggle to generalize outside the training distribution. While it may be possible to develop ML-based PDE solvers that are both fast and robust, we argue that there are serious and possibly insurmountable challenges to doing so. As a result, we conclude that ML-based PDE solvers can only be useful in applications with three conditions. However, we have not been able to identify any specific applications which satisfy these conditions. Unless an application with these conditions can be identified, we anticipate that ML will not be useful for solving PDEs except in toy problems.

There is one area of PDE solving research for which we have some optimism about the potential of ML: accelerating the solution of linear systems of equations. This is a fundamental task in linear algebra, and commonly required within standard numerical algorithms for solving PDEs. While much is still uncertain about this possibility, modest speedups have been achieved relative to state-of-the-art numerical methods, while maintaining theoretical guarantees of convergence and thus reliability. We believe the best case scenario is the development of an efficient and reliable ML-based preconditioner or iterative algorithm suitable for integration into widely used linear algebra packages; we anticipate this would be quite difficult to accomplish and propose it as a challenge problem for researchers to work towards. If achieved, it would be the first example of ML being unambiguously useful in PDE solving.

\appendix 

\chapter{Additional details of Biot-Savart line integral \label{sec:appendixbiotsavart}}
\section{Derivation of Piecewise Linear Shift \label{sec:ch3-biotsavart-appendixA}}
Here we show that by shifting each point on the filamentary coil in the outwards normal direction by an amount \unstretch{$\kappa|\delta \bm r'|^2/12$}, where $\kappa$ is the local curvature and \unstretch{$d \bm r' \equiv (d\bm r'/ds) \delta s$}, cancels the second-order error in the Biot-Savart field near the center of the coil. To do this, we compare the Biot-Savart fields for the straight segment and curved section as shown in figure \ref{fig:ch3-biotsavart-shift}. {With the origin defined as the midpoint between the two endpoints of the curved segment in figure \ref{fig:ch3-biotsavart-shift},} the straight segment can be written as \unstretch{$\bm r' = x' \hat{x} + \alpha \hat{y}$} for \unstretch{$x'\in [-L/2(1 + \alpha/R), L/2(1 + \alpha/R)]$} while the curved segment can be written as \unstretch{$\bm r' = x' \hat{x} + ({L^2}/{8R} - {x'^2}/{2R}) \hat{y}$ for $x'\in [-L/2, L/2]$}. {This quadratic parametrization of the curved segment is equal, through second order, to that of any arbitrary smooth curve.} {We compute the magnetic field at the point \unstretch{$\bm r_0 = (0, -R, 0)$} where \unstretch{$R \equiv 1/\kappa$}.} We compute the Biot-Savart fields for the straight segment and curved segment and work in the limits \unstretch{$L/R \sim \alpha/L \sim \mathcal{O}(\epsilon)$} and \unstretch{$\alpha/R \sim \mathcal{O}(\epsilon^2)$}, keeping all terms of up to second order.

The magnetic field for the straight segment at $\bm r_0$ is given by the Biot-Savart law (equation \ref{eq:ch3-filamentary_biot_savart}) and equal to
\begin{align}
    \bm B(\bm r_0) =&\nonumber\\ -&(R + \alpha) \hat{z} \int_{-L/2(1 + \alpha/R)}^{L/2(1 + \alpha/R)} \frac{dx'}{\Big(x'^2 + (R + \alpha)^2\Big)^{3/2}}.
\end{align}
Defining \unstretch{$\beta \equiv x'/ L$}, this can be written as
\begin{align}
    \bm B(\bm r_0) =&\nonumber \\ -\frac{L}{R^2}&(1+\frac{\alpha}{R}) \hat{z} \int_{-(1+\frac{\alpha}{R})/2}^{(1+\frac{\alpha}{R})/2} \frac{d\beta}{\Big((1 + \frac{\alpha}{R})^2 + \frac{L^2}{R^2}\beta^2\Big)^{3/2}}.
\end{align}
The integral can be performed analytically, it gives
\begin{equation}
        \bm B(\bm r_0) = -\frac{L}{R^2} \frac{\hat{z}}{\sqrt{\frac{L^2}{4R^2}(1 + \frac{\alpha}{R})^2 + (1 + \frac{\alpha}{R})^2}}.
\end{equation}
Expanding the denominator gives
\begin{equation}\label{eq:ch3-biotsavart-straightfinal}
    \bm B(\bm r_0) = -\hat{z}\frac{L}{R^2}\Big(1 - \frac{\alpha}{R} - \frac{L^2}{8R^2}\Big) + \mathcal{O}(\epsilon^4).
\end{equation}

The magnetic field for the curved segment is given by
\begin{equation}
    \bm B(\bm r_0) = -\hat{z}\int_{-L/2}^{L/2} \frac{R +  \frac{L^2}{8R} + \frac{x'^2}{2R}}{\Big(x'^2 + (R + \frac{L^2}{8R} - \frac{x'^2}{2R})^2\Big)^{3/2}}dx'.
\end{equation}
Again defining \unstretch{$\beta \equiv x'/ L$}, this can be written as
\begin{align}
    \bm B(\bm r_0) =&\nonumber \\ -\hat{z}\frac{L}{R^2}& \int_{-1/2}^{1/2} \frac{1 + \frac{L^2}{8R^2}(1 + 4\beta^2)}{\Big(\frac{L^2}{R^2}\beta^2 + (1 + \frac{L^2}{8R^2}(1 - 4\beta^2))^2 \Big)^{3/2}}d\beta.
\end{align}
Expanding the denominator, this gives
\begin{align}
    \bm B(\bm r_0) = &-\hat{z}\frac{L}{R^2} \int_{-1/2}^{1/2}\Big(1 + \frac{L^2}{8R^2}(1 + 4\beta^2) \nonumber\\&- \frac{3L^2}{2R^2}\beta^2 - \frac{3L^2}{8R^2}(1 - 4\beta^2)\Big) d\beta
\end{align}
\begin{equation}\label{eq:ch3-biotsavart-curvedfinal}
    \bm B(\bm r_0) = -\hat{z}\frac{L}{R^2}\Big(1 - \frac{5L^2}{24R^2} \Big) + \mathcal{O}(\epsilon^4)
\end{equation}

Comparing equations \ref{eq:ch3-biotsavart-straightfinal} and \ref{eq:ch3-biotsavart-curvedfinal}, we see that setting \unstretch{$\alpha = L^2/12R$} gives the same second-order error in $B_z$ at \unstretch{$\bm r_0 = (0, -R, 0)$}. Here, \unstretch{$R \equiv 1/\kappa$} and to fourth order \unstretch{$L^2 = |\delta \bm r'|^2 \equiv |(d \bm r'/ds) \delta s|^2$}.

\section{Minimization of Mean Squared Distance \label{sec:ch3-biotsavart-appendixB}}

With the same parameterization as above, we have that the squared distance from the curved segment to the straight segment is
\unstretch{$( {\alpha x'}/R )^2 + ( {L^2}/{8R} - {x'^2}/{2R} - \alpha )^2$} for \unstretch{$x'\in [-L/2, L/2]$}.  Taking the average over the segment gives
\begin{align}
    \langle |\Delta \bm r'|^2 \rangle = \int_{-1/2}^{1/2} \Big( \big( \frac{L^2}{8R}& - \alpha \big)^2 + \nonumber \\ \big( \frac{\alpha^2}{R^2} - \frac{1}{R}\big( &\frac{L^2}{8R} - \alpha \big) \big) L^2 \beta^2 + \frac{1}{4R^2} L^4 \beta^4 \Big) d\beta \nonumber\\
    = \Big( \frac{L^4}{64R^2} - \frac{L^2}{4R} \alpha + \alpha^2 \Big)&\nonumber \\ + \frac{1}{12}\Big( - \frac{L^4}{8R^2} + &\frac{L^2}{R} \alpha + \frac{L^2}{R^2} \alpha^2 \Big) + \frac{1}{80} \Big( \frac{L^4}{4R^2} \Big). \label{eq:ch3-biotsavart-AverageSquaredDistance}
\end{align}
The derivative with respect to $\alpha$, 
\begin{equation}
    d\langle |\Delta \bm r'|^2 \rangle/d\alpha = - \frac{L^2}{4R} + 2\alpha + \frac{L^2}{12R} + \frac{L^2}{6R^2}\alpha, 
\end{equation}
when set to zero, gives the optimal shift of the straight segment: 
\begin{equation}
    \alpha = \frac{1}{12}\frac{L^2}{R} \Big( 1 + \mathcal{O}(\epsilon^2) \Big) \label{eq:ch3-biotsavart-SquaredShiftDerivation}
\end{equation}


\section{Prescription for {Piecewise Continuous} Curvature \label{sec:ch3-biotsavart-appendixD}}
If the coil curvature is only \textit{piecewise} continuous, one should approximate each {segment} individually. {Although we have found that closed curves with continuous curvature have fourth-order error convergence using the shifted piecewise linear approach, to obtain fourth-order convergence from a non-closed segment requires a slightly modified method. We describe this method below.}

Consider approximating the magnetic field coming from one such section, parameterized by $s_i\in[0,1]$ for $i=0,1,\ldots,N$.  The procedure is the same as before, with the following modulations: 
\begin{enumerate}
    \item Set $s_0=0$ and $s_N=1$.  These points will not be shifted.  
    \item For $i\in1,2,\ldots,N-1$, we used to place $s_i$ uniformly.  Instead, spread them out slightly, such that the spacing in $s$ between ($s_0$ and $s_1$) and ($s_{N-1}$ and $s_N$) are smaller than the others by a factor of $\sqrt{2}$.  
    \item Shift the points $1,2,\ldots,N-1$ using equation \eqref{eq:ch3-biotsavart-alpha}.  For all shifts, use the $\delta s$ between the interior points, not the shorter length of the endpoints.  
\end{enumerate}

\chapter{Details of machine learned solvers \label{sec:appendix_verification_details}}
\section{Details for section \ref{sec:why_standard_dont_work}}

We set $c=1$ and use periodic BCs. We use the continuous-time FV update equation \cref{eq:fv_a}. We use a SSPRK3 ODE integrator \cite{ssprk}. We choose the timestep $\Delta t$ using a CFL condition with a CFL number of 0.3. The initial conditions for both the training set and test set are draws from a sum-of-sines distribution
\begin{equation*}
    u_0(x) = \sum_{i=1}^{N^{\textnormal{modes}}} A_i \sin{\big(2\pi k_i x + \phi_i\big)}
\end{equation*}
where $N^\textnormal{modes} \sim \{1,2,3,4,5,6\}$ and $k_i \sim \{1,2,3,4\}$ are uniform draws from a set while $A_i \sim [-1.0,1.0]$ and $\phi_i \sim [0, 2\pi]$ are draws from uniform distributions. The loss function $L$ is given by computing the mean squared error (MSE) between the predicted time-derivative and the so-called `exact' time-derivative
\begin{equation*}
    L = \frac{1}{N_x} \sum_{j=1}^{N_x} \bigg (\frac{du_j(t)}{dt} - \frac{du_j^\textnormal{exact}(t)}{dt}\bigg)^2.
\end{equation*}
Both the `exact' solution $u_j^\textnormal{exact}(t)$ and the `exact' time-derivative $\nicefrac{du_j^\textnormal{exact}}{dt}$ are coarse-grained versions of a high-resolution simulation $u^\textnormal{exact}$, i.e., $u_j^\textnormal{exact}(t) = \int_{x_{j-\nicefrac{1}{2}}}^{x_{j+\nicefrac{1}{2}}} u^{\textnormal{exact}}(x,t) dx$. 
For each sample in the training data from the distribution of initial conditions, we take 50 snapshots evenly spaced in time from $t \in [0,1]$. We draw 100 samples, for a total of $5000$ snapshots in our training dataset. For each snapshot we store the exact trajectory $u_j^\textnormal{exact}$ and the exact time-derivative $\nicefrac{du_j^\textnormal{exact}}{dt}$.
Our machine learning models are periodic convolutional neural networks (CNNs) which are given the downsampled exact trajectories $u_j^{\textnormal{exact}}(t)$ as inputs. Solvers 4, 6, and 7 output the flux $f_{j+\nicefrac{1}{2}}$, while solver 5 outputs $\alpha_{j+\nicefrac{1}{2}}$. These outputs are then used to compute the predicted time-derivative $\nicefrac{du_j(t)}{dt}$.
All models train with a batch size of 32 and use the ADAM optimizer for 100 epochs over the training dataset with a learning rate of $1 \times 10^{-3}$, followed by another 100 epochs with a learning rate of $1 \times 10^{-4}$. Solver 7 uses $G_{j+\nicefrac{1}{2}}(\bm u_j) = u_{j+1} - u_j$.

\section{Details for section \ref{sec:verification_burgers}}

Our goal is to solve the 1D Burgers' equation for $u(x,t) \in \mathbb{R}$ with diffusion and forcing:
\begin{equation}\label{eq:burgers_forcing}
    \frac{\partial u}{\partial t} + \frac{\partial}{\partial x} \bigg(\frac{u^2}{2}\bigg) = \nu \frac{\partial^2 u}{\partial x^2} + F(x, t).
\end{equation}
$x \in \Omega$ and $\Omega = [0, L]$. We set $\nu = 0.01$ and $L=2\pi$. We use periodic BCs. The initial conditions $u_0(x) = 0$. Each simulation in both the training and test data uses a randomized sum-of-sines forcing function
\begin{equation}
    F(x,t) = \sum_{m=1}^M A_m \sin{(2\pi k_m x/L - \omega_m t + \phi_m)}
\end{equation}
with $M=20$. The random variables $A_m \in [-0.5, 0.5]$, $\phi_m \in [0, 2\pi]$, and $\omega_m \in [-0.4, 0.4]$ are drawn from uniform distributions while $k_m$ is sampled uniformly from the set $\{3,4,5,6\}$. 

The machine learned solver and the standard solvers use the FV formulation. The domain $\Omega$ is divided into $N$ cells of width $\Delta x = \nicefrac{L}{N}$ and the solution average within each cell is represented by $u_j$ for $j = 1, \dots, N$. The FV update equations for \cref{eq:burgers_forcing} are
\begin{equation}
    \frac{\partial u_j}{\partial t} + \frac{\mathcal{J}_{j+\frac{1}{2}} -\mathcal{J}_{j-\frac{1}{2}} }{\Delta x} = F_j.
\end{equation}
The forcing 
\begin{equation}
    F_j(t) = \int_{x_j - \frac{\Delta x}{2}}^{x_j + \frac{\Delta x}{2}} F(x,t) \mathop{dx}
\end{equation}
is approximated using a 1-point 1st-order Guassian quadrature. For all solvers, the coefficients are advanced in time using a 3rd-order strong stability preserving Runge-Kutta \cite{ssprk} ODE integrator with a CFL factor of 0.3. Reducing the timestep further does not improve the accuracy of the baseline solvers.

The machine learned solver approximates the flux at each of the $N$ cell boundaries $\mathcal{J}_{j+\nicefrac{1}{2}}$ using the equation
\begin{equation}
    \mathcal{J}_{j+\frac{1}{2}} = \frac{1}{2}u_{j+\frac{1}{2}}^2 - \nu \bigg(\frac{\partial u}{\partial x}\bigg)_{j+\frac{1}{2}}.
\end{equation}
The reconstructed values of the solution $u_{j+\nicefrac{1}{2}}$ and its derivative $(\nicefrac{\partial u}{\partial x})_{j+\nicefrac{1}{2}}$ are approximated using the `data-driven discretization' approach introduced in \cite{bar2019learning}. The data-driven discretization approach uses learned stencils $s^d_{j+\nicefrac{1}{2},k}$ to approximate the $d$th derivative of the solution:
\begin{equation}
    u_{j+\frac{1}{2}} = \sum_{k=1}^W s^0_{j+\frac{1}{2},k} u_{j-\frac{W}{2} + k} \quad\mathrm{and}\quad \bigg(\frac{\partial u}{\partial x}\bigg)_{j+\frac{1}{2}} = \sum_{k=1}^W s^1_{j+\frac{1}{2},k} u_{j-\frac{W}{2} + k}.
\end{equation}
We set the stencil width $W=6$. The learned stencils are computed as follows. First, a CNN maps the input array $u_j$ of length $N$ to two output arrays $\tilde{s}^0_{j+\nicefrac{1}{2},k}$ and $\tilde{s}^1_{j+\nicefrac{1}{2},k}$ of shape $(N, W)$. Second, for each of the $N$ stencil coefficients $\tilde{s}$, we project the length-$W$ vector of stencil coefficients into the null space of a matrix $M^d$, resulting in intermediate stencils $\bar{s}$. These projections can be written as 
\begin{equation}
    \bar{s}^0_{j+\frac{1}{2},k} = \tilde{s}^0_{j+\frac{1}{2},k} - \frac{1}{W}\sum_{l=1}^W \tilde{s}^0_{j+\frac{1}{2},l} \quad \mathrm{and} \quad \bar{s}^1_{j+\frac{1}{2},k} = \tilde{s}^1_{j+\frac{1}{2},k} - P^1_{k,l} \tilde{s}^1_{j+\frac{1}{2},l}
\end{equation}
where the projection matrix $P^1_{k,l}$ is given by
\begin{equation}
    P^1 = (M^{1})^T(M^1 (M^1)^T)^{-1} M^1
\end{equation}
where
\begin{equation}
    M^1 = \begin{bmatrix}
        1 & 1 & 1 & 1 & 1 & 1\\
        \frac{-5}{2} & \frac{-3}{2} & \frac{-1}{2} & \frac{1}{2} & \frac{3}{2} & \frac{5}{2}
    \end{bmatrix}
\end{equation}
Third, we add a `base' stencil $\hat{s}$ to the intermediate stencil $\bar{s}$. 
\begin{equation}
    {s}^0_{j+\frac{1}{2},k} = \bar{s}^0_{j+\frac{1}{2},k} + \hat{s}^0_k \quad \mathrm{and} \quad {s}^1_{j+\frac{1}{2},k} = \frac{\bar{s}^1_{j+\frac{1}{2},k} + \hat{s}^1_k}{\Delta x}
\end{equation}
where $\hat{s}^0_k = \begin{bmatrix}0 & 0 & \nicefrac{1}{2} & \nicefrac{1}{2} & 0 & 0 \end{bmatrix}$ and $\hat{s}^1_k = \begin{bmatrix}0 & 0 & -1 & 1 & 0 & 0 \end{bmatrix}$.
These three steps preserve formal 1st-order accuracy of the stencil coefficients. 

The CNN first applies two periodic convolutions with kernel size $K=5$, 32 channels, and ReLU activation function. The CNN then applies a periodic convolution with kernel size $K=4$ and $2 \times W$ output channels. A periodic convolution involves periodically padding the array with $K-1$ pixels then applying a `valid' convolution. The CNN has a receptive field of 12, half of which are on either side of the $j+\nicefrac{1}{2}$th cell boundary. The CNN weights are initialized using the LeCun normal initialization. The biases are initialized to zero.

The training data is generated using a high-resolution simulation with $N=512$ grid points and the WENO5 \cite{weno5} flux function. We generate data from 800 simulations, storing 10 snapshots per simulation taken every $t=0.5$ units of time, starting at $t=0$. We save the solution $u_j$ as well as the downsampled high-resolution `exact' time-derivative $\nicefrac{\partial u_j}{\partial t}$. We downsample by averaging $u_j$. 

Our loss function is given by the mean squared error between the downsampled high-resolution `exact' time-derivative and the time-derivative from the `data-driven discretization' time-derivative. We train with a batch size of $128$ times the downsampling factor. We use the Adam optimizer. We train for 20,000 steps with a learning rate of $3 \times 10^{-3}$ followed by 20,000 steps with a learning rate of $3 \times 10^{-4}$. 

\Cref{fig:ml_burgers} is computed by averaging the mean absolute error over 100 simulations in the test set. The `exact' solution is given by a high-resolution solver using the WENO flux. We initialize with $u_0(x)=0$, then compute the average error with each lower-resolution method from $t=0$ to $t=15$.

While our setup is nearly identical to the setup in \cite{bar2019learning}, there are a few minor differences. First, \cite{bar2019learning} incorrectly computes the Godunov flux, which leads to numerical instability of their baseline methods and worse performance. Second, we use a SSPRK3 ODE integrator while \cite{bar2019learning} uses an order 3(2) adaptive time-stepping method from SciPy. Third, our CNN has a receptive field which is 12 points wide (symmetric around the $j+\nicefrac{1}{2}$th cell boundary) while the CNN in \cite{bar2019learning} has a receptive field which is 13 points wide. We don't believe these minor differences are the root cause of the discrepancy in \cref{fig:ml_burgers} between our replication attempt (ML, black line) and figure 3 of \cite{bar2019learning} (Bar-Sinai \& Hoyer et al., brown line). We do not know why we are unable to exactly replicate figure 3 of \cite{bar2019learning}.

\section{Details for section \ref{sec:verification_2d_euler}}
Our goal is to solve the 2D incompressible Euler's equations for $\chi(x,y,t)$ with forcing and diffusion:
\begin{align}
    \frac{\partial \chi}{\partial t} + \bm \nabla \cdot (\bm u \chi) = F(x,y,t) + \nu \nabla^2\chi \textnormal{,} && \bm u = \bm \nabla \psi \times \hat{e}_z\textnormal{,} && -\bm\nabla^2 \psi = \chi.
\end{align}
We choose periodic boundary conditions such that $x, y \in [0, L]$ and $L=2\pi$. We set $\nu = 10^{-3}$ and use the forcing function in \cite{kochkov2021machine},
\begin{equation}
    F(x,y,t) = \frac{2 \pi k}{L} \cos{\bigg(\frac{2\pi k y}{L}\bigg)} - 0.1\chi.
\end{equation}
with $k=4$. The initial conditions in both the training and test sets are random draws from the curl of the `filtered velocity field' from JAX-CFD \cite{kochkov2021machine}. 

The details of the standard solver are given by the `MUSCL' scheme described in \cref{sec:energyconservation}. The only difference between the machine learned solver and the standard solver is that the machine learned solver learns a correction to the flux at cell boundaries. This correction term is given by a simple CNN, which uses periodic convolutions and outputs 2 channels per cell. Each channel represents the flux at either the right cell boundary or the top cell boundary. The neural network uses kernels of size 5, has 64 channels per layer and 6 total layers.

We generate 100 evenly spaced data points from 100 simulations. For generation of training data, we run the `exact' high-resolution simulation until $t=20$ before sampling training data every $t=0.1$ units of time. Our loss function is again given by the mean squared error between the downsampled high-resolution `exact' time-derivative and the time-derivative from the machine learned solver. We train with a batch size of 100, use the Adam optimizer, and train for 1000 epochs over the training set. We use a learning rate of $10^{-4}$.

\Cref{fig:corr_2d_euler} is computed by computing the correlation between the downsampled high-resolution $128\times 128$ `exact' solution and the low-resolution solution(s). We run the `exact' solver until $t=20$ and use the solution at $t=20$ as the initial condition.

\section{Details for section \ref{sec:verification_compressible_euler}}

Our goal is to solve the 1D compressible Euler's equations \cref{eq:compressibleeuler} and to compare the accuracy of a standard solver, a machine learned solver, and the same machine learned solver with the error-correcting invariant-preserving algorithm \cref{eq:compressible_euler_positivity_transformation,eq:compressible_euler_entropy_transformation} applied. We train solvers at different values of $N$, where $N$ is the number of spatial grid cells. We train solvers in both periodic domains and in domains with dirichlet boundary conditions. 

Our domain is $x \in [0, L]$ with $L=1$. The initial conditions in both the training and test sets are random draws from a relatively simple distribution. This distribution has $\rho_0 = \textnormal{max}\{\rho_{\textnormal{min}}, \rho_{s}\}$, $v_0 = v_{s}$, and $p_0 = \textnormal{max}\{p_{\textnormal{min}}, p_{s}\}$ where $\rho_s$, $v_s$, and $p_s$ are draws from (different) sine waves with random amplitudes and phases. The amplitude $A$ is drawn from a uniform distribution $A \in [0, 1]$. We set $\rho_{\textnormal{min}} = 0.75$ and $p_\textnormal{min} = 0.5$. 

We generate 20 snapshots from 10,000 different runs of a high-resolution `exact' simulation, for a total of 200,000 data points in the training set. Each run takes a snapshot at $t=0$ and then another snapshop every $t=0.01$ units of time until $t=0.2$. The exact simulation has $N=256$ grid cells. We train each solver with a batch size of 64 times the upsampling factor. Each solver is trained for 100,000 training iterations, a learning rate of $10^{-4}$, and uses the ADAM optimizer. 

The machine learned solver learns a correction to the fluxes of the standard solver at each cell boundary; the standard solver is once again given by the MUSCL scheme with reconstruction in characteristic variables.  The flux correction is given by the output of a simple CNN with 5 layers of 32 channels. Each layer pads with either periodic or edge padding. Each layer uses a kernel size of 5, except for the last layer which uses a kernel size of 4 due to symmetry considerations. 

Once again, our loss function is given by the mean squared error between the downsampled high-resolution `exact' time-derivative and the time-derivative from the machine learned solver.

\Cref{fig:1d_euler_verification} is computed by computing the MSE averaged over time and space and averaged over 50 different initializations in the test set.

\chapter{Detailed explanations for weak baselines \label{sec:appendixresults}}
We now give detailed explanations for each of the 76 articles in table 1, as well as the 6 articles that claim to either underperform or have varied performance relative to a standard numerical method.

\appendixentry{\fnoli}{\citefnoli}{Fourier Neural Operator for Parametric Partial Differential Equations}{\fnolicitations}
\appendixdata{1D Burgers', 2D incompressible Navier-Stokes}
{``up to three orders of magnitude faster compared to traditional PDE solvers.''}
{No}
{Pseudo-spectral method}
{{\xmark} ``All data are generated on a $256\times256$ grid and are downsampled to $64\times64$.'' ``On a $256\times 256$ grid, the Fourier neural operator has an inference time of only 0.005s compared to the 2.2s of the pseudo-spectral method used to solve Navier-Stokes.'' This comparison is not at equal accuracy. To compare at equal accuracy, reduce the resolution of the pseudo-spectral method until the two methods have equal accuracy (as measured by table 1).  }
{{\cmark} A pseudo-spectral method or DG method is considered state-of-the-art for the 2D incompressible Navier-Stokes equations with periodic boundary conditions.}
{N/A}
{{\xmark} We replicated the primary outcome of this article using a DG method and found that the speedup of Fourier neural operator on GPU was $7\times$ faster than our laptop CPU, not three orders of magnitude faster.}

\appendixentry{\deeponetlulu}{\citedeeponetlulu}
{Learning nonlinear operators via DeepONet based on the universal approximation theorem of operators}{\deeponetlulucitations}
\appendixdata{1D advection (4 cases), 1D advection-diffusion}
{``as we show in Supplementary Table 5, the computational cost of running inference of DeepONet is substantially lower than for the numerical solver.''}
{No}
{``The reference solutions of all deterministic PDEs are obtained by a second-order finite difference method.'' ``To generate the training dataset, we solve the system using a finite difference method on a 100 by 100 grid.''}
{{\xmark} Table 5 does not list the runtime at equal accuracy, and the grid resolution of the finite difference method is not reduced to match the accuracy of the DeepONet.}
{{\cmark} A second-order finite-difference method is fairly efficient for the 1D advection equation, but note that DG methods are likely more efficient for the 1D advection equation with smooth solutions.}
{N/A}
{{\xmark} We consider the 1D advection equation (case 1). As listed in table S8, this is the linear advection equation from $x \in [0, 1]$ and $t \in [0, 1]$. The error of the $100 \times 100$ second-order finite-difference method is shown in figure S11; the runtime is $9 \times 10^{-3}$ seconds using 1 core on CPU. We instead run a DG code with quadratic basis functions with 13 grid cells with a similar initial condition, the runtime is $6 \times 10^{-5}$ seconds on my laptop which has 2 cores. In summary, we were able to achieve similar accuracy with an order of magnitude lower runtime. }

\appendixentry{\tompson}{\citetompson}
{Accelerating Eulerian Fluid Simulation With Convolutional Networks}{\tompsoncitations}
\appendixdata{2D and 3D Poisson for computer graphics (GPU-only) in real-time (low-accuracy)}
{``For Jacobi to match the divergence performance of our network, it requires 116 iterations and so is $4.1\times$ slower than our network.'' ``Note that for fair quantitative comparison of output residual, we choose the number of Jacobi iterations (34) to match the FPROP time of our network (i.e. to compare divergence at fixed compute).'' Supported by figure 5. }
{No}
{``A Jacobi-based iterative solver and a PCG-based solver (with incomplete Cholesky L0 preconditioner).'' The PCG baseline ``is orders of magnitude slower and has been omitted for clarity.'' }
{{\cmark} Compares divergence (accuracy) at constant compute (speed), and speed at constant accuracy.}
{{\cmark} For Poisson's equation, Multigrid methods (or preconditioners) are highly efficient at high accuracy, because they converge in many fewer iterations than Jacobi. See, e.g., figure 4e of article {\tang} or figure 23 of article \cheng. However, for the scenario considered in the paper of GPU-only real-time computer graphics applications, the authors argue that multigrid methods are less efficient than Jacobi iteration. This argument seems plausible, but ultimately we are unsure if it is correct. }
{N/A}
{{\cmark} }

\appendixentry{\mlacceleratedcfd}{\citemlacceleratedcfd}
{Machine learning–accelerated computational fluid dynamics}{\mlacceleratedcfdcitations}
\appendixdata{2D incompressible Navier-Stokes}
{``our  results  are  as  accurate  as  baseline  solvers  with 8 to 10$\times$ finer resolution in each spatial dimension, resulting in 40- to 80-fold computational speedups.'' Supported by figures 2a and 2b.}
{No}
{Finite-volume method based on a Van-Leer flux limiter.}
{{\cmark} See figure 2b.}
{{\xmark} Pseudo-spectral and DG methods are highly efficient for this problem. The original authors of this article replicated this result using a strong spectral baseline on TPU (see article \dresdner) and found that a pseudo-spectral baseline was much faster than the weaker FV baseline and faster than a similar ML-based solver. On GPU, we find that the PS baseline is over 80x faster than the FV baseline. We also replicated this result using a DG baseline and found that DG-based methods could solve these equations at 10 to 11$\times$ coarser resolution and (on CPU) 20-40$\times$ faster than the original baseline.}
{N/A}
{{\xmark}}

\appendixentry{\meshbasedpfaff}{\citemeshbasedpfaff}
{Learning Mesh-Based Simulation with Graph Networks}{\meshbasedpfaffcitations}
\appendixdata{2D incompressible Navier-Stokes cylindrical flow, 2D compressible Navier-Stokes airfoil wing}
{``Our method is also highly efficient, running 1-2 orders of magnitude faster than the simulation on which it is trained.'' Supported by table 1 and section A.5.1.}
{No}
{COMSOL for incompressible Navier-Stokes, SU2 for compressible Navier-Stokes. Also compares to ANSYS.}
{{\xmark} Table 1 compares the runtime between the (highly accurate) ground-truth solver and the less accurate ML-based solver. This comparison is not at equal accuracy.}
{{\cmark} Though we don't have reason to believe COMSOL and SU2 are inefficient general-purpose tools, we recommend being cautious when evaluating this comparison (see recommendation 1 in Methods).}
{N/A}
{{\xmark}}

\appendixentry{\barsinai}{\citebarsinai}
{Learning data-driven discretizations for partial differential equations}{\barsinaicitations}
\appendixdata{1D Burgers', 1D Korteweg-de Vries, 1D Kuramoto-Sivashinsky}
{``The resulting numerical methods are remarkably accurate, allowing us to integrate in time a collection of nonlinear equations in 1 spatial dimension at resolutions $4\times$ to $8\times$ coarser than is possible with standard finite-difference methods.''}
{No}
{A FV method with a fifth-order upwind-biased WENO scheme with Godunov flux}
{{\cmark} See figure 3c.}
{{\xmark} We consider the 1D Burgers' equation (in figure 3). We replicate figure 3c, and compare WENO to DG methods with polynomial orders 2 and 3. We find (see github) that, as with the ML-based solver, DG methods are able to solve the 1D Burgers' equation at resolutions $4\times$ to $8\times$ coarser than WENO. The ML-based solver is still able to solve the 1D Burgers' equation with 2-4$\times$ fewer degrees of freedom compared to the DG-based method. We give an {\xmark} for rule 2 because we were able to replicate the article's primary outcome and achieve significantly improved performance with a stronger baseline.
}
{N/A}
{\xmark}

\appendixentry{\deepfluids}{\citedeepfluids}
{Deep Fluids: A Generative Network for Parameterized Fluid Simulations}{\deepfluidscitations}
\appendixdata{2D \& 3D incompressible Navier-Stokes (smoke/graphics)}
{``Reconstructed velocity fields are generated up to 700$\times$ faster than re-simulating the data with the underlying CPU solver.'' Supported by table 1.}
{No}
{Mantaflow}
{{\xmark} The resolution of the underlying CPU solver is not reduced to match the accuracy of the ML-based solver.}
{{\cmark} Though we don't have reason to believe that Mantaflow is an inefficient general-purpose tool, we recommend being cautious when evaluating this comparison (see recommendation 1 in Methods).}
{N/A}
{\xmark}

\appendixentry{\deeponetwang}{\citedeeponetwang}
{Learning the solution operator of parametric partial differential equations with physics-informed DeepOnets}{\deeponetwangcitations}
\appendixdata{1D Burgers'}
{``up to three orders of magnitude faster compared a conventional PDE solver.'' Supported by figure 11.}
{No}
{Spectral solver (Chebfun)}
{{\xmark} Figure 11 compares the runtime of a highly accurate spectral solver to that of a less accurate ML-based solver.}
{{\cmark} A spectral solver is highly efficient for Burgers' equation, as long as the diffusion coefficient is sufficiently high (so that shocks are not too strong). }
{N/A}
{{\xmark} We replicate the 1D Burgers' setup using a FV method with Godunov flux. Using 100 gridpoints on CPU, this gives an L2 error of 1\% and a runtime of $4.2\times 10^{-4}$ seconds, over an order of magnitude faster than the ML-based solver. On GPU, we can solve 1000 PDEs in $1.2\times10^{-2}$ seconds, again an order of magnitude faster than the MLP in figure 11b.}

\appendixentry{\solverinthe}{\citesolverinthe}
{Solver-in-the-Loop: Learning from Differentiable Physics to Interact with Iterative PDE-Solvers}{\solverinthecitations}
\appendixdata{2D Burgers', 2D Poisson, 2D and 3D incompressible Navier-Stokes wake dynamics}
{``A speed-up of more than 68$\times$ [for the simulation in figure 1].'' Supported by appendix C, section titled ``runtime performance''.}
{No}
{Reference simulation (baseline appears to be the \href{https://github.com/tum-pbs/PhiFlow}{PhiFlow} library, which is based on a MAC grid data structure). See section B.5 for details about 3D setup, and section B.1 for details about 2D baseline. Baseline appears to be a FV method.}
{{\xmark} The 68$\times$ speedup compares the reference simulation with MAE of 0 to a reference simulation with MAE of 0.13. This is only a 28\% improvement over the `source' simulation, which has an MAE of 0.167. A fair comparison would be between two simulations with approximately equal MAE.}
{{\cmark} While we recommend comparing with both FV and DG methods for the Navier-Stokes equations, we consider FV highly efficient for problems with fluid-structure interaction. Though we don't have reason to believe that PhiFlow is an inefficient general-purpose tool, we also recommend being cautious when evaluating this comparison (see recommendation 1 in Methods). }
{N/A}
{\xmark}

\appendixentry{\deepmmnetcvt}{\citedeepmmnetcvt}
{DeepM\&Mnet: Inferring the electroconvection multiphysics fields based on operator approximation by neural networks}{\deepmmnetcvtcitations}
\appendixdata{2D electroconvection (steady state)}
{``The speedup of DeepONets prediction versus the NekTar simulation for forward independent conditions is about 10,000 folds.'' Not supported by any other evidence. }
{No}
{Nektar: high-order spectral element (3 modes), 5 quadrature points in each direction, with second-order stiffly stable timestepping scheme. `` }
{{\xmark} To make a fair comparison, reduce the resolution of the Nektar simulation (below $32\times 32$) until its accuracy is equal to that of the DeepONet.}
{{\cmark} Though we don't have reason to believe that Nektar is an inefficient general-purpose tool, we recommend being cautious when evaluating this comparison (see recommendation 1 in Methods).}
{}
{\xmark}

\appendixentry{\belbuteperez}{\citebelbuteperez}
{Combining Differentiable PDE Solvers and Graph Neural Networks for Fluid Flow Prediction}{\belbuteperezcitations}
\appendixdata{2D compressible Navier-Stokes airfoil wing}
{``the substantial speedup of neural network CFD predictions.'' Supported by Table 1.}
{No}
{SU2}
{{\xmark} Compares ground truth (runtime 137s, RMSE 0.0) to CFD-GCD (runtime 2.0s, RMSE $5.4\times10^{-2}$) instead of comparing at equal accuracy.}
{{\cmark} Though we don't have reason to believe that SU2 an inefficient general-purpose tool, we recommend being cautious when evaluating this comparison (see recommendation 1 in Methods).}
{}
{\xmark}

\appendixentry{\pinoli}{\citepinoli}{Physics-Informed Neural Operator for Learning Partial Differential Equations}{\pinolicitations}
\appendixdata{1D Burgers', 2D incompressible Navier-Stokes}
{``Further, in PINO, we incorporate the Fourier neural operator (FNO) architecture which achieves orders-of-magnitude speedup over numerical solvers." Supported by figure 8.}
{No}
{Same spectral solver as in article \fnoli.}
{{\cmark} See figure 8.}
{{\xmark} The transient flow problem is identical to that in article \fnoli, with one key difference: the Reynolds number is now 20, instead of $10^3$-$10^5$. Thus, the problem is now diffusion-dominated rather than advection-dominated. We use a DG solver with second-order polynomial basis functions to replicate this result, except we reduce the resolution to $3\times3$ and change the timestep accordingly. With a $3\times3$ resolution, we find an error of 2-3\% with a runtime of 0.035s, 7$\times$ slower than the PINO method with similar accuracy.}
{N/A}
{\xmark}

\appendixentry{\projectionyang}{\citeprojectionyang}{Data-driven projection method in fluid simulation}{\projectionyangcitations}
\appendixdata{3D Poisson}
{``Experimental results demonstrated that our data-driven method drastically speeded up the computation in the projection step.'' Supported by table II.  }
{No}
{Preconditioned Conjugate Gradient (PCG) linear solver}
{{\xmark} The ML-based solver doesn't use iteration. Thus, the accuracy of the ML-based solver (as evidenced by figure 4) isn't as high at that of the PCG baseline.}
{{\xmark} For Poisson's equation, Multigrid methods (or preconditioners) are highly efficient. See, e.g., figure 4e of article {\tang} or figure 23 of article \cheng.}
{}
{\xmark}

\appendixentry{\neuralconverge}{\citeneuralconverge}{Learning Neural PDE Solvers with Convergence Guarantees}{\neuralconvergecitations}
\appendixdata{2D Poisson}
{``[Our model] achieves 2-3 times speedup compared to state-of-the-art solvers.'' Supported by figure 2.}
{No}
{``The FEniCS model is set to be the minimal residual method with algebraic multigrid preconditioner, which we measure to be the fastest compared to other methods such as Jacobi or Incomplete LU factorization preconditioner.'' }
{\cmark ``We evaluate the convergence rate by calculating the computation cost required for the error to drop below a certain threshold.'' This article compares runtime at equal accuracy. }
{{\xmark} Consider figure 2a: FEniCS takes almost 20 seconds to solve a $256\times 256$ Poisson problem on a square domain. We implement Poisson's equation on a square domain using a direct solve (LU decomposition) and find that the direct solve takes 12 milliseconds, over three orders of magnitude faster than FEniCS. Multigrid is a weak baseline relative to direct methods for sufficiently small problems.}
{N/A}
{\xmark}

\appendixentry{\messagepassing}{\citemessagepassing\hspace{0.1cm}(Note: we consider version 2 of this article on ArXiv, version 3 was uploaded after private communication with the authors.)}{Message Passing Neural PDE Solvers}{\messagepassingcitations}
\appendixdata{1D Burgers', 1D wave, 2D incompressible Navier-Stokes (smoke/graphics)}
{``Our model outperforms state-of-the-art numerical solvers in the low resolution regime in terms of speed and accuracy.'' Supported by tables 1 and 2.}
{No}
{WENO5 (Burgers) and spectral (wave)}
{{\cmark} See tables 1 and 2.}
{{\xmark} It's a slow implementation of WENO5 and spectral. }
{N/A}
{{\xmark} We replicated both problems with a stronger baseline, and found that the stronger baseline was orders of magnitude faster than the ML-based solver.}

\appendixentry{\deepmmnet}{\citedeepmmnet}{DeepM\&Mnet for hypersonics: Predicting the coupled flow and finite-rate chemistry behind a normal shock using neural-network approximation of operators}{\deepmmnetcitations}
\appendixdata{2D reacting Navier-Stokes}
{``DeepONets can be over five orders of magnitude faster than the CFD solver employed to generate the training data.'' }
{No}
{CFD solver (no details given) coupled with the MUTATION library}
{{\xmark} This article compares a method with relative MSE of $1e-5$ to a CFD solver with relative MSE of 0.0. For a fair comparison, reduce the resolution of the CFD solver until the relative MSE is equal. }
{{\xmark} ``Even though variations only take place along the streamwise direction, the actual computations were performed in two dimensions as stated earlier.'' The two dimensions are streamwise ($x=x_1$) and normal ($y=x_2$). In other words, this article uses a 2D code as a baseline for a 1D problem. }
{}
{\xmark}

\appendixentry{\frameworkmishra}{\citeframeworkmishra}{A machine learning framework for data driven acceleration of computations of differential equations}{\frameworkmishracitations}
\appendixdata{1D Burgers', advection, Euler}
{``Numerical experiments involving both linear and non-linear ODE and PDE model problems demonstrate a significant gain in computational efficiency over standard numerical methods.'' Supported by tables 5, 6, and 7, as well as page 20.}
{No}
{Rusanov for Burgers' and Euler, backwards euler time-stepping for advection}
{{\cmark} Focuses on speedup at constant error (page 20) or error at constant speed (see ``gain'' on page 12).}
{{\xmark} The Rusanov scheme is a first-order scheme, and is much more diffusive and less accurate at solving the 1D Euler equations than a higher-order scheme (e.g., MUSCL scheme with reconstruction in characteristic variables). For Burgers' equation, WENO5 would be a strong FV baseline.}
{N/A}
{\xmark}

\appendixentry{\optimizemultigrid}{\citeoptimizemultigrid}{Learning to Optimize Multigrid PDE Solvers}{\optimizemultigridcitations}
\appendixdata{2D elliptic diffusion}
{``Experiments on a broad class of 2D diffusion problems demonstrate improved convergence rates compared to the widely used Black-Box multigrid scheme.'' Supported by figure 3.}
{No}
{Black box multigrid scheme}
{{\cmark} Measures number of iterations (i.e., convergence rate) at constant accuracy. See figure 3.}
{{\cmark} Multigrid is highly efficient for elliptic problems.}
{N/A}
{\cmark }

\appendixentry{\donga}{\citedonga}{Local extreme learning machines and domain decomposition for solving linear and nonlinear partial differential equations}{\dongacitations}
\appendixdata{1D Helmholtz, 1D advection, 1D diffusion, 1D non-linear helmholtz, 1D Burgers', 2D Poisson.}
{``The computational performance of the current method is on par with, and often exceeds, the FEM performance in terms of the accuracy and computational cost." Supported by tables 2, 5, 7, 10, 11, as well as figures 50 and 52.}
{No}
{FEniCS, linear elements, second-order, sparse LU decomposition for linear solver, Newtons method for non-linear equations. For 1D diffusion and 1D Burgers', uses second-order backward differentiation formula (BDF2), an implicit timestepping method. }
{{\cmark} Figures 50 and 52 make plots of speed versus accuracy, allowing readers to make comparisons at equal accuracy or equal speed.}
{{\xmark} The primary outcome (in this case defined by what is reported in the abstract) compares the ML-based solver to a weaker baseline (second-order FEM). In the appendix, a stronger baseline is compared to (higher-order FEM). Both the ML-based solver and second-order FEM underperform relative to higher-order FEM. Because the primary outcome reports performance relative to the weaker baseline rather than the stronger baseline, we determine that rule 2 is not satisfied. To satisfy rule 2, report in the abstract performance relative to the strong baseline.}
{N/A}
{\xmark}

\appendixentry{\rayb}{\citerayb}{Detecting troubled-cells on two-dimensional unstructured grids using a neural network}{\raybcitations}
\appendixdata{2D advection, Burgers', Euler}
{``Through several numerical tests, the MLP indicator has been shown to outperform the TVB indicator (with various values of M) both in terms of solution accuracy and the number of cells flagged, while maintaining a comparable computational cost.'' Supported by table 6.}
{No}
{TVB limiter}
{{\cmark} The computational time in table 6 is nearly constant. Comparing accuracy at constant runtime.}
{{\cmark} TVB limiter is highly efficient.}
{}
{\cmark }

\appendixentry{\novelcnn}{\citenovelcnn}{A Novel CNN-Based Poisson Solver for Fluid Simulation}{\novelcnncitations}
\appendixdata{3D Poisson (smoke)}
{``We have shown that our approach accelerates the projection step in a conventional Eulerian fluid simulator by two orders of magnitude.'' Supported by table 1 (page 1462) and section 7.2.}
{No}
{Preconditioned conjugate gradient, both with and without multigrid preconditioning.}
{{\xmark} Compares runtime at different accuracy. ``The accuracy of the PCG-based method [is] higher than our CNN-based solver. As such, our CNN-based linear equation solver is suitable for the simulations which are not strict with numerical accuracy.''}
{{\xmark} The primary outcome of this article (in this case defined by what is reported the abstract) compares to the weaker baseline (MIC(0)-PCG), while table 1 compares to both weak and strong baselines (MG-PCG). Because the primary outcome reports performance relative to the weaker baseline rather than the stronger baseline, we determine that rule 2 is not satisfied. To satisfy rule 2, report in the abstract performance relative to the strong baseline.}
{N/A}
{\xmark}

\appendixentry{\wandel}{\citewandel}{Learning Incompressible Fluid Dynamics from Scratch -- Towards Fast, Differentiable Fluid Models that Generalize}{\wandelcitations}
\appendixdata{2D incompressible Navier-Stokes wake dynamics}
{``The $\vec{v}$-Net as well as the $\vec{a}$-Net'' are significantly faster than PhiFlow ($11\times$ on CPU and $40\times$ on GPU)." Supported by table 1 and Appendix E.}
{No}
{Phiflow (FV method, relies on iterative conjugate gradient solver) using $100\times 100$ grid.}
{{\xmark} The loss (which in this case, doesn't measure accuracy because the pressure is evolved independently of the velocity when pressure is not independent of velocity) is compared, but even so, the runtime is not compared at equal values of the loss.}
{{\cmark} While we recommend comparing with both FV and DG methods for the Navier-Stokes equations, we consider FV highly efficient for problems with fluid-structure interaction. Though we don't have reason to believe that PhiFlow is an inefficient general-purpose tool, we also recommend being cautious when evaluating this comparison (see recommendation 1 in Methods).}
{N/A}
{{\xmark}}

\appendixentry{\shan}{\citeshan}{Study on a Fast Solver for Poisson’s Equation Based on Deep Learning Technique}{\shancitations}
\appendixdata{2D \& 3D Poisson}
{``Numerical experiments show that the same ConvNet architecture is effective for both 2-D and 3-D models\dots with a significant reduction in computation time compared to the finite-difference solver.'' Supported by last paragraph before conclusion.}
{No}
{Finite-difference method }
{{\xmark} The convolutional network presumably has lower accuracy than the finite-difference baseline, because the authors never mention the number of iterations of the finite-difference baseline. The authors admit that their ``comparison is not exactly fair.'' }
{{\xmark} For Poisson's equation, Multigrid methods (or preconditioners) are highly efficient. See, e.g., figure 4e of article {\tang} or figure 23 of article \cheng. Note also that for sufficiently small 2D problems, direct solves (such as LU decomposition) will outperform Multigrid methods. Although this article says nothing about how the system of linear equations (equation 7) is solved, the baseline takes 17s to solve the 2D problem on a $64\times64$ grid, which is orders of magnitude slower a direct solve (such as LU decomposition).}
{N/A}
{{\xmark }}

\appendixentry{\algebraic}{\citealgebraic}{Learning Algebraic Multigrid Using Graph Neural Networks}{\algebraiccitations}
\appendixdata{2D elliptic diffusion}
{``Experiments on a broad class of problems demonstrate improved convergence rates compared to classical AMG.'' Supported by table 4.}
{No}
{algebraic multigrid}
{{\cmark} Table 4: ``required to reach specified tolerance.'' Compares number of iterations at constant accuracy.}
{{\cmark} Algebraic multigrid is highly efficient for elliptic problems.}
{N/A}
{\cmark}

\appendixentry{\zhuang}{\citezhuang}{Learned discretizations for passive scalar advection in a 2-D turbulent flow}{\zhuangcitations}
\appendixdata{2D advection}
{``The method maintains the same accuracy as traditional high-order flux-limited advection solvers, while using 4× lower grid resolution in each dimension.''}
{No}
{second-order Van Leer advection scheme (FV)}
{{\cmark} See figure 8.}
{{\xmark} DG or pseudo-spectral methods are state-of-the-art for scalar advection. DG schemes in particular will solve the advection equations at much coarser resolution. }
{N/A}
{\xmark}

\appendixentry{\pathak}{\citepathak}{Using Machine Learning to Augment Coarse-Grid Computational Fluid Dynamics Simulations}{\pathakcitations}
\appendixdata{2D incompressible Navier-Stokes}
{``The ML-assisted coarse-grid evolution resulted in corrected solution trajectories that were consistent with the solutions computed at a much higher resolution in space and time.'' Supported by figures 2 and 3.}
{No}
{Dedalus, a spectral solver.}
{{\cmark} Although this article never directly reduces the runtime to make a comparison at equal accuracy, it is fair to say that the solver runs at `lower resolution.' If this article had said `$4\times$ lower resolution', that would not have been fair.}
{{\cmark} Spectral solvers are highly efficient for the incompressible Navier-Stokes equations.}
{{\cmark} Though we don't have reason to believe that Dedalus is an inefficient general-purpose tool, we recommend being cautious when evaluating this comparison (see recommendation 1 in Methods).}
{\cmark}

\appendixentry{\leoni}{\citeleoni}{DeepONet prediction of linear instability waves in high-speed boundary layers}{\leonicitations}
\appendixdata{Parabolized stability equations}
{``\dots at a very small computational cost compared to discretization of the original equations.'' Supported by subsection V.B.}
{No}
{The code in reference [5]}
{{\xmark} Section V.B compares the runtime of the highly accurate code (error 0.0) to the less accurate ML-based forward solver (error 2-5\%, see figure 10). This comparison is not at equal accuracy.}
{{\cmark} We are unsure if this code is an efficient numerical method.}
{N/A}
{\xmark}

\appendixentry{\fnodeformation}{\citefnodeformation}{Fourier Neural Operator with Learned Deformations for PDEs on General Geometries}{\fnodeformationcitations}{}
\appendixdata{2D compressible Navier-Stokes airfoil wing, 2D incompressible Navier-Stokes pipe flow}
{``Geo-FNO is $10^5$ times faster than the standard numerical solvers.'' }
{No}
{Second-order implicit FV solver.}
{{\xmark} The comparison is made between the highly accurate (but slow) ground truth solver and the less accurate ML-based solver.}
{{\cmark} Finite-volume methods are highly efficient for the 2D compressible Navier-Stokes airfoil problem.}
{N/A}
{\xmark}

\appendixentry{\stevensa}{\citestevensa}{Enhancement of shock-capturing methods via machine learning}{\stevensacitations}
\appendixdata{1D Burgers', advection, Euler}
{``We find that our method outperforms WENO in simulations where the numerical solution becomes overly diffused due to numerical viscosity.'' Supported by figure 8.}
{No}
{WENO}
{{\cmark} Faster runtime at equal accuracy (see figure 8).}
{{\cmark} WENO5 is highly efficient for problems with shocks/discontinuities.}
{N/A}
{\cmark }

\appendixentry{\illarramendi}{\citeillarramendi}{Towards an hybrid computational strategy based on Deep Learning for incompressible flows}{\illarramendicitations}
\appendixdata{2D Poisson}
{``For the same accuracy to be achieved with only Jacobi iterations, the calculation is 3.2 times slower than the hybrid method.''}
{Yes}
{Jacobi method}
{{\cmark} Compares runtime at equal accuracy.}
{{\xmark} For Poisson's equation, Multigrid methods (or preconditioners) are highly efficient. See, e.g., figure 4e of article {\tang} or figure 23 of article \cheng. Note also that for sufficiently small 2D problems, direct solves (such as LU decomposition) will outperform Multigrid methods by a large factor.}
{N/A}
{\xmark}

\appendixentry{\stachenfeld}{\citestachenfeld}{Learned Coarse Models for Efficient Turbulence Simulation}{\stachenfeldcitations}
\appendixdata{2D incompressible \& 3D compressible Navier-Stokes}
{``Broadly, we conclude that our learned simulator outperforms traditional solvers run on coarser grids.'' Supported by figure 2 and ``Running Time'' section.}
{Yes}
{Athena++ with HLLC flux}
{{\cmark} It is fair to say that the ML-based solver outperforms the Athena++ baseline (compare the accuracy in figure 2 to the numbers in the ``running time'' section), because this comparison can be done at equal accuracy. It is not fair to say that the ML-based solver is $1000\times$ faster than Athena++ (see page 8).}
{{\dangersign} Athena++ is a state-of-the-art FV software package for shock-dominated problems.
DG and spectral methods are state-of-the-art for incompressible turbulence without shocks. In the videos available at \url{https://sites.google.com/view/learned-turbulence-simulators}, the compressible decaying turbulence seems to be weakly compressible (no shocks). See, e.g., \cite{markert2022discontinuous}, which writes that ``At least for subsonic turbulence, high order DG offers significant benefits [over FV methods] in computational efficiency for reaching a desired target accuracy.'' We give a warning sign because we believe that DG and/or spectral methods are state-of-the-art for weakly compressible turbulence and would likely outperform Athena++ with HLLC flux on this test problem, but we haven't replicated the result and so we don't have enough evidence to say for sure. }
{N/A}
{{\dangersign}}

\appendixentry{\han}{\citehan}{Predicting Physics in Mesh-reduced Space with Temporal Attention}{\hancitations}
\appendixdata{2D incompressible \& compressible Navier-Stokes}
{``We compare the evaluation cost of the learned model with the FV-based numerical models in table 6, and observe significant speedups for all three datasets.'' 100, 682, and 800 speedup reported in table 6.}
{No}
{OpenFOAM, open-source FV library.}
{{\xmark} Doesn't reduce resolution to match accuracy.}
{{\dangersign} While we recommend comparing with both FV and DG methods for the Navier-Stokes equations, we consider FV highly efficient for problems with fluid-structure interaction. However, we give a warning sign for rule 2 because OpenFOAM is known to be an inefficient general-purpose tool, with high overhead and slow convergence. See, e.g., \cite{capuanocost,featoolwebsite}. Because OpenFOAM is a general-purpose tool, we also recommend being cautious when evaluating this comparison (see recommendation 1 in Methods).}
{N/A}
{\xmark}

\appendixentry{\stevensb}{\citestevensb}
{FiniteNet: A Fully Convolutional LSTM Network Architecture for Time-Dependent Partial Differential Equations}{\stevensbcitations}
\appendixdata{1D advection, Burgers', Kuramoto-Sivashinsky}
{``We train the network on simulation data, and show that our network can reduce error by a factor of 2 to 3 compared to the baseline algorithms.'' Supported by table 1.}
{No}
{WENO5 for Burgers', 4th order finite difference for KS}
{{\cmark} Compares error (accuracy) at equal resolution (a proxy for runtime).  }
{{\cmark} WENO is highly efficient for Burgers', high-order finite difference is highly efficient for KS.}
{N/A}
{\cmark}

\appendixentry{\ozbay}{\citeozbay}
{Poisson CNN: Convolutional neural networks for the solution of the Poisson equation on a Cartesian mesh}{\ozbaycitations}
\appendixdata{2D Poisson}
{``Analytical test cases indicate that our CNN architecture is capable of predicting the correct solution of a Poisson problem with mean percentage errors below 10\%, an improvement by comparison to the first step of conventional iterative methods.'' Supported by table 6, conclusion.}
{No}
{Algebraic multigrid package PyAMG with a tolerance of $10^{-10}$}
{{\cmark} Similar speed (and theoretically faster speed on large grid sizes) and superior accuracy compared to a single cycle of multigrid. Comparing accuracy at equal runtime.}
{{\cmark} This method is focused on solving 2D Poisson problems at large grid sizes, for which multigrid methods are state-of-the-art. Note that for sufficiently small 2D problems, direct solves (such as LU decomposition) will outperform Multigrid methods. }
{N/A}
{\cmark }

\appendixentry{\zili}{\citezili}
{Graph neural network-accelerated Lagrangian fluid simulation}{\zilicitations}
\appendixdata{}
{``Overall, FGN achieves $\sim$5-8$\times$ acceleration over MPS under different resolution.'' Supported by figure 10.}
{No}
{Moving particle semi-implicit method (MPS), with conjugate gradient for pressure solve. ``MPS is a numerical method based on SPH [smooth particle hydrodynamics] which prioritizes accuracy over calculation speed.'' }
{{\cmark} ``During the benchmark, we set the absolute tolerance of CG solver to be 0.1 and maximum iteration to be 10 (note that these hyperparameters are $10^{-5}$ and 100 respectively when used to generate training data).'' When comparing runtime, they reduce the tolerance (accuracy) of the CG solver. The tolerance is chosen to be the minimum iteration that the simulation can still run without throwing NaNs (private communication with authors).
}
{{\xmark} The iterative pressure projection (i.e. Poisson solve) accounts for most of the calculation time in the baseline (see table 5). The Poisson solve uses conjugate gradient (CG), which is much slower than a state-of-the-art method such as algebraic multigrid. See, for example, \cite{sodersten2019bucket}. }
{N/A}
{\xmark}

\appendixentry{\peng}{\citepeng}{Attention-Enhanced Neural Network Models for Turbulence Simulation}{\pengcitations}
\appendixdata{2D incompressible Navier-Stokes}
{``Both models provide 8000 folds speedup compared with the pseudo-spectral numerical solver.'' Supported by table II.}
{No}
{}
{{\xmark} The ML-based solvers have non-zero error (see table 1) while the spectral solver has zero error. They are not comparing runtime at equal accuracy.}
{{\xmark} A pseudo-spectral solver for the 2D compressible Navier-Stokes with high Reynolds number on a $64\times64$ grid for 10 timesteps should take much fewer than 502 seconds to run. This is a slow implementation compared to the JAX-CFD \cite{dresdner2022learning} implementation, see \url{https://github.com/nickmcgreivy/WeakBaselinesMLPDE/blob/main/article4/data/runtime\_corr.png} where a pseudo-spectral solver on a $64\times64$ grid takes 0.1s to advance forward one unit of time.}
{N/A}
{\xmark}

\appendixentry{\chen}{\citechen}{Numerical investigation of minimum drag profiles in laminar flow using deep learning surrogates}{\chencitations}
\appendixdata{2D incompressible Navier-Stokes airfoil, steady laminar flow}
{``Therefore, relative to OpenFOAM, the speed-up factor is between 600X and 300X.'' Supported by table 2.}
{No}
{SimpleFOAM (Semi-Implicit Method for Pressure-Linked Equations, or SIMPLE), a second-order FV method within OpenFOAM}
{{\xmark} The ``performance'' section compares the runtime (for 200 iterations) of the ``ground truth'' OpenFOAM simulation to that of the ML-based surrogate solvers which are ``close to the OpenFOAM result'' but still have some error in the flow (see figure 10). To satisfy rule 1, increase the minimum mesh size of the OpenFOAM simulation to match the error of the ML-based solver, then compare runtimes at equal error.}
{{\dangersign} We give a warning sign for rule 2 because OpenFOAM is known to be an inefficient general-purpose tool, with high overhead and slow convergence. See, e.g., \cite{capuanocost,featoolwebsite}. Because OpenFOAM is a general-purpose tool, we recommend being cautious when evaluating this comparison (see recommendation 1 in Methods). }
{N/A}
{\xmark}

\appendixentry{\alguacilb}{\citealguacilb}{Predicting the propagation of acoustic waves using deep convolutional neural networks}{\alguacilbcitations}
\appendixdata{2D acoustic wave propagation, derived from Boltzmann equation (see Appendix B). Low LBM viscosity, closed domain with hard reflecting walls. ``Acoustic propagation takes place in a linear regime.'' ``The fluctuating density $\rho'$ is chosen such that $\rho_0 \gg \rho'$ to avoid non-linear effects.''}
{``The combination of both strategies can achieve a speed-up of 15.5 times with respect to the LBM code.''}
{No}
{Multi-physics lattice Boltzmann solver Palabos.}
{{\xmark} Compares the runtime of the ground truth Palabos solution to that of the ML-based solver with non-zero error.}
{{\dangersign} A Lattice Boltzmann solver is an inefficient numerical method to use to solve the linear acoustic wave equation. We give a warning sign because we believe that it would likely be much more efficient to use a linear acoustic solver or an Euler solver, but we haven't replicated this PDE and so we don't have enough evidence to say for sure. Though we don't have reason to believe that Palabos is an inefficient general-purpose tool, we also recommend being cautious when evaluating this comparison (see recommendation 1 in Methods).}
{N/A}
{{\xmark}}

\appendixentry{\wandelb}{\citewandelb}{Teaching the incompressible Navier–Stokes equations to fast neural surrogate models in three dimensions}{\wandelbcitations}
\appendixdata{3D compressible navier-stokes wake flow}
{``Furthermore, the U-Net as well as the pruned U-Net are considerably faster than Phiflow since they only require one forward pass through a convolutional neural network which can be easily parallelized and Phiflow relies on an iterative conjugate gradient solver.'' Supported by table 1.}
{No}
{Phiflow (FV method, relies on iterative conjugate gradient solver) using $128\times 128 \times 64$ grid.}
{{\xmark} The runtime of the two methods are compared in table 1, and the value of the loss function of the two methods are compared in table 1, but the accuracy of the two methods are never compared. The loss (which in this case, doesn't measure accuracy because the pressure is evolved independently of the velocity when pressure is not independent of velocity) is compared, but even so, the runtime is not compared at equal values of the loss.}
{{\cmark} While we recommend comparing with both FV and DG methods for the Navier-Stokes equations, we consider FV highly efficient for problems with fluid-structure interaction. Though we don't have reason to believe that PhiFlow is an inefficient general-purpose tool, we also recommend being cautious when evaluating this comparison (see recommendation 1 in Methods).}
{N/A}
{{\xmark}}

\appendixentry{\blist}{\citeblist}{Learned Turbulence Modelling with Differentiable Fluid Solvers: Physics-based Loss-functions and Optimisation Horizons}{\blistcitations}
\appendixdata{2D incompressible Navier-Stokes cylindrical flow}
{``For the former, our model evenly matches the performance of a $4\times$ simulation for several hundred time steps, which represents a speedup of 14.4.'' ``Measures speedups of up to 14 with respect to comparably accurate solutions from traditional solvers.'' Supported by figure 21 and table 8.}
{No}
{Semi-implicit PISO scheme, a second-order FV method}
{{\cmark} Figure 21 compares the accuracy of the ML-based solver to methods with lower resolution than ground truth. The article then compares runtime at equal accuracy.}
{{\xmark} Pseudo-spectral and DG methods are state-of-the-art for 2D incompressible Navier-Stokes. Finite-volume methods are significantly less efficient.}
{N/A}
{\xmark}

\appendixentry{\cheng}{\citecheng}{Using neural networks to solve the 2D Poisson equation for electric field computation in plasma fluid simulations}{\chengcitations}
\appendixdata{2D Poisson}
{``For this configuration, the resolution time [sic] of the neural network running on A100 GPU is about a factor 2 [sic] lower than the linear system solver on 128 cores, making it a viable option in terms of performance.'' Supported by figure 17. }
{No}
{Multigrid preconditioner with PETSc (see figure 23).}
{{\cmark} Reduces tolerance (see figure 16) to match accuracy.}
{{\cmark} Uses multigrid method. Note that for small domains, direct solves (such as LU decomposition) outperform multigrid.}
{N/A}
{\cmark}

\appendixentry{\wen}{\citewen}{An edge detector based on artificial neural network with application to hybrid compact-WENO finite difference scheme}{\wencitations}
\appendixdata{1D \& 2D shallow water and Euler}
{There are lots of results in this article, so it is hard to tell which should be considered the primary outcome. We will choose the statement in the abstract ``the ANN edge detector can capture an edge accurately with fewer grid points than the classical multi-resolution analysis.'' We also considered choosing the statement ``Generally speaking, the hybrid-ANN scheme captures the shock waves and high gradients more accurate [sic] than the Hybrid-MR scheme.'' }
{No}
{Hybrid-MR (multi-resolution analysis) scheme}
{{\cmark} The accuracy is higher (see figure 12, 13, etc) while the runtime is about the same (see table 6, 7, etc) compared to the Hybrid-MR scheme. }
{{\cmark} This is probably a strong baseline, but we are unsure.}
{}
{{\cmark}}

\appendixentry{\delaraa}{\citedelaraa}{Accelerating high order discontinuous Galerkin solvers using neural networks: 1D Burgers' equation}{\delaraacitations}
\appendixdata{1D Burgers'}
{``We see a substantial increase in efficiency, leading to ratios $CHO/CLO = 75$, 22, and 59 is case 1, 2, and 3 respectively.'' Also, ``the method shows potential and significant cost savings for high-order polynomials.'' Supported by table 1.}
{No}
{DG, very high order (polynomial 5, 7, and 28)}
{{\xmark} The ratio of cost is between a highly accurate high-order simulation and a less accurate ML-corrected low-order simulation. Thus, the cost ratio is not computed at constant accuracy.}
{{\cmark} For Burgers' equation with smooth solutions, we consider very high order DG methods highly efficient.}
{N/A}
{{\xmark} We replicate the test case (see first paragraph of section 4) using our own DG code on CPU, written in Python with JAX, and find that when the error is about $6 \times 10^{-3}$ our DG baselines take about 0.05s to run. Our baseline is 4 to 10$\times$ faster than the low-order ML-based solvers listed in table 1.}

\appendixentry{\zhao}{\citezhao}{Learning to Solve PDE-constrained Inverse Problems with Graph Networks}{\zhaocitations}
\appendixdata{2D scalar wave equation, 2D incompressible Navier-Stokes}
{``The proposed method \dots is $35\times$ faster than the classical FEM solver.'' Supported by table 2.}
{No}
{Wave equation: FEniCS using Euler Method and first order elements, with GMRES and LU factorization as preconditioner. Navier-stokes: Chorin's method.}
{{\dangersign} Table 2 compares the runtime at equal resolution (a proxy for runtime) rather than at equal accuracy. We give this article a warning sign because while it does reduce the resolution (relative to the fine grid ground truth FEM solver) it doesn't reduce it enough to reach equal accuracy. The ML-based solver has an error $2.5\times$ higher than the FEM solver at equal resolution, and thus we think this comparison is likely unfair.}
{{\cmark} While we recommend comparing with both FV and DG methods for the Navier-Stokes equations, we consider FV highly efficient for problems with fluid-structure interaction. Though we don't have reason to believe that FEniCS is an inefficient general-purpose tool, we also recommend being cautious when evaluating this comparison (see recommendation 1 in Methods).}
{N/A}
{{\dangersign}}

\appendixentry{\assessments}{\citeassessments}{Performance and accuracy assessments of an incompressible fluid solver coupled with a deep Convolutional Neural Network}{\assessmentscitations}
\appendixdata{2D Poisson}
{``These networks can provide solutions 10-25 faster than traditional iterative solvers.''}
{No}
{Jacobi method}
{{\cmark} ``A fair performance comparisons is performed \dots allowing the assessment of the time of inference at a fixed error level.'' }
{{\xmark} A multigrid method or preconditioner is considered highly efficient for elliptic PDEs.}
{N/A}
{\xmark}

\appendixentry{\holloway}{\citeholloway}{Acceleration of Boltzmann Collision Integral Calculation Using Machine Learning}{\hollowaycitations}
\appendixdata{6-dimensional Boltzmann collision operator}
{``Our method demonstrated a speed up of 270 times compared to these methods while still maintaining reasonable accuracy.'' Supported by table 1.}
{No}
{DG discretization of collision operator}
{{\xmark} Compares runtime of highly accurate ground DG method to that of less accurate ML-based solver.}
{{\cmark} We believe that the DG discretization is likely highly efficient.}
{N/A}
{\xmark}

\appendixentry{\azulay}{\citeazulay}{Multigrid-augmented deep learning preconditioners for the Helmholtz equation}{\azulaycitations}
\appendixdata{2D Helmholtz}
{``We show that while our U-Net may require more FLOPs than traditional methods, it can applied efficiently on GPU hardware, and yield favorable running times.'' Supported by figure 10 and section 4.3.5.}
{No}
{Geometric multigrid preconditioner, followed by GMRES iterations }
{{\cmark} Figure 10 compares runtime at constant accuracy/error.}
{{\cmark} Multigrid preconditioner is highly efficient for elliptic PDEs.}
{}
{\cmark}

\appendixentry{\wulatent}{\citewulatent}{Learning to Accelerate Partial Differential Equations via Latent Global Evolution}{\wulatentcitations}
\appendixdata{1D Burgers', 2D incompressible Navier-Stokes, 3D incompressible Navier-Stokes cylinder flow }
{``It achieves significant [sic] smaller runtime compared to the MP-PDE model (which is much faster than the classical WENO5 scheme).'' Also, ``we see that our LE-PDE achieves a $70.80/0.084 \simeq 840\times$ speed up compared to the ground-truth solver on the same GPU.'' See supplementary material, figure S1, and table 5.}
{No}
{PhiFlow}
{{\xmark} Table 5 compares the runtime of a highly accurate PhiFlow to that of a less accurate ML-based solver.}
{{\cmark} While we recommend comparing with both FV and DG methods for the Navier-Stokes equations, we consider FV highly efficient for problems with fluid-structure interaction. Though we don't have reason to believe that PhiFlow is an inefficient general-purpose tool, we also recommend being cautious when evaluating this comparison (see recommendation 1 in Methods).}
{N/A}
{\xmark}

\appendixentry{\liu}{\citeliu}{Predicting parametric spatiotemporal dynamics by multi-resolution PDE structure-preserved deep learning}{\liucitations}
\appendixdata{2D Burgers', 2D incompressible Navier-Stokes}
{``In particular, the speedup by the PPNN varies from $10\times$ to $60\times$ without notably sacrificing the prediction accuracy.'' Supported by figure 7b.}
{No}
{Burgers': 3rd-order accurate up-wind scheme for convection, 6th order accurate central-difference scheme for diffusion term, forward-euler timestepping. Navier-stokes: PISO algorithm using OpenFOAM.}
{{\xmark} Doesn't reduce the resolution of the PISO algorithm to compare at equal accuracy.}
{{\xmark} Pseudo-spectral or DG algorithms are considered highly efficient for 2D incompressible Navier-stokes. Finite-volume algorithms like PISO are much less efficient. Furthermore, OpenFOAM is known to be an inefficient general-purpose tool, with high overhead and slow convergence. See, e.g., \cite{capuanocost,featoolwebsite}. Because OpenFOAM is a general purpose tool, we also recommend being cautious when evaluating this comparison (see recommendation 1 in Methods). }
{N/A}
{\xmark}

\appendixentry{\zhang}{\citezhang}{A Hybrid Iterative Numerical Transferable Solver (HINTS) for PDEs Based on Deep Operator Network and Relaxation Methods}{\zhangcitations}
\appendixdata{1D \& 2D Poisson, 1D 2D \& 3D Helmholtz}
{``The results show that HINTS performs consistently better than the corresponding numerical methods, with the improvement of computational efficiency being up to $\mathcal{O}(10^2)$.'' Supported by figure S9.}
{No}
{For 1D Poisson: Multigrid with damped Jacobi relaxation, either 3 or 5 grid levels.}
{{\cmark} Figure S9 measures time until convergence, which implies equal accuracy.}
{{\xmark} For small problems (such as 1D Poisson), direct methods like LU decomposition are much faster than iterative methods like multigrid. A 1D Poisson problem should take microseconds to milliseconds to solve to machine precision, not 1 to 100 seconds (as in figure S9).}
{N/A}
{\xmark}

\appendixentry{\duarte}{\citeduarte}{Black hole weather forecasting with deep learning: a pilot study}{\duartecitations}
\appendixdata{2D black hole hydrodynamics: ``Our data set was generated from two-dimensional hydrodynamical simulations of viscous accretion on to a Schwarzschild BH.''}
{``For instance, once trained the model evolves an RIAF on a single GPU four orders of magnitude faster than usual fluid dynamics integrators running in parallel on 200 CPU cores.'' Supported by table 3, but caveated by the comments in section 5.1.}
{Yes}
{PLUTO code which uses Godunov-like flux}
{{\xmark} This article compares the runtime of a highly accurate standard numerical method to that of a less accurate ML-based solver. Doesn't compare runtime at equal accuracy.}
{{\cmark} We consider FV schemes highly efficient for this problem.}
{}
{\xmark}

\appendixentry{\alguacil}{\citealguacil}{Deep Learning Surrogate for the Temporal Propagation and Scattering of Acoustic Waves}{\alguacilcitations}
\appendixdata{2D (acoustic) wave}
{``When both strategies are combined, a large acceleration factor of 141 can be achieved with respect to the MPI-based simulation.'' Supported by table 5.}
{No}
{Lattice-Boltzmann simulation Palabos, same as article \alguacilb.}
{{\xmark} Table 5 compares the runtime of the highly accurate baseline to that of the less accurate ML-based solver.}
{{\dangersign} A Lattice Boltzmann solver is an inefficient numerical method to use to solve the linear acoustic wave equation. We give a warning sign because we believe that it would likely be much more efficient to use a linear acoustic solver or an Euler solver, but we haven't replicated this PDE and so we don't have enough evidence to say for sure. Though we don't have reason to believe that Palabos is an inefficient general-purpose tool, we also recommend being cautious when evaluating this comparison (see recommendation 1 in Methods).}
{N/A}
{\xmark}

\appendixentry{\bezginb}{\citebezginb}{WENO3-NN: A maximum-order three-point data-driven weighted essentially non-oscillatory scheme}{\bezginbcitations}
\appendixdata{1D advection, 1D \& 2D Euler}
{``The WENO3-NN scheme shows very good generalizability across all benchmark cases and different resolutions, and exhibits a performance similar to or better than the classical WENO5-JS scheme.'' Supported by table 4.}
{No}
{WENO5-JS}
{{\cmark} Table 4 and various figures compares accuracy at constant resolution, a proxy for runtime.}
{{\xmark} The primary outcome (in this case defined by what is reported in the abstract) compares the ML-based solver (WENO3-NN) to a baseline (WENO5-JS). As we learn in section 4.4 and in the appendix, both WENO5-JS and WENO3-NN underperform relative to WENO5-Z for 5 or 6 out of the 6 benchmark problems. Because the primary outcome compares to the weaker baseline (WENO5-JS) rather than the stronger baseline (WENO5-Z), we determine that rule 2 is not satisfied.
}
{N/A}
{\xmark }

\appendixentry{\shang}{\citeshang}{Deep Petrov-Galerkin Method for Solving Partial Differential Equations}{\shangcitations}
\appendixdata{2D Poisson, 2D Wave}
{``This new method outperforms traditional numerical methods in several aspects: compared to the finite element method and finite difference method, DPGM is much more accurate with respect to degrees of freedom.'' Supported by tables 6 and 7.}
{No}
{FEniCS, lagrange elements $P_k$ $k=1,2,3$ for Poisson and }
{{\cmark} Compares accuracy at constant degrees of freedom (which is a proxy for runtime). }
{{\cmark} Though we don't have reason to believe that FEniCS is an inefficient general-purpose tool for these PDEs, we recommend being cautious when evaluating this comparison (see recommendation 1 in Methods).}
{N/A}
{\cmark}

\appendixentry{\kube}{\citekube}{Machine learning accelerated particle-in-cell plasma simulations}{\kubecitations}
\appendixdata{1D particle-in-cell (linear solver)}
{``We find that this approach reduces the average number of required solver iterations by about 25\% when simulating electron plasma oscillations.'' Supported by figure 2.}
{No}
{GMRES (Jacobian-free Newton-Krylov method) to solve system of non-linear equations}
{{\xmark} Although this article compares the number of iterations at equal accuracy, (based on private communication with the first author) the runtime of the ML-based solver is significantly longer than the standard GMRES solver. We gave an `X' because, in this case, the number of iterations is not a fair proxy for speed.}
{{\cmark} Unsure. }
{N/A}
{\xmark}

\appendixentry{\shi}{\citeshi}{LordNet: Learning to Solve Parametric Partial Differential Equations without Simulated Data}{\shicitations}
\appendixdata{2D Poisson, 2D incompressible Navier-Stokes}
{``For Navier-Stokes equation, the learned operator is over 50 times faster than the finite difference solution with the same computational resources.'' Supported by figure 3 and section 4.2.}
{No}
{Finite-difference scheme (FDM) with central differencing, with conjugate gradient method to solve sparse algebraic equations}
{{\xmark} Doesn't reduce resolution of FDM to match accuracy of ML-based solver.}
{{\cmark} Unsure.}
{}
{\xmark}

\appendixentry{\ranadea}{\citeranadea}{A Latent space solver for PDE generalization}{\ranadeacitations}
\appendixdata{3D steady-state electronic cooling with natural convection}
{``Thus, the hybrid solver results in a 200$\times$ speedup over Ansys Fluent in generating solutions on fine girds [sic].'' Supported by section 3.2.1.}
{No}
{ANSYS Fluent}
{{\xmark} Doesn't reduce resolution of ANSYS baseline to match accuracy of ML-based solver. }
{{\cmark} Though we don't have reason to believe that ANSYS is an inefficient general-purpose tool, we recommend being cautious when evaluating this comparison (see recommendation 1 in Methods).}
{N/A}
{\xmark}

\appendixentry{\chenb}{\citechenb}{A machine learning based solver for pressure Poisson equations}{\chenbcitations}
\appendixdata{2D Poisson}
{``The ML-block provides a better initial iteration value for the traditional iterative solver, which greatly reduces the number of iterations of the traditional iterative solver and speeds up the solution of the PPE.'' Supported by table 2 and equation 15.}
{No}
{Preconditioned conjugate gradient}
{{\cmark} Equation 15: ``to achieve the same solution accuracy''. Compares number of iterations at equal accuracy. }
{{\xmark} Multigrid methods (or preconditioners) are state-of-the-art for Poisson's equation. See, e.g., figure 4e of article {\tang} or figure 23 of article \cheng. Note also that for sufficiently small 2D problems, direct solves (such as LU decomposition) will outperform Multigrid methods.}
{N/A}
{\xmark}

\appendixentry{\ranadeb}{\citeranadeb}{A composable autoencoder-based iterative algorithm for accelerating numerical simulations}{\ranadebcitations}
\appendixdata{2D Laplace, 2D incompressible Navier-Stokes, 3D electronic cooling with natural convection, 3D steady-state channel flow (incompressible Navier-Stokes?), 3D transient flow over a cylinder with heating}
{``We observe that the CoAE-MLSim approach is about 40-50$\times$ faster in the steady-state cases and about 100$\times$ faster in transient cases as compared to commercial PDE solvers such as Ansys Fluent for the experiments presented in this work.'' Not supported by any quantitative analysis.}
{No}
{Ansys Fluent}
{{\xmark} Doesn't compare the runtime at equal accuracy, instead compares runtime of highly accurate ANSYS Fluent to less accurate ML-based solver.}
{{\cmark} Though we don't have reason to believe that Ansys is an inefficient general-purpose tool, we recommend being cautious when evaluating this comparison (see recommendation 1 in Methods).}
{N/A}
{{\xmark}}

\appendixentry{\pengb}{\citepengb}{Linear attention coupled Fourier neural operator for simulation of three-dimensional turbulence}{\pengbcitations}
\appendixdata{3D incompressible Navier-Stokes}
{``During inference, both neural network models provides [sic] 20 folds speedup compared with the DNS approach with the traditional numerical solver.'' Supported by table 4.}
{No}
{A pseudo-spectral method, $64\times 64\times 64$ box.}
{{\xmark} This article never reduces the resolution of the spectral baseline to compare at equal accuracy.}
{{\cmark} Pseudo-spectral methods are highly efficient for this problem.}
{N/A}
{\xmark}

\appendixentry{\delarab}{\citedelarab}{Accelerating high order discontinuous Galerkin solvers using neural networks: 3D compressible Navier-Stokes equations}{\delarabcitations}
\appendixdata{3D compressible Navier-Stokes in the incompressible limit}
{``The low order corrected solution is 4 to 5 times faster than a simulation with comparable accuracy (polynomial order 6).'' Supported by section 3.5, figure 12, table 2.}
{Yes}
{Very high order DG (polyOrder 5 to 6) with constant number of elements. HORSES3D, a nodal high-order DG spectral element method (DGSEM).}
{{\cmark} Compares runtime at equal accuracy (see section 3.5).}
{{\cmark} We consider DG methods highly efficient for weakly compressible flows.
}
{N/A}
{\cmark}

\appendixentry{\ranadec}{\citeranadec}{A composable machine-learning approach for steady-state simulations on high-resolution grids}{\ranadeccitations}
\appendixdata{2D Laplace, 2D Poisson, 2D non-linear Poisson, 3D electronic cooling with natural convection}
{``We observe that the CoAE-MLSim approach is about 40-50$\times$ faster as compared to commercial steady-state PDE solvers such as ANSYS Fluent for the same mesh resolution and physical domain size in all the experiments presented in this work.'' See appendix E. }
{Yes}
{Ansys Fluent}
{{\xmark} Compares runtime at constant mesh resolution (a proxy for runtime), not at constant accuracy. For a fair comparison, reduce the resolution of ANSYS Fluent to match the accuracy of the ML-based solver.}
{{\cmark} Though we don't have reason to believe that Ansys is an inefficient general-purpose tool, we recommend being cautious when evaluating this comparison (see recommendation 1 in Methods).}
{N/A}
{\xmark}

\appendixentry{\fang}{\citefang}{Immersed boundary-physics informed machine learning approach for fluid–solid coupling}{\fangcitations}
\appendixdata{2D incompressible Navier-Stokes cylindrical flow, oscillating cylinder}
{``The time consumed by the machine learning model is reduced by 38.5\% compared with IB-LBM.'' Supported by figure 12.}
{No}
{Immersed-boundary lattice boltzmann method (IB-LBM). A D2Q9 scheme (see section 2.1).  }
{{\xmark} This comparison is not done at equal accuracy. See the differences between figure 5 and figure 9, as well as table 4. Since this paper is using the drag coefficient as a measure of accuracy, a fair comparison of simulation time would be at equal drag coefficient (or equal deviation from the high-resolution simulation's drag coefficient). }
{{\cmark} Unsure.}
{N/A}
{\xmark}

\appendixentry{\shukla}{\citeshukla}{Deep neural operators can serve as accurate surrogates for shape optimization: A case study for airfoils}{\shuklacitations}
\appendixdata{2D compressible Navier-Stokes airfoil wing}
{Importantly, DeepONets exhibit almost no generalization error over the dataset, so it follows that the resulting optimized geometry is accurate and achieved 32, 253 speed-up compared to the CFD baseline.'' Supported by table 4.}
{No}
{Nektar, using DG spectral element method (DGSEM) with basis functions spanned in 2D by Legendre polynomials of the second degree. DIRK used for time integration. See section 3.1.3.}
{{\xmark} While the Nektar solution has zero error, the DeepONet surrogate model has positive error: ``a closer look reveals non-physical streamlines originating from the surface of the airfoil. This is due to the error in the velocity fields near the airfoil surface, predicted by the DeepONet.'' Thus, the comparison in table 4 is not done at constant accuracy. }
{{\cmark} DGSEM is likely highly efficient for this problem. Though we don't have reason to believe that Nektar is an inefficient general-purpose tool, we recommend being cautious when evaluating this comparison (see recommendation 1 in Methods).}
{N/A}
{\xmark}

\appendixentry{\zhangb}{\citezhangb}{Learning the elastic wave equation with Fourier Neural Operators}{\zhangbcitations}
\appendixdata{(elastic) wave equation}
{``Post-training, the FNO is observed to generate accurate elastic wave fields at approximately 10 times the speed of traditional finite difference methods.'' Supported by table 1 and conclusion.}
{No}
{``Wavefields generated with the isotropic stress-velocity wave equation, using a staggered grid finite difference method with 4th order accuracy in space as training data.'' 84 by 84 grid. }
{{\xmark} Compares the runtime of the ground truth finite difference method to that of the less accurate ML-based solver. Doesn't reduce grid resolution to compare at equal accuracy. }
{{\cmark} Unsure.}
{N/A}
{\xmark }

\appendixentry{\bezgin}{\citebezgin}{A fully-differentiable compressible high-order computational fluid dynamics solver}{\bezgincitations}
\appendixdata{2D \& 3D compressible Navier-Stokes}
{``The NN-Rusanov flux stays stable over the course of the simulation and consistently outperforms the Rusanov flux\dots The NN-Rusanov flux is less dissipative than the classical Rusanov scheme.'' Supported by figure 5.}
{No}
{Rusanov flux}
{{\cmark} Compares error (accuracy) at equal resolution (a proxy for runtime) in figure 5. }
{{\xmark} Rusanov flux is a first-order method (with ``excess numerical diffusion [that] leads to a smeared out solution'') and is not a highly efficient baseline for this problem. }
{N/A}
{\xmark}

\appendixentry{\yang}{\citeyang}{Rapid Seismic Waveform Modeling and Inversion with Neural Operators}{\yangcitations}
\appendixdata{2D acoustic wave equation}
{``We show that full waveform modeling with neural operators is nearly two orders of magnitude faster than conventional numerical methods.'' Supported by page 2: runtime of 0.02 sec versus 1.23 sec. }
{No}
{Finite difference code on a $64 \times 64$ mesh.}
{{\xmark} Doesn't reduce resolution to compare at equal accuracy. }
{{\cmark} The finite-difference baseline in reference 40 seems likely to be a strong baseline, but we aren't entirely sure.}
{N/A}
{\xmark}

\appendixentry{\tang}{\citetang}{Neural Green’s function for Laplacian systems}{\tangcitations}
\appendixdata{2D Poisson}
{``Although our method saves only about $2\times$ the number of multiply-add operations compared to MGPCG, its intrinsic parallel nature enables it to reach a speedup of up to $12\times$ at all resolutions.'' Supported by table 2 and figure 4b. }
{No}
{Multigrid and multigrid-preconditioned conjugate gradient (MGPCG) are the two strongest baselines. }
{{\cmark} Compares at equal accuracy in table 2.}
{{\xmark} Consider table 2: the strongest baseline takes 90ms, 189ms, 299ms, and 425ms for grid sizes of $33\times33$, $65\times65$, $129\times129$, and $257\times257$ respectively. We implement Poisson's equation on a square domain using a direct solve (LU decomposition) and find that the direct solve takes 0.2ms, 0.4ms, 3ms, and 12ms for grid sizes of $32\times32$, $64\times64$, $128\times128$, and $256\times256$ respectively. LU decomposition is between 500 and 35 times faster than multigrid. Multigrid a weak baseline relative to direct methods for sufficiently small problems.}
{N/A}
{\xmark}

\appendixentry{\nastorg}{\citenastorg}{DS-GPS : A Deep Statistical Graph Poisson Solver (for faster CFD simulations)}{\nastorgcitations}
\appendixdata{2D poisson}
{``By taking advantage of GPU parallelism, we observe that our method can compute the solution ten times faster than LU decomposition.'' Not supported by any evidence.}
{No}
{LU decomposition}
{{\xmark} The error of LU decomposition on the graph is zero. The error of this ML-based solver is higher than zero (see figure 1). For a fair comparison, compare the runtime of LU decomposition on a coarsened graph (so that the error on the fine graph is comparable).}
{{\cmark} For small problems, LU decomposition is highly efficient or state-of-the-art.}
{N/A}
{\xmark}

\appendixentry{\gopakumar}{\citegopakumar}{Fourier Neural Operator for Plasma Modelling}{\gopakumarcitations}
\appendixdata{2D magnetohydrodynamics (MHD)}
{``Our work shows that the FNO is capable of predicting magnetohy- drodynamic models governing the plasma dynamics, 6 orders of magnitude faster than the traditional numerical solver, while maintaining considerable accuracy (NMSE $\sim10^5$).'' Supported by table 1.}
{No}
{JOREK code. 200 by 200 bi-cubic finite-elements.}
{{\xmark} Doesn't reduce accuracy of JOREK to match accuracy of FNO. See errors in, e.g., figure 2.}
{{\dangersign} 160 CPU hours to solve a 2D advection-dominated time-dependent PDE is unusually long, even at a high resolution of $200\times200$. We believe that JOREK is likely using inefficient numerical methods for advection-dominated flows. We believe the most likely explanation for the slow implementation is that JOREK uses fully implicit timestepping, while advection-dominated PDEs are better suited for explicit timestepping. 
}
{N/A}
{\xmark}

\appendixentry{\shit}{\citeshit}{Semi-Implicit Neural Solver for Time-dependent Partial Differential Equations}{\shitcitations}
\appendixdata{2D advection-diffusion, with advection speeds in range [-2, 2] and diffusion coefficients in range [0.2, 0.8]. }
{``We observe that for a given acceptance, the neural solver is 19.2\% faster.'' Supported by figure 5.}
{No}
{Semi-implicit scheme (see equation 3), using fixed-point iteration introduced in section 2.1.}
{{\cmark} Figure 5 compares runtime at equal error.}
{{\xmark} For this mixed hyperbolic-parabolic PDE, an (explicit) super-time-stepping method (e.g., \cite{meyer2014stabilized}) would be more efficient than an implicit method. Let us explain why. The semi-implicit scheme uses a timestep of $\Delta t_{\textnormal{implicit}} = 0.2$, and each timestep requires a number of fixed point iterations (up to 25). An explicit method would (since the maximum diffusion coefficient is $\nu = 0.8$ and $\Delta x = 0.098$) have a timestep of $\Delta t_{\textnormal{explicit}} \approx \frac{(\Delta x)^2}{2\nu} = 0.006$. Since $\frac{\Delta t_{\textnormal{implicit}}}{\Delta t_{\textnormal{explicit}}} \approx 33$, but each implicit timestep takes roughly 25$\times$ more runtime, then implicit timestepping would likely be as fast or slightly faster (as a rough estimate, as much as $33/25\approx 1.3\times$ faster) than naive explicit timestepping. However, $s$-stage super-time-stepping allows for timesteps proportional to $s^2$ and a runtime speedup (relative to explicit timestepping) of $s^2/s = s$, where the optimal number of stages $s = \sqrt{\frac{\Delta t_{\textnormal{implicit}}}{\Delta t_{\textnormal{explicit}}}}$. In this case, the speedup from using super-time-stepping would be $s = \sqrt{33} = 5.75$, which would be faster than the implicit method. }
{N/A}
{\xmark}

\appendixentry{\suforecast}{\citesuforecast}{Forecasting Variable-Density 3D Turbulent Flow}{\suforecastcitations}
\appendixdata{3D compressible Navier-Stokes}
{``Across flows with different density-ratio, our method over [sic] 3 orders of magnitude faster than high-fidelity numerical simulations.'' Supported by table 1. }
{No}
{Petascale variable-density version of the CFDNS code. Spectral code: spatial derivatives are evaluated using Fourier transforms. Uses ($64^3$) resolution.}
{{\xmark} Table 1 compares the runtime of the ground truth solver to that of the less accurate ML-based solver. For a fair comparison, reduce the resolution of the CFDNS code until the accuracy matches the ML-based solver.}
{{\cmark} Spectral codes are highly efficient for weakly compressible turbulence.}
{N/A}
{\xmark}

\appendixentry{\jeon}{\citejeon}{Physics-Informed Transfer Learning Strategy to Accelerate Unsteady Fluid Flow Simulations}{\jeoncitations}
\appendixdata{2D incompressible Navier-Stokes}
{``The simulation was accelerated by 1.8 times in the laminar counterflow CFD dataset condition including the parameter-updating time.''}
{No}
{IcoFoam, which is part of OpenFOAM, which uses the PISO algorithm.}
{{\xmark} The abstract writes that ``Notably, the cross-coupling strategy with a grid-based network model does not compromise the simulation accuracy for computational acceleration.'' If this were true, then rule 1 would be satisfied. The problem is that it is the residual of the governing equations that is ``not compromise[d]'', not the accuracy of the solution. As the paper correctly points out at the top of page 3, a low residual does not imply low error. Yes, the cross-coupling ML-CFD strategy is $1.8\times$ faster than the ``ground truth'' CFD solver, but as figures 9 and 12 show the ``ground truth'' CFD solver has zero error while the cross-coupling ML-CFD strategy has positive error. Thus, rule 1 is not satisfied. To satisfy rule 1, increase the error of the CFD solver to match that of the cross-coupling strategy. One way to do this would be to reduce the resolution of the OpenFOAM solution; a second would be to reduce the residual of OpenFOAM below the default residual.
}
{{\dangersign} We give a warning sign for rule 2 because OpenFOAM is known to be an inefficient general-purpose tool, with high overhead and slow convergence. See, e.g., \cite{capuanocost,featoolwebsite}. Because OpenFOAM is a general purpose tool, we also recommend being cautious when evaluating this comparison (see recommendation 1 in Methods). While we recommend comparing with both FV and DG methods for the Navier-Stokes equations, we consider FV highly efficient for problems with fluid-structure interaction. }
{N/A}
{\xmark}

\appendixentry{\dai}{\citedai}{FourNetFlows: An efficient model for steady airfoil flows prediction}{\daicitations}
\appendixdata{2D incompressible Navier-Stokes airfoil wing}
{``We quantitatively shows [sic] the accuracy of FourNetFlows is matched with the traditional method, running four to five orders of magnitude faster.'' Supported by table 1. }
{No}
{SimpleFoam, which uses SIMPLE (semi-implicit method for pressure linked equations) and is part of OpenFOAM.}
{{\xmark} Compares the runtime of ground truth SimpleFOAM to that of the less accurate ML-based solver.}
{{\dangersign} We give a warning sign for rule 2 because OpenFOAM is known to be an inefficient general-purpose tool, with high overhead and slow convergence. See, e.g., \cite{capuanocost,featoolwebsite}. Because OpenFOAM is a general purpose tool, we also recommend being cautious when evaluating this comparison (see recommendation 1 in Methods).}
{N/A}
{\xmark}

\appendixentry{\sun}{\citesun}{Local Randomized Neural Networks with Discontinuous Galerkin Methods for Partial Differential Equations}{\suncitations}
\appendixdata{1D Helmholtz, 2D Poisson.}
{``We compare the proposed methods with the finite element method and the usual DG method. The LRNN-DG methods can achieve better accuracy under the same degrees of freedom.'' Supported by figure 4.}
{No}
{FEM and DG, up to 3rd order. Doesn't have any other details.}
{{\cmark} Compares accuracy at constant number of degrees of freedom, a proxy for runtime. }
{{\cmark} FEM and DG are efficient methods for solving Poisson's equation.}
{N/A}
{\cmark}

\appendixentry{\shao}{\citeshao}{A Poisson’s Equation Solver Based on Neural Network Precondtioned CG Method}{\shaocitations}
\appendixdata{2D Poisson}
{``Numerical examples demonstrate that compared to conjugate gradient (CG) method, NN-PCG significantly improves convergence performance on solving 2-D Poisson’s equation.''}
{No}
{Conjugate gradient method.}
{{\cmark} Figure 2 compares accuracy at constant number of iterations.}
{{\xmark} Multigrid methods are state-of-the-art for this PDE. See, e.g., figure 4e of article {\tang} or figure 23 of article \cheng. Note also that for sufficiently small 2D problems, direct solves (such as LU decomposition) will outperform Multigrid methods.}
{N/A}
{\xmark}

\appendixentry{\discacciati}{\citediscacciati}{Controlling oscillations in high-order Discontinuous Galerkin schemes using artificial viscosity tuned by neural networks}{\discacciaticitations}
\appendixdata{1D \& 2D advection, 1D \& 2D Burgers', 1D \& 2D Euler}
{``The network-based model is always at par with the best among the traditional optimized models.'' Supported by section 6 and tables 13 and 14.}
{No}
{N/A (doesn't claim superiority)}
{N/A}
{N/A}
{N/A}
{N/A}

\appendixentry{\magiera}{\citemagiera}{Constraint-aware neural networks for Riemann problems}{\magieracitations}
\appendixdata{1D Euler, 1D scalar hyperbolic PDE}
{``Naturally, the case study problems considered in Section 6 do not show a speedup of the computational time by exchanging the analytical Riemann solver by a surrogate neural network.''}
{No}
{N/A}
{N/A}
{N/A}
{N/A}
{N/A}

\appendixentry{\bezginc}{\citebezginc}{A data-driven physics-informed finite-volume scheme for nonclassical undercompressive shocks}{\bezginccitations}
\appendixdata{Cubic scalar hyperbolic conservation law}
{``For the weak shock test case which is calculated on a coarse mesh, the NN scheme is roughly 10 times slower than the WCD scheme.''}
{No}
{WCD scheme}
{N/A (doesn't claim superiority)}
{N/A}
{N/A}
{N/A}

\appendixentry{\dongb}{\citedongb}{On computing the hyperparameter of extreme learning machines: Algorithm and application to computational PDEs, and comparison with classical and high-order finite elements}{\dongbcitations}
\appendixdata{2D Poisson, 2D non-linear Helmholtz, 1D Burgers.}
{``It is shown that the current improved ELM far outperforms the classical FEM. Its computational performance is comparable to that of the high-order FEM for smaller problem sizes, and for larger problem sizes the ELM markedly outperforms the high-order FEM.'' Supported by figures 16, 27, and 36.}
{No}
{FEniCS}
{N/A (doesn't claim superiority to stronger baseline)}
{N/A}
{N/A}
{N/A}

\appendixentry{\dresdner}{\citedresdner \hspace{0.1cm}(Note: we consider version 1 of this article on ArXiv, version 2 was uploaded after our systematic review was completed but before this article was finished.)}{Learning to correct spectral methods for simulating turbulent flows}{\dresdnercitations}
\appendixdata{1D Kuramoto-Sivashinsky (KS), 1D Burgers', 2D incompressible Navier-Stokes }
{``Overall there is little potential for accelerating 2D turbulence beyond traditional spectral solvers.''}
{No}
{Spectral solver}
{N/A (doesn't claim superiority)}
{N/A}
{N/A}
{N/A}

\appendixentry{\toshev}{\citetoshev}{E(3) Equivariant Graph Neural Networks for Particle-Based Fluid Mechanics}{\toshevcitations}
\appendixdata{3D incompressible Navier-Stokes}
{``Our main findings are that while being rather slow to train and evaluate\dots'' Supported by table 1.}
{Yes}
{N/A}
{N/A (Doesn't claim superiority)}
{N/A}
{N/A}
{N/A}


\singlespacing

\cleardoublepage
\ifdefined\phantomsection
  \phantomsection  
\else
\fi
\addcontentsline{toc}{chapter}{Bibliography}

\bibliographystyle{plain}
\bibliography{thesis}

\end{document}